\documentclass[12pt]{iopart}
\usepackage{amssymb}
\usepackage{cite}
\usepackage{epsfig}  
\usepackage{bm}       
\usepackage{dcolumn}  
\usepackage{graphicx} 
\usepackage{mathrsfs}
\usepackage{multirow}
\usepackage{vruler}

\usepackage{longtable}       
\usepackage{rotating}

\usepackage{color}           

\begin{document}


\def\Erightarrow{{\dashrightarrow}}
\def\Eleftarrow{{\dashleftarrow}}
\def\Qrightarrow{{\rightarrow}}
\def\Qleftarrow{{\leftarrow}}
\def\Nrightarrow{{\Rightarrow}}
\def\Nleftarrow{{\Leftarrow}}
\newcommand{\spar}{{\ensuremath\stackrel{\Lrightarrow}{\Nrightarrow}}}
\newcommand{\sant}{{\ensuremath\stackrel{\Lrightarrow}{\Nleftarrow}}}
\def\Nuparrow{\Uuparrow}
\def\Ndownarrow{\Ddownarrow}

\def\xbj{\ensuremath{x_{\mbox{\tiny B}}}}
\def\x{\ensuremath{x}}
\def\pT{\ensuremath{k_T}}
\def\kT{\ensuremath{p_T}}
\def\PT{\ensuremath{{\bm k}_T}}
\def\KT{\ensuremath{{\bm p}_T}}
\def\phT{\ensuremath{p_{h,T}}}
\def\PhT{\ensuremath{{\bm p}_{h,T}}}
\def\PpiT{\ensuremath{{\bm p}_{\pi,T}}}

\def\g{\ensuremath{G}}
\def\Dq{\ensuremath{\Delta q}}
\def\Dqbar{\ensuremath{\Delta \bar{q}}}
\def\Dg{\ensuremath{\Delta G}}
\def\Du{\ensuremath{\Delta u + \Delta\bar{u}}}
\def\Dd{\ensuremath{\Delta d + \Delta\bar{d}}}
\def\Ds{\ensuremath{\Delta s + \Delta\bar{s}}}
\def\Dux{\ensuremath{\Delta u(x) + \Delta\bar{u}(x)}}
\def\Ddx{\ensuremath{\Delta d(x) + \Delta\bar{d}(x)}}
\def\Dsx{\ensuremath{\Delta s(x) + \Delta\bar{s}(x)}}
\def\dq{\ensuremath{\delta q}}
\def\ubar{\ensuremath{\bar{u}}}
\def\dbar{\ensuremath{\bar{d}}}
\def\sbar{\ensuremath{\bar{s}}}
\def\Dubar{\ensuremath{\Delta\bar{u}}}
\def\Ddbar{\ensuremath{\Delta\bar{d}}}
\def\Dsbar{\ensuremath{\Delta\bar{s}}}
\newcommand{\ordera}[1]{\ensuremath{\mathcal{O}(\alpha_s^{#1})}}
\def\bT{{\bm b}_\perp}
\def\DT{{\bm \Delta}_\perp}
\def\RT{{\bm R}_\perp}
\def\hatP{{\bm {\widehat P}}}

\def\csdof{\ensuremath{\chi^2/DoF}}

\def\hermes{{\sc Hermes}}
\def\compass{{\sc Compass}}
\def\belle{{\sc Belle}}
\def\star{{\sc Star}}
\def\phenix{{\sc Phenix}}
\def\rhic{{\sc Rhic}}
\def\jlab{{\sc JLab}}
\def\desy{{\sc Desy}}
\def\hera{{\sc Hera}}
\def\delphi{{\sc Delphi}}
\def\smc{{SMC}}
\def\emc{{EMC}}
\def\slac{{SLAC}}
\def\clas{{\sc Clas}}
\def\lepto{{\sc Lepto}}
\def\pepsi{{\sc Pepsi}}
\def\jetset{{\sc Jetset}}
\def\pythia6{{\sc Pythia6}}
\def\msbar{\ensuremath{\overline{\mathrm{MS}}}}

\def\eg{{\it e.g.}}
\def\ie{{\it i.e.}}
\newcommand{\de}{{\rm\,d}}
\newcommand{\desix}{{\rm\,d^6}}
\newcommand{\dxbj}{\!\!\!{\rm\,d}\xbj\,}
\newcommand{\dx}{\!\!\!{\rm\,d}x\,}

\newcommand{\be}{\begin{equation}}
\newcommand{\ee}{\end{equation}}
\newcommand{\bea}{\begin{eqnarray}}
\newcommand{\eea}{\end{eqnarray}}
\newcommand{\bml}{\begin{multline}}
\newcommand{\eml}{\end{multline}}

\def\ppol{\Uparrow(\Downarrow)}

\title[]{Spin-polarized 
high-energy scattering of charged leptons on nucleons}

\author{M Burkardt$^1$$^\dag$, C A Miller$^2$ and W-D Nowak$^3$}  
\address{$^1$ Dept. of Physics, New Mexico State Univ. Las Cruces,
NM 88003-0001, U.S.A.}
\address{$^2$ TRIUMF, Vancouver, British Columbia V6T 2A3, Canada}
\address{$^3$ DESY, 15738 Zeuthen, Germany}
\address{$^\dag$ Present Address: Thomas Jefferson National Accelerator 
Facility, Newport News, VA 23606, U.S.A.}

\eads{\mailto{burkardt@nmsu.edu}, \mailto{miller@triumf.ca}, 
\mailto{Wolf-Dieter.Nowak@desy.de}}






\begin{abstract}
The proton is a composite object with spin one-half, understood to
contain highly relativistic spin one-half quarks exchanging spin-one
gluons, each possibly with significant orbital angular momenta.
While their fundamental interactions are well described
by Quantum ChromoDynamics (QCD), our standard
theory of the strong interaction, nonperturbative calculations 
of the internal structure of the proton based directly on QCD are
beginning to provide reliable results.  Most of our present knowledge 
of the structure of the proton is based on experimental measurements
interpreted within the rich framework of QCD.
An area presently attracting intense interest, both experimental and
theoretical, is the relationship between the spin of the proton
and the spins and orbital angular momenta of its constituents.
While remarkable progress has been made, especially in the last decade,
the discovery and investigation of new concepts have
revealed that much more remains to be learned.
This progress is reviewed and an outlook for the future is offered. \\
(Some figures in this article are in colour only in the electronic version)
\end{abstract}

\pacs{13.40.-f, 13.60.-r, 14.20.Dh}

\noindent
This is an author-created un-copyedited version of an article accepted for
publication in {\em Rep. Progr. Phys.}. IOP Publishing Ltd is not responsible
for any errors or omissions in this version of the manuscript or any version
derived from it. The definitive publisher authenticated version will soon be
available online.

\tableofcontents

\title[Spin-polarized high-energy scattering of charged leptons on nucleons]{}

\maketitle

%
%
%
\section{Introduction}
%
\label{sec:Intro}
The nucleon (proton or neutron) is a suitable object for our curiosity 
in view of the fact
that it accounts for most of the directly visible mass of the universe.
Half a century ago, elastic electron-proton scattering provided direct
experimental evidence~\cite{Hofstadter:1956qs} that the proton has a 
finite size of about 1\,fm.  (With the benefit of hindsight, we now
realize that its already-known large anomalous magnetic 
moment could already 
have been interpreted to indicate that it is not a point-like 
object~\cite{Bawin:2000px}.)
The amazing variety of baryons and mesons discovered using new
multi-GeV proton accelerators in the 1950's and 1960's demanded some
explanation in terms of an underlying structure that would unify their 
observed quantum numbers and decay modes. 
Such a composition in terms of more elementary fractionally charged 
spin-$\frac{1}{2}$ particles called `quarks' was postulated in 
1964~\cite{GellMann:1964nj,Zweig}. Then the internal structure of the proton 
came under direct experimental observation in the late 1960's at the SLAC 
laboratory, where an intense electron beam became available with an energy 
large compared to the relevant scale of $\sim 1$\,GeV. 
In these now-famous Deep-Inelastic Scattering (DIS) experiments, a substantial
amount of energy and momentum is transferred to the proton
target, causing it to disintegrate.  In 1969, it was hypothesized that the
energy and momentum was being transferred to internal constituents of the 
proton, called `partons'~\cite{Feynman:1969ej}.  Information about the 
properties of the constituents can be inferred from the distribution of the 
scattering cross section in these measured transfers. In the same exciting 
year, it was realized that the data showed a property called 
`scaling'~\cite{Bjorken:1968dy}: at large enough momentum transfer, 
the kinematic dependence of the cross section resembles that for 
elastic scattering off point-like charged particles that might be identified
with those partons.  As the properties of the cross section also indicated
that the partons have spin $\frac{1}{2}$, it became accepted that the charged 
partons could be assumed to be the hypothesized quarks.  However, it emerged
that the total momentum carried by observed quarks accounts for only about half
of the proton's momentum.  This suggested that there must also be electrically
neutral partons, possibly being exchanged by quarks to provide their binding
force, leading to the name `gluons'.  Then in the early 1970's, an elegant 
field theory was devised to explain this binding, as well as the observed 
weakening of this force at short distances or large momentum transfers, known 
as `asymptotic freedom'~\cite{Gross:1973id,Politzer:1973fx}. This theory 
became known as Quantum Chromodynamics (QCD), the contemporary standard theory 
of strong interactions.

The QCD Lagrangian can be written in one line.  Nevertheless, as in
other fields of science, simple rules can give rise to complex behaviour.
The unique richness of this field theory accounts 
for the challenge presented by the study of hadron structure.   
The QCD vacuum is believed to resemble a `dual superconductor',
which, unlike conventional QED superconductors, repells
color {\em electric} fields rather than magnetic fields. Unlike in QED,
where vacuum polarization screens charges as their separation 
increases, the QCD vacuum provides anti-screening that results in
a rapidly increasing (`running') QCD coupling constant
$\alpha_s$ at larger distance scales or smaller momentum transfer.
Moreover, the analogue of the dual Meissner effect in a superconductor
leads in the QCD vacuum to
the formation of a color flux tube connecting quarks and antiquarks
in the same way as the magnetic flux connecting two magnetic
monopoles in a conventional (QED) superconductor would be
squeezed into narrow flux tubes. 
This apparently leads to the phenomenon of confinement, as free quarks are 
not observed, although no proof has been found for the theorem that 
confinement inevitably follows from QCD. In any case, the quantitative 
prediction of hadron properties in QCD is presently subject to uncontrolled 
approximations. Laborious simulations on a Euclidian lattice are 
now producing such predictions for an increasing number of observables.
In contrast, the asymptotic freedom of QCD at short distances
allows precise perturbative treatment of hard interactions.   Hence such hard
interactions with bound partons can be used to 
experimentally probe the structure of hadrons.  
A good representative for the energies that are
required to reach the perturbative regime is provided by the
scale parameter $\Lambda_{QCD}$.  While this parameter appears
also in the description of the running of the QCD coupling, it 
provides a rough estimate about the scale of many nonperturbative
effects, such as the energy density needed to break the
superconducting phase.  The numerical value of $\Lambda_{QCD}$
depends somewhat on the renormalization procedure (scheme) that is
used in its definition, 
but a value $\Lambda_{QCD} \approx 200$--300\,MeV 
provides a good scale for estimates of the relevance of
effects governed by the nontrivial QCD vacuum structure.

Quark kinetic energy together with QCD field energy accounts for
most of the mass of the nucleon, which in turn accounts for most 
of the mass of the visible universe.  Hence the nucleon is a
worthy subject of investigation.
Deep-Inelastic Scattering of leptons has continued to provide
most of our present knowledge of nucleon structure.  
\begin{figure}[htb]
\begin{center}
\epsfig{file=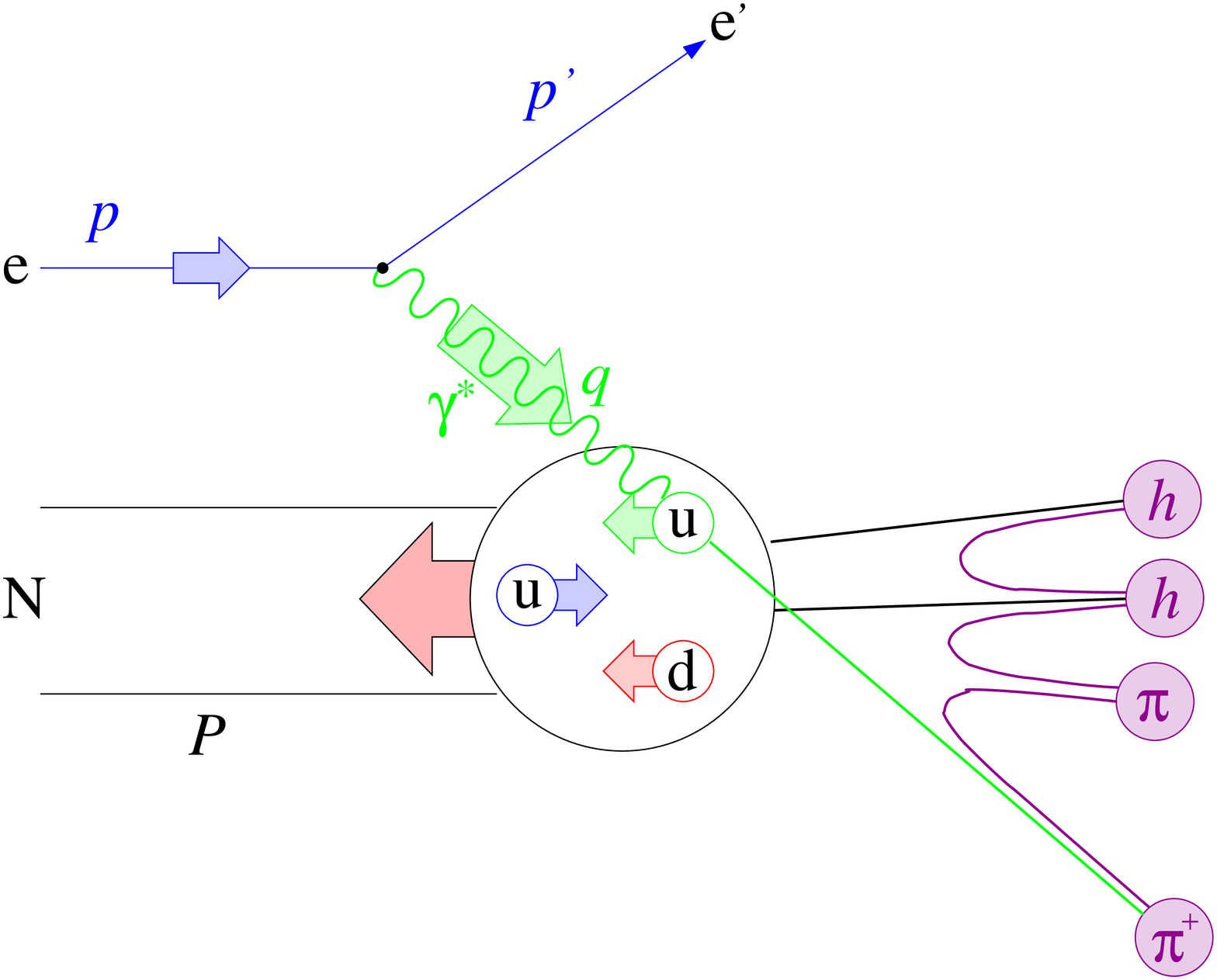,width=8cm}
\end{center}
\caption{\label{fig:SIDIS-diagram}
        Simplified schematic diagram of semi-inclusive deep inelastic scattering.  The large shaded arrows represent typical helicities, as described later in the text.}
\end{figure}
As shown in figure~\ref{fig:SIDIS-diagram}, a virtual photon $\gamma^*$ emitted 
by the incident lepton e and absorbed by a bound quark of flavour 
$up\ (u)$ or $down\ (d)$ in the nucleon N has space-like four-momentum $q$ that can be inferred
in {\em inclusive} measurements in which only the scattered lepton e' is detected.
The motion of the mostly low-mass $u$ and $d$ flavours of quarks populating the 
proton is highly relativistic.  Hence virtual quark-antiquark pairs are
abundant and may be manifest in suitable processes.  (In the general context
of DIS, the word `quarks' can be taken to also include antiquarks.) 
The relativistic motion also implies that DIS results are best interpreted
in a reference frame where the target proton has `infinite' momentum opposite 
to that of the incident virtual photon.  One may intuitively think of 
this tactic as freezing the transverse motion of the partons during 
the interaction time through time dilatation (while remembering that 
their intrinsic transverse momentum $\kT$ is of course invariant under such
a boost).  
A virtual photon has adequate spatial resolution to resolve 
the target's internal structure~\cite{Mueller:2001fv}
if its virtuality $Q^2\equiv-q^2$ is sufficiently large,
a requirement empirically found to be at least $Q^2>1$\,GeV$^2$.  
Given this condition, a key experimental 
observable is the distribution of cross sections in the Bjorken 
scaling variable $\xbj\equiv Q^2/(2P\cdot q)$.
In the infinite-momentum frame, $\xbj$ can be interpreted as the 
fraction \x\ of the proton's four-momentum $P$ that was carried by the 
quark before it was struck by the virtual photon.  Thus
we distinguish here between the experimentally obervable 
kinematic variable $\xbj$ and the theoretical concept of \x\ as a parton 
momentum fraction.  Only in the parton-model approximation 
to the DIS process can they 
be equated.  In this framework, the dependence of the DIS cross section
on $\xbj$ provides information about the $\x$-distribution of partons 
in the target (often called Parton Distribution Functions or PDFs).
For a hypothetical hadron consisting of three non-relativistic quarks 
with the same large mass, this distribution would be a $\delta$-function 
at $\x=\frac{1}{3}$.  For systems of light quarks, 
this is far from being the case.

PDFs can be considered to be intrinsic properties of hadrons
only if they are `universal',  {\em i.e.} if they can be applied in descriptions
of various hard processes involving the same hadron.  
Asymptotic freedom of QCD has provided the means to overcome its 
complicating aspects to prove profound `factorization theorems' showing that
cross sections for several experimentally important hard processes 
can be represented as a product of a cross section 
for one or more perturbatively calculable partonic subprocesses and one or more 
non-perturbative but universal parton distributions.   For the case of 
a DIS cross section $\sigma(\xbj,Q^2)$ in the Bjorken limit $Q^2 \to \infty$ 
at fixed \xbj, the quark distribution $q_f(\xbj,Q^2)$ appears in a sum over quark and
anti-quark flavours 
$f=u,d,s,\bar{u},\bar{d},\bar{s}\ldots$ weighted by the squares 
of their electric charges $e_f$ in units of the elementary charge:
\be \label{eq:sigpdf}
\sigma(\xbj,Q^2) \propto  \sum_f e_f^2 \, q_f(\xbj,Q^2)\,.
\ee
A key feature here is the incoherence between the amplitudes for the various
quark flavours.  At least in the Bjorken limit, this may be understood 
to be associated with the
possible distinguishability of hadronic final states arising from
different quark flavours, so that interference can't arise.
Parton distributions such as $q_f(\x,Q^2)$ have probabilistic interpretations
as parton number densities.  
At finite values of $Q^2$, the relationship between PDFs and observable cross
sections is only a fairly crude approximation at leading order in $\alpha_s$
as in (\ref{eq:sigpdf}), so the perturbative expansion is continued to
Next-to-Leading-Order (NLO) or even further.  Based on the extensive available 
data set for DIS as well as other processes, there is now available a precise 
(except at very small or large \x)
parameterization of $\x$ distributions of the $u$ and $d$ `valence'
quarks that determine the quantum numbers of the proton, 
and less precisely of the virtual 
`sea' quarks that can be produced by gluon splitting, 
such as $anti$-$up\ (\ubar)$, $\dbar$, and $strange\ (s,\sbar)$. 

The dependence of parton distributions on the `hard scale' $Q^2$ of the process
is weak (logarithmic), and can be understood to arise from the finer 
spatial resolution of a probe at a harder scale, which allows it 
to resolve more of the virtual partons that `dress' each quark.  
These virtual $q\bar{q}$ or gluon pairs are abundant because
their small or zero mass makes them energetically inexpensive.
With quantum fluctuations so important at short distances,
the `appearance' of the nucleon depends on how it is probed.
This might be considered to be analogous to the familiar change
in appearance of a macroscopic object when viewed in light of
different colours or wavelengths, due in part to the interactions between the
atoms of its surface.
Unlike the  $\x$ dependence, the $Q^2$ dependence of parton distributions
is perturbatively calculable in QCD at large enough $Q^2$.  
Thus a distribution that is known over the range $x_0<\x<1$  
(or alternatively $0<x<\x_0$ )
for one value of $Q^2$ can be `evolved' to the distribution
over the same \x-range at a larger (smaller) value of $Q^2$, 
using the so-called DGLAP integro-differential 
equations~\cite{Gribov:1972ri,Dokshitzer:1977sg,Altarelli:1977zs}.
The evolution kernels for several parton distributions have been computed
up to Next-to-Next-to-Leading-Order (NNLO) in $\alpha_s$~\cite{
Mertig:1995ny,Vogelsang:1995vh,Vogelsang:1996im,Vogt:2008yw},
meaning that participation of gluons in hard QCD interactions
are taken into account.  Gluons may be created in these hard interactions,
or they may be found in the target nucleon.  Hence,
at order $\alpha_s$ or higher, the DGLAP equations couple the 
distributions for quarks and gluons.
This feature has been exploited in global fits of all available
inclusive DIS data, which cover a range in $Q^2$ from 1 to 50\,GeV$^2$,
to extract information about
the gluon distribution $\g(\x)$.  From this and other types of data, 
$\g(\x)$ is now quite well determined except at small \x\ with
moderately small $Q^2$.
For convenience in the rest of this paper,
we suppress the $Q^2$ dependence of distributions in our notation.

Parton distributions depend on the order of perturbation theory
relating them to experimental observables.  Beyond leading
order in $\alpha_s$, they lose their probabilistic interpretation 
as number densities, and can even become negative.  However, at each
order, they remain universal among various observable processes in 
which they participate, and hence are still much more than
just parameterizations.  Furthermore, their basic
construction is motivated by simple semiclassical parton-model
ideas, but implemented to account for the intricacies of a quantum
system.  

For values of $Q^2$ below $\sim10$\,GeV$^2$, there may be additional
contributions to the cross section that scale as $1/Q^n, n>0$.  
These may be corrections arising because the mass of
the target nucleon is not negligible compared to $Q^2$, or they may be
due to `higher-twist' subprocesses, in which
an additional parton in the target nucleon experiences a hard 
interaction.  (Some authors apply the term `higher twist' for all 
`power-suppressed' terms, even those suppressed by factors of 
$1/Q^n$ arising only because of kinematic relationships.  
In this case, contributions involving
an additional parton are designated as `interaction dependent' or
`dynamical higher twist'.)
For DIS processes, leading twist is twist-2, with no such contributions,

Baryons are the fermionic bound states of the spin-$\frac{1}{2}$ quarks.  
All known
baryons have quantum numbers consistent with those of a set of three quarks
of particular flavours.  (Bound states of 5 or more quarks are theoretically 
possible, and in fact such wave function components may contribute to the 
ground states of ordinary baryons such as the proton.)  
Soon after the conception of quarks as the constituents of the nucleon,
nonrelativistic models were constructed in which the mass the the nucleon
was directly attributed to three massive quarks.
One of the successes of these models was considered to be the remarkable 
consistency of the measured magnetic moments of many baryons with a
combination of Dirac magnetic moments of their three `constituent quarks'.

\begin{figure}[htb]
\begin{center}
\epsfig{file=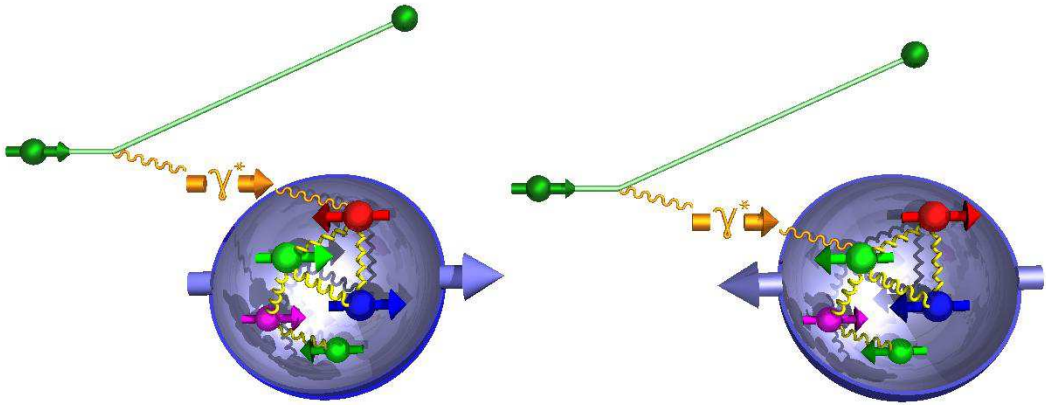,width=0.9\textwidth}
\end{center}
\caption{\label{fig:SIG3half-SIG1half}
Visualization of scattering of longitudinally polarized 
leptons and protons.}
\end{figure}

Then in the 1970's, the internal spin structure of the proton came under
investigation via DIS of helicity-polarized lepton beams on hydrogenous
targets containing protons polarized with the beam axis as quantization axis,
{\em i.e.} longitudinal polarization.
The exchanged virtual photon inherits the helicity of the beam lepton,
to a degree that depends on the lepton kinematics.  Due to conservation
of helicities in hard or short-distance interactions, only quarks
with the opposite spin direction ({\em i.e.} same helicity) as the spin-1 
photon can absorb it in the leading-order process $\gamma^* q \to q$ 
(see figure~\ref{fig:SIG3half-SIG1half}).  
Hence the `polarized virtual-photon beam' selects quarks of one helicity.  
From the difference in cross sections with same or opposite polarizations of 
beam and target, it is therefore possible to extract the number densities 
of quarks having the same ($q_f^{\stackrel{\Qrightarrow}{\Nrightarrow}}$) 
or opposite ($q_f^{\stackrel{\Qrightarrow}{\Nleftarrow}}$)
helicity as the target proton in the infinite momentum frame.  Here 
the symbol $\Qrightarrow (\Nrightarrow)$ designates the helicity of the quark
(target proton).  The quark helicity distribution $\Dq_f(\x)$ is then defined as
$
\Dq_f(\x) = q_f^{\stackrel{\Qrightarrow}{\Nrightarrow}}(\x) - q_f^{\stackrel{\Qrightarrow}{\Nleftarrow}}(\x),
$
while the polarization-averaged distribution $q_f(\x)$ introduced above can be written
$
q_f(\x) = q_f^{\stackrel{\Qrightarrow}{\Nrightarrow}}(\x) + q_f^{\stackrel{\Qrightarrow}{\Nleftarrow}}(\x).
$ 
From these definitions, the obvious `positivity limit' arises:
$
|\Dq_f(\x)| < q_f(\x)
$
The distribution $q_f(\x)$ [$\Dq_f(\x)$] is sometimes written as $f_1^q(\x)$ [$g_1^q(\x)$],
although this is arguably confusing, because the spin structure functions $f_1$ and
$g_1$ are defined to be functions of the observable kinematic quantity \xbj, 
while the parton distributions are defined to be functions of the parton 
momentum fraction \x, with the relations between them being subject to approximations
such as single-photon exchange and the quark-parton model.
For convenience hereafter, we use the term `(un)polarized' to describe measurements,
cross sections or parton distributions, meaning (polarization averaged) 
polarization dependent.  
Polarized PDFs evolve logarithmically with $Q^2$ according to a set 
of DGLAP equations that differs somewhat from those for the unpolarized case.

Certain $\x$-moments of parton distributions (integrals over \x\ weighted with
$x^n$) have important physical interpretations. 
First $(n=0)$ moments can be conveniently written as 
{\em e.g.} $\Dq_f \equiv \int~\dx \Dq_f(\x)$.
A quantity that has played a central role in the discussion of 
nucleon spin structure in recent decades
is the flavour sum $\Delta\Sigma\equiv\sum_f \Dq_f$, 
as it can be interpreted as the net
contribution of the quark helicities to the spin of the nucleon.  
(At the present precision of the field, the proton and neutron 
can be considered to be related by isospin symmetry.)
Note that contributions of individual quark flavours may tend to
cancel one another in this sum.  A pivotal event
in this field in the late 1980's was the DIS measurement with longitudinal
polarization of both muon beam and hydrogenous target by the EMC experiment~\cite{EMC:1988,EMC:1989} 
at the CERN
laboratory, which for the first time extended over a wide enough range in $\xbj$
that  $\Delta\Sigma$ could be inferred, albeit under the assumption
of SU(3) flavour symmetry (neglecting the effects of the $\sim100$\,MeV mass 
of the strange quark), which allows
use of additional constraints from measured weak decay lifetimes of hyperons.  
The surprisingly small resulting value suggested that quark helicities make little net contribution 
to the nucleon spin, in contrast to naive expectations based on the
baryon magnetic moments mentioned above.  In language that with the 
benefit of hindsight may now seem extravagant,
this finding was called a `spin crisis', and inspired more than a thousand
theoretical papers attempting its interpretation, and several new experiments 
with the aim of confirming the measurement with higher precision, 
and extending it to include the `neutron' target.

Because of the relativistic nature of bound systems of light quarks, 
there exists a third independent flavour-set of quark 
distributions that have a probabilistic interpretation
as number densities.  These quark {\em transversity}
distributions $\dq_f(\x)$ are needed to describe the transverse ($\uparrow$) 
polarization of quarks in a nucleon 
polarized transversely ($\Uparrow$) with respect to its infinite momentum:
$\dq_f(\x) \equiv q_f^{\uparrow\Uparrow}(\x) - q_f^{\uparrow\Downarrow}(\x)$. 
Their probabilistic interpretation applies only in a basis of 
transverse spin eigenstates, not in the helicity basis relevant
for the $\Dq_f(\x)$. 
The distribution $\dq_f(\x)$ is sometimes written as $h_1^q(\x)$,
although this might be misleading because no spin structure function 
$h_1(\xbj)$ is defined.  There exists little experimental information 
about transversity, because the coupling of hard vector bosons to quarks 
is insensitive to the transverse polarization of the quark,
unless the transverse polarization of the struck quark
in the final state is `measured'.
However, some traces of this transverse polarization appear 
in the structure of the `jet' of final-state
particles produced by the energetic struck quark. 
Information about transversity has begun to emerge from
{\em semi-inclusive} measurements with transverse target polarization,
which include the detection of one or more
hadrons in addition to the scattered lepton (see figure~\ref{fig:SIDIS-diagram}).

All parton distributions
actually have a two-dimensional dependence on both \x\ and the 
intrinsic quark momentum component $\kT$ transverse to the nucleon's
`infinite' momentum.  This two-dimensional dependence
does not factorize.  The three flavour-sets of distributions mentioned 
above that have probabilistic interpretations are considered basic
in the sense that they survive integration over $\kT$, and it is
this integral that is signified when only the \x\ argument appears.  Five
other sets of distributions have been identified that also have probabilistic 
interpretations but do not survive such integration.  Some of these
have been found to be particularly interesting, as in DIS they embody the
effects of a `final-state interaction' between the departing struck
quark and the remainder of the target nucleon.  The contribution
of such a distribution to the cross section depends on polarization and
momentum vectors in a manner that is called `T-odd', although this does
not refer to fundamental time symmetry.  The study of these T-odd
effects has provoked a reconsideration of the meaning of parton
distributions and their universality.  Experimental evidence for
T-odd effects has now been observed in the same semi-inclusive
data from which information about transversity has been extracted.

The dependence of the elastic lepton-nucleon cross section on the momentum
transfer is represented by the elastic `form factor', as its Fourier
transform can be interpreted as the spatial charge distribution 
in the nucleon.
Up until about a decade ago, information about hadron structure embodied
in elastic form factors and in PDFs were interpreted separately.
It was then discovered that these concepts could be unified in what became
known as Generalized Parton Distributions (GPDs), also called off-forward
or skewed parton distributions.  Even more profound was the finding that
the dependence of these GPDs on three kinematic variables encodes much more 
detailed information, about {\em e.g.} parton 
orbital angular momentum~\cite{Ji:1996ek} and correlations between transverse
position and longitudinal momentum fraction of 
partons~\cite{Burkardt:2000za,Diehl:2002he,Belitsky:2003nz,Ralston:2001xs}.  
Of particular interest is that a second moment of a certain combination of GPDs 
for each parton flavour was found to represent the total (helicity plus orbital) 
contribution of that flavour to the spin of the nucleon~\cite{Ji:1996ek}.
This offers the first opportunity for experimental access to parton
orbital angular momentum, as it was discovered that the
additional information could in principle be obtained from experimental studies
of hard exclusive processes involving additional hard interaction(s), yet
`replacing' the struck quark in the target nucleon to leave it intact.
Most prominent among these processes is Deeply Virtual Compton Scattering (DVCS),
the hard exclusive production of an energetic real photon.
Within the last decade, rapid theoretical progress has been made in
understanding this new subject, and several DVCS observables have been measured.

\begin{figure}[h] 
\begin{center} 
\begin{minipage}[b]{.3\linewidth}
\epsfig{file=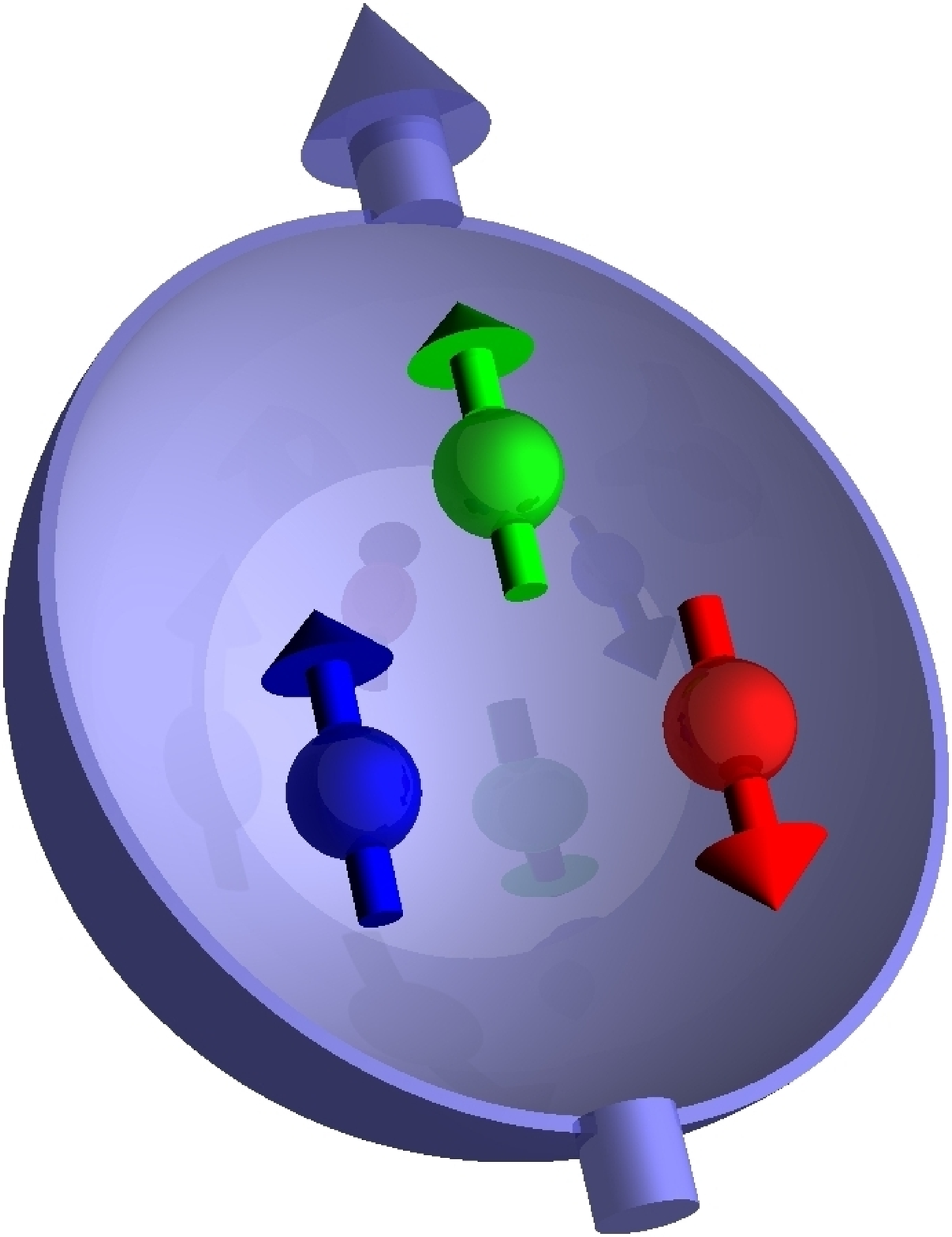,width=4.75cm}
\end{minipage} \hspace*{.01\linewidth}
\begin{minipage}[b]{.3\linewidth}
\epsfig{file=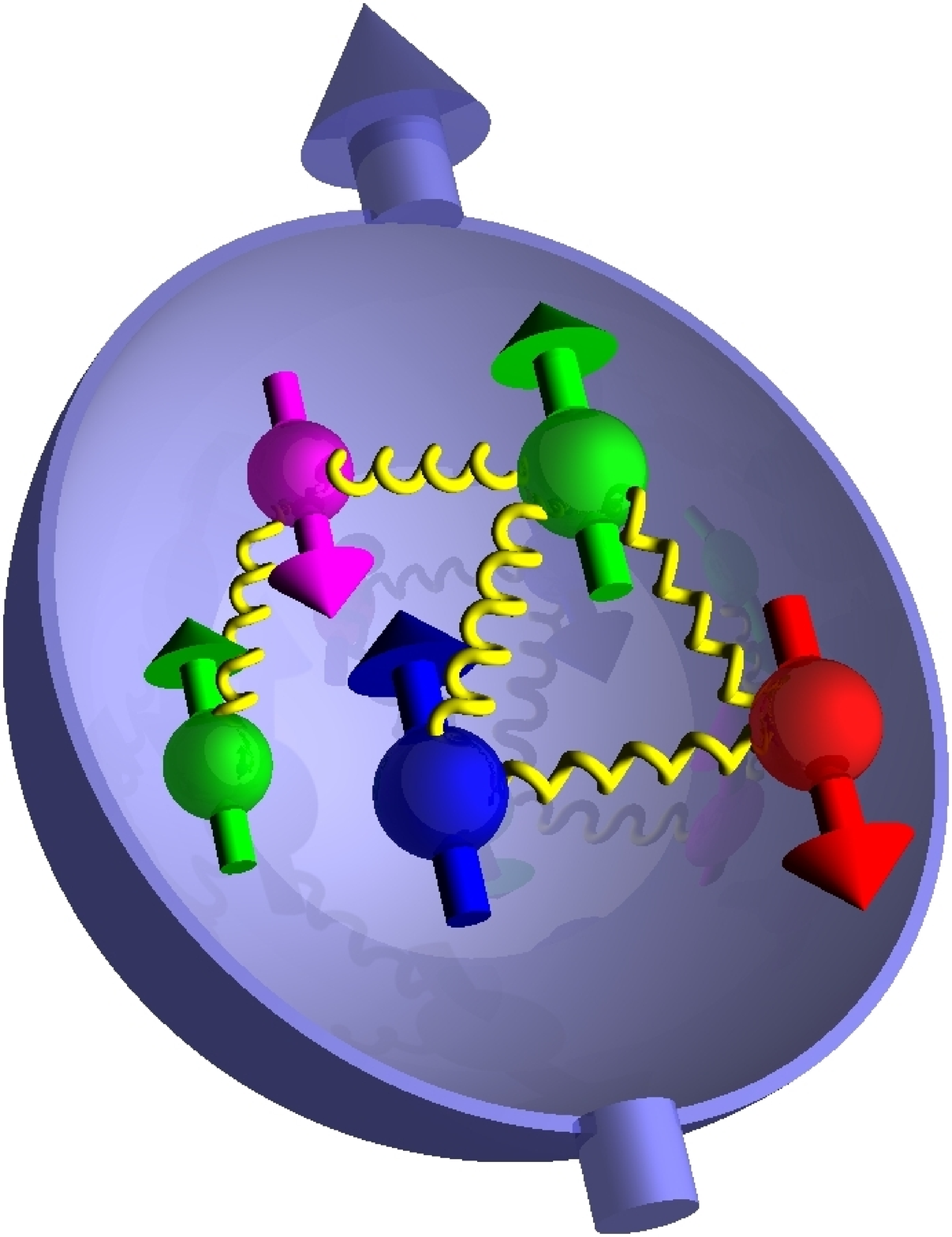,width=4.75cm}
\end{minipage} \hspace*{.01\linewidth}
\begin{minipage}[b]{.3\linewidth}
\epsfig{file=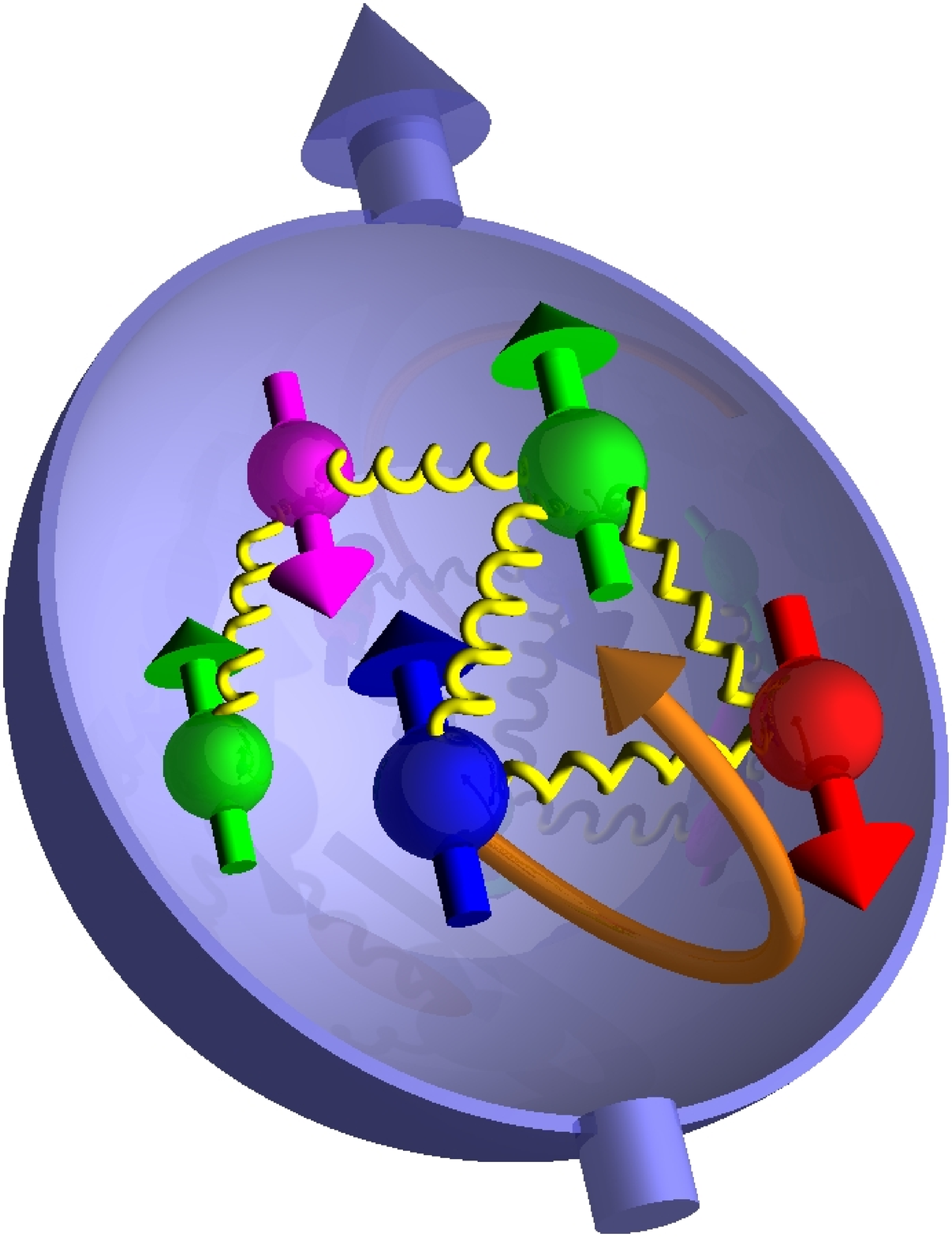,width=4.75cm}
\end{minipage}
\end{center}
\caption{\label{fig:Cartoon3}
        Illustration of nucleon structure. Left: valence quarks only.
        Middle: Gluons and quark-antiquarks in addition. Right:
        Orbital angular momentum of quarks.} 
\end{figure}

Thus the evolution of our understanding of nucleon spin structure might 
be imaginatively summarized by figure~\ref{fig:Cartoon3}.
%

%
%
\section{The longitudinal spin structure of the nucleon}
\label{sec:long}
%
%
\subsection{Introduction}
\label{subsec:LongIntro}
%
As mentioned above, parton distributions
have been investigated experimentally using DIS of
lepton beams.  In order to measure
helicity distributions $\Dq_f(\x)$, both beam and target must be
longitudinally polarized.
The relevant observable is the difference in cross sections with same 
or opposite polarizations of beam and target:
\be \label{eq:sigLL}
\sigma_{LL}(\xbj) \equiv \frac{1}{2} [\sigma^{\stackrel{\rightarrow}{\Nrightarrow}}(\xbj) - \sigma^{\stackrel{\rightarrow}{\Nleftarrow}}(\xbj)]\,.
\ee
However, measurements
of absolute cross sections are technically difficult with polarized targets.  Hence
the experimental results are typically extracted from the data as `double-spin' {\em asymmetries}:
\be \label{eq:Apar}
A_{||}(\xbj) \equiv \frac{\sigma_{LL}(\xbj)}{\sigma_{UU}(\xbj)} 
\ee
where $L$ $(U)$ designates Longitudinal polarization (Unpolarized)
appearing in pairs corresponding to beam followed by target.
The polarization-averaged cross section
\be
\sigma_{UU}(\xbj) \equiv \frac{1}{2} [\sigma^{\stackrel{\rightarrow}{\Nrightarrow}}(\xbj) + \sigma^{\stackrel{\rightarrow}{\Nleftarrow}}(\xbj)]
\ee
has been previously measured precisely using unpolarized targets.
Using these known values and measured values of $A_{||}$, $\sigma_{LL}$ can be extracted.
This approach based on {\em cross sections} avoids confusions 
that sometimes arise.  For example, it has been proposed that, if the
values of $\sigma_{LL}$ are interpreted theoretically in leading order, 
then the values of $\sigma_{UU}$ should be calculated in leading order
from a theoretical representation in terms of parton distributions,
neglecting the portion involving so-called `longitudinal' virtual 
photons!\footnote{Longitudinal virtual photons are those having `magnetic' 
quantum number $m=0$  with respect to the quantization axis along 
the photon momentum direction, while photons with `magnetic' 
quantum number $m=\pm1$ are called `transverse'.
Real photons are always transverse.  Sometimes the word `polarization'
is included in this unfortunate nomenclature, even though the
distinction has nothing to do with the longitudinal or transverse
direction of a spin quantization axis.}
This misses the key point that $\sigma_{UU}$ became involved only
because of the experimental convenience of asymmetry measurements, 
and is irrelevant to the theoretical intepretation of spin observables.
Hence it should be treated empirically in terms of experimental data,
with no influence from
theoretical concepts such as PDFs or order in $\alpha_s$.  Another point
of confusion is that the extraction has often been formulated in a way 
influenced by the technical convenience that precisely measured 
values of $\sigma_{UU}$
are typically represented in terms of an unpolarized structure 
function called $F_2$ and the ratio $R$ of the cross sections for longitudinal
and transverse photons.  Unfortunately, various values $R$ were
used to extract the values of $F_2$, so the world data set for $F_2$
contains inconsistencies.  This confusion can be avoided by using 
in equation~\ref{eq:Apar} the original experimental values of 
$\sigma_{UU}$, reconstructed if necessary
using the values of $R$ employed in the original extraction of $F_2$.

If the measurement of $A_{||}$ is {\em inclusive}, {\em i.e.}, 
only the scattered lepton is detected while ignoring the produced hadrons,  
$\sigma_{LL}$ is linearly related via a 
kinematic factor that includes the QED hard scattering cross section
to the {\em spin structure function} $g_1(\xbj)$.
(There is a small contribution to $\sigma_{LL}$ from another structure function $g_2$, which 
has no probabilistic interpretation, and is measured using a transversely polarized target.)
At leading order in $\alpha_s$, $g_1$ for the proton (p) or neutron (n) target
has a simple probabilistic interpretation in terms of quark helicity densities for flavour $f$:
\bea \label{eq:g1Dq}
g_1^{p,n}(\xbj) &=& \frac{1}{2} \sum_f e_f^2 \, \left(\Dq_f^{p,n}(\xbj)+\Dqbar_f^{p,n}(\xbj)\right)\,.
\eea
The same linear combination relates the first moments $\Dq_f$ of the 
helicity densities to the moments $\Gamma_1^{p,n}$ of $g_1$:
\be
\Gamma_1^{p,n} \equiv \int_0^1\dxbj g_1^{p,n}(\xbj) 
= \frac{1}{2} \sum_f e_f^2 \, \left(\Dq_f^{p,n}+\Dqbar_f^{p,n}\right)\,.
\ee
Since the quark charges enter only as squares, inclusive DIS measurements
cannot distinguish quarks from antiquarks, and hence can constrain only
combinations of $\Du$, $\Dd$ and $\Ds$.
As mentioned earlier, charge symmetry is a good approximation, implying 
$\Delta u^p + \Delta \ubar^p =\Delta d^n + \Delta \dbar^n$ and  
$\Delta u^n + \Delta \ubar^n =\Delta d^p + \Delta \dbar^p$,
and $\Ds$ is the same in the neutron and proton.  (By convention, when the
superscript is absent, the proton is implied.)
Hence measurements of $\Gamma_1^p$ and $\Gamma_1^n$ (inferred from data on
proton and {\em e.g.} deuteron targets) provide constraints on two linear combinations
of these three unknowns.  

A third such constraint can be derived from measurements of the lifetimes of
weakly decaying hyperons, under the assumption of the SU(3) flavour symmetry
among quark distributions that would apply in the limit of no mass difference
among quark flavours.  This symmetry has been found to be valid at the few 
percent level in a study of the relationship between the lifetimes of hyperons.
The additional third constraint allows all three unknown moments to be extracted,
and also their flavour-singlet combination 
\be \label{eq:defDSigma}
\Delta \Sigma\equiv (\Du)+(\Dd)+(\Ds).  
\ee
As described below, it has been inferred from a series of
DIS experiments of ever increasing precision  
that $\Delta \Sigma = 0.2\ldots 0.4$ and $\Ds$ is significantly negative.
This implies that there must be substantial contributions from quark 
orbital angular momentum $L^q$
and/or gluon total angular momentum $J_\g$, in order to complete the 
`proton spin budget'.
The experimental investigation of these contributions is the subject of following sections of this article.  Subtle issues about the conceptual basis of spin decompositions of a system of interacting constituents are discussed in detail in section~\ref{subsec:Decompose}.

\subsection{Axial charge of the nucleon }
\label{subsec:axial}

The axial charge $g_A$ of the nucleon is obtained from the forward matrix
element of the operator $\psi^\dagger\gamma^\mu \gamma^5 \psi$ in the
same way as the familiar (vector) charge $g_V$ is obtained from the
forward matrix element of the operator $\psi^\dagger\gamma^\mu \psi$.
In terms of quark helicity eigenstates, the contribution of each quark
flavour $f$ to the axial charge of the nucleon involves
the same linear combination
$q_f^{\stackrel{\Qrightarrow}{\Nrightarrow}} - q_f^{\stackrel{\Qrightarrow}{\Nleftarrow}}
+\bar{q}_f^{\stackrel{\Qrightarrow}{\Nrightarrow}} - \bar{q}_f^{\stackrel{\Qrightarrow}{\Nleftarrow}}$
that also enters the polarized structure function
$g_1(\xbj)$. As a result, the integral of $g_1(\xbj)$
yields the quark-charge-squared-weighted axial charge $g_A$
of the nucleon.
Since the isovector axial charge of the nucleon also describes the
$\beta$-decay of the neutron, one can thus relate the integral of
the isovector polarized structure function $g_1(\xbj)$ to the
neutron $\beta$-decay constant.  In the scaling limit of infinite $Q^2$,
this relationship between the spin structure functions of the proton
and neutron is:
\be \label{eq:bjsr}
\int_0^1\dxbj \left[g_1^p(\xbj)-g_1^n(\xbj)\right] = \frac{1}{6}\frac{g_A}{g_V}.
\ee
While we presented here only a simple plausibility argument, 
this famous Bjorken Sum Rule (BSR) was derived~\cite{bjsr} 
in the Bjorken limit of infinite $Q^2$ even before the advent of QCD, 
using only Gell-Mann's current algebra relations~\cite{Gell-Mann62,Gell-Mann64,Feynman:1964fk}
together with a reasonable assumption about smooth asymptotic 
high-energy behaviour of the virtual Compton amplitude.
Within the modern context of QCD where the sum rule can be evolved to 
finite $Q^2$, it can be tested precisely by practical experiments.  
If a violation were discovered, the implications would be profound.
This is expressed eloquently in \cite{Jaffe-EIC2}:
``Occasionally it is worth reminding ourselves what it means to `understand' something in 
QCD. In the absence of fundamental understanding we often invoke `effective descriptions' 
based on symmetries and low-energy expansions. While they can be extremely useful, we 
should not forget that a thorough understanding allows us to relate phenomena at very 
different distance scales to one another. In the case of Bjorken's sum rule, the operator 
product expansion, renormalization group invariance and isospin conservation combine to 
relate deep inelastic scattering at high $Q^2$ to the neutron's $\beta$-decay axial charge measured at 
very low energy. Even target mass and higher twist corrections are relatively well understood.''

\subsection{Inclusive DIS experiments}
\label{subsec:LongExp}
Experimental information about quark helicity densities in the nucleon has so 
far come only from DIS of polarized high-energy charged leptons on targets
containing polarized protons, or polarized light nuclei containing neutrons.  
The technologies for polarizing beam 
and target are challenging, and have been the subject of intense efforts
over several decades. A high beam polarization of 85\% was achieved in even 
the first experiments~\cite{hughes}, done at SLAC in the 1970's using a 
6--23\,GeV beam of electrons with polarization inherited in the 
photo-ionization of lithium atoms selected via the Stern-Gerlach method.  
However, the beam intensity was quite limited, and this approach was 
eventually replaced in modern SLAC 
experiments~\cite{e142:1993,Abe:1998wq,Abe:e143a2,Anthony:1999rm,Anthony:2000fn,e155:2003}
using electron beam energies up to 50\,GeV, polarized at a level of typically 
85\% in the electron source by photoelectric emission from a gallium arsenide 
surface.  Beam intensities increased
until they were limited by the heat tolerance of the cryogenic targets.  
Measurement times are of order a few weeks.

Meanwhile, a series of even higher energy experiments at 
CERN~\cite{SMC:1994p,smc:1995,SMC:1997p,smc:1997d,Adeva:1998vv,SMC:1999pdlox,Adeva:2000er,smc-deltaq} 
used 100--200\,GeV beams of muons naturally polarized to a 
level of 80\% by the weak decay of the precursor $\pi$ or $K$ mesons. 
These tertiary beams
had limited intensities of typically $4.5\times10^7$ per spill of 2.4 s
duration, with a spill period of 14.4\,s, but the more massive muons suffer 
less radiative energy loss in the target materials than electrons, allowing 
much larger target thicknesses of order meters instead of centimeters.  
Nevertheless, measurement times are of order a year.

Most targets used at SLAC and CERN are
cryogenic solids of butanol, ammonia, or more recently $^6$LiD, in a strong
magnetic field, with most protons or deuterons in the target
polarized through the 
process of Dynamic Nuclear Polarization (DNP).  For DNP to proceed, a 
suitable material doped with a 
paramagnetic compound is cooled to temperatures less than about 1\,K. 
In the high magnetic field (2--6.5\,T), the unpaired 
electron of the dopant is polarized to almost 100\%, while the proton or 
deuteron has a polarization of less than 1\%.  The process of DNP causes 
most of the electron polarization to be transferred to that of the protons by 
microwave irradiation of the sample.  By this means, proton polarizations of 
more than 95\% can be achieved. Typical deuteron polarizations are 40--55\%; 
recently though, deuteron polarizations of 80\% have been obtained.
However, most of the material encountered by the beam in 
such solid polarized targets is non-hydrogenous.  Hence the asymmetry of
interest is strongly diluted by the resulting background events, 
by a factor of up to about 7.  This constrains the statistical precision
that can be achieved, and requires careful background corrections.

As spin structure functions must be determined for both the proton and 
neutron, and targets of free neutrons are still infeasible, 
measurements are made using targets containing either deuterons or
$^3$He.  The structures of these light nuclei are such that the spin 
of the neutron is strongly correlated with that of the nucleus.
Protons can be replaced by deuterons in solid-state targets without
fundamental changes in the technology.  In contrast, the very different
polarized $^3$He targets
consist of a low pressure gas polarized via spin exchange with
alkaline atoms, the polarization of which is pumped by an infrared 
laser.  Such targets were used in a pair of experiments at SLAC.
The dilution in these targets is contributed by the essentially
unpolarized protons in $^3$He and the thin target windows, and hence 
is smaller than in the cryogenic solids.

A completely different experimental approach was followed at the
DESY laboratory, where the 27.5\,GeV HERA storage ring contained beams
of typically 40\,mA of electrons or positrons.  Such stored high-energy 
electron beams may become
spontaneously polarized via a small polarization asymmetry in the
emission of synchrotron radiation by the beam particles as they
are deflected by the magnetic fields of the ring~\cite{sokolov-ternov}.  
(Several MW of power was 
radiated via synchrotron radiation, and was continuously replenished
by a system of accelerating radio-frequency cavities.)  The
polarization rise time was of order $\frac{1}{2}$ hour, and
polarizations as large as 60\% were achieved.  To avoid disrupting
the stored beam, only very thin targets were possible.  These
were tenuous gases of nuclear-polarized atoms of hydrogen 
or deuterium partially contained in an cylindrical open-ended storage
cell coaxial with the beam.  The polarized atoms were injected 
into the cell from an Atomic Beam Source based on Stern-Gerlach
polarization filtering and radio-frequency transitions
between atomic substates in a magnetic field.  The atoms bounced
off the cell walls of order 100 times before escaping from the
ends, after which they were differentially pumped away by a large system of
turbo-pumps.  Polarized hydrogen target thicknesses of about
$10^{14}$\,nucleons/cm$^2$ were achieved, and there was no target 
dilution from non-hydrogenous material.
Measurement times were of order a year.

\subsection{Quark helicity densities from inclusive DIS data}
\label{subsec:Dq-incl}
\begin{figure*}
\begin{center}
\includegraphics[width=\textwidth]{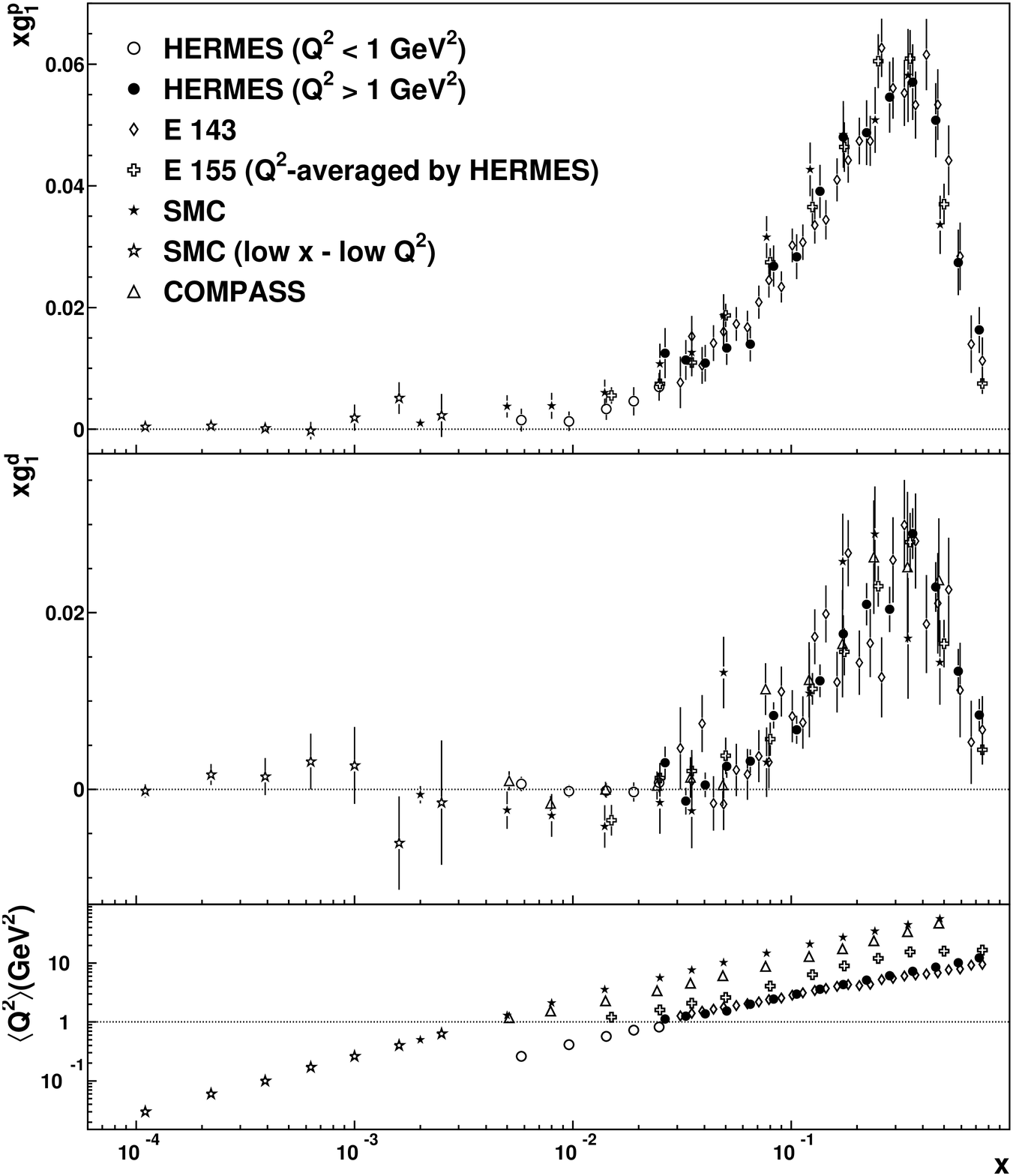}
\end{center}
\caption{\label{fig:g1pd}  Published experimental results on 
$\xbj\cdot g_1^p$ and $\xbj\cdot g_1^d$ versus $\xbj$, shown on separate 
panels, from \hermes~\cite{HERMES_g1pd} at DESY, SMC~\cite{Adeva:1998vv,SMC:1999pdlox,Adeva:2000er}
at CERN,
E143~\cite{Abe:1998wq} and  E155~\cite{Anthony:2000fn,Anthony:1999rm} at SLAC,
and COMPASS~\cite{compass} at CERN.
Error bars represent the sum in quadrature of statistical  and
systematic uncertainties. The \hermes\ data points shown are
statistically correlated by  unfolding 
QED radiative and detector smearing effects; the statistical
uncertainties shown are  obtained from   the {\it diagonal} elements of their
covariance matrix, only. The E143 and E155 data
points are correlated through QED radiative  corrections. 
The lower panel shows the $\xbj$ dependence of the mean values 
$\langle Q^2\rangle$ for the various experiments.
The figure is taken from \cite{HERMES_g1pd}.
}
\end{figure*}

\begin{figure}
\begin{center}
\includegraphics[width=0.63\columnwidth]{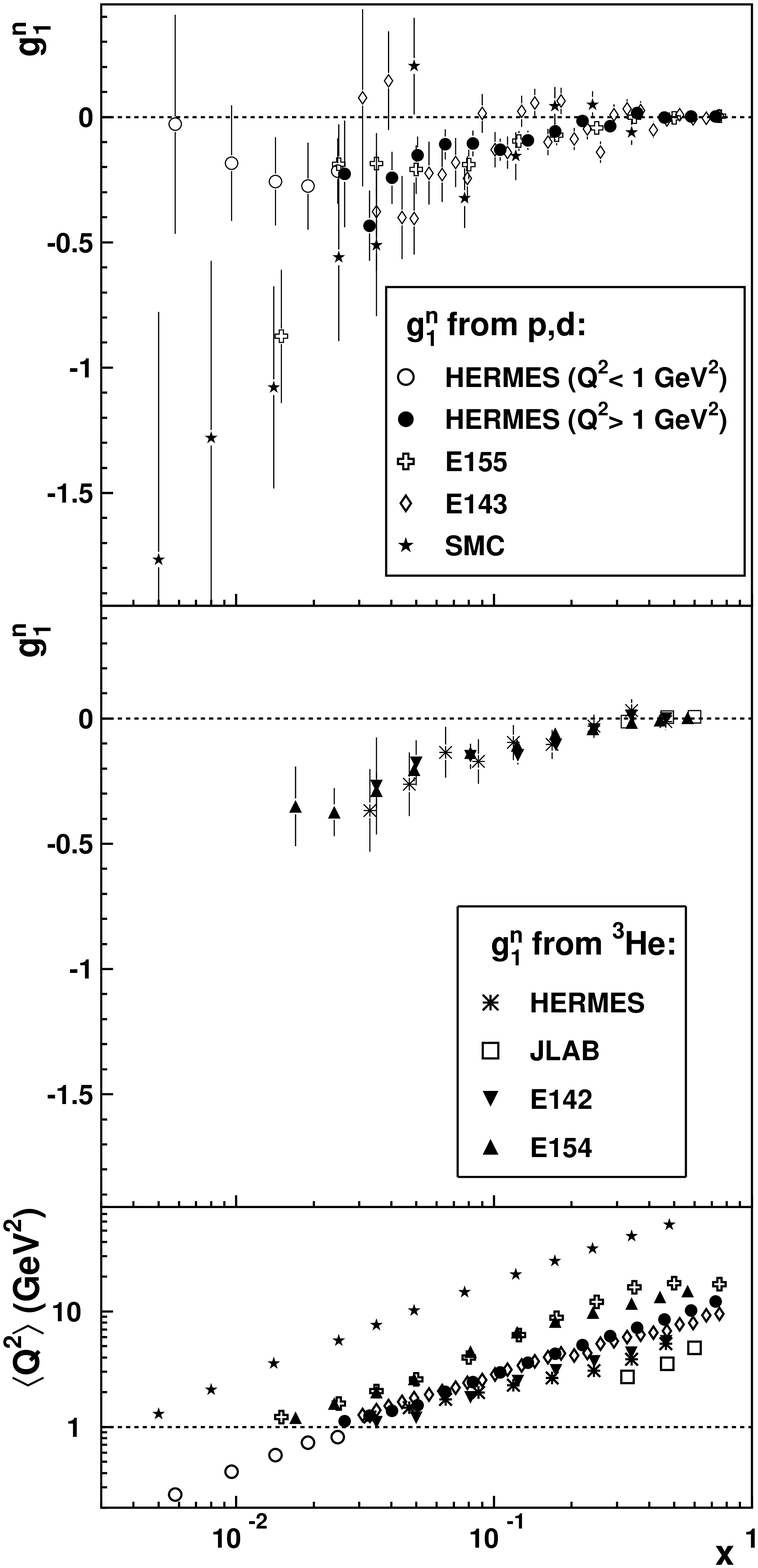}
\end{center}
\caption{\label{fig:g1n}
Top panel: the structure function $\xbj\cdot g_1^n$ obtained from
 $g_1^p$   and  $g_1^d$, from \hermes~\cite{HERMES_g1pd},
SMC~\cite{Adeva:1998vv,SMC:1999pdlox,Adeva:2000er},
E143~\cite{Abe:1998wq} and E155~\cite{Anthony:2000fn,Anthony:1999rm}. 
Second panel from the top:  $\xbj\cdot g_1^n$ obtained from a  
$^3$He target~\cite{jlab:g1n,g1n_hermes,E142n,E154n}.
Total error bars are shown, obtained by
combining statistical and systematic uncertainties in quadrature.
The bottom panel shows the $\langle Q^2\rangle$ of each data point in the top 
two panels. E155 data have been averaged over their $Q^2$ bins for
visibility.
The figure is taken from \cite{HERMES_g1pd}.
}
\end{figure}

The data from the first polarized DIS measurements at SLAC were used 
to test simple theoretical expectations
of the time, but did not cover a large enough $\xbj$ range to permit
the inference of the first moment of $g_1$.  The first opportunity for this
appeared in the late 1980's with the EMC experiment at CERN, which
employed 120--200\,GeV beams of muons.
As already mentioned, the $\Gamma_1^p$ moment extracted from this data
was interpreted to imply a surprisingly small net contribution of
quark helicities to the proton spin, including a negative contribution
from the strange sea.  The resulting excitement led to a series
of experiments of ever-higher precision at 
SLAC~\cite{e142:1993,Abe:1998wq,Anthony:1999rm,Anthony:2000fn,e155:2003},
CERN~\cite{SMC:1994p,smc:1995,SMC:1997p,smc:1997d,Adeva:1998vv}, 
and DESY~\cite{g1n_hermes,HERMES:g1p,HERMES_g1pd}.  Figure~\ref{fig:g1pd}
shows published data for the spin structure functions $g_1(\xbj)$
of the proton and deuteron, while figure~\ref{fig:g1n} 
shows the results for the neutron inferred from the data in 
figure~\ref{fig:g1pd}, or from measurements with a $^3$He target.
This information is best displayed as
$\xbj\cdot g_1(\xbj)$ versus $\log \xbj$, which conserves contributions
to the area under the data representing the first moment $\Gamma_1$.

A key issue to be addressed using the data sets containing 
measurements on both proton and `neutron' targets is the BSR, 
introduced in section~\ref{subsec:axial}.  To test this Rule, 
the first moments $\Gamma_1^p$ and $\Gamma_1^n$ must be extracted 
from the data at some fixed value of $Q^2$.
However, measurements can never cover the complete
range $0<\xbj<1$, so the inference of  $\Gamma_1$ requires 
some estimate of the contributions to the
moments from the unmeasured regions 
at the extremes of this range.
These estimates are based on some assumption 
or ansatz motivated by reasonable expectations.  Another 
problem is that these data were all produced by fixed-target 
experiments, resulting in a strong kinematic correlation 
between $\xbj$ and the hard scale $Q^2$, so that each data point 
corresponds to a different value of $Q^2$, as shown in the 
bottom panels of figures~\ref{fig:g1pd} and \ref{fig:g1n}.  
This implies that it is impossible to directly calculate
the first moment of $g_1(\xbj)$ from experimental data.
Hence, some assumption or ansatz
is typically employed to approximately `evolve' the data from
that experiment to a common value of $Q^2$.  (The resulting
model uncertainty has not always been estimated and included
with the final result.)  Applying these approaches to the
results of individual experiments has resulted in agreement with
the BSR within the experimental uncertainty,
which typically amounts to about 9\% of the Sum~\cite{HERMES_g1pd}.
  
Important information about the polarized parton distributions
can be extracted from the measured structure functions via
(\ref{eq:g1Dq}).  The most efficient use of the data
from the various experiments done at different beam energies
and kinematic conditions is achieved by incorporating all available data in
a global fit based on the evolution predicted by QCD in
Next-to-Leading Order (NLO) in $\alpha_s$.  This approach has
the important advantage that it also provides information about
the polarization of the gluons contained in the nucleon. 
In the order $\alpha_s$ version of the DGLAP
evolution equations, gluons couple to quarks
via `splitting': $g \rightarrow q\bar{q}$.  Hence the
evolution of the $g_1$ structure functions with $Q^2$
contains information on the helicity density of the gluons.
However, in NLO some choice
must be made between various `factorization schemes'
for sharing a dependence on $Q^2$ between the parton
distributions and the coefficient functions relating them
to the structure functions.  This introduces an
arbitrary scheme dependence to the results of the fit,
although results obtained in any one scheme can be transformed
into any other scheme.  The results presented here were
obtained in the so-called \msbar\ scheme~\cite{msbar}.

The fitted parameters 
appear in some functional form for the $\x$ dependence of
each of several parton helicity densities, including the
gluon helicity density $\Dg(\x)$, at some
arbitrarily chosen `starting' value $Q^2_0$.  (While this
functional form is chosen with the benefit of some physical
insight, the choice introduces some model dependence into
the results from the fit.)  At each step in the parameter
search, the $g_1(\xbj)$ function defined by the current values 
of the parameters is evolved using the DGLAP equations to
the $Q^2$ value of each experimental data point, and the
discrepancy is taken as a $\chi^2$ contribution. 

Since the inclusive cross section is sensitive to only the square 
of the quark charges and not to their signs, inclusive
measurements cannot distinguish quarks from antiquarks.
Because of this as well as the limited precision and $Q^2$ range 
of the existing
data, a useful fit is impossible without further assumptions.
The global fits done up to now include the previously mentioned
constraint based on hyperon beta decay lifetimes together with the
assumption of SU(3)
flavour symmetry, and sometimes also impose the BSR. 
The strong coupling constant $\alpha_s$ may be taken from
other experimental information, or may be also fitted as a
parameter, in which case it is found to be consistent with 
previous knowledge.  Some fits also impose an (arbitrary) symmetry
condition among the sea quarks: 
$\Delta \bar{u} = \Delta \bar{d} =\Delta s = \Delta \bar{s}$.
This is only a device for defining valence quark densities
$\Dq_{f_v}\equiv \Dq_f-\Dqbar_f$,
and does not affect the fitted strange sea quark and gluon
densities.

\begin{figure}
\begin{center}
\includegraphics[width=0.47\columnwidth]{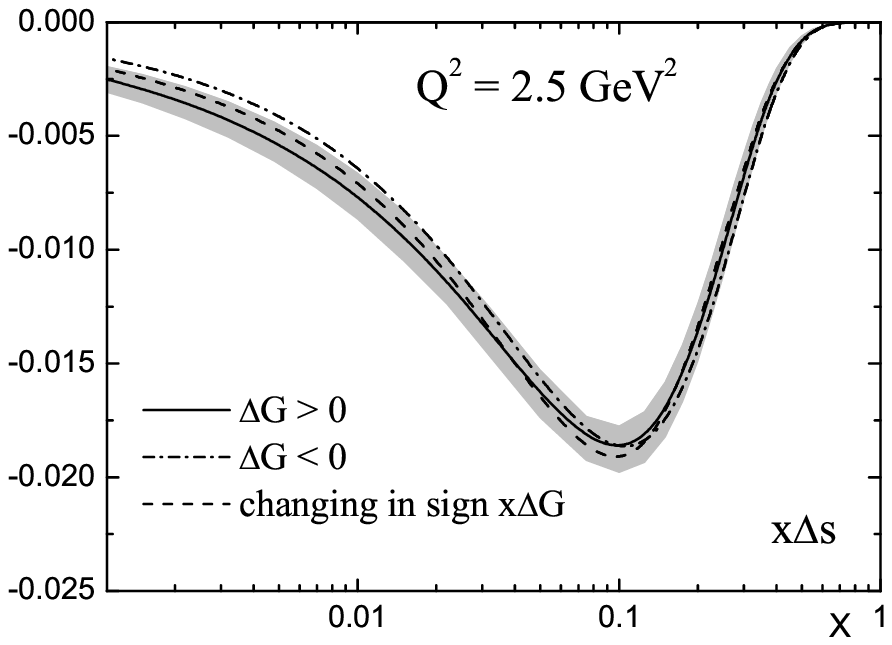}
\includegraphics[width=0.47\columnwidth]{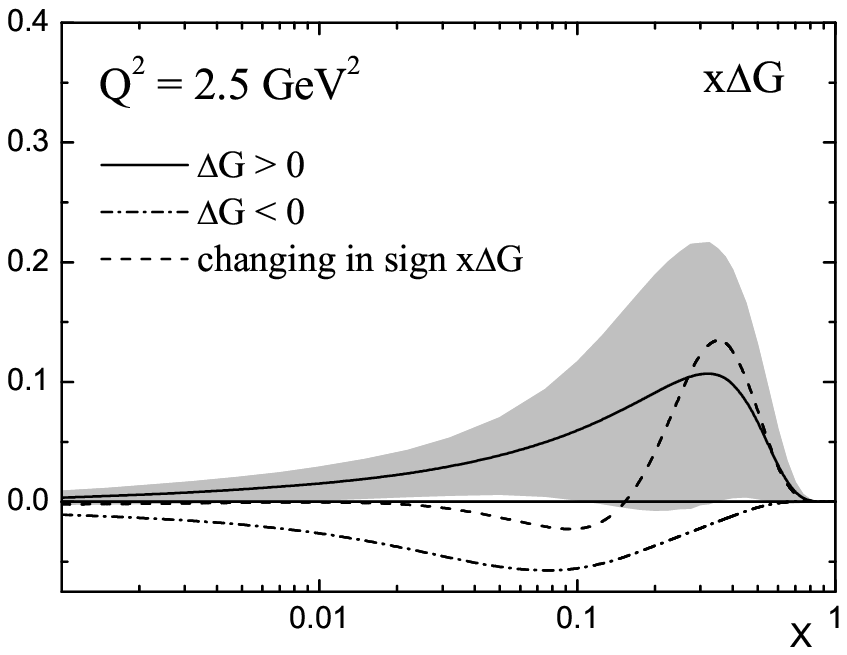}
\end{center}
\caption{\label{fig:QCDfit}
Helicity densities of the strange sea $\Dsx$ (left) 
and of the gluons (right), at $Q^2=2.5$\,GeV$^2$, for three different
choices for the shape of the gluon densities, all with a similar quality
of fit to the $g_1$ data.
The figure is taken from \cite{lss06}.
}
\end{figure}
Many groups have presented results of NLO global fits of all
available $g_1$ data available at that time.  Relatively recent examples
are reported in \cite{Anthony:2000fn,GRSV-std,BB,dFS:2005,AAC06,lss06}.
Figure~\ref{fig:QCDfit} shows the results of one recent fit~\cite{lss06}
that accounts for contributions to $g_1$ that scale as $1/Q^2$, 
which are significant
for the data measured at $Q^2$ less than a few GeV$^2$.
As the photoabsorption cross section is independent of the sign 
of the quark charge, $\Delta \bar{u}$ and $\Delta \bar{d}$ are unconstrained
by inclusive data.
Hence for this fit the convention was adopted of a flavour symmetric sea:
\begin{equation}
\Delta u_{sea}=\Delta\bar{u}=\Delta d_{sea}=\Delta\bar{d}= \Delta
s=\Delta\bar{s} \label{SU3sea}\,.
\end{equation}
It is clearly
determined that $up$ quarks have a substantial positive polarization over
the entire measured $\x$ range, while $down$ quarks have a somewhat
smaller and negative polarization.  The polarization of the strange quark sea
is found to be small but clearly negative.
The gluon helicity distribution $\Dg(x)$ is still poorly constrained,
even though many new precise data have recently been added
to the global set.  In fact there are two alternative solutions (local minima
in $\chi^2$) of equal quality with opposite signs of the gluon density, 
and yet another of similar quality using a bipolar shape for the $x$ dependence. 
The reason for the large uncertainty is that the
$g_1$ data set is still far less extensive and precise than the
unpolarized set, and covers a much more limited range in $Q^2$
(see section~\ref{subsec:DeltaG}).
\begin{table*}
\caption{\label{tab:moments} 
First moments of the helicity densities resulting from the NLO QCD fit to
all available $g_1$ data of reference~\cite{lss06} in the ($\rm\overline{MS}$) 
factorization scheme, at the input scale of $Q^2=1$\,GeV$^2$. 
The uncertainties shown are total (statistical and systematic combined in quadrature). 
The parameters marked by (*) are fixed by the assumption of SU(3) flavour
symmetry mentioned in the text. The Bjorken sum rule is also enforced.
The table is taken from \cite{lss06}.
}
\begin{center}
\lineup
\begin{tabular}{ccccc} \hline
 &~~~$\Dg > 0$~~~&~~~$\Dg < 0$~~~ \\ \hline
 $DoF$  &     $826-20$   &  $826-20$    \\
 $\chi^2$  &      721.7     &    722.9      \\
 \csdof  &      0.895     &    0.897       \\  \hline
 $\Delta u_v\equiv\Delta u-\Dubar$ &    0.926$^*$  &~~0.926$^*$ \\
 $\Delta d_v\equiv\Delta d-\Ddbar$ &    $-0.341^*$      &   $-0.341^*$ \\
 $\Delta \Sigma$  & $0.207 \pm 0.040$ & $0.243 \pm 0.065$ \\ 
 $\Ds$ &$-0.063\pm0.005$~&$-0.057\pm0.010$   \\
 $\Dg$ &~~$0.129\pm0.166$~&$-0.200\pm0.414$   \\
\hline
\end{tabular}
\end{center}
\end{table*}
The first moments of the helicity densities determined
from the fit are shown in table~\ref{tab:moments}.  (These moments
have some sensitivity to the assumed functional form of the 
densities, especially in the unmeasured small-$\x$ region.)
The flavour-singlet moment $\Delta \Sigma$ is found to be 
in the range 0.2 to 0.25,
implying that the total $\frac{1}{2}(1-\Delta \Sigma)$
of the remaining contributions to the spin of the nucleon
from the gluons and/or quark orbital angular momenta
is substantial.
The gluon moment $\Dg$ is still poorly determined,
but it appears that large values are disfavoured.

This collection of increasingly precise data from all of the experiments
also offers the opportunity to more severely test the BSR.
It is difficult to efficiently employ {\em all} of the data while
computing the Sum directly from measured values of $g_1^p$ and 
$g_1^n$ (or $g_1^d$) using (\ref{eq:bjsr}), because kinematic
interpolation and evolution to a common value of $Q^2$ is needed.
This in turn requires a paramerization of the $\xbj$ dependence
at some `starting' value of $Q^2$, just as in the fits described 
above.  In fact, those fits can be used to test the BSR,
simply by omitting the usual BSR constraint from the fit,
and computing the Sum from the resulting $\Dq(\x)$ distributions
produced by the fit (while accounting for uncertainty correlations).
The relevant combination providing the structure function difference
is the `flavour non-singlet' distribution:
\be \label{eq:DqNS}
\Dq_{NS}(\x) \equiv \sum_f (e_f^2/\langle e^2\rangle -1) \,
\left(\Dq_f(\x)+\Dqbar_f(\x)\right).
\ee
The removal of the BSR constraint from the fit may result in
large uncertainties for individual densities, but their
uncertainty correlations should result in a smaller uncertainty
for $\Dq_{NS}$.
This approach might at first seem bizarre, because the sum rule
is believed to be much more reliable than the assumption of
SU(3) flavour symmetry used in the fits!  However, it can be
expected that $\Dq_{NS}$ will be insensitive to the constraint
from hyperon decay data if the proton and deuteron data have
comparable precision and kinematic coverage, which is the case.
(This can easily be verified by varying this input to the fit to
study the sensitivity.)  Hence for this purpose, the fitting 
process provides mainly the evolution of the data points to a 
common value of $Q^2$, as well as the model used to extrapolate 
to $\x=0,1$.  This approach leads to good agreement with the BSR
within the total uncertainty from the fit, which amounted 
to about 10\% of the sum, using the data available in 1998~\cite{smcqcd}. 
Much more precise data are available today.

There are several technical issues arising in the QCD fits
that can be kept in mind when considering their results.
First, typically the values taken from experiment are those 
for $g_1/F_1$, even though the unpolarized structure function
$F_1$ has no relevance, and 
the systematic uncertainties for $g_1/F_1$ include a significant
contribution from the poorly known ratio $R=\sigma_L/\sigma_T$ of cross sections
for longitudinal and transverse photons, which has no effect on
$g_1$ itself.  The reason given for this choice is the fact 
that experimental groups may have extracted $g_1$ from the 
measured asymmetries effectively using values for the unpolarized cross 
section that were calculated using various values of $R$.
However, as it is the ratio $g_1/F_1$ that
is sensitive to $R$ and not $g_1$, it appears that this choice 
aggravates the problem.
(Fortunately, the effects of this may be small.)
Secondly, it can be difficult to obtain reliable results from
multidimensional fits of data with limited statistical accuracy.
Evidence for such difficulty here is the irregular shape of the dependence of
$\chi^2$ on some parameters that are reported.  Up to three
minima have recently been seen with similar values of $\chi^2$~\cite{lss06}.
A common method of testing the stability and accuracy of algorithms
for extracting results from experimental data is to analyze
Monte Carlo pseudo-data.  The situation here could be clarified if 
$g_1$ pseudo-data were
generated from a chosen set of polarized parton distributions, which
could be compared to the results of the fit.  
Thirdly, two groups report statistical uncertainties that are inflated 
by approximately the square root of the number of fitted 
parameters (assuming that the $\chi^2$ surface is quadratic in a
substantial region near the minimum), so that their uncertainties
cannot be directly compared to those of other groups.  These inflated
uncertainties do not correspond to what is understood by most of the 
community as a statistical uncertainty.  (One way of defining this
conventional uncertainty is the standard deviation of the
distribution of the results that would be derived from similarly fitting a
large number of Monte Carlo data sets resembling the global experimental
data set, but with each data point fluctuating independently
according to its experimental statistical uncertainty.)  
Fourthly, some groups account for the (substantial) experimental normalization
uncertainty that is common to an entire data set from one experiment
(arising from {\em e.g}. uncertainty in polarization of beam or target)
by adding that common component of the systematic uncertainty in 
quadrature to the statistical uncertainty of each data point.  
This incorrectly deprecates the information available in those data.
It is straight-forward to more appropriately account for the normalization
uncertainty by means of a $\chi^2$ penalty term driven by
a renormalization factor for that data set. 
Fifthly,  few groups (see \eg\ \cite{smcqcd,BB}) estimate 
`theoretical' uncertainties
for their results, arising from {\em e.g.} the somewhat arbitrary choice 
of functional forms to parameterize the helicity densities, and 
the dependence of the results on factorization and renormalization scales, 
indicating incomplete convergence of the perturbative expansion.

We have seen that the global data set for the spin structure
function $g_1$ on the proton and neutron target has firmly
established that quark helicities account for only a fraction
of the proton spin,
leaving substantial contributions to be made by gluons and
quark orbital angular momenta.
The large uncertainty in the moment $\Dg$ from global fits
to $g_1$ data has motivated several major efforts to
directly measure $\Dg(\x)$ using processes in which gluons enter in 
leading order, as described in section~\ref{subsec:DeltaG}.
Furthermore, information about quark orbital angular momentum is beginning
to emerge from both calculations on the lattice and measurements of 
cross section differences or asymmetries for the deeply virtual Compton 
scattering process, as described in sections~\ref{sec:GPDs} and 
\ref{subsec:Implications}.

\subsection{Helicity densities of sea quarks from semi-inclusive DIS data }
\label{subsec:sidis}

As mentioned above, quarks and antiquarks of the same flavour cannot be distinguished
in inclusive DIS measurements, since the contributions of the various
quark flavours to the cross sections scale as the square
of the quark charges and hence are independent of the sign.  
Furthermore, the strange sea can be
separated only in terms of first $\x$-moments, because of the
necessary constraint from hyperon beta decay data.  In order to obtain
information about the $\x$ dependence of the helicity densities
of the various flavours, particularly of sea quarks, it is
necessary to use additional types of experimental information.  In
{\em semi}-inclusive scattering (see figure~\ref{fig:SIDIS-diagram}), 
an energetic hadron is detected in
coincidence with the scattered lepton.  If suitable kinematic
criteria are applied in the selection of the hadrons,
they are likely to be members of the `jet' of hadrons produced in the
{\em fragmentation} of the struck quark, the non-perturbative process
in which the energy of the quark is shared among hadrons created
through the excitation of quark-antiquark pairs from the vacuum.
For our purposes here, a key point is that the identities of the hadrons
are statistically correlated with the flavour of the struck quark.
For example, a $\pi^+$ meson having valence composition
$u\bar{d}$ is more likely to appear in the fragmentation of a
$u$ or $\bar{d}$ struck quark, and kaons are more likely to
arise from strange quark fragmentation.  However, these probabilities
are also modulated by the square of the charges of the various flavours
and their (known) spin-averaged quark densities.
Hence the exploitation of this method is limited to some degree by
the `$u$-quark dominance' arising from both of these effects.
For example, $\pi^-=d\bar{u}$ mesons and even $K^-=\bar{u}s$ mesons, 
which provide good 
sensitivity to sea quarks, arise more often from fragmentation of
$u$ quarks.  Hence a quark flavour decomposition based on data for
various identified hadrons requires the solution of 
a strongly coupled system of relationships, 
which in turn requires primary data of high precision.

Given an adequate understanding of the fragmentation process, a
complete flavour decomposition of the quark {\em and antiquark} helicity 
distributions can be extracted from sufficiently precise measurements on 
both proton and `neutron' targets of double-spin asymmetries
in the cross sections for leptoproduction of various types of hadrons.
The extraction requires knowledge of the
probabilities for the various types $h$ of hadrons to appear
in the fragmentation of a struck quark of a given flavour $f$.
These probabilities are embodied in the (familiar unpolarized) 
fragmentation functions $D_f^h(z)$, where $z\equiv
E_h/\nu$ and $\nu$ and $E_h$ are the energies
in the target rest frame of the absorbed virtual photon
(and hence of the struck quark) and of the detected hadron.
While these fragmentation
functions have typically been extracted~\cite{Hirai:2007cx,Albino:2008fy} 
from mostly high energy $e^+ e^-$ collider data, 
a recent extraction~\cite{frag:dFSS} includes 
preliminary semi-inclusive DIS multiplicities
of identified hadrons~\cite{Hil05}.

For leptoproduction of a hadron $h$, the semi-inclusive cross section 
difference that is analogous to the inclusive case of (\ref{eq:sigLL}) 
can be factorized in leading order in $\alpha_s$ as
\be \label{eq:sidis}
\frac{\de [
\sigma_h^{\stackrel{\Erightarrow}{\Nrightarrow}}(\xbj,z) 
  - \sigma_h^{\stackrel{\Erightarrow}{\Nleftarrow}}(\xbj,z)]}
{\dxbj \de z}
 \propto \sum_f e_f^2 \ \Dq_f(\xbj) D_f^h(z) \,.
\ee
If, for example, there are available experimental measurements 
of these cross section differences for production of 
$\pi^+$, $\pi^-$, $K^+$ and $K^-$ mesons on both proton and
`neutron' targets, this system of eight over-constrained
equations can in principle be solved for the helicity densities
of up to six light quark flavours.  As mentioned above, in practice
$u$-quark dominance limits the sensitivity to some sea quark
flavours.   
\begin{figure*}
\begin{center}
\includegraphics[width=\textwidth]{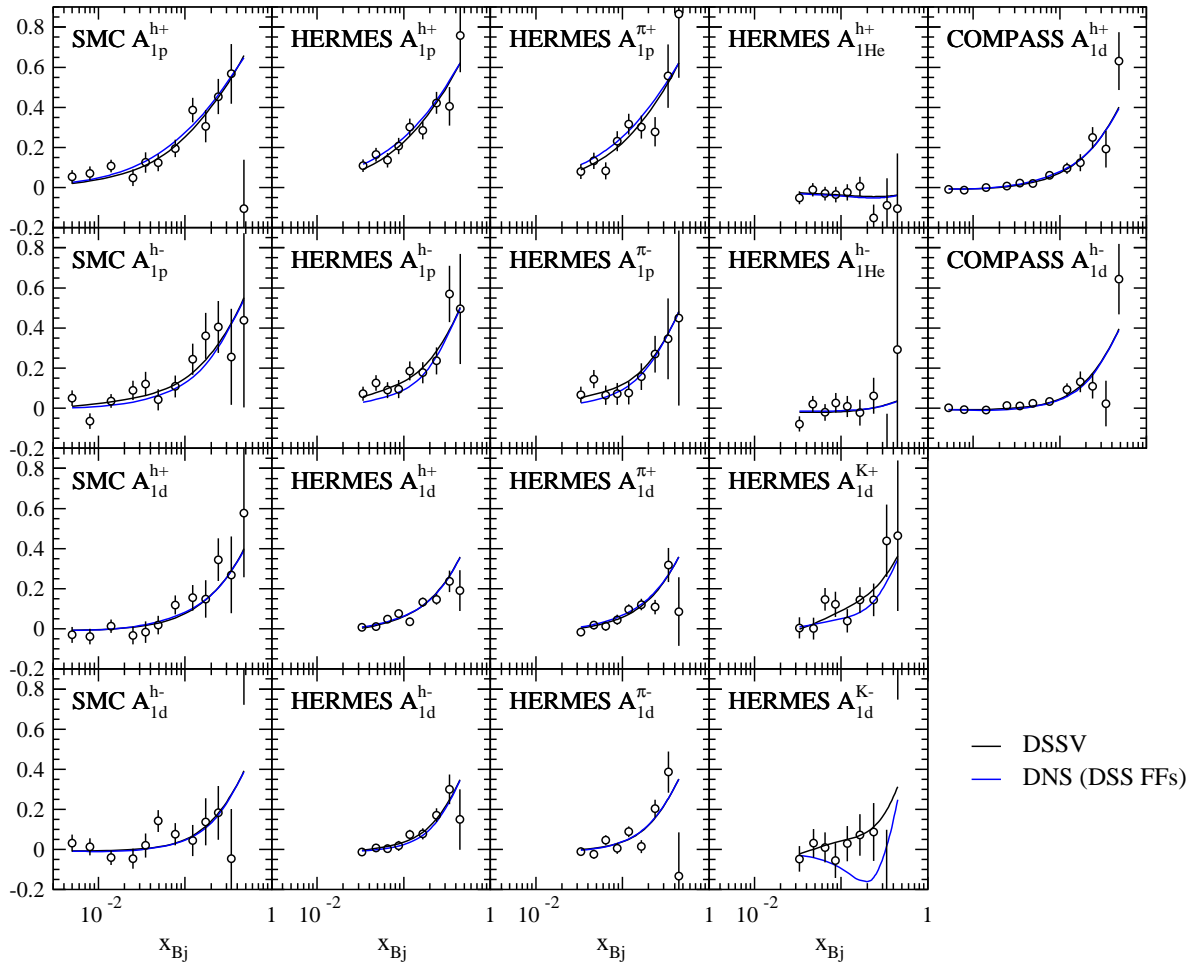}
\end{center}
\caption{\label{fig:deFlorian1} 
Semi-inclusive asymmetries for virtual-photon-nucleon DIS on both the proton (p)
and deuteron (d), producing the type of hadron labelled in the
superscript on the symbols in the figure (`h' means unidentified hadrons).  
These data from SMC~\cite{smc-deltaq}, \hermes~\cite{hermesdeltaq} and
\compass~\cite{Alekseev:2007vi} are compared to curves computed from 
helicity densities fitted to these and other data in NLO~\cite{DSSV}.  
The acronym `DNS' refers to the previous fit by this 
group~\cite{dFS:2005,Navarro:2006bb} using older information~\cite{Kretzer} 
about fragmentation functions. 
The figure is taken from \cite{DSSV}.
}
\end{figure*}

As in the case of inclusive measurements, absolute cross section
differences are difficult to measure, especially with polarized
targets.  Hence cross section asymmetries are extracted from
the data.  
Figure~\ref{fig:deFlorian1} shows all of the measured semi-inclusive
double-spin asymmetries on both hydrogen and deuterium targets.  
It includes data from SMC for unidentified hadrons~\cite{smc-deltaq},
from \hermes\ for production of 
identified $\pi^+$, $\pi^-$, $K^+$ and $K^-$ mesons~\cite{hermesdeltaq},
and from \compass\ for unidentified hadrons~\cite{Alekseev:2007vi}.
Each collaboration extracted helicity densities in leading order in
$\alpha_s$.  In the case of \hermes, all three sea quark flavours
could be distinguished, while fragmentation functions were effectively
fitted to hadron multiplicities measured by the same experiment.
It was found that the helicity densities of all three sea quark
flavours are consistent with zero within the measured \x\ range.
Further analysis of all these data were later done in NLO, including 
data from other processes.  This is described in the next section.

\subsection{Helicity densities from a global analysis in NLO}
\label{subsec:GlobalNLO}

\begin{figure*}
\begin{center}
\includegraphics[width=0.75\textwidth]{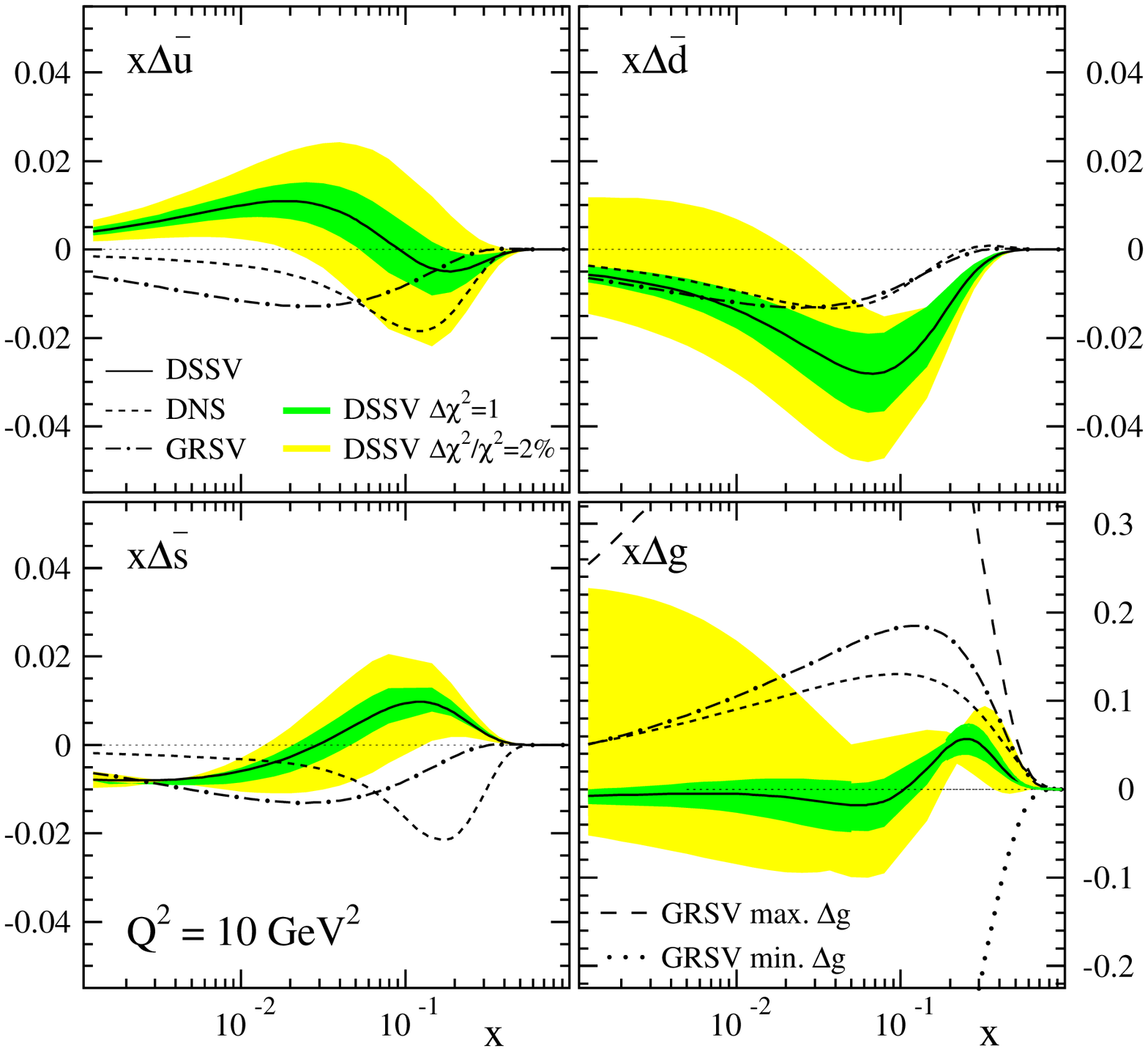}
\end{center}
\caption{\label{fig:deFlorian2} 
The continuous and dashed curves represent
parton helicity densities multiplied by $\x$, extracted in NLO from 
double-spin DIS and $p^\Nrightarrow + p^\Nrightarrow$ data and evaluated 
at $Q^2=10$ GeV$^2$, with their inner and outer uncertainty bands 
corresponding to $\delta \chi^2=1$ and $\delta \chi^2=2\%$ of $\chi^2$, 
respectively.  The dotted and long-dashed curves represent 
the previously known unpolarized densities, which act as positivity limits
to the helicity density limits through the requirement that all cross sections
be non-negative.  The dashed (dot-dashed) curves represent the results of the 
previous fits of \cite{dFS:2005} (\cite{GRSV-std}).
The figure is taken from \cite{DSSV}.
}
\end{figure*}
The most complete extraction of helicity densities from all
relevant published data appears in \cite{DSSV}.
It is a global NLO QCD analysis like that described in
section~\ref{subsec:Dq-incl} for inclusive DIS data, 
but with the crucial inclusion of semi-inclusive DIS data 
to provide sensitivity to individual sea quark flavours,
and for the first time RHIC  $p^\Nrightarrow + p^\Nrightarrow$ data 
for inclusive production of neutral pions and jets, providing 
much stronger constraints on the gluon helicity.
Another new feature of the fit is the use of the above-mentioned set of 
fragmentation functions~\cite{frag:dFSS}
based on not only $e^+ e^-$ data but also preliminary semi-inclusive 
DIS multiplicities of identified hadrons.

The extensive data set allows some freedom in the application of the 
constraint of neutron [hyperon] beta decay data through the assumption of 
SU(2) [SU(3)] flavour symmetry.  Penalty terms were added to 
$\chi^2$ to limit violations of SU(2) and SU(3) symmetries,
but their values at the solution are negligible compared to unity,
suggesting that, at least in terms of first moments, the data
are fully compatible with these symmetries.  Because the data 
are still unable to distinguish $s$ and $\bar{s}$ flavours, the
assumption is made that $\Delta s = \Delta \bar{s}$.
(A previous analysis~\cite{hermesdeltaq} in LO found that only $\Delta s$ 
is constrained by the \hermes\ data, and the above assumption
has negligible effect on that analysis.)

Well-defined unique minima appear in the dependence of $\chi^2$ 
on each of the three sea quark flavours, and the quality of the
fit is excellent: $\csdof = 0.88$.  
Figure~\ref{fig:deFlorian1} shows a comparison of the resulting
fits with the available semi-inclusive asymmetries.
Asymmetries calculated from the same fragmentation functions but with 
helicity densities from a previous fit based on an older 
parameterization~\cite{Kretzer} 
of fragmentation functions are also shown, illustrating the sensitivity
of the kaon asymmetries and hence sea densities on this input to the fit.

\begin{table}[h]
\caption{\label{tab:DSSV-1stmoments} Truncated first moments 
$\int_{x_{\min}}^1 \dx \Delta q(x,Q^2)$
at $Q^2=10\,\mathrm{GeV^2}$, taken from \cite{DSSV}.
It is important to note the factor of two relating 
$\Delta\bar{s}$ in this table (assuming $\Delta s = \Delta \bar{s}$)
to \Ds\ in table~\ref{tab:moments}.
The uncertainties for $\Delta \chi^2=1$ account only for experimental
statistical and systematic uncertainties added in quadrature, and not
for `theoretical' or model uncertainties.
}
\begin{center}
\lineup
\begin{tabular}{crrr}
\hline
& $x_{\min}=0$ & \multicolumn{2}{c}{$x_{\min}=0.001$} \\
& best fit &\multicolumn{1}{c}{$\Delta \chi^2=1$}& \multicolumn{1}{c}{$\Delta \chi^2/\chi^2=2\%$} \\ \hline
$\Delta u + \Delta\bar{u}$  & 0.813  &  0.793 $^{+0.011}_{-0.012}$ & 0.793 $^{+0.028}_{-0.034}$ \\
$\Delta d + \Delta\bar{d}$  & -0.458 & -0.416 $^{+0.011}_{-0.009}$ &-0.416 $^{+0.035}_{-0.025}$ \\
$\Delta\bar{u}$             & 0.036  &  0.028 $^{+0.021}_{-0.020}$ & 0.028 $^{+0.059}_{-0.059}$  \\
 $\Delta\bar{d}$            & -0.115 & -0.089 $^{+0.029}_{-0.029}$ &-0.089 $^{+0.090}_{-0.080}$ \\
$\Delta\bar{s}$             & -0.057 & -0.006 $^{+0.010}_{-0.012}$ &-0.006 $^{+0.028}_{-0.031}$ \\
$\Dg$                  & -0.084 &  0.013 $^{+0.106}_{-0.120}$ & 0.013 $^{+0.702}_{-0.314}$  \\ 
$\Delta\Sigma$              & 0.242  &  0.366 $^{+0.015}_{-0.018}$ & 0.366 $^{+0.042}_{-0.062}$ \\
\hline
\end{tabular}
\end{center}
\end{table}
Figure~\ref{fig:deFlorian2} shows the resulting parton helicity
densities with uncertainty bands, while table~\ref{tab:DSSV-1stmoments}
shows their (truncated) first moments.  
The densities $\Delta u + \Delta \bar{u}$ and $\Delta d + \Delta \bar{d}$
are not shown because they were already well constrained by 
the inclusive data.  The semi-inclusive
data constrain the sea densities, especially $\Delta s$, where a solution
to a long-standing mystery appears.  Leading-order analysis of semi-inclusive
data has 
indicated that the strange helicity density is small or even positive
at large $x$, whereas all analyses of only inclusive data imposing SU(3)
symmetry clearly favour negative values of the first moment of \Ds.  These
findings are now reconciled by a bipolar shape of \Dsbar\ that is negative
at small $x$ and positive at large $x$.  However, figure~\ref{fig:deFlorian2}
shows that the first moment \Dsbar\ is negative only because of the
substantial contribution from the region $x<0.01$, and 
table~\ref{tab:DSSV-1stmoments} shows that even $x<0.001$ contributes strongly.
It may seem mysterious that the fit forces 
\Dsbar\ to be negative in a region of small $x$ devoid of data, while the
penalty term for SU(3) violation is so small at the solution point.  However,
the explanation may be that the \Dsbar\ distribution is completely
unconstrained in this $x$ range by anything but the penalty term,
so a very small value of the penalty is sufficient.

The gluon helicity density is constrained, primarily by the addition 
of the RHIC data, to be significantly smaller than previous uncertainties, 
in the experimentally accessible $x$ range.  Table~\ref{tab:DSSV-1stmoments}
shows that the first moment \Dg\ of the possibly bipolar distribution
is consistent with zero, still with rather large uncertainties.  

A quantity of particular interest that is accessible to
semi-inclusive DIS data is the flavour asymmetry  
$\Delta\bar{u}(x) - \Delta\bar{d}(x)$.  Naively, one might
expect that since $u$ and $d$ quarks have similarly small
masses and they both are produced by gluon `splitting'
($g \rightarrow q\bar{q}$), both their densities and 
polarizations would be similar.  However, the difference in
their unpolarized densities $\bar{u}(x) - \bar{d}(x)$ is
experimentally well established~\cite{Towell:2001nh}, 
and is explained by various
non-perturbative models that also predict an asymmetry in
the helicity densities.  Such a delicate imbalance in
nucleon structure provides an excellent test of such models.
The flavour asymmetry $\Dubar - \Ddbar$ in the first moments can
be compared to the predictions
of two representative models.  One is the chiral quark soliton 
model ($\chi$QSM)~\cite{Dressler:1999zg}, which is an effective 
theory where baryons appear as soliton solutions of the chiral 
Lagrangian, and the other is a `meson cloud' model~\cite{Cao:2003zm} 
that considers quantum fluctuations of the spin-$\frac{1}{2}$ nucleon
into a virtual pion with spin 1 plus a $\Delta$ hadron with 
spin $\frac{3}{2}$.
The value $\Dubar - \Ddbar = 0.15^{+0.11}_{-0.10}$ from the fit 
is consistent with the predictions of both models, but with a preference
for the $\chi$QSM.

%
%
\subsection{Comparison of first moments of helicity densities}
\label{subsec:moments}
Two previous sections discussed three significantly different
but not independent sources of information about first moments
of helicity distributions: the direct interpretation of the
first moment of the spin structure function $g_1^d$ of the deuteron,
NLO QCD fits of parameterized helicity densities to all $g_1$ data 
(see section~\ref{subsec:Dq-incl}),
and global fits of helicity densities to all available relevant data,
including double-spin asymmetries for both semi-inclusive DIS and 
inclusive production of neutral pions and jets in
$p^\Nrightarrow + p^\Nrightarrow$ 
(section~\ref{subsec:GlobalNLO}).
The first moments from these analyses are compared in 
table~\ref{tab:compmoments} on page~\pageref{tab:compmoments}.
The comparison of results from the NLO QCD fits should be undertaken with caution, 
as the analyses differ
in several respects.  The LSS fit uses only inclusive DIS data, including the
\compass\ data on the deuteron, which leads to two equally probable
solutions.  They are merged into a range of values given in the table rather
than central values with standard deviations.  The AAC fit does not yet include the
\compass\ data, but includes some double-spin asymmetries for inclusive production 
of neutral pions in $p^\Nrightarrow + p^\Nrightarrow$.  Finally the DSSV fit 
includes all inclusive and semi-inclusive DIS data, as well as 
double-spin asymmetries for both semi-inclusive DIS and 
inclusive production of neutral pions and jets in
$p^\Nrightarrow + p^\Nrightarrow$.  Another important difference between
the analyses is the treatment of experimental statistical uncertainties,
indicated in the first column.
The LSS uncertainties are based on the standard treatment specified by the PDG~\cite{pdg},
while the AAC group adopts a different treatment that increases 
their uncertainties.  The DSSV group provides the results from two treatments.
Finally, the common elements of the data sets used for the various 
results imply that the uncertainties are correlated, so that they
should in principle be in better agreement than indicated by their
uncertainties.

The values for $\Delta \Sigma$ are all consistent, with the possible
exception that the DSSV result for the `standard' treatment of
statistical uncertainties $(\Delta\chi^2=1)$ appears to be somewhat smaller.  
However it is important to bear in mind that the contribution $-0.124$ to that 
result from the unmeasured region $x<0.001$ is much larger than this
`discrepancy'.

Having accounted for the factor of two relating
the DSSV value for $\Delta\bar{s}$  (assuming $\Delta s = \Delta \bar{s}$)
to the values for \Ds\ from the other two fits, all of the results again
are consistent, again with the possible exception that the DSSV result for 
$\Delta\chi^2=1$ appears to be somewhat more negative.  However, 
essentially all of this value is contributed by the unmeasured region $x<0.001$. 
The uncertainty of the DSSV result is much larger than that of AAC, in spite
of the additional data used in the DSSV fit, presumably because
of the relaxation of SU(3) flavour symmetry.

Finally, all the results for \Dg\ are consistent with each other and with zero.
The additional RHIC data included in the DSSV fit resolves the
ambiguity arising in the LSS fit, and much improves the precision.
Here the contribution to that result from the unmeasured region is about $-0.07$.
Other (leading order) experimental constraints on \Dg\
are discussed in the next subsection.
\begin{sidewaystable}[h]
{\renewcommand{\baselinestretch}{1.}\caption{ \label{tab:compmoments}
Recent published values of first moments of helicity densities, 
separated into experimental evaluations based on the first moment of
$g_{1d}$, and evaluations from NLO QCD fits of various overlapping data sets.
The DSSV values for \Ds are computed from their fit result for $\Delta\bar{s}$  using their assumption $\Delta s = \Delta \bar{s}$.
The fit uncertainties for $\Delta \chi^2=1$ account only for experimental
statistical and systematic uncertainties added in quadrature, and not
for `theoretical' or model uncertainties.
The DSSV uncertainties apply only to the {\em partial} moments over the range 
$x>0.001$, and hence should be interpreted with care.
All evaluations are in the \msbar\ scheme.
}}
\lineup
\begin{tabular}{l|c|c|l|l|l}
Analysis            &  year & $Q^2$  & $\Delta \Sigma$ & \Ds & \Dg \\        
                    &       &(GeV$^2$)&                 &      &      \\
\hline
\hline
\multicolumn{4}{c}{~~~~~~~~~~~~~~~~~~~~~~~~~~Direct Experimental evaluations}\\
\hline
\hermes~\cite{HERMES_g1pd}\ (d target) & 2006 & 5  & $0.330 \pm 0.011\mathrm{(theor.)}$ & $-0.085 \pm 0.013\mathrm{(theor.)}$ & $-$ \\ 
&& & $\pm 0.025\mathrm{(exp.)}\pm 0.028$(evol.) & $\pm 0.008\mathrm{(exp.)}\pm 0.009$(evol.) & \\ 
\compass~\cite{Alexakhin:2006vx}~~~~~``     & ``   & 3  & $0.35 \pm 0.03\mathrm{(stat.)}$ & $-0.08 \pm 0.01\mathrm{(stat.)}$ & $-$ \\ 
&& & $\pm 0.05\mathrm{(syst.)}$ & $\pm 0.02\mathrm{(syst.)}$ & \\ 
\hline
\multicolumn{3}{c}{~~~~~~~~~~~~~~~~~~~~~~~~~~NLO QCD fits}\\
\hline
AAC~\cite{AAC06} $(\Delta\chi^2=12.65)$ & 2006 & 1 & $0.27 \pm 0.07$ & $-0.10\pm0.02$ & $0.31\pm0.32$  \\
LSS\cite{lss06} $(\Delta\chi^2=1)$ & 2006 & 1 & 0.17\ldots 0.31 & -0.068\ldots -0.047 & -0.6\ldots 0.3  \\
DSSV\cite{DSSV} $(\Delta\chi^2=1)$ & 2008 & 1 & $0.242^{+0.015}_{-0.018}$ & $-0.114\pm0.02$ & $-0.084^{+0.106}_{-0.120}$ \\
DSSV\cite{DSSV} $(\Delta\chi^2=0.02,\ \chi^2=7.85)$ & 2008 & 1 & $0.242^{+0.042}_{-0.062}$ & $-0.114\pm0.06$ & $-0.084^{+0.702}_{-0.314}$ \\
\end{tabular}
\end{sidewaystable}

\subsection{Gluon helicity distribution}
\label{subsec:DeltaG}
The nucleon spin can be decomposed into contributions of quarks and gluons,
as will be explained in section~\ref{subsec:Decompose}. A decomposition in 
which all terms have a probabilistic interpretation as parton densities
in the helicity basis can be written as:
\be \label{eq:spinsumGluon}
\frac{1}{2} = \frac{1}{2}\Delta \Sigma  + \Dg + \cal{L},
\ee
where $\cal{L}$ is the total orbital angular momentum of partons. The 
intrinsic gluon contribution $\Dg$ is accessible to experiment.
Until recently the only existing knowledge about the gluon helicity 
distribution $\Dg(x)$ was derived as its first moment
$\Dg \equiv \int_0^1 \de x~ \Dg(x)$, from NLO pQCD fits to the 
spin-dependent structure function $g_1(x,Q^2)$ measured in double-spin 
lepton-nucleon scattering experiments (see section~\ref{subsec:LongExp}).
However, this method is subject to large uncertainties, as $g_1(x,Q^2)$ 
is only weakly sensitive to $\Dg(x)$ via the $Q^2$ evolution of parton 
distribution functions. Very recently, better precision was achieved in 
a global analysis of DIS and $pp$ data, as explained in 
section~\ref{subsec:GlobalNLO}.

\begin{figure*}
\begin{center}
\includegraphics[width=.8\textwidth]{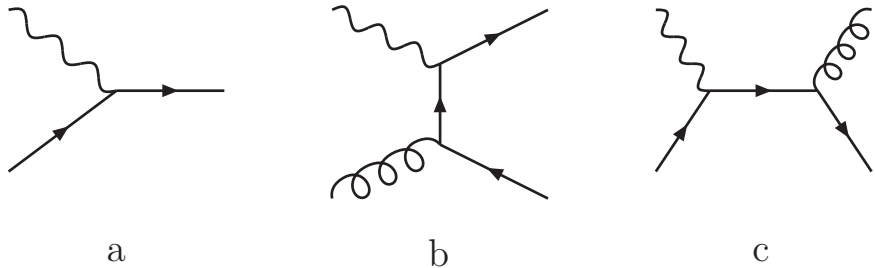}
\end{center}
\vspace*{-1.5cm}
\caption{\label{fig:DGG_subproc} 
Feynman diagrams for hard subprocesses in lepton-nucleon scattering: \\ 
a) \ordera{0} DIS 
($\gamma^* q \rightarrow q$), b) \ordera{1} Photon-Gluon Fusion
($\gamma^* g \rightarrow q\bar{q}$), c) \ordera{1} QCD Compton 
scattering ($\gamma^* q \rightarrow q g$).}
\end{figure*}

More experimental data are now emerging from the study of the
subprocess in which gluons appear in leading order in double-spin
lepton-nucleon scattering: Photon-Gluon Fusion (PGF), see 
figure~\ref{fig:DGG_subproc}(b). Here, the gluon emits a $q\bar{q}$ 
pair in which each quark inherits the helicity of the gluon. As explained in 
section~\ref{sec:Intro}, the (anti-)quark can absorb the photon only if they 
have the same helicity. Given positive (negative) values of $\Dg(x)$, the
PGF process has a negative (positive) asymmetry, while the major background 
processes DIS (figure~\ref{fig:DGG_subproc}(a)) and QCD Compton 
(figure~\ref{fig:DGG_subproc}(c)) have always a positive asymmetry.

The PGF process produces a $q\bar{q}$ pair resulting in 2 jets of hadrons.
At fixed-target energies, one has to use a leading hadron to represent a jet,
plus possibly another hadron separated by large transverse momentum $\pT$ to 
represent the other jet. In order to suppress background processes, final 
states are selected which
contain either large transverse momenta or charm. These signatures are also
exploited to access $\Dg(x)$ in double-spin proton-proton scattering
in the experiments \star\ and \phenix\ at \rhic. A cleaner channel in 
these experiments will be single direct photon production.
\begin{figure*}
\begin{center}
\includegraphics[width=\textwidth]{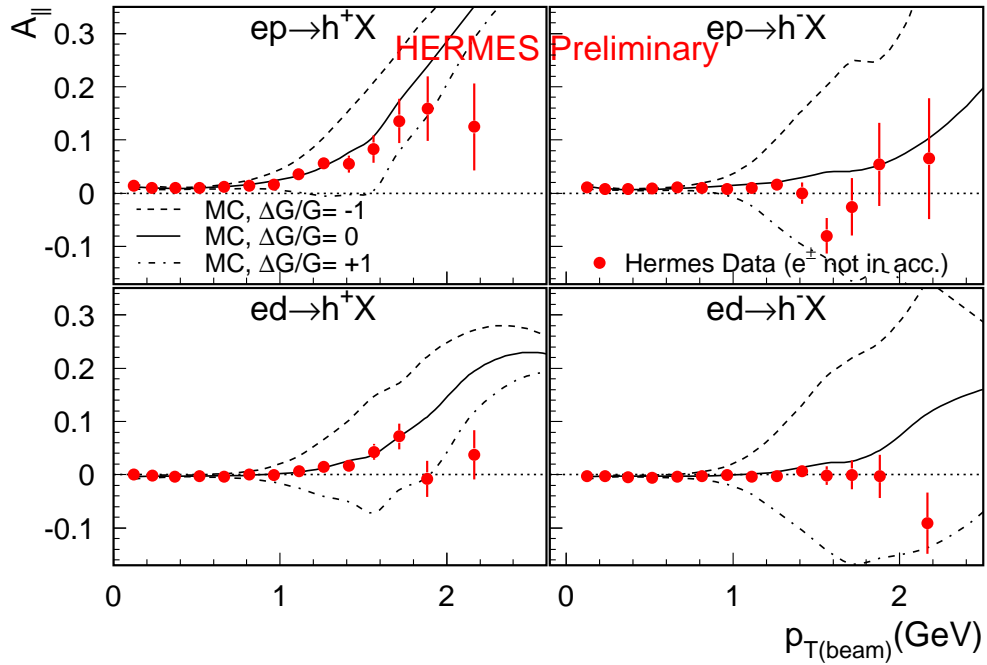}
\end{center}
\vspace*{-2cm}
\caption{\label{fig:DGG_asy}
Asymmetry measured at {\sc Hermes} for positive (left) and negative (right)
inclusive charged hadrons on the proton (top) and deuteron (bottom), as a
function of $p_{T(beam)}$. The errors are statistical only. There is an overall
experimental normalization of 5.2\% (3.9\%) for proton (deuteron) data. The
curves show the asymmetries as expected from the 
{\sc Pythia}~\cite{Sjostrand:2000wi,Patty_thesis} Monte Carlo
simulation for three different assumptions on the gluon polarization. 
This figure is taken from \cite{Liebing:2007bx}.}
\end{figure*}

The gluon polarization $\frac{\Dg}{\g}(x)$ is extracted in a
leading-order analysis from the measured 
double-spin asymmetry $A_{||}$ in the hadron production cross section. 
Results on $\frac{\Dg}{\g}(x)$ are available from \smc, \compass, and 
\hermes. In particular, \compass\ uses charmed meson decay products and 
pairs of 
high-$\pT$ hadrons, while \hermes\ uses single charged high-$\pT$ hadrons. 
The hard scale is given by the charm quark mass in the former, and by $\pT$
in the latter case. Using as an example \hermes\ data for $h^{\pm}$-production 
on the proton and deuteron, respectively~\cite{Liebing:2007bx}, $A_{||}$ is 
shown in figure~\ref{fig:DGG_asy} as a function of hadron transverse momentum 
$p_{T(beam)}$ that is measured with respect to the beam axis. Also shown is
the dependence expected from a Monte Carlo simulation for several values
of $\frac{\Dg}{\g}$ ($0, \pm1$), which demonstrates that for
$p_{T(beam)}>1$ GeV the asymmetry clearly starts to depend on the
underlying value for $\frac{\Dg}{\g}$. Comparing it to the data 
reveals at the level of the asymmetry, that small values of 
$\frac{\Dg}{\g}$ are prefered. In order
to enhance the sensitivity, the `detected' fragmenting quark should 
have the largest possible transverse momentum. This is then reflected in large 
$p_{T(beam)}$ values for the outgoing hadron(s). 

\begin{figure*}
\begin{center}
\includegraphics[width=0.8\textwidth]{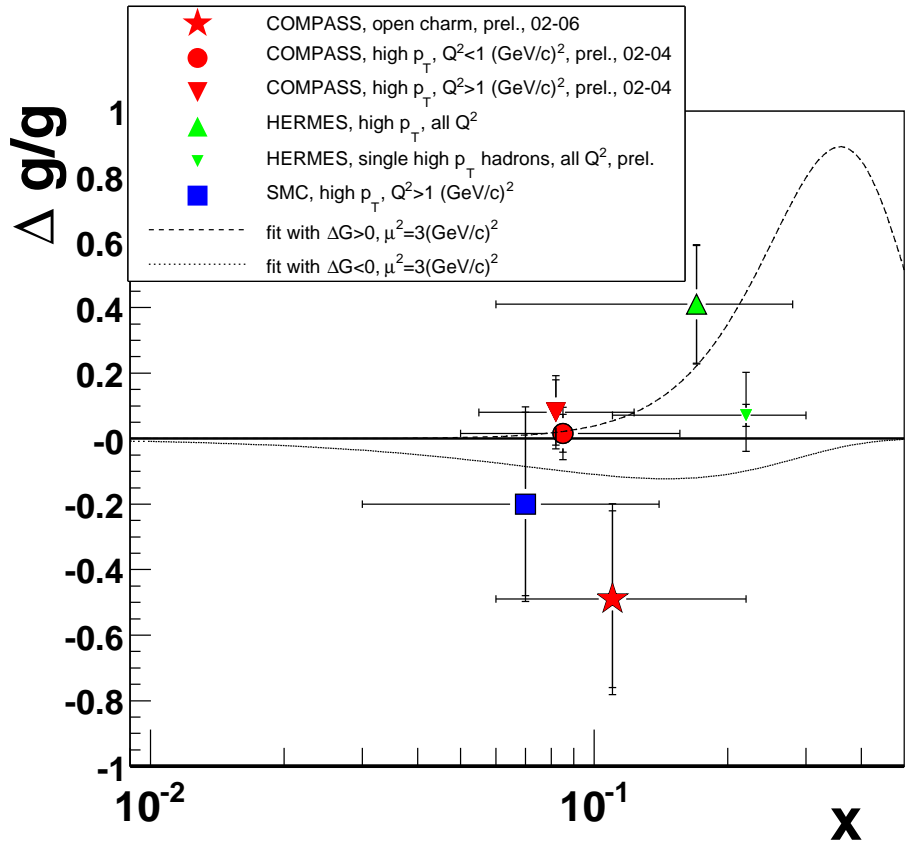}
\end{center}
\caption{\label{fig:DGGvsX}
Leading-order results on $\frac{\Dg}{\g}(x)$, published or 
preliminary, from 
\compass~\cite{Alekseev:2008cz,Ageev:2005pq,Stolarski:2008jc}, 
\hermes~\cite{Airapetian:1999ib,Liebing:2007bx}, 
and SMC~\cite{Adeva:2004dh}. The sequence of references corresponds 
to the order of measurements shown in the legend, where also the
analysed final state is indicated. Outer error bars, if shown, 
represent statistical and systematic uncertainties combined in 
quadrature. Inner error bars represent statistical uncertainties
only. Additionally, two equivalently good QCD NLO fits of the 
structure function $g_1^d$ are shown~\cite{Alexakhin:2006vx}, 
obtained with the constraints $\Dg<0$ and $\Dg>0$, respectively. 
The figure is taken from \cite{Alekseev:2008cz}, updated 
with~\cite{Stolarski:2008jc}.}
\end{figure*}
In figure~\ref{fig:DGGvsX} the above mentioned results on 
$\frac{\Dg}{\g}(x)$ are shown. The three statistically most precise 
ones are from \hermes\ on single hadron production on the deuteron 
at very low $Q^2$~\cite{Liebing:2007bx}
and from \compass\ on high-transverse-momentum hadron-pair production 
on the deuteron at $Q^2 < 1$ GeV$^2$~\cite{Ageev:2005pq} as well as at
$Q^2 > 1$ GeV$^2$~\cite{Stolarski:2008jc}. For the \hermes\ result,
very detailed studies were performed to estimate systematic 
uncertainties associated with Monte Carlo models of the background
processes. The major contribution here stems from the 
uncertainties in the fractional contribution from soft processes, which 
are poorly known. Among the \compass\ results on high-$p_t$ hadron pairs,
the most recent also includes a study of systematic uncertainties
associated with the Monte Carlo model.

From the three high-$p_t$ hadron results, it can be concluded that 
$\frac{\Dg}{\g}(x)$ at medium-$x$ (0.06...0.3) is about 0.1 with a total 
uncertainty of the same magnitude. This leading-order result is in
fair agreement with results of NLO fits of $g_1$ from
\cite{Alexakhin:2006vx} shown as curves in figure~\ref{fig:DGGvsX}.
The value of the unpolarized gluon distribution from the CTEQ6 set 
of PDFs~\cite{Lai:1999wy} is $G(x)|_{x=0.1} \approx 10$, so that for 
the gluon helicity distribution $\Dg(x)|_{x=0.1} \approx 1 \pm 1$ is 
obtained. Expressed as $x \cdot \Dg(x)$, the LO result from high-$p_t$ 
hadrons is $0.1 \pm 0.1$, in agreement with the less model-dependent
result from the 
NLO QCD fits of \cite{DSSV} discussed in the previous section (see 
bottom right panel of figure~\ref{fig:deFlorian2}). We note that any
comparison of next-to-leading-order to leading-order results has to be
done with caution since $k$ factors, defined as their ratio, are known 
to be large. In the kinematic region considered here, $k$ factors were
found~\cite{Jager:2005uf} of up to 5 for cross sections and up to 2 
for asymmetries that are used to extract information on the polarized
gluon distribution.
  
If and when it becomes 
possible to experimentally constrain the {\em full} first moment $\Dg$ with 
sufficient accuracy, it will be possible to determine the total orbital 
angular momentum ${\cal{L}}$ of partons according to 
(\ref{eq:spinsumGluon}).

%
%
\section{The transverse spin structure of the nucleon}
\label{subsec:TransIntro}
\subsection{Transverse spin in deep-inelastic scattering}
The quark transversity distribution $\dq_f (\x)$ was defined in 
section \ref{sec:Intro} as the difference between
number densities of quarks with  transverse polarization
in the same and opposite direction as that of the nucleon spin, in a 
nucleon that is polarized transverse to its `infinite' momentum.
It may at first appear mysterious that it is necessary to 
distinguish between $\delta q_f(\x)$ and the helicity 
distribution $\Delta q_f(\x)$, since
there would be no need to distinguish between them in a 
nonrelativistic system. 

All twist-2 parton distributions are defined as light-cone Fourier
transforms of products of quark fields $\psi_f$ (or gluon fields). 
The only difference
between $q_f(x)$, $\Delta q_f(x)$, and $\delta q_f(x)$ are the 
Dirac matrices that 
appear between $\psi_f^\dagger$ and $\psi_f$. For $\Delta q_f$ the 
Dirac matrices
are (in Dirac representation, where the lower components vanish
in the nonrelativistic limit)
\be
\gamma^0\gamma^i\gamma^5 = 
\left(\begin{array}{cc} 1 & 0 \\ 0 & -1 \end{array}\right)
\left(\begin{array}{cc} 0 & \sigma_i \\ -\sigma_i & 0 \end{array}
\right)
\left(\begin{array}{cc} 0 & 1 \\ 1 & 0 \end{array}\right)
=
\left(\begin{array}{cc} \sigma_i & 0 \\ 0 & \sigma_i 
\end{array}\right),
\ee
while for $\delta q_f$ they read
\be
i\gamma^0\sigma^{0i}\gamma^5 = \gamma^i\gamma^5 =
\left(\begin{array}{cc} 0 & \sigma_i \\ -\sigma_i & 0 \end{array}
\right)\left(\begin{array}{cc} 0 & 1 \\ 1 & 0 \end{array}\right)
=\left(\begin{array}{cc} \sigma_i & 0 \\ 0 & -\sigma_i 
\end{array}\right).
\ee
So the only difference between $\Delta q_f$ and $\delta q_f$ is the 
minus sign in
front of the lower components. In the nonrelativistic limit, the
lower components vanish like $p/m$, so the difference between 
$\Delta q_f$
and $\delta q_f$ vanishes like $v^2/c^2$.

Transversity  loses its probabilistic interpretation as a 
difference in number density distributions 
when expanded in a helicity basis. Every observable in quantum 
mechanics has a probabilistic interpretation as the sum of 
eigenvalues of the observable weighted by the probability to be in 
each eigenstate.\footnote[1]{This is one of the postulates
of quantum mechanics.} 
However, if basis functions are used in which the observable is not 
diagonal, the expectation value of the observable becomes
a sum over products of amplitudes and the probabilistic interpretation
is `lost', or at least not apparent. In the present context 
this elementary principle
appears most clearly using Pauli matrices and 
identifying eigenstates of $\sigma_z$ or $\sigma_x$ with
helicity eigenstates $|\!\leftrightarrow\rangle$ or
transversity eigenstates $|\!\updownarrow\rangle=
\frac{1}{\sqrt{2}}\left(  |\rightarrow\rangle+e^{i\varphi}
|\leftarrow\rangle\right)$,  respectively, 
where $\varphi$ specifies the orientation of the transverse
polarization direction.
Obviously $\sigma_x$ is diagonal when expressed in terms of its own 
eigenvectors, but loses this property when expressed in terms
of the eigenvectors of $\sigma_z$:
\be
\sigma_x = 
\left|\uparrow \right\rangle \!\left\langle \uparrow \right|\,-\,
\left|\downarrow\right\rangle\! \left\langle \downarrow \right|\,
=\, \left|\leftarrow \right\rangle \!\left\langle \rightarrow \right|
\,+\,
\left|\rightarrow\right\rangle \left\langle \leftarrow \right| .
\label{off}
\ee
Therefore the expectation value of the `transversity operator'
$\sigma_x$ in some state
$\psi$ becomes a sum of products of amplitudes in the helicity basis:
\be
\langle \psi|\sigma_x|\psi\rangle 
= \left| \langle \psi |\uparrow\rangle \right|^2
- \left| \langle \psi |\downarrow\rangle \right|^2
= \langle \psi |\leftarrow\rangle \langle \psi |\rightarrow\rangle^*
+ \langle \psi |\rightarrow\rangle \langle \psi |\leftarrow\rangle^*
\label{probilost}.
\ee
The right hand side of (\ref{probilost}) is not a sum over 
absolute squares and hence the probabilistic interpretation for 
$ \langle \psi|\sigma_x|\psi\rangle$ is no longer apparent.

Despite the loss of a direct probabilistic interpretation, there 
are (at least) two reasons for discussing
transversity in a helicity basis. First, that basis has 
already been used to study the more familiar
longitudinal spin polarization, and
secondly only helicity and not transversity is conserved in 
hard processes.
The second statement is based on the observation that
for negligible quark masses, both the QCD and QED Lagrangians
are invariant under a `chiral'  transformation of the quark 
spinors:
\be
\psi_f(x) \longrightarrow e^{i\varepsilon \gamma^5}\psi_f(x).
\label{eq:chiral}
\ee
As a consequence of this `chiral symmetry', the chirality of a 
massless quark (the eigenvalue of $\gamma^5$) is conserved. 
It turns out that for a massless quark its chirality agrees 
with its helicity, but for an antiquark with the opposite 
sign.\footnote[2]{This minus sign for antiquarks also
illustrates that, even for massless quarks,
chirality and helicity are not the same in the sense of being 
described by the same operator, which is one reason why it is necessary to 
introduce the concept of chirality in addition to helicity.
Another is that the helicity of a {\em massive} particle is frame dependent, 
whereas chirality is always Lorentz-invariant.}
Chiral symmetry thus implies the conservation of the
helicity of a massless quark in hard scattering processes. 
However, even though helicity is conserved, transversity is not, 
as the scattering amplitudes for
the two helicity states may differ in their phase.
An analogue from optics is the propagation of light through an
optically active medium where the refraction indices for the
two circular polarization states are different.
While the circular polarization does not change as the
beam propagates through this medium, the linear polarization,
which is related to the longitudinal polarization through a simple
linear combination, precesses around the beam axis since the
waves describing the two circular polarization states acquire
different phases in such a medium.

An important distinction must be made between transversity and
transverse spin.  Transverse spin is the expectation value of the operator
$\psi^\dagger\gamma^\perp\gamma_5\psi$ and is chiral-even, whereas 
chiral-odd transversity is the expectation value of the operator 
$\psi^\dagger\sigma^{\mu\nu}\gamma_5\psi$,
which has matrix elements between states of opposite quark helicity in
the proton rest frame.  Transverse spin does not have a partonic
interpretation in the light-cone framework and is associated with the 
(twist-3) spin structure function $g_2$, whose effects vanish like 
$1/\sqrt{Q^2}$ in the Bjorken limit, whereas transversity does not
vanish in this limit.  For a detailed
discussion of these subtle issues see~\cite{JaffeErice}.

\begin{figure}[htb]
\begin{center}
\epsfig{file=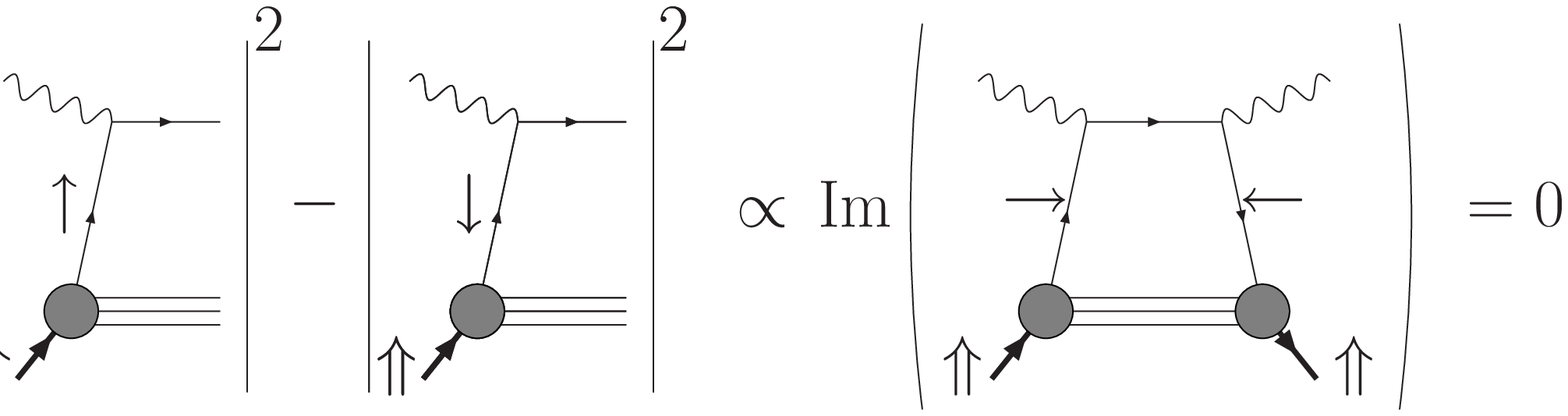,width=12cm}
\end{center}
\caption{\label{fig:transtheorem}
        When expressed in terms of quark helicity
states, the dependence of the DIS cross section on the transversity
of the quarks is proportional to the imaginary part of the
quark helicity-flip forward Compton amplitude which
vanishes at large $Q^2$.}
\end{figure}

Since hard electromagnetic (or weak) processes conserve the helicity
of the quark, inclusive DIS experiments cannot be used to
measure the transversity of the quarks. In order to see why, we 
first use the optical theorem in a transversity basis 
and then express transversity eigenstates in terms of 
helicity eigenstates.
The difference between the contribution to the cross section from
quarks with the same and with opposite transverse polarization
as the nucleon would be proportional to the imaginary part of a
forward scattering amplitude involving quark helicity flip (see 
figure \ref{fig:transtheorem}). 
However, since absorption of a hard vector boson ({\it e.g.} photon)
cannot flip the helicity of a massless quark, this
forward scattering amplitude must be zero and therefore inclusive
DIS
cannot be sensitive to the transverse spin of the quarks.
Since this argument does not involve the polarization of the
photon in DIS, this result also illustrates why a transversely
polarized charged lepton beam does not help to determine the
transverse polarization of the quarks in inclusive DIS. 
In semi-inclusive DIS, the meson production vertex {\em can}
flip quark helicity, which is why the latter process can be
sensitive to transverse quark polarization.

It is instructive to compare DIS mediated by vector boson
exchange, where quark helicity is conserved, with a Gedanken DIS
experiment 
involving scalar bosons, where quark helicity flips at the hard
quark-scalar vertex. If only
scalar bosons are exchanged, then the quark helicity flips
at both vertices in the forward Compton amplitude analogous to
figure~\ref{fig:transtheorem} and such a process would not be 
sensitive to quark 
transversity either. Only the interference term between
vector and scalar boson exchange provides the single helicity
flip that is needed to make this amplitude non-vanishing,
{\it i.e.} the interference between photon exchange and the
exchange of a hypothetical scalar boson (such as the Higgs boson)
would in principle be suitable for accessing transversity
experimentally using inclusive DIS \cite{Ratcliffe:2004we}.

\subsection{Tensor Charge of the Nucleon}
%
The tensor charge is obtained from the forward matrix element of
the operator $\psi_f^\dagger \sigma^{\mu \nu}\psi_f$, analogously to the
vector and axial vector charge (see section \ref{subsec:axial}).
In terms of quark transversity eigenstates, the tensor charge involves
the linear combination 
$q_f^{\uparrow \Uparrow} - q_f^{\downarrow \Uparrow}
-\left(\bar{q}_f^{\uparrow \Uparrow} - \bar{q}_f^{\downarrow \Uparrow}
\right)$,
{\it i.e.} the transversity of the quarks {\em minus} the transversity
of the antiquarks. The fact that the tensor charge contains the 
difference between quark and antiquark transversity is convenient
as the contributions from short-lived 
virtual $q\bar{q}$ pairs cancel, and it can be identified
with the transversity of valence quarks only.
Moreover, the tensor charge can be computed in lattice gauge theory,
and with better precision than the axial charge (see section
\ref{sec:GPDLattice}).

In a helicity basis, transversity involves both a quark and a
nucleon helicity flip, {\it i.e.} it involves the overlap between a
nucleon `state' that had positive helicity and from which a quark 
with positive helicity has been removed and another nucleon `state' 
that had negative helicity and from which a quark 
with negative helicity has been removed. The Cauchy-Schwarz 
inequality tells us that this `scalar product' is less than the
norms of the states involved.
Since the latter represent the probability to find a quark with
the same helicity as the nucleon spin, one thus finds
the Soffer inequality \cite{Soffer:1994ww}
\be
2|\delta q_f (x)| \leq 2 
q_f^{\stackrel{\rightarrow}{\Rightarrow}}(x) =
q_f(x) + \Delta q_f(x),
\ee
which holds independently for each quark and antiquark flavor.
It provides useful constraints on parameterizations of 
transversity.

Attempts have been made to develop a `transversity sum rule' 
\cite{Bakker:2004ib,Leader:2008hd} that is analogous to the `Ji 
decomposition' of the nucleon spin
(\ref{eq:SpinDeco}) but involves the net transversity of
quarks (plus antiquarks).
However, one problem with any transversity sum
rule is that quarks and anti-quarks contribute with opposite signs
to the tensor charge (first moment of transversity), 
while they contribute to either 
longitudinal or transverse nucleon spin with the same sign. 
Therefore, transversity cannot contribute 
to such a sum rule in terms of the tensor charge, but rather in
terms of the tensor charge for the quarks {\sl plus} the
anti-quarks --- a combination that cannot be represented in terms
of a local operator. Therefore lattice calculations will be unable
to calculate the transversity contribution to a transversity
sum rule.
Furthermore, in the transversity sum rule in
\cite{Bakker:2004ib,Leader:2008hd}, neither an
operator definition nor an experimental procedure for determining
the orbital angular momentum contribution has been provided. 
Thus this sum rule cannot be tested and is unfalsifiable.

\subsection{Transversity in the Drell-Yan reaction}
\label{sec:DY}
%
The conceptually simplest practical way to measure transversity is 
to exploit the Drell-Yan reaction in doubly transversely
polarized hadron-hadron 
collisions $h_1^{\ppol}+h_2^{\ppol} \longrightarrow \mu^+ \mu^-+ X$, 
involving the hard scattering process
\be
q+ \bar{q} \longrightarrow \gamma^* \longrightarrow
\mu^+ \mu^-.
\ee
The high-energy cross section for this reaction is obtained
by convoluting the quark distribution in one of the hadrons 
with the antiquark distribution in the other hadron
and with the cross section for $q+\bar{q} \longrightarrow
\mu^+ \mu^-$, where the latter can be calculated in 
perturbation theory (one photon in the intermediate state 
plus higher-order corrections). 
Transversity can be studied in Drell-Yan experiments with
transversely polarized (anti-) protons because the quark and 
antiquark  preferentially annihilate into a vector boson
when their transverse spins are parallel. 
Hence the double-spin asymmetry  
$A_{TT}$ is capable of
measuring the correlation between the spin of the quark in one
of the hadrons with that of the antiquark in the other hadron. 
This double-spin asymmetry is therefore sensitive to the product of 
the transversity distributions for quarks and antiquarks
carrying momentum fractions $x_1$ and $x_2$ respectively:
\be
A_{TT}(x_1,x_2) = \frac{
\sigma^{\Uparrow \Uparrow}-\sigma^{\Uparrow \Downarrow}}
{\sigma^{\Uparrow \Uparrow}+\sigma^{\Uparrow \Downarrow}}
\propto 
{\sum_f e_f^2 \dq_{f/h_1} (x_1)\delta \bar{q}_{f/h_2} (x_2)}.
\ee
The momentum fractions of the quark and antiquark can be 
reconstructed from the invariant mass and net longitudinal 
momentum of the muon pair into which they annihilate. The struck 
$\bar{q}$ in a Drell-Yan reaction can be either a sea 
or valence quark in an incident antiproton (or meson or proton).
In the interpretation of the data for $p+p\longrightarrow
\mu^+\mu^-+ X$, one $\x$ value can be 
associated with the $\bar{q}$ by considering asymmetric events where
the momentum fraction of one of the initial quarks was
so high that it is very unlikely to have been an antiquark
since the latter carry only small momentum fractions.
As the Drell-Yan cross section depends on products of PDFs, 
determining the
overall normalization may in general be ambiguous. In the case of
$\bar{p}+p\longrightarrow \mu^+\mu^-+X$ there is no such ambiguity as 
the same valence PDF of the proton appears as a product with itself. 
Once this normalization has been determined, it can be used to 
determine the normalization of sea quarks in $pp$ scattering.
Plans are being laid at several laboratories to exploit this
reaction to measure transversity~\cite{Efremov:2004qs,CompassPlans}.

All existing experimental constraints on transversity are based on
the sensitivity of fragmentation to the transverse polarization of
the struck quark in processes other than Drell-Yan, 
as manifested in single-spin
asymmetries. These will be discussed in Section 3.5.

\subsection{Intrinsic parton transverse momentum}
\label{sec:SSA}
Partons can also have momentum $\KT$ transverse to the
infinite momentum of the hadron. Hence
parton distributions can be introduced that depend not 
only on $\x$ but also on 
$\KT$ \cite{Mulders:1995dh,Bacchetta:2006tn} and have interpretations as
number densities or differences between number densities
\cite{Bacchetta:1999kz}.
Only three of these distributions  
survive integration over $\KT$ leading
to the familiar PDFs that depend only on $x$.
These have already been discussed
in section 1. They are the spin-averaged distribution
$q_f(x,\KT)$, the helicity distribution 
$\Delta q_f(x,\KT)$ and the
transversity distribution $\delta q_f(x,\KT)$.
However, many other twist-2 distributions exist and they have
probabilistic interpretations similar to that of integrated
PDFs \cite{Bacchetta:1999kz}.
These distributions\footnote[3]{In the `Amsterdam notation', distributions that do not `survive' integration over ${\rm d}^2\KT$ are
signified by the superscript $\perp$.} 
are sometimes known as 
Transverse-Momentum-Dependent (TMD) parton distributions
(TMD-PDFs, or simply TMDs).
In the rest of this subsection we take the unpolarized case as an 
example.

When a quark in a nucleon absorbs a hard virtual photon,
the resulting momentum of the quark is the sum of its initial
momentum and the momentum transfered from the lepton. 
In leading order $\alpha_s$, any  momentum component transverse to 
the virtual photon direction thus originates
from the intrinsic transverse momentum $\KT$
of the parton before the 
scattering, described by $q_f(\x,\KT)$.

The distribution of hadrons $h$ in a jet originating from a
fast quark of flavor $f$ is described by fragmentation
functions $D_{f/h}(z,\PT)$. Here $\PT=\PhT-
z\KT $ is the transverse momentum of the hadron relative to the
momentum of the fragmenting quark. In DIS,
the factorized semi-inclusive cross section for producing
a hadron with transverse momentum $\PhT$ with respect to
the virtual photon momentum is obtained
by convoluting\footnote{Only for Gaussian dependences can these
convolutions be factorized.}  the transverse-momentum-dependent 
parton distributions $q_f(\x,\KT)$ with these fragmentation
functions:
\be
\sigma^{ep\rightarrow ehX}(\xbj,z,\PhT)
\propto \sum_f q_f(\xbj,\KT) \,\otimes \, D_{f/h}(z,\PT) 
\,\otimes \,\delta (z\KT+\PT-\PhT).
\label{frag}
\ee
These fragmentation functions are expected to be universal, {\it i.e.}
independent of both process and target \cite{Collins:2004nx}.
For example, the distribution of hadrons in jets produced in 
$e^+e^-\rightarrow hadrons$ 
is described by the same fragmentation functions that
describe jets in DIS. However, as discussed in
section \ref{sec:sivers}, initial-state interactions (ISIs) and 
final-state interactions (FSIs) need to be considered before the 
concept of universality can be literally applied.

\subsection{Semi-inclusive DIS and the Collins effect}
\begin{figure}
\begin{center}
\epsfig{file=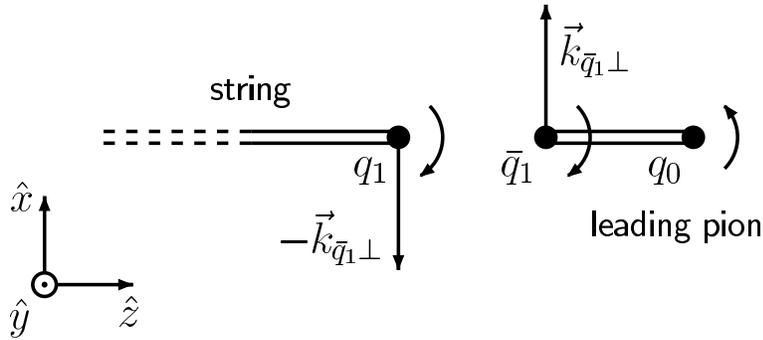,width=10cm}
\end{center}
\caption{\label{fig:Artru} Behind the leading quark ($q_0$) 
moving along $\hat{z}$ a string
forms, which eventually breaks through the formation of the
$\bar{q}_1q_1$ pair. In the $^3P_0$ model, the $\bar{q}q$ pair has
orbital angular momentum in the opposite direction as the spin of the
$\bar{q}q$ pair, i.e. in the case of fragmentation into pions 
(spin $0$) in the same direction as the spin of the leading quark.
Since the antiquark from the pair is expected to be produced
closer to the leading quark as the produced quark, this implies
that the produced antiquark inherits a positive transverse momentum 
${\vec k}_{\bar{q}_1 \perp}$
in the $\hat{y}$ direction from the orbital motion (for a
leading quark polarized in the $\hat{y}$-direction).
}
\end{figure}

%
\label{sec:TransvSpin_Collins}
\label{sec:Artru}
When a transversely polarized quark fragments into hadrons,
the structure of the jet is sensitive to the polarization of the
quark. In particular, the orientation of the transverse spin 
direction together with the quark's momentum defines a plane 
relative to which the momentum distribution of 
hadrons may not be left-right symmetric. To understand how this may
occur, we consider a simple model for the case of
fragmentation into pions: in this process 
for example a quark with spin pointing up needs to pick up from 
the vacuum an anti-quark with spin down in order to form a spin
zero state. 
The $^3P_0$ model for quark anti-quark pair creation 
\cite{LeYaouanc:1972ae} postulates that the $q\bar{q}$ pair is created
with `vacuum quantum numbers' $J^P=0^+$,  the simplest configuration
of which is $S=1$, $L=1$ and no gluon. This model
is based on the idea that as the QCD string between the fragmenting
quark and the target remnant is stretched and stores energy,
a $q\bar{q}$ pair can tunnel from the vacuum to remove some of the
stored energy and the string breaks.
Due to angular momentum conservation, this model suggests
that the $q\bar{q}$ pair is thus produced with orbital angular 
momentum in the same direction (i.e. up in this case)
as the spin of the fragmenting
quark in the produced pion (see figure~\ref{fig:Artru}). 
The antiquark that merges with the 
fragmenting quark should retain some of that transverse
orbital angular momentum, causing it to move to the left
(when looking into the
direction of motion of the fragmenting quark with spin up) 
\cite{Artru:1995bh}.
The pion inherits the transverse momentum carried 
by the antiquark.

\begin{figure}
\begin{center}
\includegraphics[height=10cm,angle=-90]{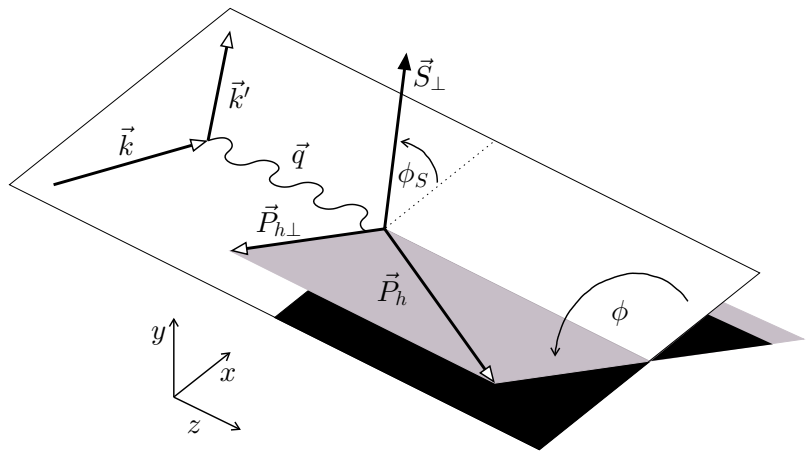}
\end{center}
\caption{\label{fig:phis} The definitions of the azimuthal
angles of the hadron production plane and the axis of the 
relevant component $\vec{S}_\perp$ of the target spin,
relative to the scattering plane that contains the momenta 
$\vec{k}$ and $\vec{k}'$ of the incident and scattered leptons.}
\end{figure}

The fragmentation function $H_{1,f/h}^\perp(z,\pT)$ describing the 
left-right asymmetry in the fragmentation of a transversely 
polarized quark of flavor $f$ into a hadron $h$ was identified
by Collins \cite{Collins:1992kk}, and found to be chirally odd.
Its combination with transversity $\delta q_f(\x)$ that
describes the distribution of transversely polarized quarks
thus gives rise to a correlation between the transverse target 
spin and the transverse momentum of the produced hadron(s).
The associated asymmetry is proportional to 
$\sin \left(\phi-\phi_q\right)$, where
$\phi_q$ is the azimuthal angle of the polarization vector of the 
fragmenting quark about the virtual photon direction
and $\phi$ is the azimuthal angle  of the 
trajectory of the produced hadron $h$.
However, the polarization direction of the struck quark and that 
of the fragmenting quark are not identical. In QED, when the
electron emits a high-momentum virtual photon, the electron
helicity is conserved and therefore the orientation of the
electron spin is rotated in the lepton scattering plane. The virtual
photon inherits the change in angular momentum and is thus
linearly polarized in the lepton scattering plane. 
When a quark absorbs this linearly polarized photon,
its spin component along the polarization direction of the
photon tends to flip. For a transversely 
polarized quark, this implies that its polarization direction
is on average tilted symmetrically with respect to the normal
of the lepton scattering plane. For azimuthal angles measured
relative to the lepton scattering plane, as illustrated in 
figure~\ref{fig:phis}, 
the tilt results in 
$\phi_q=\pi-\phi_S$, where $\phi_S$ denotes the polarization
direction of the target nucleon, since the polarization of the
quark before the photoabsorption is proportional to the product
of the target 
nucleon polarization and the transversity distribution.
The resulting $\sin \left(\phi+\phi_S\right)$ angular dependence
can thus be used to extract the product of transversity and the 
Collins fragmentation function.

However, this `Collins effect' is not the only mechanism 
that can give rise to a 
correlation between the target spin and
the momenta of final state hadrons. A correlation between
transverse momenta of quarks in the target and the target nucleon
spin may give rise to a $\sin \left(\phi-\phi_S\right)$ asymmetry, 
as we will now discuss.

\subsection{Sivers effect}
\label{sec:sivers}
%
In a target that is polarized transverse to the virtual-photon
direction ({\it e.g.} up), the quarks in the target 
nucleon can exhibit a (left/right) asymmetry of the distribution 
$f_{q/p^\uparrow}(x,\KT)$ in their transverse 
momentum $\KT$ 
\be
f_{q/p^\uparrow}(x,\KT) = f_1^q(x,\kT^2)
-f_{1T}^{\perp q}(x,\kT^2) \frac{ (\hatP
\times \KT)\cdot {\bm S}}{M},
\label{eq:sivers}
\ee
where ${\bm S}$ is the spin of the target nucleon and
$\hatP$ is a unit vector opposite to the direction of the
virtual-photon momentum. The fact that such a term
may be present in (\ref{eq:sivers}) is known as the Sivers effect
and the function $f_{1T}^{\perp q}(\x,\kT^2)$
is known as the Sivers function \cite{Sivers:1989cc}.
The latter vanishes in a naive parton 
picture since $(\hatP \times \KT)\cdot {\bm S}$ 
is odd under naive time reversal (a property known as naive-T-odd), 
where one merely reverses
the direction of all momenta and spins without interchanging the
initial and final states. 
In fact, until a few years ago, it was believed that full
time reversal invariance of QCD forbids the existence
of the Sivers effect.  However, then it was realized that a FSI 
mediated by soft gluon(s) between the struck quark and
the remainder of the target nucleon, before the quark
fragments into hadrons, could avoid this restriction~\cite{Collins:2002kn}.
This led to a generalization of our understanding of what is
meant by a parton distribution and its universality, and how it 
may appear in expressions for matrix elements of processes.
The momentum fraction $x$, which is equal to $\xbj$ in DIS experiments, 
represents the longitudinal momentum of the quark {\it before}
it absorbs the virtual photon, as it is determined solely from the 
kinematic properties of the virtual photon and the target nucleon. 
In this new view, in DIS the transverse momentum $\KT$
represents the asymptotic transverse momentum
of the active quark after it has absorbed the virtual photon and
then left the target but still before it
fragments into hadrons.  Thus the Sivers function for semi-inclusive
DIS includes the FSI 
between struck quark and target remnant, and
time reversal invariance no longer requires that it vanishes.
Indeed, time reversal not only reverses the signs of all
spins and momenta, but also transforms FSIs into ISIs. 
It has been shown that the Sivers 
function relevant for SIDIS and that relevant for 
Drell-Yan (DY) processes must have opposite signs 
\cite{Collins:2002kn},
\be
f_{1T}^\perp(\x,\kT^2)_{SIDIS} =
- f_{1T}^\perp(\x,\kT^2)_{DY} ,
\label{SIDISDY}
\ee
where the asymmetry in DY arises from the ISI between the 
incoming antiquark and the target.
The experimental verification of this relation 
would provide a test of the current understanding of the Sivers 
effect within QCD. 
While the argument leading to (\ref{SIDISDY}) is 
nonperturbative, it is instructive to elucidate its physical
origin in the context of a perturbative picture:
for instance, when the virtual photon in a DIS process hits a red 
quark, the spectators must be collectively anti-red in order to
form a color-neutral bound state, and thus attract
the struck quark. In DY, when an anti-red antiquark annihilates with
a target quark, the target quark must be red in order to merge
into a photon, which carries no color. Since the proton was 
colorless before the scattering, the spectators must be anti-red
and thus repel the approaching antiquark \cite{Burkardt:2007rv}.

\begin{figure}
\begin{center}
\epsfig{file=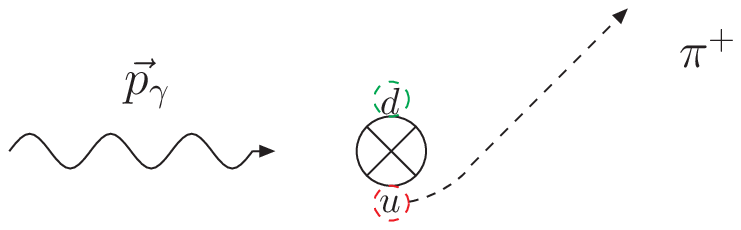,width=10cm}
\end{center}
\caption{\label{fig:gpdssa}
        A correlation between the nucleon spin and the quark orbital
        angular momentum, in combination     
        with an attractive final state interaction, gives rise to a 
        Sivers effect for $u$ ($d$) quarks imparting a 
        transverse momentum that is on average directed upward 
        (downward).}
\end{figure}

Another reason why the Sivers effect is attracting attention
is that it relates to the correlation between the orbital angular 
momentum of partons
and the spin of the parent nucleon. A semi-classical
argument can help to understand how.  
Should quarks have orbital angular momentum, it would be correlated
with the spin of the nucleon, as the latter defines the only 
preferred direction. When viewed in the nucleon rest frame,
quarks orbiting in a transversely polarized nucleon move 
towards the virtual photon on one
side of the nucleon but away from it on the opposite side.
Upon boosting the nucleon to the infinite-momentum frame, 
the quarks on one side of the nucleon with a particular 
momentum fraction $x$ thus appear to have
larger longitudinal momenta than on the other side. Since parton 
distributions are rapidly falling functions of $x$, shifting them 
towards larger
(smaller) $x$ results in an increased (decreased) probability of 
interaction when viewed at a fixed value of $x$.  The result
is that, at a given value of $x$, more quarks absorb photons
on the left or right side of the nucleon when the nucleon spin
is pointing \eg\ up.  As the struck quark moves away, it experiences
an attractive FSI that tends to `focus' it towards the axis
through the center of the nucleon along the direction of
the virtual photon (see figure~\ref{fig:gpdssa}).  
Thus the excess of interactions on one side results in
a tendency of the struck quark and hence the hadron jet to
be deflected towards the opposite side, producing a sinusoidal
azimuthal distribution of the hadrons relative to the transverse
polarization direction of the target nucleon.  Quarks of different
flavours may have different spin-orbit correlations, and also
tend to produce different populations of hadron types.  Hence
different hadron types may have different azimuthal distributions.

The type of deflection shown in figure~\ref{fig:gpdssa} also 
arises in optics when a
convex lens is illuminated asymmetrically, \eg\ when only one side
of the lens is illuminated with a laser beam parallel to the axis
of the lens. The beam is then deflected towards the lens axis and
the outgoing beam has a left-right `momentum asymmetry'
\cite{Burkardt:2002ks}.
Because of this analogy, the above mechanism has been called
`chromodynamic lensing' \cite{Burkardt:2003uw}.

As the asymmetry is caused by the orbital motion of the quarks,
it is a (left-right) asymmetry relative to the direction of the
(vertical) transverse target
polarization, {\it i.e.} proportional to $\sin(\phi-\phi_S)$ in
SIDIS. This differs from the $\sin(\phi+\phi_S)$
angular dependence caused by the Collins effect, 
thus permitting the two contributions to be separated through the
use of transverse target polarization.

The Sivers function not only provides a clue about the
orbital angular momentum structure of the nucleon wave function, 
but in addition
carries important information about the complex phase structure
of the FSI \cite{Brodsky:2002cx}.

\subsection{Boer-Mulders effect}
\label{sec:BM}
%
Even when the target nucleon is unpolarized, there can exist a
correlation between the distribution in transverse momentum 
and transverse polarization of its quarks.
The Boer-Mulders effect is a possible dependence of the distribution 
$f_{q^\uparrow/p}(\x,\KT)$ of 
transversely polarized quarks in an unpolarized target
on the transverse polarization ${\vec S}_q$ of the quarks:
\be
f_{q^\uparrow/p}(\x,\KT) = \frac{1}{2}\left[f_1^q(\x,\kT^2)
-h_{1}^{\perp q}(\x,\kT^2) \frac{ (\hatP
\times \KT)\cdot { {\bm S}_q}}{M}\right] ,
\label{eq:BM}
\ee
and the Boer-Mulders function $h_{1}^{\perp q}(\x,\kT^2)$
embodies this correlation \cite{Boer:1997nt},
which can naturally arise from non-perturbative spin-orbit
interactions. 

\begin{figure}
\begin{center}
\epsfig{file=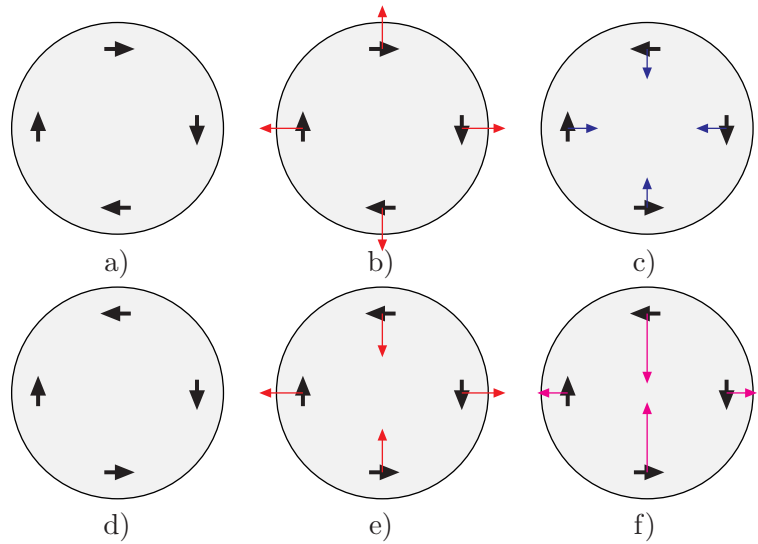,width=10cm}
\end{center}
\caption{\label{fig:bm} Illustration for the interplay between
        transverse momenta from the Boer-Mulders and the Collins 
        effect, leading to a $\cos 2\phi$ asymmetry:
        a) `primordial' transverse spin distribution in an
             unpolarized target,
        b) transverse momentum from the Collins effect for 
            hypothetical SIDIS without transverse spin flip,
        c) transverse momentum from the attractive FSI,
        d) after the virtual photon absorption, the transverse spin
            component in the (horizontal) lepton scattering plane 
            has flipped,
        e) transverse momentum from the Collins effect including
            the transverse spin flip,
        f) net transverse momentum from FSI and Collins effect.
 }
\end{figure}

In the case of the Sivers effect, the relevant spin
is that of the target nucleon, whereas here it is that of the
struck quark.  However, both effects can be understood
to arise in part from the preference of the virtual photon to absorb
on a quark with excess momentum fraction associated with its  
orbital motion, because of the monotonic decrease with \x\ of the
unpolarized PDF.  Thus a quark found on the top
(bottom) [left] \{right\} side of the nucleon, 
as seen by the virtual photon, 
will tend to have orbital angular 
momentum pointing right (left) [up] \{down\}  (on first reading it may
be helpful to consider only the first option and compare to 
figure \ref{fig:bm}).  If the quark spin
is correlated with its orbital angular momentum as represented by
the Boer-Mulders function, then its
polarization will have a similar (or opposite) spatial pattern
(figure \ref{fig:bm}a).
As described in section~\ref{sec:TransvSpin_Collins}, the
quark polarization affects the transverse momentum of produced
hadrons via the Collins effect.  Quarks with polarization
pointing right (left) [up] \{down\}  may be deflected \eg\ up (down)
[left] \{right\} , or the opposite depending on the sign of
the Collins fragmentation function.  We first consider the
simpler case of exchange of a scalar boson rather than 
the vector photon, where the former cannot carry any
information about the orientation of the lepton scattering plane 
so that the initial quark transverse
polarization direction persists through the photoabsorption process. 
Then the combined effect would be a net `focussing' or `defocussing' 
of the produced hadrons relative to the direction of the
exchanged boson, according to the product of the Collins function 
with the Boer-Mulders function representing the original 
spin-orbit correlation of the quark (figure \ref{fig:bm}b).  
However, we must beware 
that this term `(de)focussing' can be misleading in this context.  
Since one side of the nucleon is not experimentally distinguished
from the other, 
and the target nucleon is uniformly illuminated by the parallel
`beam' of virtual photons, the effect only randomly scatters or
smears the direction of the hadron.  At each azimuthal angle 
of the hadron direction about the virtual photon direction there
is a net smearing effect, while
the mean change of direction due to this `(de)focussing' is zero.

An additional independent axially symmetric `focussing' action 
is provided by the attractive FSI that is crucial also for the Sivers 
effect (figure \ref{fig:bm}c).  
(The Sivers effect itself plays no role here because 
the target nucleon is unpolarized.)  This FSI effect either enhances 
or competes with that in figure \ref{fig:bm}b to produce a net 
`(de)focussing' action on the direction of the produced meson relative
to that of the virtual photon.   Thus for scalar exchange, the net 
effect may be a change in the cross section for hadron production 
that is independent of azimuth.  For example, for observed 
values of the meson momentum component $\phT$ transverse to the 
direction of the virtual photon that are larger than the most
probable value, this smearing would increase the cross section.

On the other hand, as also described in 
section~\ref{sec:TransvSpin_Collins}, 
for vector (photon) exchange, the initial transverse quark 
polarization 
does not always persist through the photoabsorption process.  
Because of the linear polarization of the virtual photon 
in the lepton scattering plane, the component of the
quark polarization in that plane can be flipped 
(figure \ref{fig:bm}d), 
inverting the  effect in figure \ref{fig:bm}b 
only for hadron transverse momentum components normal to that plane
(figure \ref{fig:bm}e).  
Now this effect tends to enhance the FSI effect in one plane
and cancel it in the orthogonal plane (figure \ref{fig:bm}f).  
Since the net smearing
effect is proportional to the {\em magnitude} of the net
`(de)focussing' action, the smearing is different in the lepton
scattering plane from that in the orthogonal plane.  Hence
the effect on the cross section can be different for hadrons
produced in the lepton scattering plane compared to the
orthogonal plane.  
This amounts to a $\cos(2\phi)$ dependence of the hadron azimuthal 
distibution, relative to the lepton scattering plane.  In the absence
of the FSI, the inversion of the Collins effect  in the 
lepton scattering plane (illustrated
in figure \ref{fig:bm}e) would not change the `(de)focussing' {\em magnitude},
which is all that matters for the cross section.  The necessary role
of the FSI is associated with the T-odd nature of this Boer-Mulders effect. 

A more formal argument is also included here for the sake of 
completeness, as it may not be easily available in the literature.  
The Collins effect acts as a polarimeter with analyzing power
$\propto H_1^\perp(x,\pT)(\hatP
\times \PT)\cdot { {\bm S}_q^\prime}$,
which tags  the transverse polarization ${\bm S}_q^\prime$ of the
quark {\it after} it has absorbed the virtual photon and its 
component in the lepton scattering plane has flipped compared to
$ {\bm S}_q$. Here $\PT$ is the transverse momentum 
of the outgoing hadron relative to the momentum direction of the
fragmenting quark. The resulting asymmetry in the hadron distribution
relative to the lepton scattering plane is obtained by convoluting
the Boer-Mulders function with the Collins function to
obtain the distribution of the transverse (with respect to the
virtual photon direction) hadron momentum
$\PhT$ in the final state. (For simplicity, the dependence
of these functions on the longitudinal momentum is not shown here.)
In addition, the result must be averaged over the (unmeasured)
transverse polarization direction $\phi_S$ of the quark in the 
initial state:
\bea
A(\PhT) &\propto& \int_0^{2\pi}\!\!\!\!{\rm d}\phi_S\!
\int\! {\rm d}^2 \KT \!
\int\! {\rm d}^2 \PT\,\,
h_{1}^{\perp q}(\kT^2)( \KT\times { {\bm S}_q})_z\\
& &\quad\quad\quad \times
H_1^\perp(\pT^2)(\PT\times { {\bm S}_q^\prime})_z
\delta(\KT+\PT-\PhT).\nonumber
\eea
As $\PhT$ is the only vector on which the transverse 
momentum integral can depend, rotational invariance implies 
\bea
\int\! {\rm d}^2 \KT \!\!\int\! {\rm d}^2 \PT\,
h_{1}^{\perp q}(\kT^2)\,k_{T,i}\, H_1^\perp(\pT^2)\, p_{T,j}\,
\delta(\KT+\PT-\PhT)& &\\
\quad \quad \quad
= a(\phT^2) \delta_{ij} + b(\phT^2) \phT^i \phT^j,
& &\nonumber
\label{eq:integral}
\eea
with some suitable functions $a(\phT^2)$ and $b(\phT^2)$, 
yielding
\be
\!\!\!\!\!\!\!\!
A(\PhT) \propto \int_0^{2\pi}\!\!\!{\rm d}\phi_S\, 
\left[ a(\phT^2)
{\bm S}_q \cdot {\bm S}_q^\prime + b(\phT^2) 
\left({\bm S}_q \times \PhT\right)_z
\left({\bm S}_q^\prime \times \PhT\right)_z\right].
\ee
The transverse polarization
of the active quark does not always flip, but events where it
does not flip do not produce any asymmetry. For the determination
of the asymmetry we thus focus on those events where the transverse
spin component in the lepton scattering plane flips.
Taking this plane to lie in the $\hat{x}$-$\hat{z}$
plane, the transverse polarization vectors take on the form
${\bm S}_q = (\cos \phi_S, \sin \phi_S)$ and
${\bm S}_q^\prime = (-\cos \phi_S, \sin \phi_S)$, yielding
\bea
\!\!\!\!\!\int_0^{2\pi}\!\!\!{\rm d}\phi_S\, {\bm S}_q\cdot {\bm S}_q^\prime
&=& 0 \\
\!\!\!\!\!\int_0^{2\pi}\!\!\!{\rm d}\phi_S\, 
\left({\bm S}_q\times \PhT\right)_z
\left({\bm S}_q^\prime \times \PhT\right)_z 
&=& \frac{\pi}{2}\left(p_{h,T,x}^2
-p_{h,T,y}^2\right) = \frac{\pi}{2}\phT^2 \cos 2\phi ,  
\eea 
where $\phi$ specifies the transverse direction of the outgoing
hadron momentum $\PhT$. In combination with 
(\ref{eq:integral}) this
implies a $\cos 2\phi$ asymmetry for the transverse
momentum distribution of the produced hadrons
\be
A(\PhT) \propto \phT^2\, b(\phT^2) 
\cos 2\phi
\ee
which involves a convolution $b(\phT^2)$ of the Boer-Mulders 
and Collins functions. We note here that a flip of 
the quark polarization direction relative
to the normal of the lepton scattering plane is essential to
provide a reference plane (the lepton scattering plane)
for the transverse momentum of the outgoing hadrons and thus
making the Boer-Mulders effect observable, just as it was
essential for separating the Collins from the Sivers effect in SIDIS.
We also note that the Boer-Mulders effect is a spin effect, even
though no spin is being measured.

Alternatively, one can also study the $\cos{2\phi}$ 
asymmetry in the unpolarized Drell-Yan process, in which case
$\phi$ is the azimuthal angle of the $\mu^+\mu^-$ plane
about the virtual photon axis with respect to the incident proton
trajectory.

While ISI/FSI are obviously essential features for the Sivers and 
Boer-Mulders functions, they affect all twist-2 $\KT$-dependent
parton distributions. However, this does not spoil standard
universality,
as all other $\KT$-dependent PDFs are (naive) T-even and thus
the ISI in DY has the same effect as the FSI in SIDIS.
Nevertheless, their $\KT$ dependence is expected
to differ from the `intrinsic' $\KT$ dependence that one would
for example obtain theoretically through the evaluation
of a matrix element of a local operator, as is usually done
in lattice gauge theory (see section \ref{sec:GPDLattice}).

\subsection{Recent semi-inclusive data with transverse target polarization}
\label{subsec:A_UT-data}

Almost 20 years ago, surprisingly large single-spin asymmetries were 
observed in inclusive production of pions by beams of polarized protons 
and anti-protons~\cite{E704}.  (Recently this phenomenon was also observed
at the higher energies of RHIC~\cite{Adams:2003fx,Nogach:2006gm}, 
where the cross section for these
reactions can be understood within the framework of perturbative QCD.)
Various theoretical interpretations
were proposed, including the Collins and Sivers effects.  However,
no way of disentangling these effects on those data could be found.  
A similar ambiguity plagued the interpretation of the first substantial
single-spin asymmetries, with respect to longitudinal target polarization,
observed in the semi-inclusive leptoproduction of mesons by the \hermes\ 
experiment~\cite{hermes-sidis-AUL}.  When the
target is polarized along the lepton beam axis, both Collins and
Sivers mechanisms produce a common sinusoidal behavior in the 
azimuthal angle $\phi$ of the detected hadron about the direction of the virtual
photon, with respect to the lepton scattering plane.
On the other hand, when the target polarization is orthogonal to
the beam axis (more precisely, when it is orthogonal
to the direction of the virtual photon), the azimuthal angle $\phi_S$ 
of the axis of transverse target polarization resolves the effects
of the two mechanisms into distinctive signatures:
$\sin(\phi-\phi_S)$ for the Sivers mechanism, and
$\sin(\phi+\phi_S)$ for the Collins mechanism~\cite{Boer:1997nt}.
In DIS, the Collins effect serves as a `polarimeter' sensitive to 
transverse quark polarization to reveal its correlation with 
transverse target polarization, which is known as transversity.  
As described in section~\ref{sec:SSA}, the Collins mechanism depends on the
influence of the struck quark's polarization on the
transverse momentum $\PT$ acquired by the produced hadron in the 
fragmentation process, in particular on the component of $\PT$ that is 
orthogonal to the quark's transverse polarization.
This `measurement' of the struck quark's polarization occurs {\em after} 
its spin component in the lepton scattering plane
has been flipped by the photo-absorption.  
The orientation of the lepton scattering plane is `remembered'
by the virtual photon because that is the plane of its linear polarization.  
Hence the Collins mechanism is sensitive to the orientation of the 
lepton scattering plane from which the azimuthal angles are measured,
as indicated by the dependence on $\phi+\phi_S$.
In contrast,  as described in section~\ref{sec:sivers}, 
the Sivers effect can be understood in terms of
retention by the struck quark of the $\KT$ that it had in the target
due to a correlation with transverse target polarization.
This transverse momentum tends to be
inherited by a forward hadron that may `contain' this quark.
Hence in either case, the hadron $\PhT$ is correlated with
transverse target polarization.  However, the different azimuthal
dependences allow their experimental separation, and semi-inclusive 
DIS measurements with a transversely polarized target were eagerly 
anticipated.  

\begin{figure}[h]
\begin{center}
\includegraphics[height=0.75\columnwidth,angle=-90]{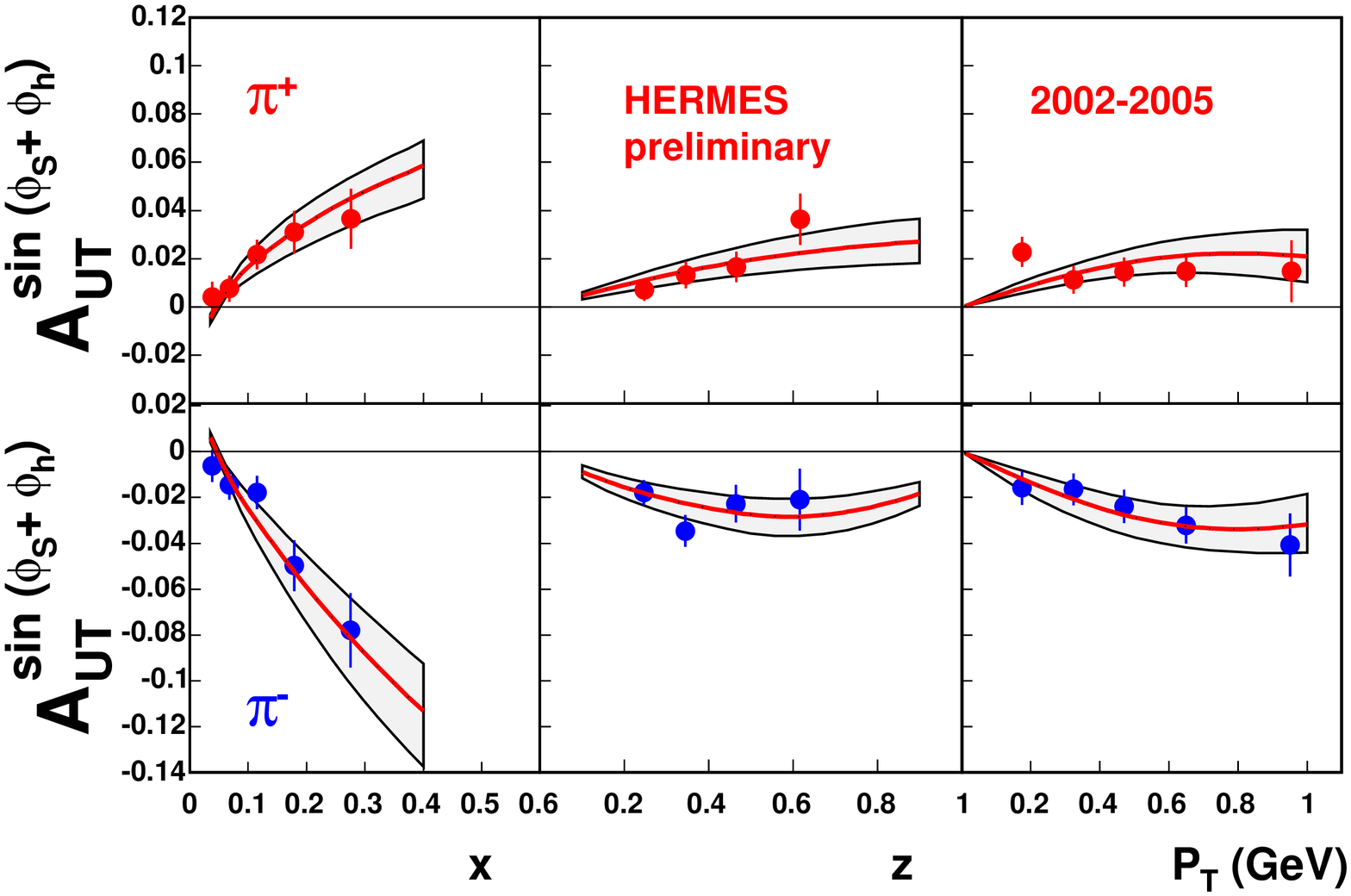} 
\includegraphics[height=0.75\columnwidth,angle=-90]{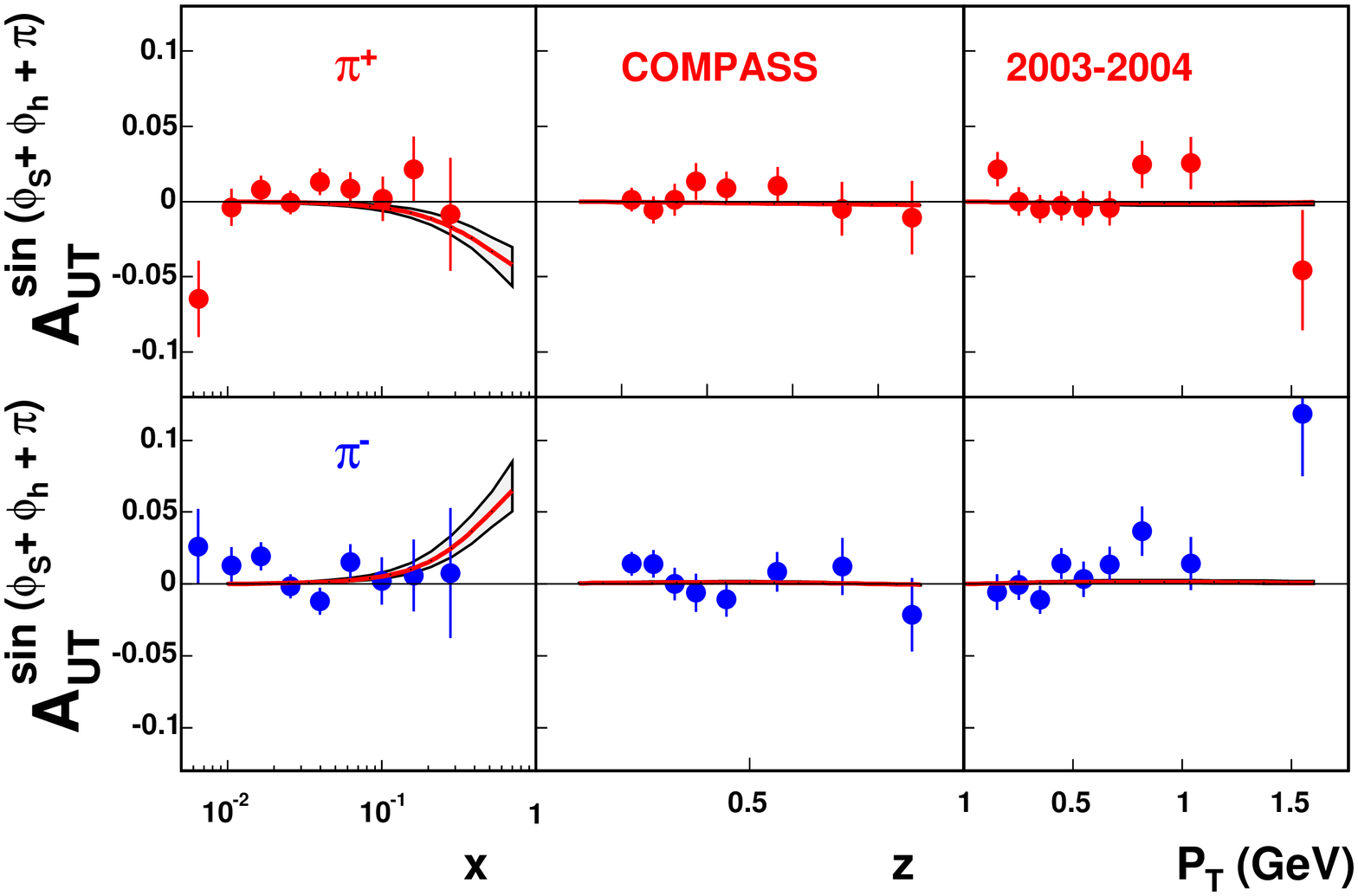} 

\end{center}
\caption{\label{fig:Collins} 
Collins sinusoidal amplitude of the lepton-beam asymmetry 
with respect to transverse polarization
of a proton~\cite{hermes-AUT-2} (deuteron~\cite{Alekseev:2008dn}) target, 
for the semi-inclusive production by positrons (muons) of 
identified charged mesons 
as labelled in the upper (lower) panel, 
as a function of $x$, $z$ or $\phT$ (the latter appearing as $P_T$ in the figure).
The sign of the amplitudes is according 
to the Trento Conventions~\cite{Trento},
although for the deuteron target the sign is reversed
because of the extra value of $\pi$ added to the argument of the $\sin$ function.
The error bars represent the statistical uncertainties, while
for the proton target there is an additional common 8\% scale uncertainty.
The curves are fits to the data, as described in the 
text~\cite{Anselmino:2008sj}.
The error bands on these curves arise from the statistical uncertainties of the data,
but the procedure chosen implies that they are several times
larger than the conventional definition given by, e.g., the Particle Data Group~\cite{pdg}.
On the other hand, uncertainties related to model assumptions are not estimated.
The figure is taken from \cite{Anselmino:2008sj}.
}
\end{figure}
\begin{figure}[h]
\begin{center}
\includegraphics[height=0.48\columnwidth,angle=-90]{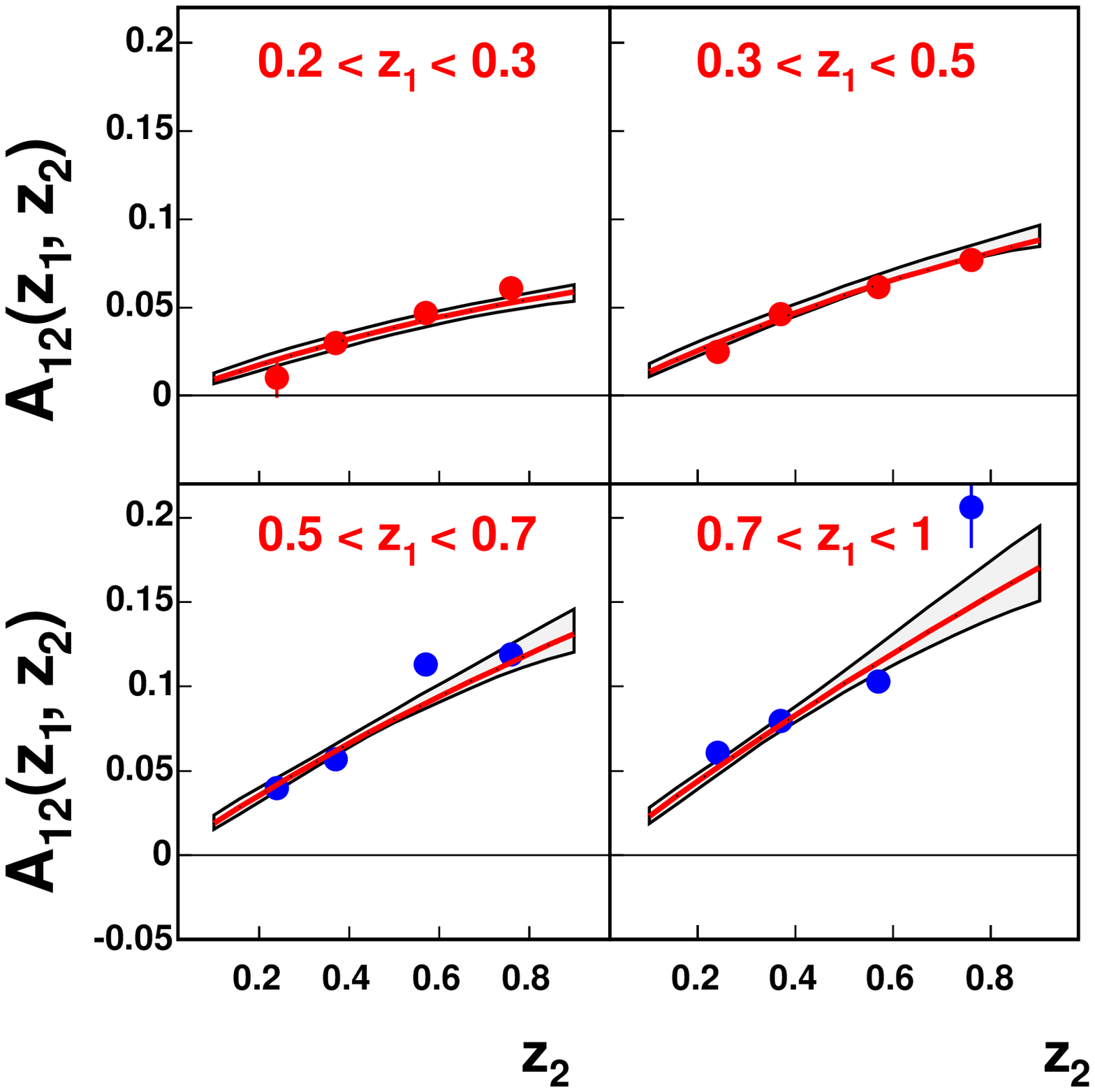}
\includegraphics[height=0.48\columnwidth,angle=-90]{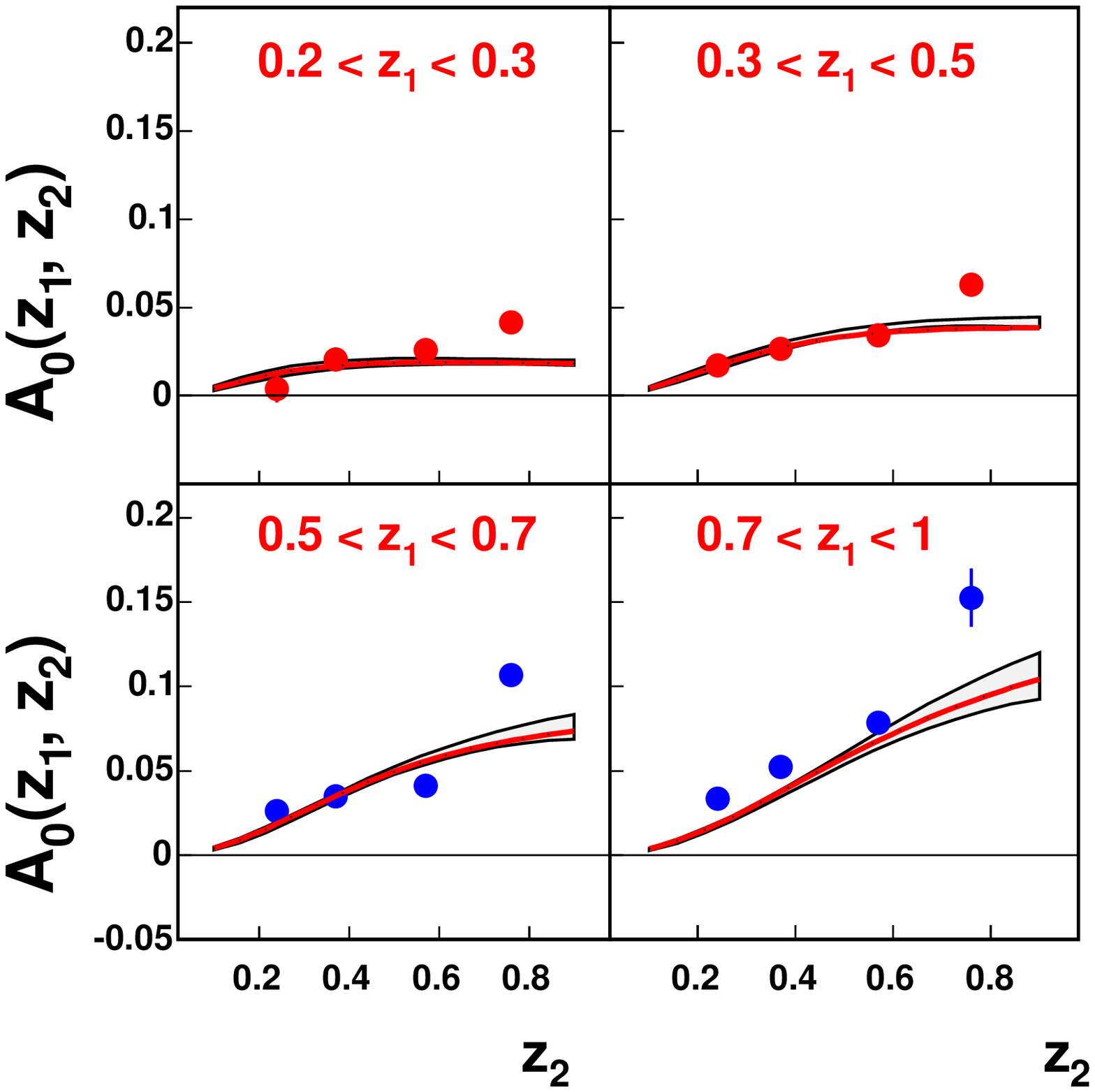} 
\end{center}
\caption{\label{fig:Belle} 
Two alternative jet asymmetries extracted from \belle\ data~\cite{Belle} 
for $e^+e^- \rightarrow 2~jets$,
compared to a fit~\cite{Anselmino:2008sj} of parameterized Collins functions.
The figure is taken from \cite{Anselmino:2008sj}.
}
\end{figure}
\begin{figure}[h]
\begin{center}
\includegraphics[height=0.75\columnwidth,angle=-90]{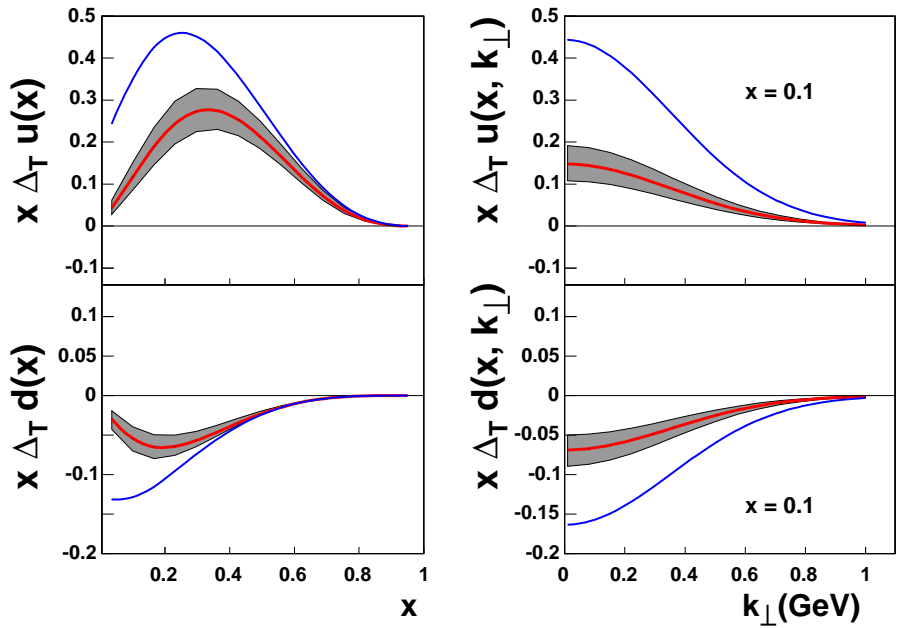}
\end{center}
\caption{\label{fig:transversity} 
The (red) curves within the error bands represent values of the 
transversity distributions at $Q^2=2.4$\,GeV$^2$ for $u$ (upper panels) 
and $d$ (lower panels) quarks
as a function of $x$ and $\kT$ (labelled $k_{\perp}$
in the figure), from the fit~\cite{Anselmino:2008sj} 
to the pion data shown in figure~\protect\ref{fig:Collins} 
together with $e^+e^-$ data from \belle~\cite{Belle}.
The $\kT$ dependence in the right panels was chosen to be the
same as that used for the unpolarized distributions, which was
derived from fitting the azimuthal dependence of unpolarized SIDIS
cross sections. The `Torino notation' used in the figure for the 
transversity distributions is 
equivalent to the `Amsterdam notation' used elsewhere in this paper.
(This is not true for, \eg, TMDs~\cite{Trento}.) The outermost (blue) 
curves represent the Soffer positivity bounds~\cite{Soffer:1994ww}.
The error bands have the same meaning as in figure~\protect\ref{fig:Collins}.
The figure is taken from \cite{Anselmino:2008sj}.
}
\end{figure}
\begin{figure}[h]
\begin{center}
\includegraphics[height=0.75\columnwidth,angle=-90]{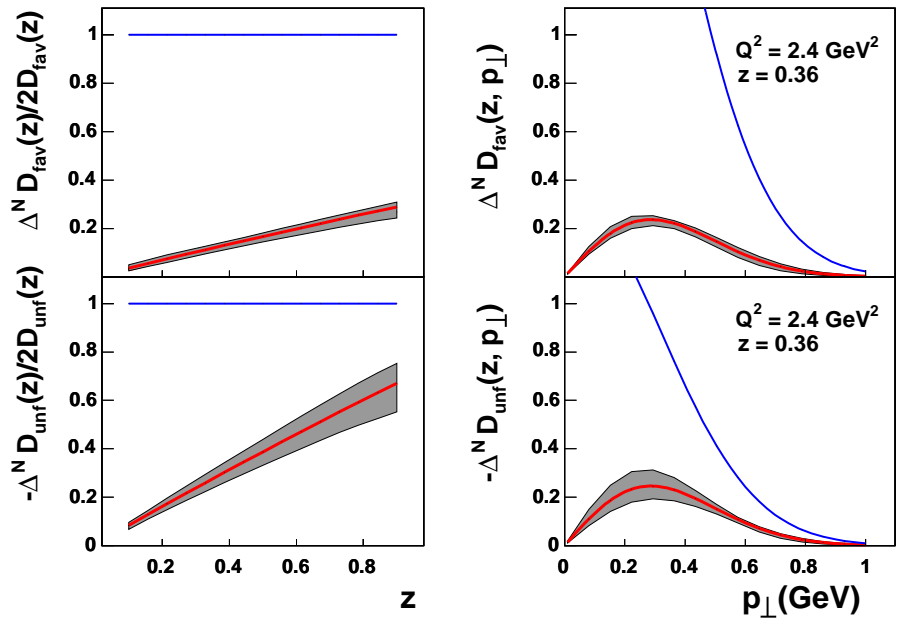} 
\end{center}
\caption{\label{fig:collins} 
The (red) curves within the error bands represent
values of the favoured 
and disfavoured Collins fragmentation functions at $Q^2=2.4$\,GeV$^2$ 
as functions
of $z$ or $\pT$ (the latter appearing in the figure as $p_{\perp}$),
from the fit~\cite{Anselmino:2008sj} to the pion data shown in
figure~\protect\ref{fig:Collins},
together with $e^+e^-$ data from \belle~\cite{Belle}.
In the left panels, the $\pT$-integrated functions are shown as ratios to twice
the corresponding unpolarized fragmentation functions, while
in the right panels, the functions are evaluated at fixed values of $z$.
The Collins function is represented in the Torino notation, which is
related to the Amsterdam notation used elsewhere in this paper by
$\Delta^N D_{h/q^{\uparrow}}(z,\phT^2) =
\frac{2 |\PhT|}{z M_h} \,  H_1^{\perp q}(z, \phT^2)$~\cite{Trento}.
The outermost (blue) curves represent the positivity limits.
The error bands have the same meaning as in figure~\protect\ref{fig:Collins}.
The figure is taken from \cite{Anselmino:2008sj}.
}
\end{figure}
\begin{figure}[h]
\begin{center}
\includegraphics[height=0.48\columnwidth,angle=-90]{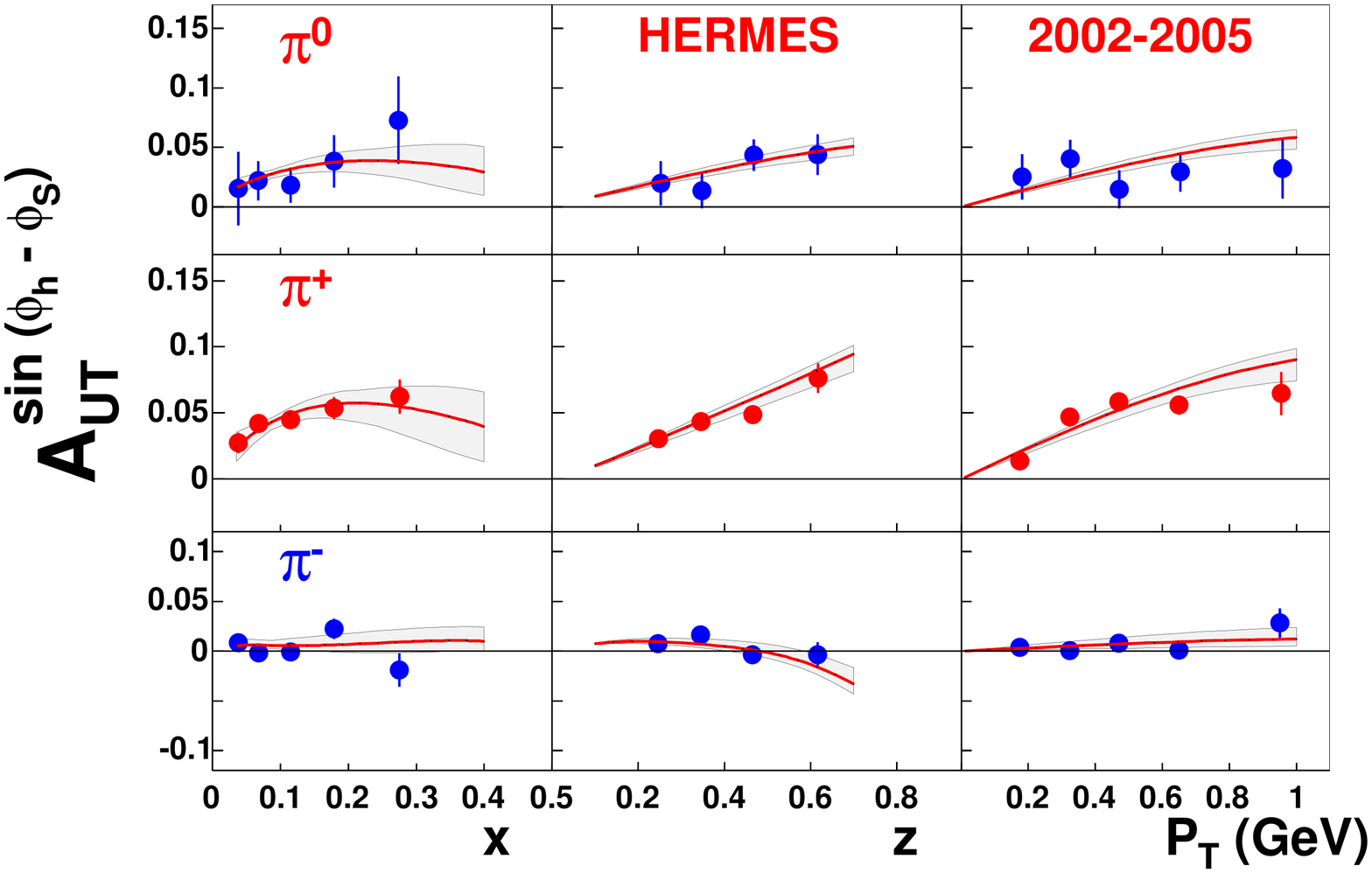}
\includegraphics[height=0.48\columnwidth,angle=-90]{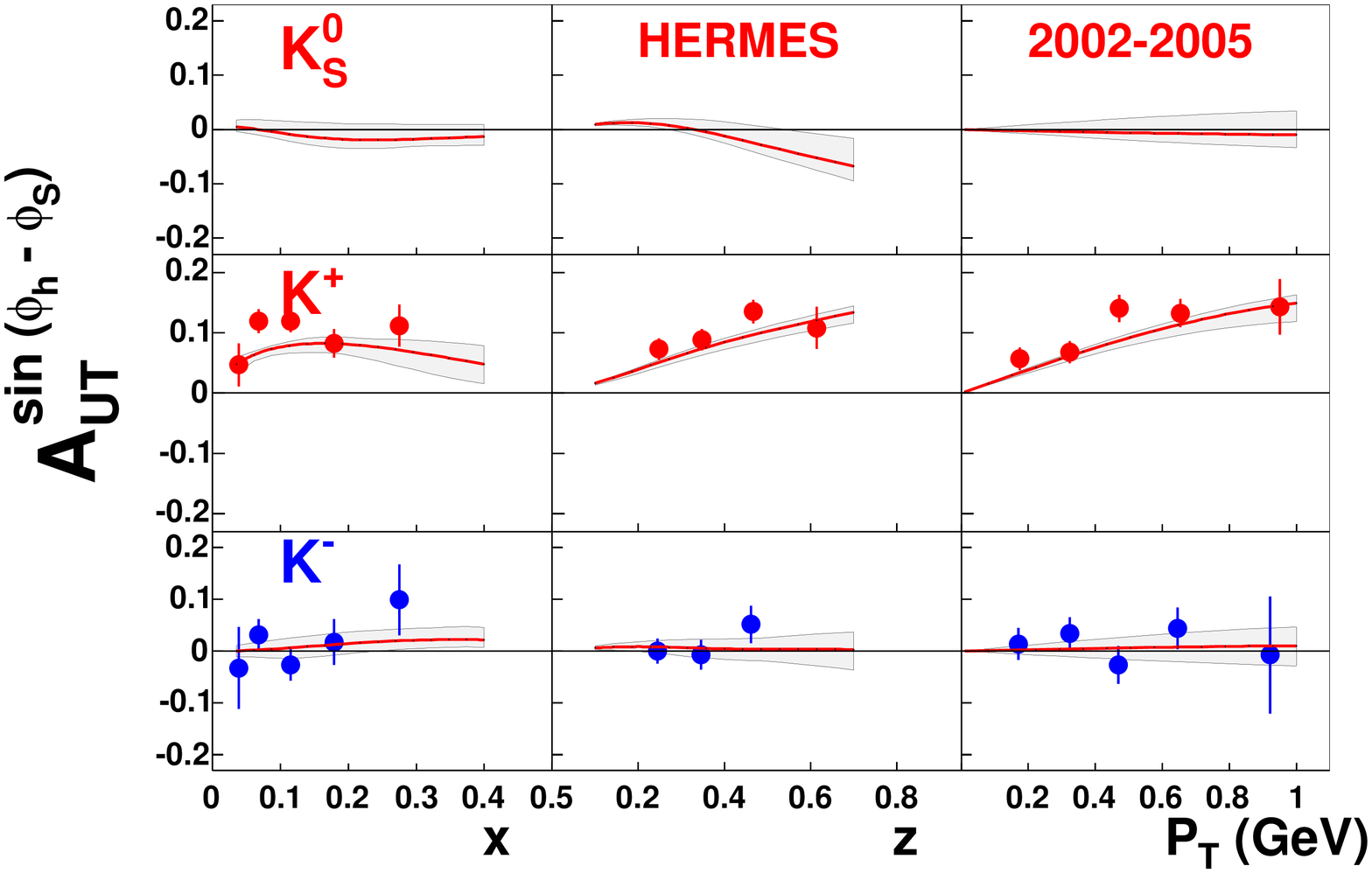}
\includegraphics[height=0.48\columnwidth,angle=-90]{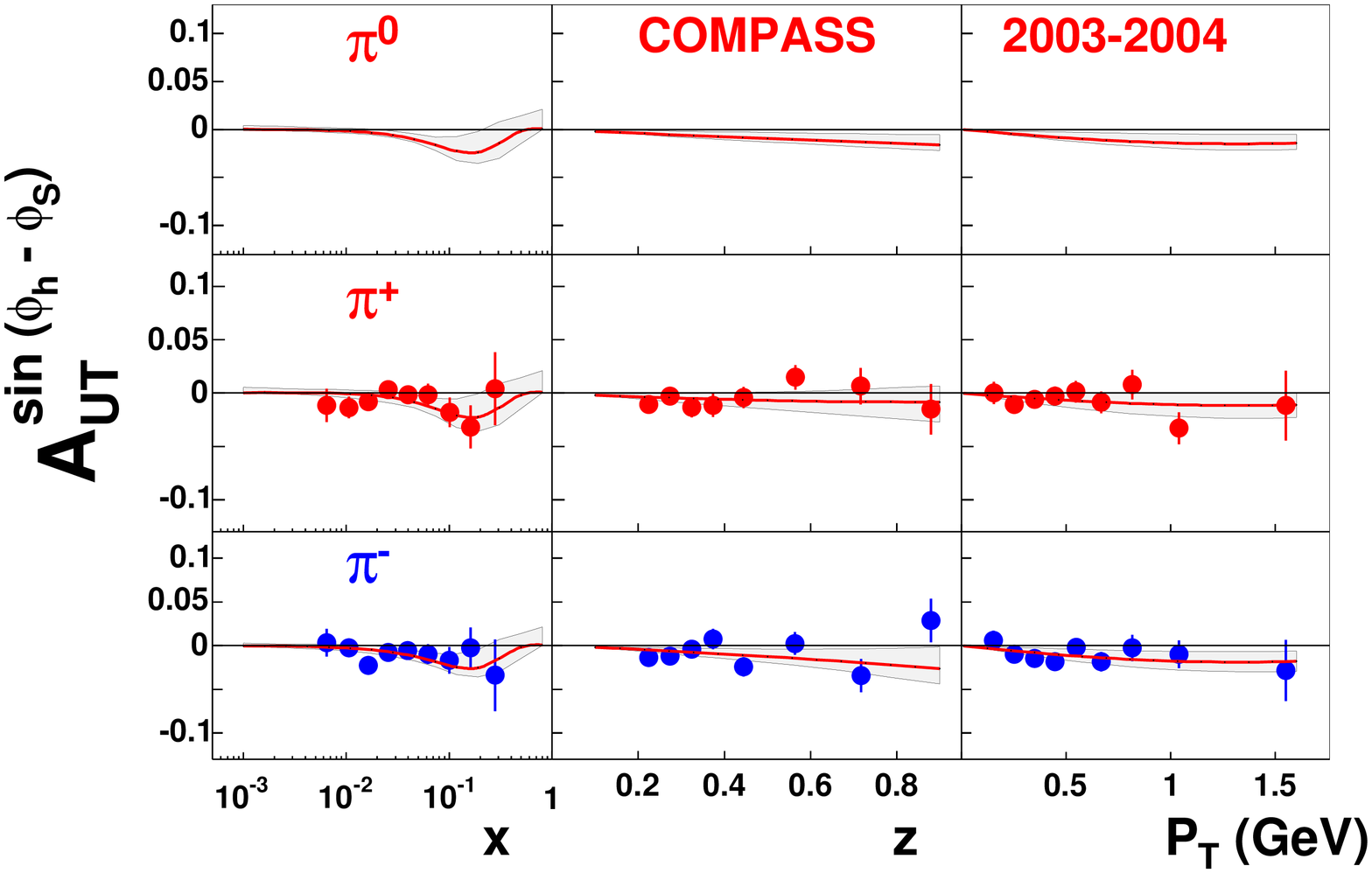}
\includegraphics[height=0.48\columnwidth,angle=-90]{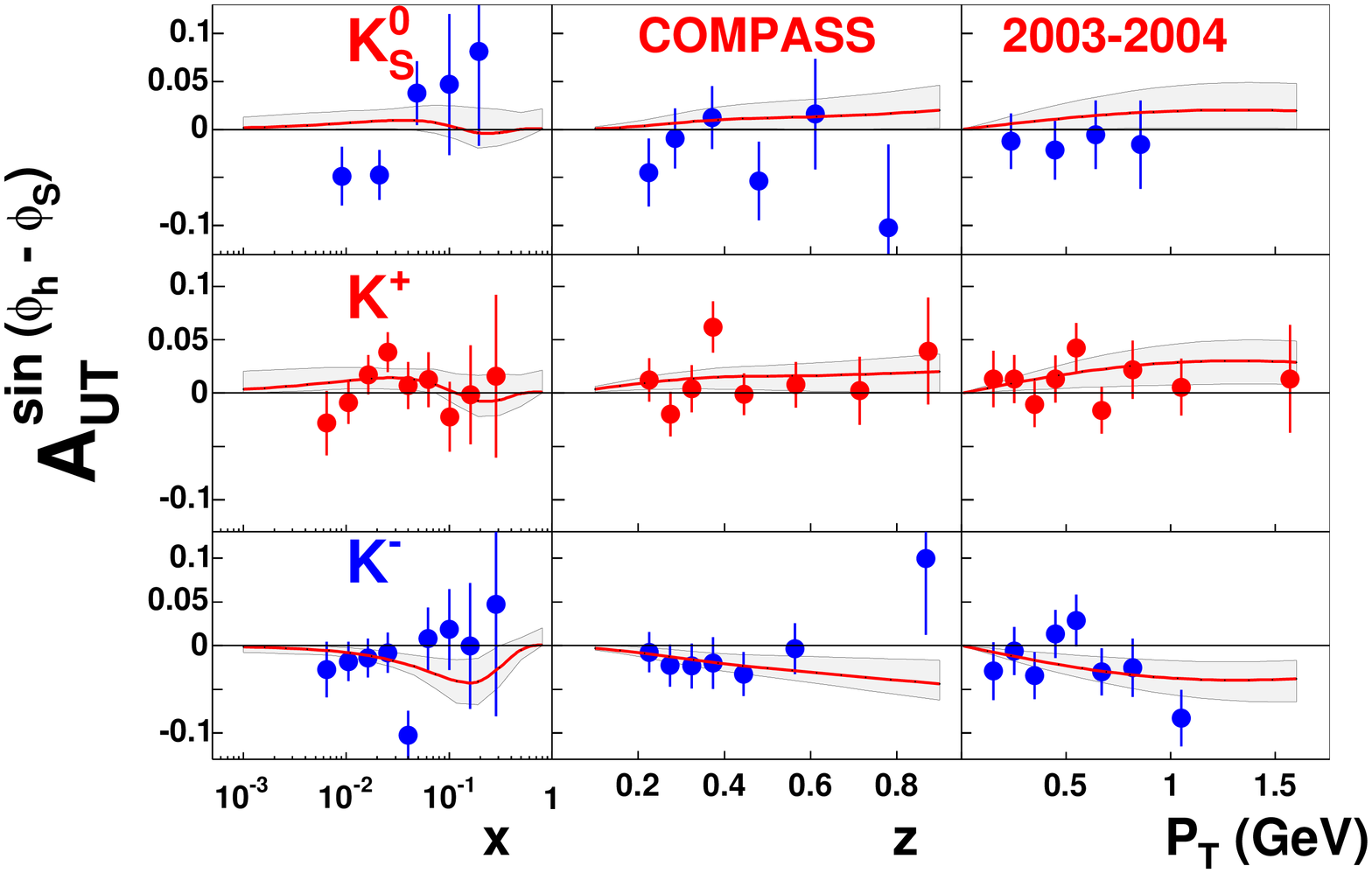}
\end{center}
\caption{\label{fig:Sivers} 
Sivers sinusoidal amplitude of the asymmetry with respect to transverse polarization
of a proton~\cite{hermes-AUT-2} (deuteron~\cite{Alekseev:2008dn}) target, 
for the semi-inclusive production by positrons (muons) of 
identified charged mesons as labelled in the upper (lower) panels, 
as a function of $x$, $z$ or $\phT$ (the latter appearing as $P_T$ in the figure).
The sign of the amplitudes is according to 
the Trento Conventions~\cite{Trento}.
The error bars represent the statistical uncertainties, while
for the proton target there is an additional common 8\% scale uncertainty.
The curves are fits to the data, as described in the 
text~\cite{Anselmino:2008sga}.
The error bands on these curves arise from the statistical uncertainties of the data,
but the procedure chosen implies that they are several times
larger than the conventional definition given by, e.g., the Particle Data Group~\cite{pdg}.
On the other hand, uncertainties related to model assumptions are not estimated.
The figure is taken from \cite{Anselmino:2008sga}.
}
\end{figure}
\def\cmh{\mbox{\large$\bm{\langle}$}\ensuremath{\sin(\phi+\phi_S)}
        \mbox{\large$\bm{\rangle}$}\ensuremath{_{UT}^h}}
\def\smh{\mbox{\large$\bm{\langle}$}\ensuremath{\sin(\phi-\phi_S)}
        \mbox{\large$\bm{\rangle}$}\ensuremath{_{UT}^h}}
The first such data
were reported a few years ago by \hermes~\cite{hermes-AUT-1}.
The cross section asymmetry with respect to
target polarization was extracted as a two-dimensional distribution 
in $\phi$ versus $\phi_S$.
The Collins azimuthal moment $\cmh$ and Sivers moment $\smh$
of the virtual-photon asymmetry were extracted in the fit
\begin{equation}
\hspace*{-1cm}
\frac{A^h_{UT}(\phi,\phi_S)}{2} = \cmh \sin(\phi+\phi_S) 
+ \smh \sin(\phi-\phi_S)\,.
\end{equation}
This simultaneous extraction of both 
contributions was shown by detailed Monte Carlo
simulations to avoid significant cross-contamination, even when they have
very different magnitudes in the context of a limited detector acceptance.  
These initial data on the proton target provided evidence for both 
Collins and Sivers signals.  
On the other hand, asymmetries measured by \compass\ using 
higher energy muons on a deuteron target are consistent with 
zero~\cite{Alexakhin:2005iw}.
As illustrated in figures~\ref{fig:Collins} and \ref{fig:Sivers}, 
the nonzero results for the proton are now firmly established by 
recent more precise preliminary data~\cite{hermes-AUT-2},
while the most recent results for
the deuteron remain consistent with zero~\cite{Alekseev:2008dn}.
For the proton target, the Collins azimuthal amplitude for $\pi^+$ 
is clearly positive and non-zero, 
while it is negative and at least as large for $\pi^-$.  The latter 
finding was a surprise because of the contrast with the double-spin
longitudinal asymmetries shown in figure~\ref{fig:deFlorian1}.
In that case asymmetries for pions of both charges are positive,
even though $\Delta u$ is positive and $\Delta d$ is negative, albeit
smaller in magnitude.  The reason there is that both charges are produced
mostly from photoabsorption by $u$ quarks~\cite{hermesdeltaq},
but not as dominantly for $\pi^-$.  Lattice calculations and models 
predict that transversity densities resemble the helicity densities,
at least at large $\x$ where valence quarks dominate, and where
the  $\pi^+/\pi^-$ contrast is most apparent in figure~\ref{fig:Collins}.
Hence efforts to find an explanation focused on the Collins
fragmentation function.

The explanation that emerged for the larger negative $\pi^-$ amplitudes is
a substantial magnitude for the disfavoured 
Collins function describing {\em e.g.} the fragmentation of 
$u$ quarks to $\pi^-$ mesons, and with a sign opposite to that of the 
favoured function.  Opposite signs could be expected in the light of the 
string model of fragmentation discussed in section~\ref{sec:Artru}. 
If a favoured pion forms as the string end created by
the first break, a disfavoured pion from the next break will
inherit transverse momentum from the first break in the opposite
direction from that acquired by the first pion.  Such a $\PpiT$ 
anticorrelation between favoured and disfavoured pions is demonstrated
by the \jetset\ simulation~\cite{Sjostrand:1993yb}, which is based on a string
fragmentation model.  Hence any correlation between $\PpiT$ and 
another kinematic or spin observable should have the opposite
sign for favoured and disfavoured pions.

This expectation for the Collins function has recently been confirmed
by results for both favoured and disfavoured functions from a simultaneous fit
of these DIS data together with information about the azimuthal structure
of jets in $e^+e^-$ collisions from the \belle\ collaboration~\cite{Belle}.
Two alternative jet asymmetries extracted from the \belle\ data are shown in
figure~\ref{fig:Belle}.
The most recent such fit~\cite{Anselmino:2008sj} is shown as the
curves in figures~\ref{fig:Collins} and \ref{fig:Belle}.  Such fits 
of data from these three experiments produced the first experimental 
determination of the transversity distributions of $u$ and $d$ quarks, 
the most recent example of which is shown in figure~\ref{fig:transversity}.  
At this stage, the quark sea is neglected.  As is the case for the helicity 
distributions, transversity is found to be positive for $u$ quarks
and negative and smaller in magnitude for $d$ quarks.  For both
flavours, the magnitudes are of order half of the Soffer positivity 
bound~\cite{Soffer:1994ww}.  Especially for $u$ quarks and for larger
values of $x$ (bearing in mind that there are no data for $x>0.3$), 
the extracted density is smaller than existing model 
predictions~\cite{Wakamatsu:2000fd,Schweitzer:2001sr,Pasquini:2005dk}. 
This fit also yields values for
the Collins fragmentation function, shown in figure~\ref{fig:collins}.
They confirm that indeed the disfavoured Collins function is opposite in
sign and larger in magnitude than the favoured function.
The initially surprising features of the Collins asymmetry data 
now appear to be well understood.  A further confirmation of this
recently appeared in the form of preliminary single-spin asymmetry data
for unidentified hadrons from the proton by the \compass\ 
collaboration~\cite{Levorato:2008tv}.
These data for the Collins asymmetry are consistent with predictions 
based on this fit of earlier data from \hermes\ on the proton and from
\compass\ on the deuteron.

\begin{figure}[h]
\begin{center}
\includegraphics[width=0.75\columnwidth]{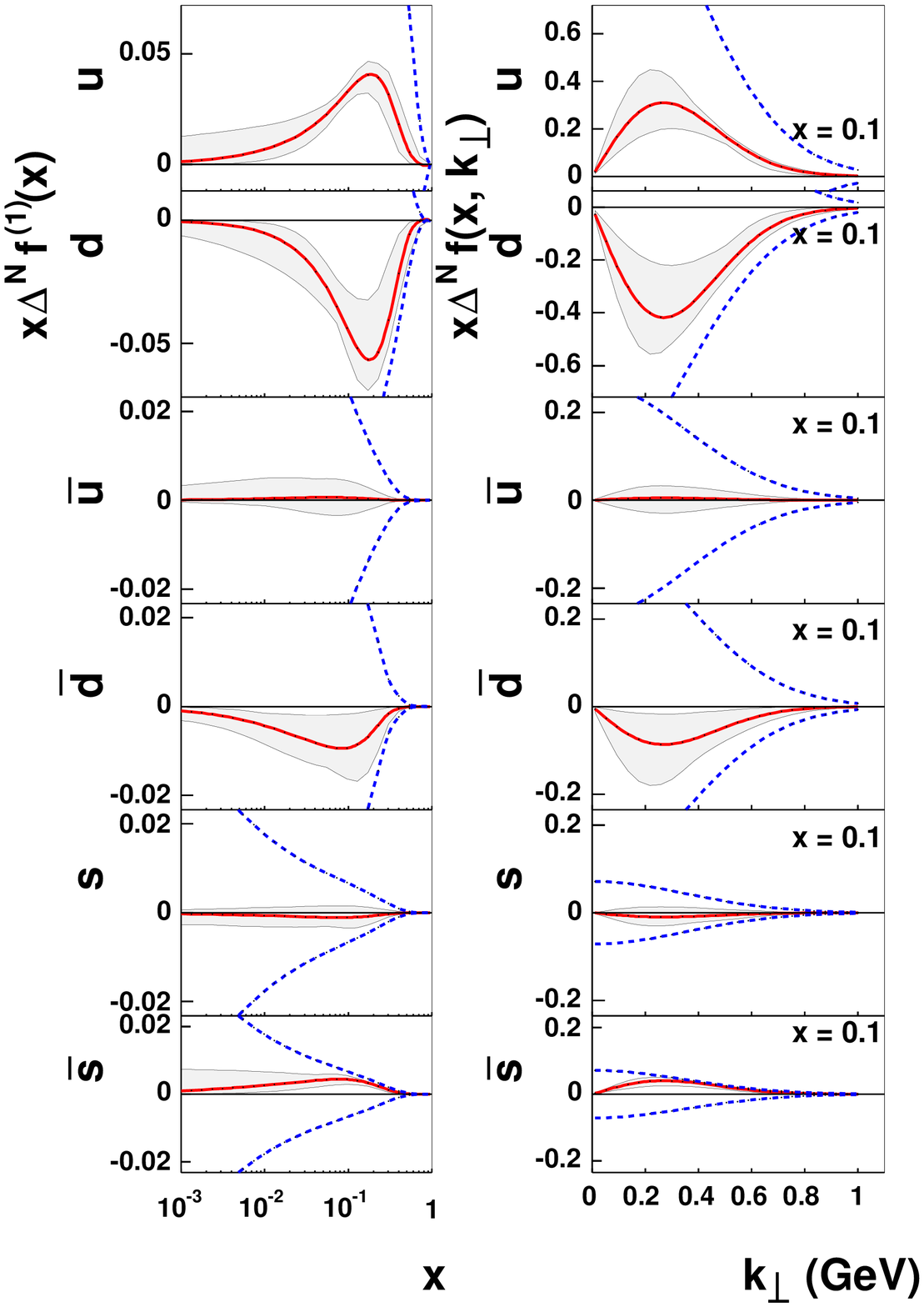} 
\end{center}
\caption{\label{fig:Mauro-Sivers} 
The left [right] panels show as continuous (red) curves within the 
error bands values of the 
first $\kT$ [$x$] moment of the Sivers function as functions
of $x$ [$\kT$, appearing in the figure as $k_{\perp}$],
from a fit~\cite{Anselmino:2008sga} to the 
data shown in figure~\protect\ref{fig:Sivers}.
The Sivers function is represented in the Torino notation, which in this case is 
opposite in sign and equal in magnitude to that in the Amsterdam notation 
used elsewhere in this paper.
The error bands have the same meaning as in figure~\protect\ref{fig:Sivers}.
The figure is taken from \cite{Anselmino:2008sga}.
}
\end{figure}
The Sivers azimuthal amplitudes for production of both pions and kaons from the proton
shown in the upper
panels of figure~\ref{fig:Sivers} are clearly positive and nonzero 
for the positive charges.  This constitutes the first evidence from 
leptoproduction for a nonzero T-odd parton distribution function.
We note here that these Sivers amplitudes receive no contributions 
that are suppressed by $1/Q$ (twist-3)~\cite{Bacchetta:2006tn}.   
Since the production of both $\pi^+$ and $K^+$ is dominated by $u$ quarks,
this positive value with the definition of azimuthal angles used here
implies a negative value for the Sivers function $f_{1T}^{\perp}$ of this 
flavour~\cite{Trento}.
The amplitudes for the negatively charged mesons are consistent with zero.
Their interpretation is complicated by competition between fragmentation
of quarks of at least two flavours. 
Remarkably, for the proton target, the $K^+$ amplitudes are on average much larger 
than those for $\pi^+$, even though one might suppose that they are similarly dominated by 
photoabsorption by $u$ quarks.
The amplitudes for the deuteron target measured by \compass\ are all consistent with zero 
(as are their preliminary Sivers amplitudes on the proton for unidentified 
hadrons~\cite{Levorato:2008tv}).

The curves on these figures from \cite{Anselmino:2008sga} represent 
an 11-parameter fit to data for pions and kaons from  \hermes\ on
the proton and from \compass\ on the 
deuteron, extracting representations of the Sivers functions for 
all six quark flavours.  The shapes of all four sea flavours \ubar, \dbar, $s$ and \sbar\ 
were constrained to be the same.  The success in explaining the large
$K^+$ amplitudes for the proton depends on the use of
a recent fit of unpolarized fragmentation functions~\cite{DSS}, 
which differs from all previous such fits in that 
$D_{\bar s}^{K^+}(z) \gg D_{u}^{K^+}(z)$ over the whole $z$ range.
This feature implies that $K^+$ production need not be dominated
by $u$ quarks, allowing \sbar\ quark distributions to be probed. 

Values of the Sivers functions resulting from the fit are shown in 
figure~\ref{fig:Mauro-Sivers}.
The Sivers function for $d$ quarks
is found to be at least as large in magnitude as that for $u$ quarks.
Their signs agree with expectations based on an attractive 
final-state interaction between the struck quark and spectator diquark
(see section~\ref{sec:SSA}).  Such a relatively large magnitude 
of the Sivers function for $d$ quarks could be explained by a
similar trend in the known relative contributions of $u$ and $d$ quarks to
the anomalous magnetic moment of the proton, as explained in 
section~\ref{subsec:IPDPD}.  Hence the data 
appear to support a profound connection between static and partonic 
properties of the proton.  Furthermore, there is evidence for a 
significant role of sea quarks.  In considering the statistical
significance of this finding, it is useful to bear in mind that
the method of assigning the error bands results in values much larger 
than those that would correspond to the definition of \eg, the Particle 
Data Group~\cite{pdg}. On the other hand, uncertainties related to model 
assumptions are not estimated.

We now look forward to measurements
of single-spin asymmetries for the Drell-Yan process, from which values
of the Sivers function can be extracted that are predicted on the basis
of fundamental properties of QCD to be equal
in magnitude but opposite in sign to these from DIS, as in (\ref{SIDISDY}).  
This will test
our basic understanding of T-odd parton distribution functions within
the context of QCD.  Thus continues the exciting story of a feature
of proton structure that only a few years ago was (incorrectly) asserted to be
forbidden by fundamental time reversal invariance~\cite{Collins:2002kn}.

%
%
\section{Towards a 3-dimensional picture of the nucleon}
\label{sec:GPDs}
%
\subsection{Generalized parton distributions}
\begin{figure}
\unitlength1.cm
\begin{center}
\epsfig{file=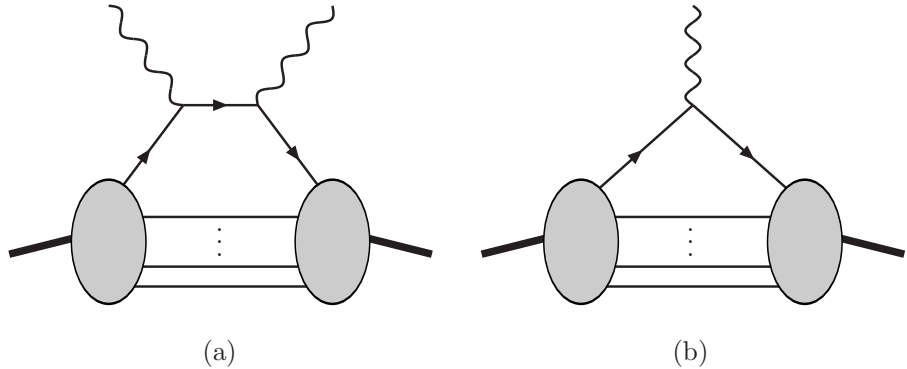,width=12.cm}
\end{center}
\caption{\label{fig:comptFF} Comparison between a) the Compton amplitude with two photon
vertices and a quark propagating between them and b)
a form factor with just one photon vertex.}
\end{figure}

%
\label{subsec:GPD-GPDs}
In section \ref{sec:Intro} it was explained that
factorization can be applied to the theoretical interpretation 
of DIS cross sections. Using the optical theorem, the inclusive DIS
cross section can be related to the imaginary part of the
forward Compton amplitude, which can also be factorized into
a `hard part' describing the vertices of the active quark with
the photons and the quark propagating between these two 
vertices, and a soft part describing the quark correlation 
function in the hadron (see figure \ref{fig:comptFF}a). 
A similar factorization of the scattering amplitude 
occurs in Deeply Virtual Compton Scattering (DVCS) 
\cite{Collins:1998be}, 
$\gamma^*p\longrightarrow \gamma p$,
where $\gamma^*$ is a space-like photon with virtuality
$q^2=-Q^2<0$. 
As in DIS, the process is best interpreted in a frame in which the
nucleon has infinite momentum opposite to that of the virtual photon.
In the Bjorken limit, and for a momentum transfer $t$ to the
proton much less than $Q^2$, the DVCS amplitude factorizes
into a convolution of the Compton amplitude for scattering
off a quark constituting the hard part, and a quark correlation 
function, constituting the soft part of the amplitude.
After factorization has been applied to the DVCS amplitude, its
matrix element has much in common with the matrix element describing
the form factor (figure \ref{fig:comptFF}b). 
Indeed, the soft part
entering the DVCS amplitude becomes identical to the soft part
entering the form factor. However, the DVCS process provides 
more `surgical' access.
In the case of the DVCS amplitude, the hard part has a dependence
on the quark momentum fraction appearing in the quark propagator,
which is not present in the form factor. 
The additional kinematic degree of freedom provided by the
hard part in the DVCS process provides
access to the dependence of the soft part on the momentum fraction 
$x$ of the active quark. 
This motivates the concept of Generalized
Parton Distributions (GPDs) as a description of the soft
part that enters the DVCS amplitude \cite{Ji:1996nm}. 
What is gained by
introducing this concept is that universality as introduced
in section \ref{sec:Intro} applies with an extended scope, 
now also encompassing form factors and hard exclusive processes
--- the same GPDs that describe
elastic form factors as well as the soft part of DVCS should also 
appear in other hard exclusive reactions. In each of these reactions,
they appear convoluted with the appropriate coefficient
function representing the hard part. In section 
\ref{subsec:GPD_Access}ff, it is
discussed how and to what extent exclusive reactions
can be used to constrain GPDs experimentally.

In elastic scattering, the photon can couple to any of the quark
lines connecting the two `soft blobs' in figure \ref{fig:comptFF}b,
leaving the nucleon intact.
The form factor 
is the coherent superposition of such diagrams for all the 
quarks of all flavors in the nucleon.
The experimental decomposition of the nucleon form factor into 
the contributions from various quark flavors has been partially 
accomplished through the
comparison of proton and neutron form factors using isospin symmetry,
and the interpretation of parity-violating electron scattering. 
Two flavor combinations $(2u-d-s)$ and $2(d-u-s)$ 
are obtained from the proton and neutron electromagnetic
form factors using charge symmetry. A further combination is
accessible in parity violating electron scattering, but the
corresponding constraints are not very strong yet \cite{Diehl:2007uc}.

However, what is still needed for a more comprehensive understanding
of the microscopic physics that underlies form factors is
their decomposition with respect to the momentum fraction of the
quark that absorbs the photon. Generalized parton distributions 
(GPDs) embody this 
information. By definition, they provide a decomposition of form 
factors evaluated at a given value of the invariant momentum transfer
$t=(p_{fin}-p_{ini})^2$ on the target, with respect to the average 
momentum fraction $x = \frac{1}{2}\left(x_{ini}+x_{fin}\right)$ 
of the active quark of flavor $f$:
\bea
\int\!\!\de x\, H^f(x,\xi,t) &=& F^f_1(t)
\quad \quad \int\!\!\de x\, E^f(x,\xi,t) = F^f_2(t)
\label{eq:GPD1}\\
\int\!\!\de x\, \widetilde{H}^f(x,\xi,t) &=& G^f_A(t)
\quad \quad 
\int\!\!\de x\, \widetilde{E}^f(x,\xi,t) = G^f_P(t)
\label{eq:GPD2} ,
\eea
where $x_{ini}$ and $x_{fin}$ are the 
initial and final momentum fractions of the active quark
with respect to 
the average momentum $\frac{1}{2}\left(p_{ini}+p_{fin}\right)$,
and the dimensionless `skewness' 
$\xi=\frac{1}{2}\left(x_{ini}-x_{fin}\right)$ 
represents their difference. 
Thus the linear combinations
$x\pm\xi$ represent the momentum fraction carried by the quark
before/after the hard interactions.
The functions
$F^f_1(t)$, $F^f_2(t)$, $G^f_A(t)$, and $G^f_P(t)$ are the 
contributions to the Dirac, Pauli, axial, and pseudoscalar form 
factors from quark flavor $f$, respectively, and the
functions $H^f$, $E^f$, $\widetilde{H}^f$, and $\widetilde{E}^f$
are the quark GPDs providing their momentum decomposition.
One can thus imagine the form factor being the result of adding (or
integrating over) the contributions from all possible quark momentum
fractions $x$ (here $x$ is not to be identified with
the Bjorken variable $\xbj$ defined in
section 1). 

Just like the ordinary (forward) PDFs, GPDs are
subject to higher-order corrections in $\alpha_s$ and are thus also
dependent on the scale $Q^2$ (and implicitly also
on the factorization scheme used) at which they are probed.
Nevertheless, as $F_1^f(t)$ and $F_2^f(t)$ are defined as
matrix elements of a conserved current, they are $Q^2$ independent
and hence the $Q^2$ dependence must disappear in the lowest moment
of the corresponding GPDs (\ref{eq:GPD1}).

\begin{figure}
\begin{center}
\epsfig{file=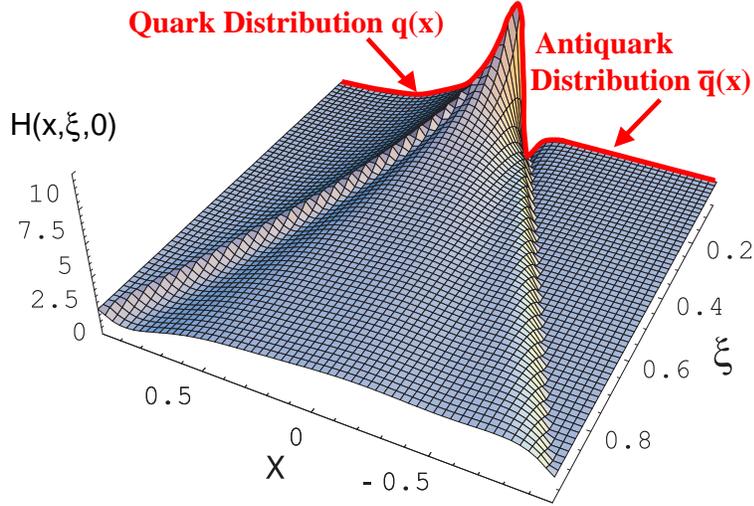,width=10cm}
\end{center}
\caption{\label{fig:vdh}
Model calculation for $H^q(x,\xi,0)$ from \cite{Goeke:2001tz}.
The thick (red) line represents the forward limit $\xi=0$. Accessible to the
DVCS process is the `outer' or `DGLAP' region $|x|>\xi$ that represents
quarks for $x>\xi$ and antiquarks for $x<-\xi$. In the `inner' or `ERBL'
region $|x|<\xi$, the quark GPD behaves like a meson distribution
amplitude in deeply virtual meson production. The figure is taken from 
\cite{Goeke:2001tz}.}
\end{figure}

Decomposing the form factors
with respect to the quark momentum fraction $x$ 
involves the infinite-momentum direction. Hence it can
make a difference whether the momentum transfer to the
target is parallel, or perpendicular to this momentum. The GPDs must 
therefore depend on an additional variable, here
called $\xi$. Its relationship to $t/M^2$ characterizes
the direction of the momentum transfer relative to the 
infinite momentum direction.  
The case $\xi=0$ corresponds to the 
case where those two momenta are perpendicular to each other (which 
is not directly experimentally accessible in DVCS), 
while nonzero $\xi$
corresponds to the situation where the momentum transfer to the
target has a component parallel to the virtual photon direction.
Very little is known about the full $x$ vs. $\xi$ dependence
of GPDs, except that they
might possibly be quite complex, as indicated for example by 
the model calculation shown in figure \ref{fig:vdh}.
In (\ref{eq:GPD1}) and (\ref{eq:GPD2}), the integration over $x$
removes all reference to the infinite momentum direction, with respect
to which $x$ and $\xi$ are defined. Lorentz invariance thus implies
that the $\xi$ dependence disappears in these integrals.

One constraint on quark GPDs comes from the fact that in the limit
of vanishing $t$ and $\xi$, the nucleon helicity non-flip quark
GPDs must reduce to the respective PDFs:
\bea
H^f(x,0,0)=q_f(x), & ~~\widetilde{H}^f(x,0,0)=\Delta q_f(x) & 
{\rm ~~~for~ } x>0 \\
H^f(x,0,0)= -\bar{q}_f(-x), & ~~\widetilde{H}^f(x,0,0)=
\Delta \bar{q}_f(-x) & {\rm ~~~for~} x<0.
\label{eq:GPDtoPDF}
\eea

GPDs became the focus of intense interest after the discovery of the
`Ji-relation', connecting a second moment of GPDs with the total angular 
momentum of quarks with flavor $f$ or gluons, which is the respective 
contribution to the nucleon spin $\frac{1}{2}$ \cite{Ji:1996ek}:
\be
\frac{1}{2} = \sum_f J_f + J_\g,
\ee
where
\bea
J_{f} = \frac{1}{2} \lim_{t \rightarrow 0} \int_{-1}^1\!\!\de x\,x
\left[H^{f}(x,\xi,t)+E^{f}(x,\xi,t)\right], 
\label{eq:Ji_SumRule}\\
J_{\g} = \frac{1}{2} \lim_{t \rightarrow 0} \int_0^1\!\!\de x
\left[H^{\g}(x,\xi,t)+E^{\g}(x,\xi,t)\right] .
\label{eq:Ji_SumRuleGlue}
\eea
Note that no $\xi$-dependence remains in (\ref{eq:Ji_SumRule}) and
(\ref{eq:Ji_SumRuleGlue}) after integration over
$x$, although the two unpolarized quark GPDs $H^f$ and $E^f$ have to be 
evaluated at some common value of $\xi$. 
Unlike the lowest moments (\ref{eq:GPD1}), these second moments and hence 
$J_{f,G}$ depend on $Q^2$ (as well as order in $\alpha_s$ and factorization 
scheme), but this dependence is usually suppressed for simplicity.

Equation (\ref{eq:Ji_SumRule}) is usually applied to the
$\hat{z}$ component of the quark total angular momentum in a 
longitudinally polarized nucleon. 
However, in principle (\ref{eq:Ji_SumRule}) 
can be applied 
for any polarization direction of the nucleon target.
Nevertheless, (\ref{eq:Ji_SumRule}) finds its most important
application in the helicity sum rule through 
\be
J_f = \frac{1}{2}\left(\Delta q_f +\Delta \bar{q}_f \right) + L_f,
\label{eq:SpinDeco}
\ee
since in that case at least the
quark helicity contribution $\Delta q_f+ \Delta \bar{q}_f$ in 
$J_f$ has a partonic interpretation. (An attempt to associate
a partonic interpretation of (\ref{eq:Ji_SumRule}) 
in the case of transversely polarized nucleons can be found in 
\cite{Burkardt:2005hp}.)

The gauge invariant orbital angular momentum $L_f$ of
quarks is defined through the expectation value of 
$\psi_f^\dagger\,{\vec r}\times \left({\vec p}-g{\vec A}\right)\psi_f$.
No process has been identified to measure $L_f$ directly, but it
can be evaluated indirectly through (\ref{eq:SpinDeco}) as
\be 
L_f = J_f- \frac{1}{2}\left(\Delta q_f +\Delta \bar{q}_f
\right).
\ee
The issues arising in a decomposition of the nucleon spin are
further discussed in section \ref{sec:SpinBudget}.

%
\subsection{General properties of GPDs}
\label{subsec:GPD_Properties}
%
GPDs must satisfy several relations arising from fundamental symmetries 
or conservation laws:
\begin{itemize}
\item Invariance under time reversal (for both quark and gluon GPDs):
   \be
       F(x,-\xi,t) = F(x,\xi,t),
   \ee
\item Polynomiality property of GPDs~\cite{Polyakov:1999gs} as a 
      consequence of Lorentz invariance:
   \bea
   \label{eq:GPD-Poly_H}
       \!\!\!\!\!\!\!\!\!\!\!\!
\int_{-1}^1 \de x \, x^{n-1} H^f(x,\xi,t) = \sum_{\stackrel{k=0}{\rm even}}^{n-1} 
       (2\xi)^k A^f_{n,k}(t) + {\rm mod}(n-1,2)(2\xi)^n C^f_n(t), \\
   \label{eq:GPD-Poly_E}
       \!\!\!\!\!\!\!\!\!\!\!\!
\int_{-1}^1 \de x \, x^{n-1} E^f(x,\xi,t) = \sum_{\stackrel{k=0}{\rm even}}^{n-1} 
       (2\xi)^k B^f_{n,k}(t) - {\rm mod}(n-1,2)(2\xi)^n C^f_n(t).
   \eea
Polynomiality means that the entire $\xi$ dependence of the $n$'th $x$-moment 
$(n>0)$ of a GPD is given by a polynomial in $\xi$ of order at most $n$.  
Polynomiality is satisfied, \eg, by the representation of GPDs in terms of 
double distributions (DDs) $F^f_{DD}(\beta,\alpha,t)$~\cite{Mueller:1998fv,
Radyushkin:1996ru,Radyushkin:1997ki}, which can 
be considered as spectral functions of nonforward matrix elements.  In terms of
a DD, a generic quark GPD $F$ can always be written as:
\bea
\!\!\!\!\!\!\!\!\!\!\!\!\!\!\!\!\!\!\!\!\!\!\!\!\!\!\!\!\!
F^f(x,\xi,t)&=&\int_{-1}^1 d\beta \int_{-1+|\beta|}^{1-|\beta|} d\alpha \,
             \delta(x-\beta-\xi\alpha)\,F^f_{DD}(\beta,\alpha,t) \nonumber \\
            &\rule{2mm}{0mm}&+\theta(\xi-|x|) \; D^f(\frac{x}{\xi},t).
\label{eq:DDansatz1}
\eea
The step function $\theta(\zeta)$ is zero for $\zeta<0$ and unity
for $\zeta>0$.  Thus the so-called `D-term' $D^f$~\cite{Polyakov:1999gs}
contributes only in the `ERBL' region $-\xi<x<\xi$,
where quark GPDs have the character of distribution amplitudes for the
creation of a quark/antiquark pair.  It does not contribute in
the complementary so-called `DGLAP' region $|x|>\xi$, where quark GPDs
describe the emission and reabsorption of an (anti-)quark in
the infinite momentum frame, thereby having properties analogous
to the familiar (anti-)quark distribution functions.  The D-term
provides a convenient means of representing this profound
difference in GPD properties between the two regions, which would
otherwise require singularities in the DD having a severity
beyond represention by delta functions or their derivatives.
Another perspective is that the maximum power of $\xi$ provided by the DD
itself is one less than that allowed by Lorentz invariance. 
Expressions analogous 
to (\ref{eq:DDansatz1}) hold for the polarised quark GPDs 
$\widetilde{H}$ and $\widetilde{E}$, for which the D-term  vanishes.  
The mapping between GPDs and DDs is not one to one; each GPD can be
represented by any of an infinite family of DDs.
The quantity $C_n^f(t)$ in (\ref{eq:GPD-Poly_H},\ref{eq:GPD-Poly_E}) is the 
$n$'th $x$-moment of $D^f(\frac{x}{\xi},t)$, which  has the same magnitude 
for the quark GPDs $H^f$ and $E^f$ but enters with the opposite 
sign.  Hence the second moment of the sum of the unpolarized quark GPDs
$H^f(x,\xi,t)+E^f(x,\xi,t)$ (`Ji integral' (\ref{eq:Ji_SumRule})) is 
independent of $\xi$.
\end{itemize}

\noindent Several sum rules apply at $\xi=0$ and $t=0$~\cite{Diehl:2003ny}. 
Suppressing the kinematic arguments:
\begin{itemize}
\item Quark and gluon GPDs are related by the conservation of momentum,
   \be
      1 = \int_{-1}^1 \de x \, x \sum_f H^f + \int_0^1 \de x \, H^\g,
   \ee
      and of the proton's angular momentum,
   \be
      \frac{1}{2} = \frac{1}{2} \int_{-1}^1 \de x \, x \sum_f (H^f + E^f)
                  + \frac{1}{2} \int_0^1 \de x \, (H^\g + E^\g).
   \ee
      Combining them leads to:
   \be
      0 = \int_{-1}^1 \de x \, x \sum_f E^f + \int_0^1 \de x \, E^\g
   \ee
\item For each quark flavour, the GPD $E^f$ is constrained by the 
      contribution $\kappa_f$ of flavour $f$ to the proton anomalous 
      magnetic moment $\kappa$:
   \be
      \int_{-1}^1 \de x E^f = F_2^f(t=0) = \kappa_f
   \ee
\end{itemize}

\noindent As quark GPDs are not generally even or odd in $x$, it is often 
useful~\cite{Diehl:2003ny} to consider the GPD combinations
\bea
\label{eq:GPD-H-comb}
H^{f(\pm)}(x,\xi,t) \equiv H^f(x,\xi,t) \mp H^f(-x,\xi,t) \\
\label{eq:GPD-Htilde-comb}
\widetilde{H}^{f(\pm)}(x,\xi,t) \equiv 
     \widetilde{H}^f(x,\xi,t) \pm \widetilde{H}^f(-x,\xi,t),
\eea
and their analogues for $E^f$ and $\widetilde{E}^f$. Here the upper (lower) 
sign corresponds to the exchange of a $C=+1$ ($C=-1$) object in the $t$ 
channel, and denotes the `singlet' (`non-singlet' [or valence]) 
combination. These names are taken from the corresponding forward limits 
given by: 
\bea
\label{eq:GPD-PDF-H}
H^{f(\pm)}(x,0,0) = q_f(x) \pm \bar{q}_f(x) \\
\label{eq:GPD-PDF-Htilde}
\widetilde{H}^{f(\pm)}(x,0,0) = \Delta q_f(x) \pm \Delta \bar{q}_f(x),
\eea
for $x>0$. While the singlet (C-even) combination of quark GPDs is probed 
in DVCS (see (\ref{eq:CFF1a})) or exclusive vector meson production, the 
non-singlet (C-odd) one is selected in pseudoscalar meson production. 

Gluons are their own antiparticles so that gluon GPDs are either even in 
$x$ ($H^\g, E^\g$) or odd ($\widetilde{H}^\g,\widetilde{E}^\g$). Their forward 
limits involve an additional factor of $x$, so that (for $x>0$):
\be
H^\g(x,0,0)=x \g(x), \;\;\;\;\;\;\; \widetilde{H}^\g(x,0,0)=x \Dg(x).  
\ee
%
%

%
\subsection{Impact-parameter-dependent parton distributions}
\begin{figure}
\begin{center}
\epsfig{file=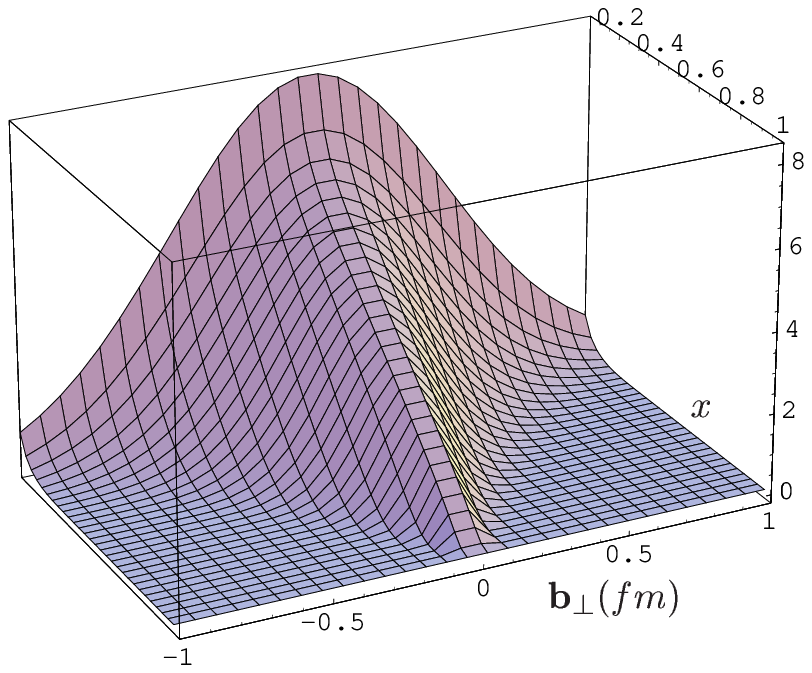,width=10cm}
\end{center}
\caption{\label{fig:ipdpd} Qualitative shape of $q(x,\bT)$.
}
\end{figure}

%
\label{subsec:IPDPD}
A particularly simple physical interpretation for GPDs can be
developed in the limiting case $\xi=0$, where the quark momentum
fraction is the same in the initial and final state.
So far we have introduced GPDs as momentum-disected form factors.
Hence a GPD evaluated at a certain value of $x$ can 
be considered to 
be the form factor for only those quarks in the nucleon that carry 
momentum fraction $x$. In analogy with nonrelativistic
form factors, whose Fourier transform provides information on the
charge distribution in position space, one would thus expect
that the Fourier transform of the $t$ dependence of GPDs for 
fixed $x$ provides information
on the position space distribution of quarks carrying a certain
momentum fraction $x$. Indeed, it can be shown that the
distribution of partons in impact parameter space can be
obtained from GPDs via \cite{Burkardt:2000za}:
\bea
q_f(x, \bT) &=& \int \frac{{\rm d}^2\DT}{(2\pi)^2 }
e^{-i\DT\cdot \bT} H^f(x,0, - \DT^2),
\label{IPD}
\eea
where $-\DT^2 = t$ for $\xi = 0$. Analogously,
the parton helicity distribution in a longitudinally polarized target
is obtained by replacing $q_f \rightarrow \Delta q_f$ and 
$H^f \rightarrow \widetilde{H}^f$.
Note that the Heisenberg uncertainty principle
forbids a simultaneous exact measurement of $x$ and the longitudinal
({\it i.e.} along the direction of the virtual photon)
position of the quarks \cite{Belitsky:2003nz}. 
Therefore, only the impact parameter 
$\bT$, {\it i.e.} the position transverse to the nucleon's momentum
can be determined simultaneously with $x$. For the determination 
of the $\bT$ distribution, the momentum transfer should be purely
perpendicular, which implies the restriction to $\xi=0$.
The reference point for the impact parameter $\bT$ is the transverse
center of longitudinal momentum $\RT$, which plays a role similar
to the center of mass in a nonrelativistic system. In analogy to
that case, it is obtained as an average of the
positions ${\bm r}_{\perp,i}$ of all the partons, 
except that the weight factor 
is not the mass fraction, but the momentum fraction $x_i$
carried by each parton \cite{Soper:1972xc}:
\be
\RT =  \sum_{i= q,g} x_i\, {\bm r}_{\perp,i}.
\label{eq:Rperp}
\ee
For $x\rightarrow 1$, the active quark dominates the sum in 
(\ref{eq:Rperp}) and $q(x,\bT)$ must necessarily become narrow 
in this limit (see figure \ref{fig:ipdpd}). 

In analogy to the inclusive PDFs that are measured in DIS, given
that we have restricted ourselves to $\xi=0$, 
these Impact-Parameter-Dependent Parton Distributions
(IPD-PDFs or simply IPDs) 
have a probabilistic interpretation and hence they
satisfy a variety of inequalities, such as 
\cite{Burkardt:2001ni,Pobylitsa:2002iu}:
\bea
q_f(x,\bT) &\geq& \left| \Delta q_f(x,\bT) \right|  \geq 0
\quad \mbox{for}
\quad x>0, \\
-q_f(x,\bT) &\geq& \left| \Delta q_f(x,\bT) \right|  \geq 0
\quad \mbox{for}
\quad x<0 \nonumber.
\eea

Remarkably, (\ref{IPD}) is 
not limited to the nonrelativistic domain of $|t|<m^2$, in contrast
to the interpretation of form factors  as {\it three}-dimensional
Fourier transforms of charge distributions. Its only limitation is
$|t|\ll Q^2$ as $1/Q$ represents the spatial
resolution of the virtual photon.
Since the electromagnetic form factors are obtained as integrals
of GPDs (see (\ref{eq:GPD1}) and (\ref{eq:GPD2})), 
this feature of (\ref{IPD}) implies that in the 
infinite momentum frame, the identification of the 
{\it two}-dimensional Fourier transform of 
form factors with the charge distribution in the transverse plane
is also not limited to $|t|<m^2$. 

For a longitudinally polarized or unpolarized target, the distribution
of quarks in the transverse plane is axially symmetric. However,
when the target is {\it transversely} polarized, the polarization
direction of the quark in the transverse plane breaks this axial
symmetry. For example, for a target polarized in the ${\hat {\bm x}}$ 
direction, the quark distribution $q_f(x,\bT)$
is deformed along the ${\hat {\bm y}}$ direction and the 
sign and magnitude of this deformation are embodied in the GPD 
$E^f(x,0,-\DT^2)$ \cite{Burkardt:2002hr}:
\bea 
\!\!\!\!\!\!\!\!\!\!\!\!\!\!\!\!\!\!\!\!
\label{deformation}
\label{deform}
&&q_f(x,\bT)=\\
\!\!\!\!\!\!\!\!\!\!\!\!\!\!\!
&& \int \frac{\de^2\DT}{(2\pi)^2 }
e^{-i\DT\cdot \bT} H^f(x,0, - \DT^2)+
\frac{1}{2M}\frac{\partial}{\partial b_y}
\int \frac{\de^2\DT}{(2\pi)^2 }
e^{-i\DT\cdot \bT} E^f(x,0, - \DT^2).
\nonumber
\eea
Such deformations can be interpreted semi-classically as
an enhanced probability of photoabsorption occuring on
the left or right of the nucleon as seen by the incident photon,
when the nucleon spin is pointing \eg\ up.
As was explained in section~\ref{sec:sivers}, this suggests
an intimate relationship between these deformations,
parton orbital angular momentum, and the Sivers effect.

Experimentally, only the $x$-integral of the  
GPD $E^f(x,0,-\DT^2)$, i.e. the form factor
$F_2(-\DT^2)$, is known (see (\ref{eq:GPD1})).
Any additional constraints come only from lattice calculations,
which will be discussed in the following section, as well as
model studies.
As a result, the details of the above deformation are not very well
known. In the considered case of a transversely
polarized target, this deformation implies the
existence of a transverse flavor dipole moment. 
The integrated deformation can
be related model-independently to the anomalous magnetic moment
of the nucleon.
For example, when the target is polarized in the ${\bm {\hat x}}$
(transverse)
direction, the average deformation ${\cal D}_f$ of the distribution 
for quarks of flavor $f$ in the ${\bm {\hat y}}$ direction reads
\cite{Burkardt:2002hr}:
\be
{\cal D}_f\equiv \int_{-1}^1\!\!\de x \int \!\!\de^2\bT \,
q_f(x,\bT) \, {\rm b}_y = \frac{\kappa_{f/p}}{2M}.
\ee
The $\kappa_{f/p}$ are the contributions from quark flavor $f$
to the anomalous magnetic moment of the proton with the
quark charge factored out:
$\kappa_p = \frac{2}{3} \kappa_{u/p} -
\frac{1}{3} \kappa_{d/p}+ \ldots\approx 1.79$. 
A flavor decomposition of the
anomalous magnetic moment, using $\kappa_n\approx -1.91$, 
charge symmetry, and
neglecting the small contribution from $s$ quarks yields 
$\kappa_{u/p} \approx 1.67$ and $\kappa_{d/p}\approx-2.03$,
which translates into average deformations of 
${\cal D}_u\approx +0.16$ fm and ${\cal D}_d\approx -0.20$ fm. 

Such a significant transverse deformation of quark distributions
should have observable consequences, {\it e.g.} in SIDIS on a 
transversely polarized target. 
Armed with the above quantitative information, the semiclassical
arguments of section~\ref{sec:sivers} about the transverse 
deformation of PDFs as a cause of single-spin asymmetries
can be made more specific.
When a virtual photon collides with
a proton that has transverse spin down, 
it `sees' more $u$ quarks on the right side of the proton and 
more $d$ quarks
on the left side, with the signs being determined by the signs
of $\kappa_{q/p}$ (see figure \ref{fig:gpdssa}).
Since the final-state interaction  acting
on the escaping quark is on average expected to be attractive,
more ejected $u$ 
quarks are expected to be deflected to the left than to the right.
For $d$ quarks, the opposite is predicted. The predicted
negative sign of the Sivers function (\ref{eq:sivers})
for $u$ quarks $f_{1T}^{\perp u/p}$ \cite{Burkardt:2002ks}, 
has been confirmed by a recent 
analysis of SIDIS data for pion production on the proton
\cite{hermes-AUT-1,Alexakhin:2005iw,hermes-AUT-2}. Here
the experimental result for $f_{1T}^{\perp d/p}$ is less clear
as its contribution to pion production is strongly suppressed. 
The asymmetry on the deuteron is consistent
with zero \cite{Ageev:2006da,Vossen:2007mh}. A small Sivers asymmetry 
could arise from cancellation between
the contributions from the proton and neutron. Using charge symmetry
this implies
$f_{1T}^{\perp u/n}\approx -f_{1T}^{\perp d/p}$, which is
consistent with the prediction based on the transverse deformation
of IPDs. However, it is also possible that at very small $x$, 
both $f_{1T}^{\perp u/p}$ and $f_{1T}^{\perp d/p}$ are small.

\subsection{QCD simulations on a space-time lattice}
\label{sec:GPDLattice}
%
Ideally, in order to describe properties and interactions of 
hadrons theoretically, hadronic matrix elements should 
be evaluated in terms of QCD wave functions. However, this is not 
possible as long as wave functions in QCD cannot be calculated from
first principles. QCD calculations on a Euclidean lattice 
(also known as {Lattice QCD} or {Lattice Gauge Theory}) 
are currently the most promising tool to 
overcome this problem. Presently there are a large number of workers
active in this field  --- resulting in many types and flavors of 
lattice QCD. The proceedings of the yearly lattice symposia 
(currently published in PoS) provide a good overview of the state of 
the field; specifically concerning hadron structure, the two most
recent summaries were reported in \cite{Orginos:2006zz,Hagler:2007hu}.

The inevitable truncations for numerical
computations in QCD are less problematic on a space-time lattice,
where space rather than momentum is discretized. After discretizing 
space-time on a 4-dimensional
grid, approximate solutions for hadron masses and matrix 
elements are obtained numerically using Monte-Carlo techniques, 
because the latter are capable of dealing with the truly astronomical
numbers of degrees of freedom (Contemporary lattices typically have
about $32^3\times 64$ discretization points. Including spin and color
degrees of freedom, this implies a total of more than $>10^7$
degrees of freedom for the entire lattice) that are necessary for an 
adequate approximation to the continuum.
These Monte-Carlo techniques are used to evaluate a so-called
`path integral', which sidesteps a direct computation of
wave functions. For example, a pion form factor can be computed
without having to know what a pion is, only its quantum numbers
\cite{Brommel:2006ww,Kaneko:2007nf,Boyle:2007wg,Simula:2007fa}.
The evaluation of these path integrals would be impractical
working with a real time variable, as amplitudes in quantum mechanics
oscillate as a function of time and would thus lead to poor
statistical precision.
These oscillations are avoided by introducing an imaginary
time $t\longrightarrow -i\tau$, which renders the $\tau$ime 
({\it sic})
evolution operator real and positive:
$\exp(-iHt) \longrightarrow \exp(-H\tau)$,
and thus suitable for Monte Carlo sampling. The variable $\tau$ is
often referred to as `Euclidean time', as the Minkowskian metric
becomes Euclidean when expressed in terms of $(\tau,{\vec r})$.
This last step requires an analytical continuation to 
translate the results into a Minkowskian space-time.

It is impossible to apply lattice QCD to directly calculate DIS cross
sections,
since this would require a summation over all possible
final states. Instead one can use the optical
theorem and evaluate the imaginary part of the forward Compton
amplitude. 
This requires the calculation of a correlation function 
$\left\langle P\left|\psi_f^\dagger(0)\gamma^+\psi_f(z^-)\right|P\right\rangle$
between nucleon states, where $\psi_f$ is the quark field operator. 
Here the light-cone variables are defined as
$z^\pm = \frac{1}{\sqrt{2}}\left( z^0 \pm z^3\right)$, 
where the presence of $z^0$ makes the
correlation function non-local in time. Such a space-time structure 
arises since the Compton amplitude involves two vertices 
in figure \ref{fig:comptFF}a coupling
the nucleon to the virtual photon, 
separated in a light-like direction,
as the struck quark moves with nearly the speed of light.
This causes additional difficulties in the translation from
imaginary $\tau$ime to real time. This problem restricts lattice gauge
theory calculations of PDFs to only their Mellin moments. Such moments
are accessible because they can be represented by matrix
elements of local (i.e. containing products of field operators
and their derivatives evaluated at the same space-time point)
operators for which this analytical continuation
is trivial. For PDFs, which can apparently be parameterized by
relatively simple functions, only a few moments may be sufficient
to constrain these functions, except very close to the endpoints
$x\longrightarrow 0,1$. In higher moments the region
$x\longrightarrow 0$ is suppressed and they thus
only indirectly help constrain this region. 
Even though higher moments emphasize the $x\longrightarrow 1$ region,
PDFs are very small for very large $x$, and only very high
moments yield useful constraints on the 
$x\longrightarrow 1$ behavior of PDFs.

Parameterizing GPDs is significantly more difficult as there
are two longitudinal variables instead of one, and the $t$-dependence
adds yet another variable. Furthermore, it is not yet
known if GPDs can be described by similarly simple functions as PDFs
and hence it is not obvious how many moments will be needed to
generate a reliable parameterization for GPDs. Indeed, both
phenomenological model calculations (see figure~\ref{fig:vdh}) as
well as perturbative QCD evolution \cite{Kirch:2005tt} 
suggest a `kink', {\it i.e.} a discontinuous first derivative,
for GPDs at $x=\pm\xi$, whose adequate parameterization may 
therefore demand a 
larger number of moments than was the case for PDFs.

Remarkable progress in lattice calculations has been
made over the last three decades. One source of systematic 
uncertainties is the coarseness (spacing) of the lattice.
Improved algorithms now permit improved precision at the
same lattice spacing \cite{Symanzik:1983gh,Weisz:1983bn,Itoh:1985wk,Takaishi:1996xj,Hasenfratz:1993sp}. 
Another source of uncertainties arises from the
finite size of the lattice (physical volume). Ideally, 
it should be large enough to contain the hadron 
state completely. Both because improved algorithms 
allow larger lattice spacings at the same precision 
and because increased computing
power allows more discretization points, large lattices with
several fm diameter are now possible.
In the past, the omission of sea quarks in `quenched calculations'
gave rise to additional
uncertainties. Steady increases in computational resources and
improved algorithms make 
it now possible to also include sea quarks (`full' lattice QCD)
and unquenched calculations are now routine.
Nevertheless, calculating observables that require inserting
operators into disconnected quark loops is at present still
prohibitively expensive in most cases; some exceptions which are 
tractable have been studied, e.g., in \cite{Michael:2007vn}.
Usually, contributions from disconnected diagrams are omitted, 
resulting in an uncontrolled
approximation that affects most flavour-singlet observables.
Lattice calculations of flavour non-singlet observables are not 
affected by this omission and hence are much more reliable.
For the Dirac form factor, the contribution from disconnected quark
loops has been estimated to be small 
\cite{Dong:1998tj,Mathur:2000cf,Leinweber:2004tc,Leinweber:2006ug}, 
consistent
with the small influence from strange quarks on the charge radius 
of the nucleon \cite{Armstrong:2005hs,Aniol:2005zf}.
For most other observables, it is not clear whether the disconnected
quark loop contribution is small, but steady advances in computer
technology will allow including those contributions in the near 
future.

The masses of light quarks ($f=u,d$) are of the order of a few MeV.
Calculations are presently restricted to
much higher values, corresponding to pion masses of 
$m_\pi \approx 250$
MeV or even higher.\footnote{Since the quark mass itself is not an 
observable, results are usually presented as a function of the pion 
mass, which strongly depends on the masses of the quarks.} 
This limitation is due not only to the increase of statistical 
fluctuations in the Monte Carlo algorithms for small quark masses 
but also to the increase in the required overall lattice volume: a 
proper description of the virtual pion fluctuations around a hadron 
demands lattices that are several pion Compton wavelengths in 
diameter.
As a result of the restriction to performing calculations
in an `unphysical universe' in which the light quarks are too heavy, 
another significant source of uncertainty arises from 
the extrapolation to small quark masses
and the difficulties involved in restoring the chiral symmetry 
(\ref{eq:chiral}) that was broken by an expedient necessary in
discretizing the Dirac equation.
Recently, new methods for approximating
fermion fields on a lattice have been applied which
provide significantly better approximations of chiral symmetry
\cite{Kaplan:1992bt,Shamir:1993zy,Neuberger:1997fp}.
The latter advances now allow more meaningful calculations 
at smaller quark masses, where chiral symmetry becomes increasingly
important. 
Fortunately, it is not necessary to perform lattice
calculations at quark masses for $u$ and $d$ quarks of a few MeV.
Chiral Perturbation Theory ($\chi$PT) is an effective field theory
for QCD at small quark masses and momentum transfers.
It can be used to extrapolate
results from larger quark masses down to these values.
However, a reliable extrapolation first requires lattice simulations for 
quark masses where $\chi$PT can be trusted, a matter of controversy.

\begin{figure}
\unitlength1.cm
\begin{picture}(12,10)(-1,-10.5)
\includegraphics{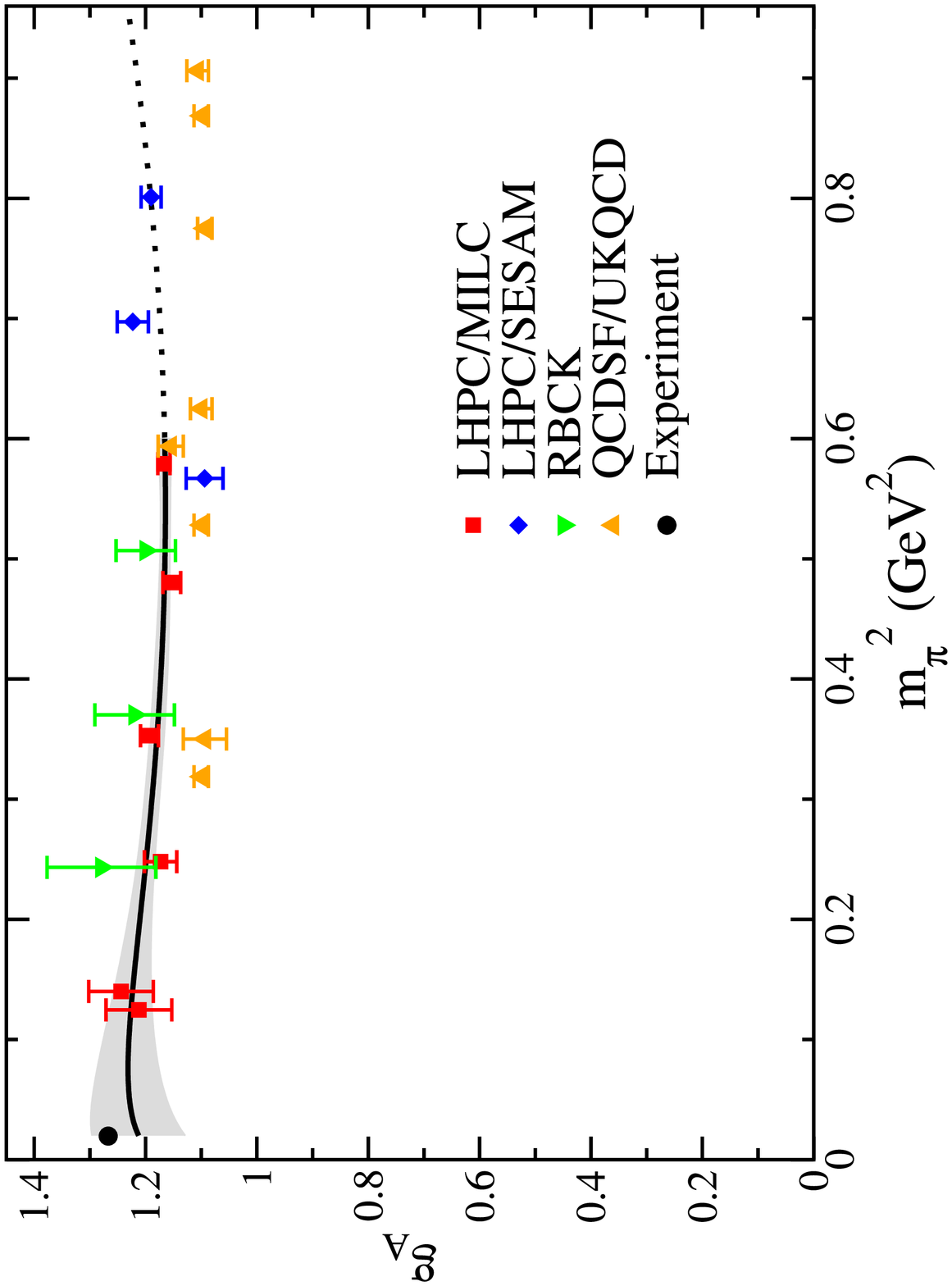}
\end{picture}
\caption{\label{fig:gA} 
Comparison of all presently existing unquenched lattice QCD 
calculations of the nucleon's axial charge $g_A$
(LHPC/MILC \cite{Edwards:2005ym}, LHPC/SESAM \cite{Dolgov:2002zm},
RBCK \cite{Ohta:2004mg}, QCDSF/UKQCD \cite{Khan:2004vw}). The
solid/dotted line and error band denote a 
chiral perturbation theory fit to the red squares only.
The error bars
reflect only statistical uncertainties and thus the
differences between results from different collaborations
gives an indication of the systematic uncertainty in these
calculations. The figure is taken from \cite{Edwards:2005ym}.
}
\end{figure}

Lattice QCD calculations have now been used to evaluate a large 
variety 
of observables, and due to lack of space we present only a few
selected cases. Figure \ref{fig:gA} shows lattice results
for the nucleon's axial charge $g_A$ \cite{Edwards:2005ym}, 
which is related to the spin content of the nucleon
by Bjorken's sum rule (\ref{eq:bjsr}). 
These results illustrate not only the wide range of quark masses
that is now accessible, but also
the accuracy that can be achieved when 
extrapolating results to the physical value $m_\pi = 139$ MeV.
For the tensor charge one finds after extrapolation to the
chiral limit $\delta u -\delta \bar{u}=0.857\pm
0.013$ and $\delta d - \delta \bar{d}=0.212 \pm 0.005$ 
\cite{Gockeler:2005cj}.

Using lattice QCD calculations to determine the second moment
of GPDs entering the Ji-relation (\ref{eq:Ji_SumRule}) yields
$J_u=0.214\pm 0.027$ and $J_d=-0.001\pm 0.027$ 
\cite{Hagler:2007xi,Brommel:2007sb}.
Combining these results with lattice calculations
for $\Delta q_f+\Delta \bar{q}_f$ one thus finds
\bea
L_u^{latt.} &=& J_u^{latt.} - \frac{1}{2}\left(
\Delta u^{latt.} +\Delta \bar{u}^{latt.}\right)= -0.195
\pm  0.044 
\label{eq:Lulatt}
\\ 
L_d^{latt.} &=& J_d^{latt.} - \frac{1}{2}\left(\Delta d^{latt.} 
+\Delta \bar{d}^{latt.}\right)= 0.200
\pm 0.044
\label{eq:Ldlatt}
\eea
at the physical pion mass
where the uncertainties are only statistical since some of the 
systematic uncertainties are difficult to quantify.
In these calculations, disconnected quark loops have been omitted
in the calculation of the matrix elements. 
The perhaps surprising result that the net orbital angular
momentum carried by the quarks is close to zero may thus be
subject to large corrections.
However, the isovector combination 
$L_u^{latt.}-L_d^{latt.}\approx -0.4$
is not affected by the omission of disconnected quark loops,
and therefore more reliable. Note that in all quark models, where the
orbital angular momentum arises from the lower component
of a confined relativistic quark wave function, such as the MIT bag model,
$u$ quarks have a positive orbital angular momentum, while
for $d$ quarks the orbital angular momentum is negative, yielding
$L_u-L_d>0$. Moreover, the sign of the Sivers function,
both in SIDIS and in purely hadronic reactions strongly suggests
$L_u>L_d$. QCD evolution of the quark orbital angular momentum
from very low $Q^2$, where the quark models are appropriate, to
the lattice scale of about $4\,{\rm GeV}^2$ has been suggested as a 
possible solution for this discrepancy \cite{Thomas:2008ga}.
However, in considering this apparent discrepancy, one
should bear in mind that there are different possibilities for 
defining what is orbital angular momentum, 
which is discussed in detail in section
\ref{subsec:Decompose}. 

\begin{figure}
\unitlength1.cm
\begin{picture}(12,12)(-.7,7.5)
\includegraphics{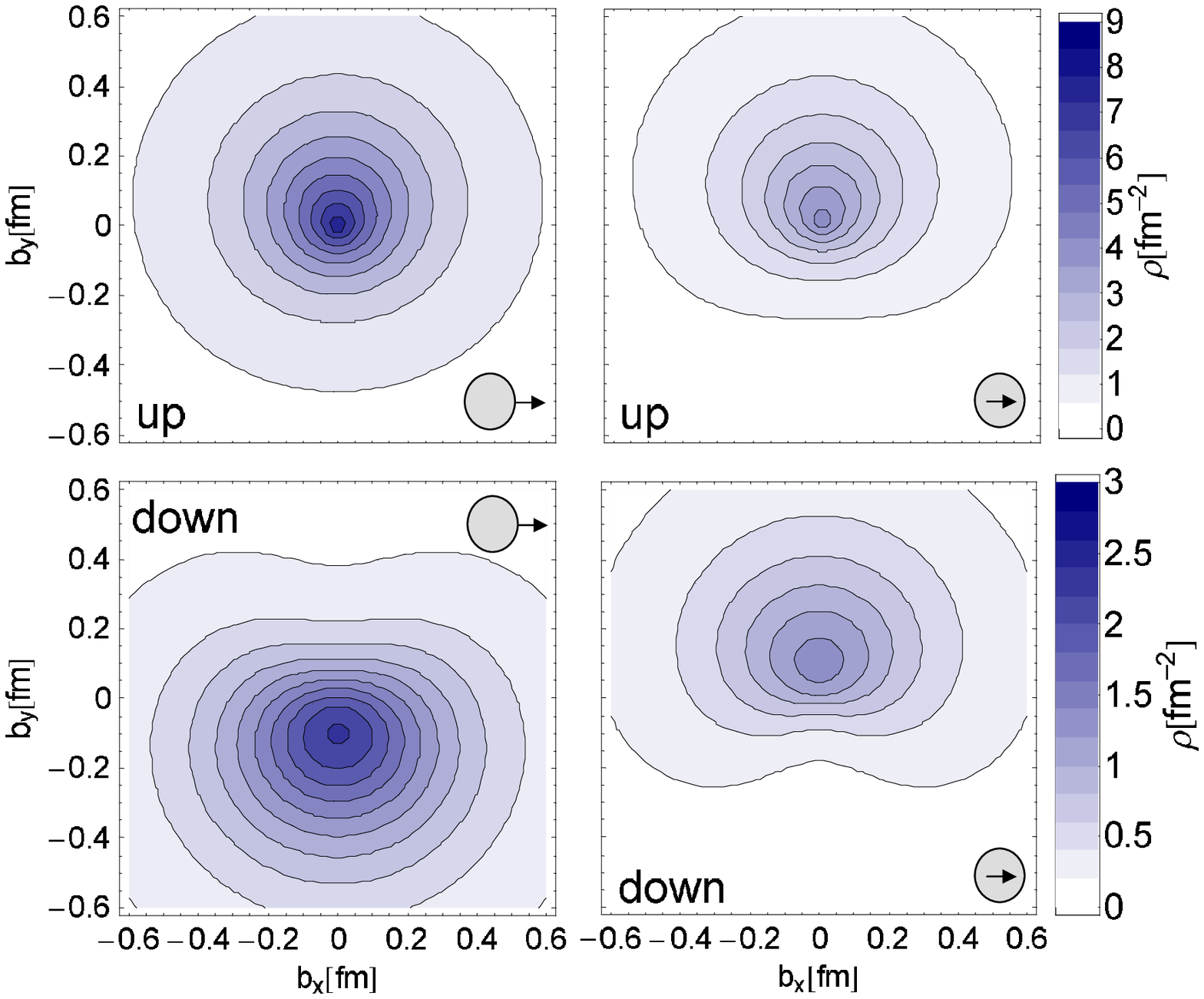}
\end{picture}
\caption{\label{fig:IPD} 
Lowest moment of the IPDs for unpolarized quarks in a 
transversely polarized nucleon (left) and transversely polarized 
quarks in an unpolarized nucleon (right) for $u$ (top)
and $d$ (bottom) quarks. In the inserts, the quark spins 
(inner arrows) and 
nucleon spins (outer arrows) are oriented in the transverse plane
as indicated. The figure is taken from~\cite{Gockeler:2006zu}.}
\end{figure}

Figure \ref{fig:IPD} shows results for the lowest moment
$\int {\rm d}x\, q(x,\bT)$ of IPDs.
The magnitude of the deformation in these panels 
(i.e. the deviation from axial symmetry) is a measure for
the correlation between the nucleon spin and the quark orbital 
angular momentum (left) and between the 
quark spin and quark orbital angular momentum (right).
Notice that the average density on the left is twice as large as on
the right since both quark polarizations are summed over on the left.
The figures clearly indicate that the correlation between quark
angular momentum and the nucleon spin (left)
has opposite signs
for the two quark flavors, while they have the same sign for
the correlation with the quark spin (right). Applying the 
chromodynamic lensing mechanism (section \ref{subsec:IPDPD}), this
is consistent 
with opposite sign Sivers functions $f_{1T}^{\perp q}$
(see section \ref{sec:sivers})
for $u$ and $d$ quarks. Using the same reasoning for transversely
polarized quarks in an unpolarized target, this leads to same-sign 
Boer-Mulders functions $h_1^{\perp q}$
(see section \ref{sec:BM}) and the sign is
the same as the sign of the Sivers function for $u$ quarks
(negative).
The correlations on the right also appear somewhat stronger than
on the left, suggesting $\left|h_1^{\perp q}\right|>
\left|f_{1T}^{\perp q}\right|$.

Lattice QCD calculations play an important role in complementing
both experiments and phenomenological models. 
For selected hadron structure observables, 
lattice calculations are reaching a 
precision that will allow  the comparison of QCD predictions to 
experimental data. Lattice calculations also complement
experiments because there are many interesting observables
which are difficult to probe experimentally, but which are
easily accessible on a lattice. Examples include most chirally odd  
distribution functions and form factors. Nevertheless, calculations
on the lattice will never be able to fully replace experiments 
in the exploration of hadron structure.
For example, no local manifestly gauge invariant operator \footnote{ 
An operator that can be written as the same expression in all gauges is called
{\em manifestly gauge invariant}.}\label{page:manifestly}
exists for $\Dg$,
which is therefore inaccessible in lattice QCD calculations. 
Simulation of scattering reactions and investigation of unstable 
particles is enormously difficult.
Already the determination of low energy scattering phase shifts
requires a sequence of calculations on lattices of various
volumes \cite{Luscher:1985dn,Luscher:1986pf,Luscher:1990ux}, 
thus multiplying the computational time. Only those resonances
(unstable particles) the mass of which is close to their decay 
threshold are accessible by this technique. 
Furthermore, the necessary use of Euclidean $\tau$ime in lattice 
calculations makes those observables inaccessible that require
calculations with a real time variable. 
For those high-energy
scattering processes where initial and/or final state interactions
are essential, such as SIDIS, no useful algorithm has been
developed yet and the interpretation of experiments still has to rely
on phenomenological models. Even after factorization has been 
applied in SIDIS, calculating the final-state interaction phase
that is crucial for TMDs would again require a real time variable. 
This is the main reason why no one has found a direct way for
calculating the Sivers 
(\ref{eq:sivers}) or Boer-Mulders (\ref{eq:BM}) functions on the
lattice. 
However, ISI/FSI affect the $\KT$ dependence of all PDFs
and since the lattice presently can only calculate intrinsic 
distributions (without ISI/FSI)
a direct comparison to experimentally measured $\KT$-dependent 
distributions (which include ISI/FSI) thus requires 
modeling of those effects.
Likewise, no algorithm exists for calculating
fragmentation functions on the lattice.

As a result of these limitations, there has also been a significant 
synergy between lattice calculations and phenomenological models in 
recent years. Even though lattice calculations cannot be used for 
direct calculations in dynamical processes, they can play a useful 
role in providing phenomenological models with input parameters,
thus effectively increasing the predictive power of these models.
The above-mentioned predictions for the Boer-Mulders functions is
an excellent example for such a synergy between lattice and
phenomenology: in the `chromodynamic lensing' model for SSAs
presented in section \ref{subsec:IPDPD}, IPDs are needed as
input. For the Boer-Mulders functions, the relevant GPDs are 
chirally odd
and no experimental constraints on them exist, but the lattice
calculations of these GPDs \cite{Gockeler:2006zu}
leading to the deformations
shown in figure \ref{fig:IPD}
allowed predictions 
to be made for the signs of the Boer-Mulders functions for
$u$ and $d$ quarks \cite{Burkardt:2005hp,Burkardt:2007xm}.

%
\subsection{Access to quark GPDs through exclusive lepton-nucleon scattering}
\label{subsec:GPD_Access}
%
Presently the experimentally most accessible GPDs are the chirally-even 
quark GPDs $F^f(x,\xi,t)$ ($F=H,\widetilde{H},E,\widetilde{E}$ and $f=u,d$). 
In order to constrain their non-forward ($\xi \neq 0$) behaviour, 
measurements can be performed
of hard exclusive leptoproduction of a photon or meson, leaving the target 
nucleon intact. Useful observables are cross sections and cross section
differences or asymmetries with respect to the polarization of the beam or 
target, or the beam charge. In these observables, GPDs typically appear 
convoluted with hard scattering amplitudes, so that they cannot be 
extracted directly from the measured data. Instead, their functional 
dependence on $x$, $\xi$, and $t$ is parameterized in theoretical models
(see section~\ref{subsec:GPD_Models}), the parameters of which have to be 
constrained by comparison to experiment (see 
section~\ref{subsec:GPD_ExpResults}). For various final states,
the convolution integrals appear combined additively or multiplicatively
in various ways, with various kinematic prefactors. Hence eventually the
extraction of quark GPDs will require a (presently not yet existing) global 
fit to data from various experiments, as has been customary for decades in 
the extraction of ordinary PDFs.

%
%
\subsubsection{Deeply virtual Compton scattering}
\label{subsec:GPD_Access_DVCS}
%
The exploitation of exclusive production 
of a real photon, {\em i.e.}, Deeply Virtual Compton Scattering 
e p $\rightarrow$ e p $\gamma$, has two advantages: \\
i) among the hard exclusive processes that can be accessed
experimentally in the foreseeable future, it is considered to be
the one with the most reliable theoretical interpretation \\
ii) effects of next-to-leading 
order~\cite{Belitsky:1997rh,Ji:1998xh,Mankiewicz:1997bk} and 
sub-leading twist (see section~\ref{sec:Intro}) 
corrections ~\cite{Anikin:2000em,Radyushkin:2000ap,Belitsky:2001ns} 
are under theoretical control. For suitably chosen observables
the lowest correction is of order $1/Q^2$ instead of $1/Q$.

In the generalized Bjorken limit of large photon virtuality $Q^2$ at fixed
$x_B$ and $t$, the dominant pQCD subprocess of DVCS is described by the 
`handbag' diagram shown in the left panel of 
\begin{figure}[htb]
\includegraphics[width=15cm]{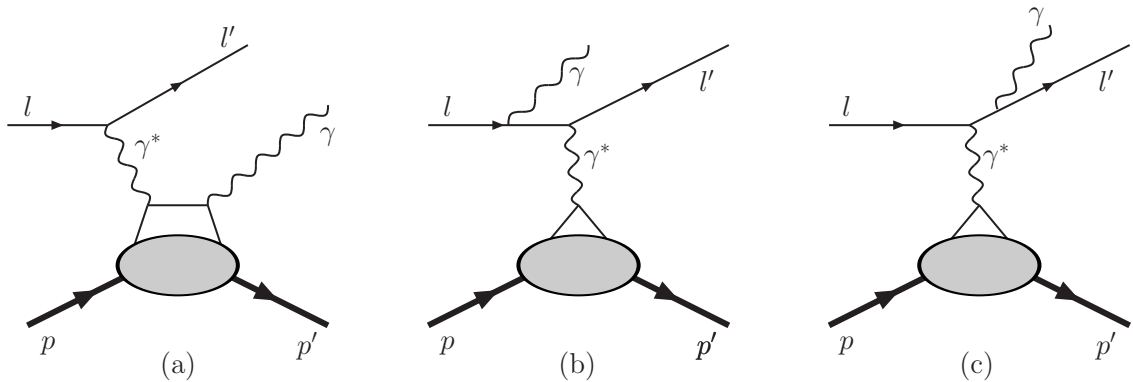}
\centering
\caption{\label{fig:handbagNEW} Leading-order processes for leptoproduction 
of real photons. Left: Deeply Virtual Compton Scattering. 
Middle and right: Bethe-Heitler process.}
\end{figure}
figure~\ref{fig:handbagNEW}. The skewness parameter $\xi$ is given by
$\xi \simeq \frac{x_B}{2-x_B}$ in this limit. Pictorially, the parton
(of flavor 
$q$) taken out of the proton carries the longitudinal momentum fraction 
$x+\xi$ and the one put back into the proton carries the fraction $x-\xi$. 
The quark GPD $F^f(x,\xi,t)$ can then be considered as describing the 
correlation between these two quarks at the given value(s) of $t$ (and $Q^2$).

The $t$-dependence of quark GPDs is directly accessible in DVCS although high
experimental precision, {\em i.e.}, high statistical accuracy in 
conjunction with sufficient resolution of the experimental apparatus, is 
required to extrapolate to the limit $t \rightarrow 0$. The latter is of 
particular importance for the evaluation of the second $x$-moment of the 
sum of the two `unpolarized' quark GPDs, $H^f+E^f$, which is related to
the total angular momentum $J_f$ of the parton species $f=(u,d,g)$, at a given 
value of $Q^2$~\cite{Ji:1996ek,Ji:1996nm} (see (\ref{eq:Ji_SumRule})).

\begin{figure}[htb]
\includegraphics[width=8cm]{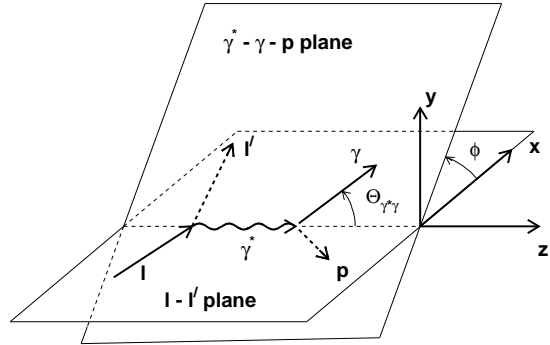}
\centering
\caption{\label{fig:kinPlane} Definition of the azimuthal angle 
$\phi$ in DVCS in the target rest frame.}
\end{figure}

The Bethe-Heitler (BH) process, or radiative elastic scattering, is 
illustrated in the two right panels of figure~\ref{fig:handbagNEW}.
Its final state is indistinguishable from that of the DVCS process
and these two mechanisms have to be combined on the level of the process
amplitudes. The differential cross section for leptoproduction of real 
photons is written in terms of the BH and DVCS process amplitudes as:
\begin{eqnarray}
\label{eq:DVCS-BH-amplitude}
\frac{d\sigma(ep \rightarrow ep\gamma)}{dx_BdQ^2d|t|d\phi} & 
\propto & |\tau^{}_{BH}|^2 + |\tau^{}_{DVCS}|^2 + 
\underbrace{\tau^{}_{DVCS} \tau^{*}_{BH}+\tau^{*}_{DVCS} \tau^{}_{BH}}_{\mathrm{I}}.
\end{eqnarray}
Here $\phi$ is the azimuthal angle between the scattering plane 
spanned by the trajectories of the incoming and outgoing leptons and the
production plane spanned by the virtual photon and the produced real photon 
(see figure~\ref{fig:kinPlane}). The BH process amplitude is calculable 
using the electromagnetic proton form factors, which are well measured 
at small $t$ (see, {\em e.g.}, \cite{Mergell:1995bf}). The pure DVCS
contribution, $|\tau^{}_{DVCS}|^2$, can then be extracted by integrating
over the azimuthal dependence of the cross section. In this situation the 
interference term I vanishes to leading order in $1/Q$; 
its total contribution at collider kinematics was estimated to be
at the percent level~\cite{Belitsky:2001ns}.

The interference term $\mathrm{I}$ is of high interest, as the measurement 
of its azimuthal dependence opens experimental access to the 
{\em complex-valued} DVCS process amplitude. In this way, {\em both} their 
real and imaginary parts 
(magnitude and phase) become accessible~\cite{Diehl:1997bu}. This method 
to study DVCS can be compared to holography~\cite{Belitsky:2002ep} in the 
sense that the phase of the process amplitude of DVCS is measured against 
the known `reference phase' of the BH process. Moreover, at fixed-target 
energies, the more abundant BH process can be considered as an 'amplifier' 
to study the rare DVCS process.

An expansion in $1/Q$ of the electroproduction cross section 
(\ref{eq:DVCS-BH-amplitude}) reveals distinctive kinematic signatures
of DVCS~\cite{Diehl:1997bu,Belitsky:2000gz,Belitsky:2001ns}. For the 
interference term I this expansion reads:
\begin{eqnarray}
\label{eq:ITtoAmpl}
\! \! \! I \propto e_l \left[\sum_{n=1,3} k^c_n \; \mathrm{Re} \widehat{M}_n
              \cos{n \phi} 
          + P_{beam} \sum_{n=1,2} k^s_n \; \mathrm{Im} \widehat{M}_n
              \sin{n \phi} 
          + {\cal{O}}(1/Q) \right],
\end{eqnarray}
where $e_l = \pm 1$ is the charge of the lepton beam and the $k^{c,s}_n$
are kinematic factors. The $\widehat{M}_n$ represent linear combinations of
helicity (Compton) amplitudes describing the DVCS process. The subscript
$n=1,2,3$ labels the helicities for initial and final state photons:
$n=1~(++)$ means helicity conservation, $n=2~(+0)$ single-helicity flip 
and $n=3~(+-)$ double-helicity flip.

In general, the portion of the interference term containing real parts of 
Compton amplitudes can be isolated by forming a cross section difference or 
asymmetry with respect to the charge of the lepton beam~\cite{Brodsky1972}, 
while that containing imaginary parts can be accessed 
by forming single-spin differences or asymmetries with respect to the 
polarization of the lepton beam~\cite{Kroll:1995pv} or of the 
target~\cite{Belitsky:2001ns,Diehl:2003ny}. In the interpretation of
experimental results in the subsequent sections, always only the 
helicity-conserving twist-2 Compton amplitude 
$\widehat{M}_1 \equiv \widehat{M}_{++}$ is considered. In the convention 
of Ref.~\cite{Belitsky:2001ns}, this amplitude
can be written as a linear combination of products of the Dirac and Pauli 
elastic nucleon form factors $F_1$ and $F_2$ with Compton form factors (CFFs)
$\mathcal{F}$ ($\mathcal{F} = \mathcal{H},\mathcal{E},\widetilde{\mathcal{H}},
\widetilde{\mathcal{E}}$).
For the three different target polarizations (U for unpolarized, L (T) for 
longitudinally (transversely) polarized target) the expressions are:
\begin{eqnarray} 
\label{eq:Munpol}
\widehat{M}_U & \propto &  \frac{\sqrt{t_0-t}}{2m} \,
                 \left[F_1 \, {\cal H} + \xi \, (F_1 + F_2) \, 
                 \widetilde {\cal H} - \frac{t}{4 m^2} F_2 \, {\cal E}
                \right], \\
\label{eq:LTSA} 
\widehat{M}_L & \propto & \frac{\sqrt{t_0-t}}{2m} \;
                   \left[ \xi\left( F_1+F_2 \right) \mathcal{H}
                   + F_1 \widetilde{\mathcal{H}} -  
                   \left( \frac{\xi}{1+\xi} F_1 + \frac{t}{4m^2} F_2 \right) 
                    \, (\xi \widetilde{\mathcal{E}}) \right], \\
\label{eq:TTSA_N}
\widehat{M}_{T} & \propto & \frac{t_0-t}{4m^2} \; \left[ 
                        F_2 \mathcal{H} - F_1 \mathcal{E} 
            + \xi (F_1+F_2) \, (\xi \, \widetilde{\mathcal{E}}) \right],
\end{eqnarray}
where terms proportional to $\xi^2$ and higher are neglected in 
(\ref{eq:LTSA}) and (\ref{eq:TTSA_N})~\cite{Belitsky:2001ns,Diehl:2003ny}.
Here $m$ is the nucleon mass and $-t_0=4\xi^2m^2/(1-\xi^2)$ is the minimum 
possible value of $-t$ at a given $\xi$.

The complex-valued CFFs $\mathcal{F}(\xi,t,Q^2)$ are flavor sums of 
convolutions of the corresponding 
leading-twist quark GPDs with the functions $C_f^{\mp}$ describing the hard
$\gamma^*q$ Compton scattering. The latter are available up to NLO in 
pQCD~\cite{Belitsky:1997rh,Ji:1998xh,Mankiewicz:1997bk}:
\begin{eqnarray}
\label{eq:CFF}
\mathcal{F}(\xi,t,Q^2) = \sum_{f} \int_{-1}^1 \de x \; 
                             C^\mp_f(\xi,x,\log Q^2)\, F^f(x,\xi,t,Q^2).
\end{eqnarray}
Here the $-(+)$ sign in the superscript applies to the CFFs
$\mathcal{F}=\mathcal{H},\mathcal{E} \;
(\widetilde{\mathcal{H}},\widetilde{\mathcal{E}})$, corresponding to the GPDs 
$F^f=H^f,E^f \; (\widetilde{H}^f,\widetilde{E}^f)$. At ${\cal{O}}(\alpha_s)$,
also contributions from $F^G(\widetilde{F}^G)$ exist in 
(\ref{eq:CFF})~\cite{Diehl:2003ny}. Except when explicitly needed, the 
dependence on $Q^2$ is omitted in the following.

Only at {\em leading order} in $\alpha_s$ there exist relatively simple 
relationships between the real or imaginary parts of a CFF and (the flavor 
sum over) the respective quark GPDs that it embodies~\cite{Belitsky:2001ns}.
Using the same sign convention, the real part of (\ref{eq:CFF}),
\be
  \label{eq:CFF2a}
  \! \! \! \mathrm{Re} \; \mathcal{F}(\xi,t) \stackrel{\rm LO}{=} 
  \sum_{f} e_q^2 \left[ \mathcal{P} \int_{-1}^1 \de x \ F^f(x,\xi,t) 
    \left( \frac{1}{x-\xi} \mp \frac{1}{x+\xi} \right) \right],
\ee
becomes at given skewness $\xi$ a flavor sum of integrals over 
$-1 < x < 1$ of the respective quark GPDs, where $\cal{P}$ denotes Cauchy's 
principal value. The imaginary part of (\ref{eq:CFF}),
\be
\label{eq:CFF1a} 
\mathrm{Im} \; \mathcal{F}(\xi,t) \stackrel{\rm LO}{=}   \pi 
\sum_{f} e_f^2\left( F^f(\xi,\xi,t) \mp F^f(-\xi,\xi,t)\right),
\ee
becomes a {\em direct} flavor sum of quark GPD singlet 
combinations $F^f(\xi,\xi,t) \mp F^f(-\xi,\xi,t)$. Singlet 
combinations are also probed in exclusive vector meson production while in 
exclusive pseudoscalar meson production nonsinglet combinations 
$F^f(\xi,\xi,t) \pm F^f(-\xi,\xi,t)$ are accessed~\cite{Diehl:2003ny}.

A recent theoretical development exploits dispersion relations for CFFs. 
A CFF $\cal{F}$ satisfies a single-variable dispersion relation showing 
that  the entire $\xi$ dependence of the real part is given by a
term that can be constructed from only the imaginary 
part~\cite{Teryaev:2005uj,Kumericki:2007sa,Diehl:2007jb}:
\be   \label{eq:CFF3}
  \mathrm{Re} \; \mathcal{F}(\xi,t) =
  \frac{1}{\pi} \mathcal{P} \int_0^1 \de \xi' \, 
  \mathrm{Im} \; {\mathcal{F}}(\xi',t) 
  \left(\frac{1}{\xi-\xi'} \mp \frac{1}{\xi+\xi'} \right) 
  + {\mathcal{C_F}}(t).
\ee
At leading order, the imaginary part depends only on the GPDs on the 
trajectories $x=\pm\xi$. The D-term form factor is then given by:
\be 
{\mathcal{C_F}}(t) = \sum_{f} e_f^2 
   \int_0^1 \de x \; D^f(x, t) \left( \frac{1}{1-x} \right),
\ee
where the D-term $D^f$ was introduced in (\ref{eq:DDansatz1}) of 
section \ref{subsec:GPD_Properties}.
This form factor vanishes for the CFFs 
$\widetilde{\cal{H}}$ and $\widetilde{\cal{E}}$ and is identical up to the 
sign  for $\cal{H}$ and $\cal{E}$~\cite{Kumericki:2007sa,Diehl:2007jb}. 
Equation (\ref{eq:CFF3}) together with (\ref{eq:CFF1a}) imply that at leading 
order and without considering effects of evolution,  the real part of the CFF 
can be decomposed into a term that shares the well-known property of
the imaginary part in depending on the value of the GPDs only along the
one-dimensional trajectories $x=\pm\xi$, and another term that depends only 
on $t$ and $Q^2$, and not on $\xi$.  Thus, in a leading-order analysis of DVCS data, 
parametric models for quark GPDs can be constrained only in the subspace 
$(x=\pm\xi,t)$. Beyond this scenario, a full $(x,\xi)$ mapping of quark GPDs 
is still possible, at least in principle:

\noindent i) when a large enough range in $Q^2$ is covered by DVCS
measurements, the known $Q^2$-evolution of quark GPDs can be used to 
constrain their $x$-dependence. 

\noindent ii) in hard exclusive leptoproduction of a {\em virtual} photon
(double DVCS or DDVCS), its virtuality, {\em i.e.} the invariant mass 
of the produced lepton pair, is an additional variable that facilitates 
a complete mapping of quark GPDs~\cite{Guidal:2002kt,Belitsky:2002tf}. 
However, the DDVCS cross section is suppressed by an additional factor 
$\alpha_{em}$, thereby making this reaction practically inaccessible 
using present facilities.

Even if complete DVCS data sets were available for both proton and neutron targets,
it would still be a very intricate theoretical problem to deconvolute the 
quark GPDs $F^f$ from the CFF $\cal{F}$. The variable $t$ 
does not pose any problem as it does not appear in the kernel; \ie, the 
deconvolution would have to be done at a given value of $t$.

\subsubsection{Hard exclusive meson production}

Additional information on GPDs is available from hard exclusive 
lepto-production of a pseudoscalar or vector meson M:
e p $ \rightarrow$ e~p~M. To leading order in $1/Q$ and 
$\alpha_s$, the meson structure itself enters only through a few 
constants~\cite{Diehl:2003ny}. (In higher order the meson distribution 
amplitude has to be explicitly taken into account.) The meson production 
cross section receives contributions from both longitudinal (L) and 
transverse (T) virtual photons:
\begin{eqnarray}
\label{eq:mesonXsection}
\frac{d\sigma}{dx_BdQ^2d|t|} & 
 \propto & \sum_{spins} |{\mathcal{A}}_T|^2 
+ \epsilon \sum_{spins} |{\mathcal{A}}_L|^2,
\end{eqnarray}
where $\epsilon$ is the ratio of longitudinal and transverse fluxes of 
the virtual photon. The dependences of the scattering
amplitudes $\cal{A}$ on the hadron polarizations is summed or averaged 
over in $\sum_{spins}$. A factorization theorem has been proved to date 
only for the longitudinal part~\cite{Collins:1996fb}. 

The contribution by {\it transverse} photons to the cross section for meson
production appears at subleading order in 
$1/Q$~\cite{Collins:1996fb,Diehl:2003ny}.
In case of production of a vector meson its spin-density matrix can
be measured, thus facilitating the longitudinal-transverse separation.
In case of pseudoscalar meson production only data at sufficiently large 
$Q^2$ may be used where the contribution of transverse virtual photons 
can be neglected. 

It appears that for meson production by {\it longitudinal} photons, at 
leading order in $1/Q$ and for an unpolarized 
target~\cite{Collins:1996fb,Diehl:2003ny},
only one pair of quark GPDs is involved in the relevant combination, 
controlled by the type of spin-parity exchange in the meson production,
so that by measuring various meson types different pairs of quark GPDs 
are filtered out.
For natural-parity exchange, i.e. for the production of e.g. vector 
mesons such as $\rho^0$, the quark-helicity-conserving GPDs $H^f$ 
and $E^f$ are involved:
\begin{equation} 
\label{eq:MesonGPDsNP}
\frac{1}{2} \sum_{spins} |{\mathcal{A}}_L|^2 =  
   (1-\xi^2) |{\mathcal{H}}|^2 - (\xi^2+\frac{t}{4m^2}) |{\mathcal{E}}|^2
   - 2 \xi^2 Re({\mathcal{E}}^* {\mathcal{H}}).
\end{equation}
For unnatural-parity exchange, i.e. for the production of e.g.
pseudoscalar mesons such as pions, the quark-helicity-flip GPDs 
$\widetilde{H}^f$ and $\widetilde{E}^f$ are involved:
\begin{equation} 
\label{eq:MesonGPDsUNP}
\frac{1}{2} \sum_{\lambda' \lambda} |{\widetilde{\mathcal{A}}}_L|^2 =  
   (1-\xi^2) |{\widetilde{\mathcal{H}}}|^2 - \xi^2 \frac{t}{4m^2} 
   |{\widetilde{\mathcal{E}}}|^2
   - 2 \xi^2 Re({\widetilde{\mathcal{E}}}^* {\widetilde{\mathcal{H}}}).
\end{equation}

Recently, detailed numerical studies were presented of the perturbative 
convergence of the longitudinal cross section for the production of 
$\rho^0, \omega, \phi$ and $\pi^+$ mesons~\cite{Diehl:2007hd}. At 
intermediate to large \xbj, typical for fixed-target experiments, NLO 
corrections of up to 100\% were found, which somewhat decrease in size when 
going from $Q^2 = 4$ GeV$^2$ to 9 GeV$^2$. 
At lower \xbj, typical for collider kinematics, NLO corrections
are huge even for $Q^2$ well above 10 GeV$^2$. Such an increased 
sensitivity to higher-order corrections is due to the above shown 
{\em quadratic} dependence on GPDs. It also exists in DVCS 
(see (\ref{eq:DVCS-BH-amplitude})) for the $|\tau_{DVCS}|^2$ term,
but not for the interference term I. We note that huge NLO corrections 
up to 200-300\% appear also in other, mainly collider processes. Here
in most cases the full NLO result is available and used, and typically 
the NNLO corrections are not again 100\% of the NLO ones.
%
%
\subsection{GPD parameterizations}
\label{subsec:GPD_Models}
%
The application of the Ji relation (\ref{eq:Ji_SumRule}) to evaluate the total 
angular momentum of quarks of flavour $f$ is a challenging task as it
requires, at some fixed common value of $\xi$, knowledge of both $H^f$ and 
$E^f$ over the entire $x$ range in the limit of small $t$. In 
section~\ref{subsec:GPD_Access_DVCS}, it was explained that the best presently 
practical probe to constrain quark GPDs is DVCS, which unfortunately does not 
directly provide all of the required information because of its 
kinematic limitations. While $H^f(x,0,0) = q_f(x)$ is well known, 
$E^f(x,\xi,0)$ is poorly constrained at $\xi=0$, because there it has no influence 
on experimental observables aside from its first moment, \ie\ the Pauli form 
factor. Hence the Ji relation must be evaluated in the nonforward regime.
Furthermore, the integral (\ref{eq:Ji_SumRule}) must be evaluated in the
limit $t \rightarrow 0$, even though the crucial dependences of DVCS cross
sections on beam charge and polarisations of the beam and target disappear
in this limit. Thus GPD parameterizations constrained by DVCS data must be
used to {\em extrapolate} as $t \rightarrow 0$, which might seem to 
introduce some degree of arbitrariness. However,
GPDs are far from arbitrary functions of the three variables $x,\xi,t$ (or 
four, when including $Q^2$), as they must satisfy several relations as 
described in section~\ref{subsec:GPD_Properties}. Considerable ingenuity is 
therefore required in finding analytic representations or phenomenological
parameterizations of quark GPDs that embody the constraints of various 
symmetries and principles, while conveniently parameterizing the remaining 
degrees of freedom. The discussion of {\em dynamical} models, a recent review 
of which is given in \cite{boffi:2007yc}, is beyond the scope of this paper.

\begin{figure}[t]
\begin{center}
\includegraphics[width=0.47\textwidth,bb=105 350 455 690]{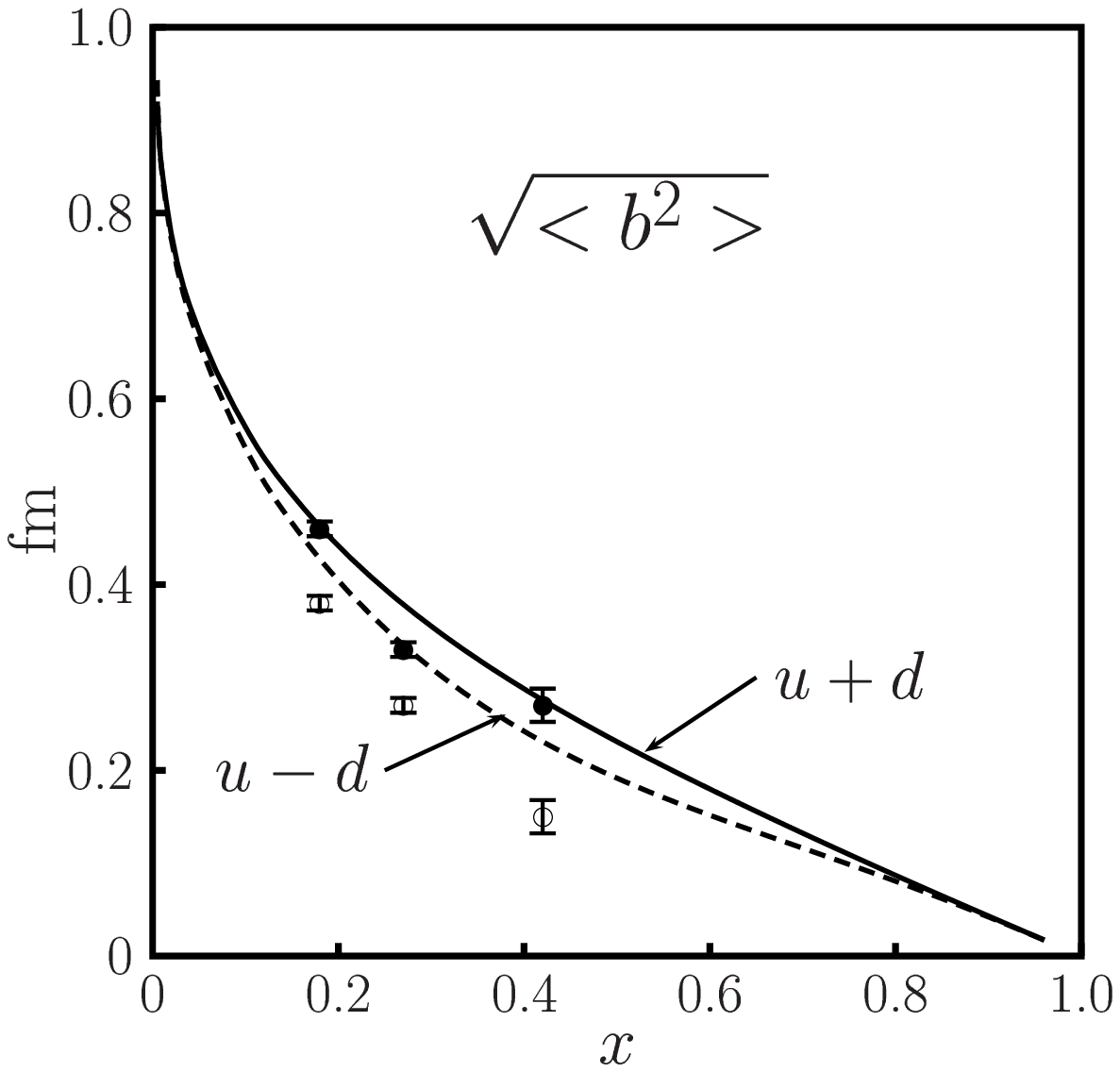}
\end{center}
\caption{\label{fig:n2_bf} Square root of the mean-square impact
parameters $\langle \bT^2 \rangle$ for the sum and the difference 
of $u$ and $d$ quark distributions from the fit to elastic form factors, 
compared with lattice QCD results from~\protect\cite{Negele:2004iu}.  
The figure is taken from \cite{Diehl:2004cx}.
}
\end{figure}

The simplest ansatz that one might consider for the $t$ dependence of quark
GPDs is a `factorized' form:
\be
H^f(x,\xi,t)=h^f(x,\xi) \, F^f_1(t).
\ee
However, there is evidence from phenomenological 
considerations~\cite{Goeke:2001tz,Diehl:2004cx,Guidal:2004nd} and lattice 
QCD calculations~\cite{Hagler:2007xi,Gockeler:2006zu} that the $t$ 
dependence is entangled with both $x$ and $\xi$, 
so that factorized ans\"atze are disfavoured. 
As discussed in section~\ref{subsec:IPDPD}, $t$ is conjugate to the 
impact parameter $\bT$, so that this entanglement corresponds to the
interplay between longitudinal and transverse degrees of freedom
in the nucleon, sometimes called `nucleon tomography'.
Measured electromagnetic form factors together with some physically
motivated functional forms for the limits $x\to 0,1$ have provided some 
information about the entanglement of $x$ and $t$ for `valence' quark 
flavours in the limited regime $\xi=0$~\cite{Diehl:2004cx}.  
In this work, an exponential ansatz was chosen for this entanglement
in the valence GPDs:
\be
 H^{f(-)}(x,0,t) = q_{fv}(x) \exp{[t f_f(x)]}\,
  \label{eq:ExpEntangl}
\ee
where the unpolarized valence PDF $q_{fv}(x) \equiv q_f(x) - \bar{q}_f(x)$ 
is known.
Although the dependence on $Q^2$ has been suppressed here for convenience,
this exponential form is approximately stable under QCD evolution, 
more so at small $x$.  It corresponds to an IPDPDF with guaranteed
positivity and a plausible Gaussian shape with an $x$-dependent width:
\be
  \label{eq:ExpImpact}
q_{fv}(x,\bT) = \frac{1}{4\pi}\, \frac{q_{fv}(x)}{f_f(x)}\,
  \exp\Bigg[ -\frac{\bT^2}{4 f_f(x)} \Bigg]\,,
\qquad \qquad
\langle \bT^2 \rangle = 4 f_f(x) \,.
\ee
The choice for the function $f_f(x)$ is motivated by simple physical
ideas.  Partons with small $x$ can be considered to arise from
a cascade of branching processes, sometimes called 
`Gribov diffusion'~\cite{Gribov:1973jg}.  This sort of random walk
corresponds to the mean-square impact parameter $\langle \bT^2 \rangle$ 
growing at small $x$ as $\log(1/x)$, a behavior that is provided by the 
`Regge-motivated' form
\be
\hspace*{-1cm}  H^{f(-)}(x,0,t) \propto
e^{(B_0/2) t} \left(\frac{1}{x}\right)^{\alpha' \,t}\!\!\!\!\cdot q_{fv}(x)
= e^{\alpha' \, t \log(1/x)+(B_0/2) t} \, q_{fv}(x)\,.
\label{eq:unfactorized_Ansatz}
\ee
\begin{figure}
\begin{center}
\leavevmode
\includegraphics[width=0.4\textwidth]{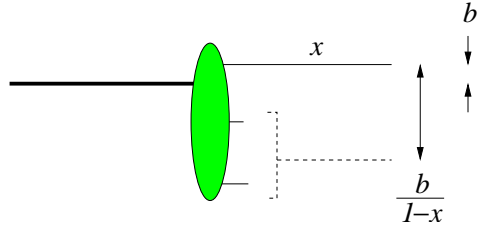}
\end{center}
\caption{\label{fig:ImpactParam} Impact parameter relationship
between a struck quark with longitudinal momentum fraction $x$ and the
collection of `spectator' partons.  The dashed line indicates the
center of momentum of the spectators and the thick continuous line
the center of momentum of the proton.
The figure is taken from \cite{Diehl:2004cx}.}
\end{figure}
The choice for the function $f_f(x)$ is also influenced by plausible
expectations for the opposite situation with $x$ approaching unity.
As was mentioned in section~\ref{subsec:IPDPD}, a struck quark with large
$x$ is expected to have a small impact parameter $\bT$ with respect to
the nucleon centre of momentum given by (\ref{eq:Rperp}).  With
the center of momentum of the spectator collection given by
$-\bT x/(1-x)$, the separation $\bT/(1-x)$ between quark and 
spectators can be interpreted as a lower limit on the transverse
size of the nucleon, as illustrated in figure~\ref{fig:ImpactParam}.
If the size of this confined system is not to diverge as $x\to 1$,
then $\langle \bT^2 \rangle$ must vanish at least as rapidly as 
$(1-x)^2$~\cite{Burkardt:2004bv,Burkardt:2001ni}.  
This implies that $f_f(x) \to A_f(1-x)^n$ as $x\to1$, 
with $n\geq 2$.  With $n=2$, this as well as the above constraint
for small $x$ are respectively satisfied by the third and first
terms in the form
\be
        \label{eq:GPDFFn2}
f_f(x) = \alpha' (1-x)^{3} \log\frac{1}{x} + B_f (1-x)^{3} 
        + A_f \, x (1-x)^{2}\,,
\ee
where the second term provides a smooth interpolation between them,
and $\alpha'$, $A_f$ and $B_f$ are free parameters to be fitted.
Using in (\ref{eq:ExpEntangl}) and (\ref{eq:GPD1}) this form for 
$H^{f(-)}(x,0,t)$, or a similar form for $E^{f(-)}(x,0,t)$
together with a simple ansatz for its unknown forward distribution
constrained by the known magnetic moments of the proton and neutron,
a good fit is obtained of existing data for both Dirac and Pauli elastic 
form factors of the nucleon.  For $x<0.8$, $H^{u(-)}(x,0,t)$ is well
determined, thereby determining the mean square impact parameter $\bT$
of $u$ quarks as a function of $x$ in this range. As illustrated
in figure~\ref{fig:n2_bf}, there are clear
indications that the transverse distribution of $d$ quarks is 
broader than that of $u$ quarks at large $x$.  While the fits
for $E^{f(-)}(x,0,t)$ are ambiguous due to the unknown forward
distributions, their second $x$-moments appearing in the Ji relation
(\ref{eq:Ji_SumRule}) are reasonably well determined within the
context of the assumed ansatz.  It is found that the contributions of
the orbital angular momenta of $u$ and $d$ valence quarks to the 
nucleon spin almost cancel, leaving a small net contribution.
In summary, a remarkable amount of information about `nucleon tomography'
was gleaned from only data for nucleon elastic form factors.

In the following, two specific parametric models for quark GPDs that have 
been most frequently used (in several variants) for comparisons to DVCS data
are briefly described as examples. Both are based on quark GPD 
parameterizations that might in principle be employed at any order in
$\alpha_s$, while the phenomenology has been done at only leading order. 
The remainder of this section has limited goals as a reference tool for 
continuing students of the field, and might be omitted on a first reading.
The first example, the `VGG-model'~\cite{Vanderhaeghen:1999xj,Goeke:2001tz}, 
is based on an ansatz that factorizes the dependences on $\xi$ and $t$ in
the framework of double distributions including the earlier mentioned D-term 
(see section~\ref{subsec:GPD_Properties}). Here the flavour dependence of the 
D-term is taken from the chiral quark soliton model~\cite{Petrov:1998kf},
resulting in a somewhat modified expression for a generic unpolarised 
quark GPD $F$ $(F=H,E)$:
\be
\!\!\!\!\!\!\!\!\!\!\!\!\!\!\!\!\!\!\!\!\!\!\!\!\!\!\!\!\!
F^f(x,\xi,t)=\int_{-1}^1 \de\beta \int_{-1+|\beta|}^{1-|\beta|} \de\alpha \,
             \delta(x-\beta-\xi\alpha)\,F^f_{DD}(\beta,\alpha,t)
             \pm \frac{1}{N_f} \theta(\xi-|x|) \; D(\frac{x}{\xi},t),
\label{eq:DDansatzVGG}
\ee
where the upper (lower) sign applies for the GPD $H$ ($E$).
The  model ansatz proposed in \cite{Musatov:1999xp} is invoked to write 
the DD in (\ref{eq:DDansatzVGG}) as a product
\be
F_{DD}^f(\beta,\alpha,t)=h^{(b)}(\beta,\alpha)\ F^f_{\rm forw}(\beta,0,t)
\label{eq:DDansatz2}
\ee
of a `forward' ($\xi$-independent) quark GPD $F^f_{\rm forw}(\beta,0,t)$
with a `profile function'
\be
h^{(b)}(\beta,\alpha) \propto \frac{\left[ 1-|\beta|^2-\alpha^2 \right]^b}
                              {(1-|\beta|)^{2b+1}}
\label{eq:DDansatz3}
\ee
that controls the skewness dependence through the 
(free) profile parameter $b$, with $b \rightarrow \infty$ approaching the 
forward limits (\ref{eq:GPD-PDF-H}) and (\ref{eq:GPD-PDF-Htilde}) for 
$H^f_{DD}(x,0,t)$ and $\widetilde{H}_{DD}^f(x,0,t)$, respectively. 
Often profile parameters are chosen separately for valence and sea quarks. We 
note that for $b \rightarrow \infty$ this ansatz for $F_{DD}(x,\xi,t)$ is 
singular at $x=0$ for finite $\xi$, which is physically implausible. While the 
clear advantage of the ansatz (\ref{eq:DDansatz2},\ref{eq:DDansatz3}) is its 
simplicity, it is considered to be too restrictive for flexible modelling 
of the $(\xi,t)$ dependence of quark 
GPDs~\cite{Polyakov:2002wz,Kumericki:2008di}, especially at small values of 
$\xi$.  The $(x,t)$ dependence is based on either a factorized or the above-mentioned
Regge-inspired ansatz.

The quark GPD $\widetilde{E}^f$ is evaluated from the `pion pole', 
which is an ansatz based on the assumption that the dominant contribution
at small $|t|$ arises from the emission of a virtual pion by the target
nucleon~\cite{Mankiewicz:1998kg,Penttinen:1999th}, which for small $|t|$ 
is not far off-shell because of the small mass of the pion.  
The `pole' in the pion-exchange $t$ dependence $1/(m_\pi^2-t)$ appears
in the unphysical region at $t=m_\pi^2$.  This ansatz contributes to only the
real part and only in the ERBL kinematic region.  Also the pion quantum
numbers limit this contribution to only the C-even part of the isovector
combination  $\widetilde{E}^{u-d}$ and to the corresponding proton-neutron
transition GPDs.  For the unknown forward (spin-flip) PDF of the 
GPD $E^f$, a parametric combination of valence and sea quark `PDFs' is used 
with a functional form inspired by results obtained in \cite{Goeke:2001tz} 
using the chiral quark soliton model~\cite{Diakonov:1987ty}. In this 
parametric 
combination, the quark total angular momenta $J_f$ of quarks and antiquarks 
of flavour $f$ ($f=u,d$) enter as model parameters, so that the derivation of 
model-dependent constraints on $J_f$ becomes possible by comparing predictions 
to experimental data, as described in 
section~\ref{subsubsec-exp:trans-target-spin}. The VGG model, after its
implementation in the `VGG code'~\cite{Vanderhaeghen:1999xj}, was widely 
compared to polarised cross sections and asymmetries in beam spin and charge 
at fixed-target energies and found to describe essential features of the data,
especially when using the Regge-inspired ansatz for the $(x,t)$ dependence.
(see section~\ref{subsubsec-exp:low-energy-cross-section}). As this model 
inherited the above mentioned limitations of the adopted DD ansatz, it cannot
describe unpolarized cross sections at collider energies where the skewness 
is small. It was not studied how seriously its limitations may affect the 
various interpretations of experimental data in terms of this model. For a 
similar model~\cite{Belitsky:2001ns}, the effects of NLO corrections to the 
quark CFFs were found to be significant but moderate, so that they need to be 
included when extracting model parameters from precise data, while LO 
calculations were considered to be sufficient for the purpose of estimating 
observables.

The so-called `dual-parameterization' model, presented as the second example, 
is based on the idea of duality in soft hadron-hadron interactions, which is 
the assumption that the 2 \(\rightarrow\) 2 scattering amplitude in the 
$s$-channel can be represented as an infinite sum of $t$-channel 
exchanges~\cite{deAlfaro:1973}. In the case of quark GPDs, this idea is 
implemented in the form of a $t$-channel partial wave expansion of the Mellin 
moments~\cite{Polyakov:1998ze}.
In addition, by using Gegenbauer moments instead of Mellin
moments, one can diagonalize the leading-order QCD evolution kernel, 
which is a major advantage of this approach. After resumming the partial
wave expansion for the Gegenbauer moments, one arrives at an expansion for a 
quark GPD $F^f$ in which each subsequent term is suppressed by two powers of 
the skewness parameter $\xi$:
\be
F^f(x,\xi,t) = K_0 Q^f_0(x,t) + \xi^2 K_2 Q^f_2(x,t) +
\xi^4 K_4 Q^f_4(x,t)+ \ldots
\label{eq:GPDexpansion}
\ee
Here the $K_k$ represent parameter-free $\xi$-dependent (but 
$t$-independent) integral transformations acting on the variable $x$. The 
functions $Q^f_k(x,t)$ are called `forward-like' functions as their LO 
evolution is governed by the DGLAP evolution equations, just as for PDFs. 
The parameters of the model appear in these functions. In the ($t$-dependent) 
`forward' limit $\xi=0$, (\ref{eq:GPDexpansion}) simplifies to only the first 
term and the $K_0$ becomes simple enough to invert. In the case of the quark
GPD $H^f$, the functions $Q^f_0(x,t)$ can thus be related to 
$q_f(x,t) \equiv H^{f(+)}(x,0,t)$: 
\be
Q^f_0(x,t) = (q_f(x,t)+\bar{q}_f(x,t))-\frac{x}{2} 
             \int_x^1 \frac{\de \zeta}{\zeta^2} (q_f(\zeta,t)+\bar{q}_f(\zeta,t)),
\ee
and $Q^f_0(x,0)$ is thus constrained in terms of the PDFs. An analogous 
expression holds for the GPDs $E^{f(+)}(x,\xi,t)$ with ($t$-dependent) 
`forward' distributions $e_f(x,t)$. The expansion in (\ref{eq:GPDexpansion}) is
introduced in such a way that as $k$ increases their terms are increasingly 
suppressed for small skewness $\xi$ and fixed $x$. It remains to be shown 
that this expansion in $\xi$ gives a good approximation for fixed $x/\xi$ 
when truncated at finite $k$~\cite{Diehl_PrivComm}.

The leading-order DVCS amplitude relates to quark GPDs through the convolution 
integral:
\be
{\cal A}^f(\xi,t) \stackrel{\rm LO}{=} \int_0^1 \de x \, F^{f(+)}(x,\xi,t)\left[
\frac{1}{x-\xi+i0}+\frac{1}{x+\xi-i0}\right].
\label{eq:singular}
\ee
Another major advantage of the dual parameterization is that this
singular integration can be performed analytically
in terms of the $Q_k^f$ \cite{Polyakov:2002wz}:
\bea
\label{eq:singular2}
\!\!\!\!\!\!\!\!\!\!\!\!\!\!\!\!\!
{\rm Im} \, {\cal A}^f(\xi,t) &\stackrel{\rm LO}{=}& -2 
\int_{\bar{x}}^1 
\frac{\de x}{x}N^f(x,t)  
\frac{1}{\sqrt{\frac{2x}{\xi} - x^2 - 1}},\\
\!\!\!\!\!\!\!\!\!\!\!\!\!\!\!\!\!
{\rm Re} \, {\cal A}^f(\xi,t) &\stackrel{\rm LO}{=}& 
-2 \int_0^{\bar{x}}
\frac{\de x}{x}N^f(x,t)\left[ \frac{1}{1-\sqrt{\frac{2x}{\xi} + x^2}}
+ \frac{1}{1+\sqrt{\frac{2x}{\xi} + x^2}} - \frac{2}{1+x^2}\right]
\nonumber\\
\!\!\!\!\!\!\!\!\!\!\!\!\!\!\!\!\!
& &-2\int_{\bar{x}}^1
\frac{\de x}{x}  N^f(x,t)
\left[\frac{1}{1+\sqrt{\frac{2x}{\xi} + x^2}}- \frac{2}{1+x^2}\right]
- 4 {\cal D}^f(t),
\label{eq:singular3}
\eea
where $\bar{x} =\frac{1-\sqrt{1-\xi^2}}{\xi}$,
$N^f(x,t)\equiv \sum_{k=0}^\infty x^k Q_k^f(x,t)$ and ${\cal D}^f(t)$ is the
$D$-term form factor. The integral transform in (\ref{eq:singular2}) can be 
inverted analytically, yielding
\bea
\!\!\!\!\!\!\!\!\!\!\!\!\!\!\!\!\!\!\!\!
N^f(x,t) = \frac{1}{\pi} \frac{x(1-x^2)}{\left(1+x^2\right)^{3/2}}
\int_{x_0}^1 \frac{\de \xi}{\xi^{3/2}} \frac{1}{\sqrt{\xi- x_0}}
\left[ \frac{1}{2} {\rm Im} {\cal A}^f(\xi,t) - \xi \frac{\de}{\de \xi}{\rm Im} 
{\cal A}^f(\xi,t)\right],
\label{eq:singularinverse}
\eea
with $x_0 =\frac{2x}{1+x^2}$.
This result can be summarized in the following statements that all hold to
leading order only:
\begin{itemize}
\item The imaginary part of the DVCS amplitude alone (but for all $\xi$)
is sufficient to determine $N^f(x,t)$. If data are available only above a 
certain value of $\xi$, one can still reconstruct $N^f(x,t)$ above a certain 
value of $x$. In principle, \ie\ given ideal data (negligible uncertainties), 
this reconstruction can be done exactly.
\item Upon inserting $N^f(x,t)$ from (\ref{eq:singularinverse}) into 
(\ref{eq:singular3}), data for the real part of the Compton amplitude
can be used to determine the D-term form factor. Given ideal data with 
full coverage in $\xi$, the only new information contained in 
${\rm Re}\,{\cal A}^f(\xi,t)$ that was not already contained in ${\rm Im}\,{\cal A}^f(\xi,t)$ 
is the D-term form factor. Of course, as ideal data are unfortunately
unavailable, (\ref{eq:singular3}) also provides useful constraints on $N^f(x,t)$ 
as the D-term form factor does not depend on $\xi$.
\item At fixed $Q^2$, $N^f(x,t)$ and the D-term form factor constitute the only 
information about quark GPDs that can be obtained from DVCS. However, QCD 
evolution can be a viable tool for the extraction of quark GPDs, as the LO 
evolution of the $Q_{2k}^f$ is known to be governed by
the LO DGLAP evolution equations. Since the $Q_{2k}^f(x,t)$ enter $N^f(x,t)$
multiplied by $x^{2k}$, the $n$'th Mellin moment of $N^f(x,t)$ will depend
on the $(n+2k)$'th Mellin moment of the $Q_{2k}^f$. Since the anomalous
dimensions that govern the LO evolution depend on the order of the moment
(and only on the order), this implies that at different $Q^2$ the Mellin
moments of $Q_{2k}^f$ (at some renormalization scale $Q^2_0$) enter 
$N^f(x,t)$ with different relative weights. At least in principle, DVCS data 
taken at a set of different values of $Q^2$ thus provide a system of linear equations 
for the Mellin moments of the  $Q_{2k}^f$ that (after suitable truncation) can 
be solved, which in turn can be used to reconstruct the quark GPD.
\end{itemize}
Unfortunately, not only is there a lack of ideal data that are required for 
the above procedure to work, but also LO evolution is likely to be 
insufficient at presently accessible values of $Q^2$. An approach that is 
closely related to \cite{Polyakov:2002wz} is that of 
\cite{Mueller:2005ed,PassekKumericki:2008sj}, which uses conformal symmetry to 
diagonalize the evolution up to NLO and may thus overcome the above 
discussed limitations of the LO treatment.

The dual parameterization does not model the $t$ dependence of the
GPDs, which must be specified separately.
Present applications of the model do not include the quark GPDs 
$\widetilde{H}$ and $\widetilde{E}^f$ nor higher-twist contributions. The model
is worked out only to leading-order accuracy, as only then the DVCS amplitude 
has a simple form in terms of the $Q^f_k(t)$, and also their evolution can 
be handled easily. The generalization from LO to NLO is far from obvious. 
On the other hand, an 
advantage of this approach is the clear separation of the ($\xi$-independent) 
`forward' distributions $q_f(x,t)$ [$e_f(x,t)$] of the quark GPDs 
$H^f(x,\xi,t)$ 
[$E^f(x,\xi,t)$] from genuine non-forward effects encoded in the distribution 
of strength among the functions $Q^f_2, Q^f_4, \ldots$ in (\ref{eq:GPDexpansion}).
Restricting oneself to the two lowest functions ($k=0,2$) yields the so-called 
`minimal version' of the model~\cite{Polyakov:1998ze}.

Also assuming in this `minimal version' $Q_2^f(x,0) \propto Q_0^f(x,0)$, 
it was initially found that the $Q^2$ and $W$ dependences of the 
low-$x$ HERA collider data were fairly well described~\cite{Guzey:2005ec}, 
until a missing factor was noticed, leading to a cross section magnitude 
discrepancy of about a factor of four~\cite{Guzey:2008ys}. 
In \cite{Polyakov:2008xm}, calculations in the minimal version using 
{\em only} the lowest function $Q_0^f$ (called `zero-step' approach) as well 
as calculations of the VGG model were recently compared to \jlab\ data on 
cross sections and asymmetries (see section \ref{subsubsec-exp:beam-spin},
\eg\ figure \ref{fig:Polyakov+Vand_A_LU_JLab_data}).  These comparisons
are unaffected by that missing factor.

No global fits to all existing DVCS data yet exist. Future algorithms must
necessarily include enough orders beyond LO to ensure convergence. The 
only existing approach fulfilling this requirement is a recently 
developed formalism~\cite{Kumericki:2007sa} based on dispersion relations 
discussed in section~\ref{subsec:GPD_Access_DVCS}. In this new formalism, 
dispersion relation and operator expansion techniques are combined leading to 
convergence of the conformal partial wave expansion of the VCS amplitude. 
This formalism was successfully applied to fit DVCS observables in NNLO at 
low $x_B$~\cite{Kumericki:2007sa}, as described in 
section~\ref{subsubsec-exp:high-energy-cross-section}. In a first 
attempt, this fit was very recently extended to include also data from 
fixed-target experiments~\cite{Kumericki:2009uq}.

%
\subsection{DVCS experimental results}
\label{subsec:GPD_ExpResults}
%
\subsubsection{DVCS cross section at high energies}
\label{subsubsec-exp:high-energy-cross-section}

The DVCS cross section (\ref{eq:DVCS-BH-amplitude}) integrated over 
its azimuthal dependence has been measured in hard exclusive
electroproduction of photons by the experiments {\sc H1} and 
{\sc Zeus} at the {\sc Hera} collider. For the $\xbj$ range 
accessible at collider kinematics (${\cal{O}}(10^{-2}$ \ldots 10$^{-4}$)), 
two-gluon exchange plays a major role besides the quark-exchange 
handbag diagram of figure~\ref{fig:handbagNEW}, so that 
{\em both} gluon and quark GPDs are probed simultaneously. This is 
however limited to small skewness values below 10$^{-2}$.

The analysis method used by the two experiments is similar due to their 
similar geometries. As the outgoing proton
remains undetected in the beam pipe, the event topology is
defined by two electromagnetic clusters, the outgoing lepton and 
the produced real photon, at most one of which is associated with a 
charged track. Two events samples are selected:

\noindent i) in the `DVCS-enriched' sample an energetic photon is required. 
With respect to the incoming lepton it is required to be detected 
at a large scattering angle while the outgoing lepton is measured at
a small angle. Still, a high enough virtuality $Q^2 > 4$~GeV$^2$ is ensured 
by requiring a large energy of the scattered lepton ($> 15$~GeV).

\noindent ii) in the Bethe-Heitler dominated `reference sample', the 
radiatively produced photon is emitted at a small angle with respect to the 
incoming lepton, and the outgoing lepton at a large angle. 

A Monte Carlo simulation of the completely known BH process, validated by
the reference sample, is used to subtract the BH contribution from 
the DVCS-enriched sample. The remainder of the spectrum is due to DVCS and 
possible additional background; no contribution from the interference term 
exists at leading twist, as the data are integrated over the azimuthal angle. 

\begin{figure}[htb]
\vspace*{3mm}
\begin{center}
\begin{minipage}[b]{.48\linewidth}
\epsfig{file=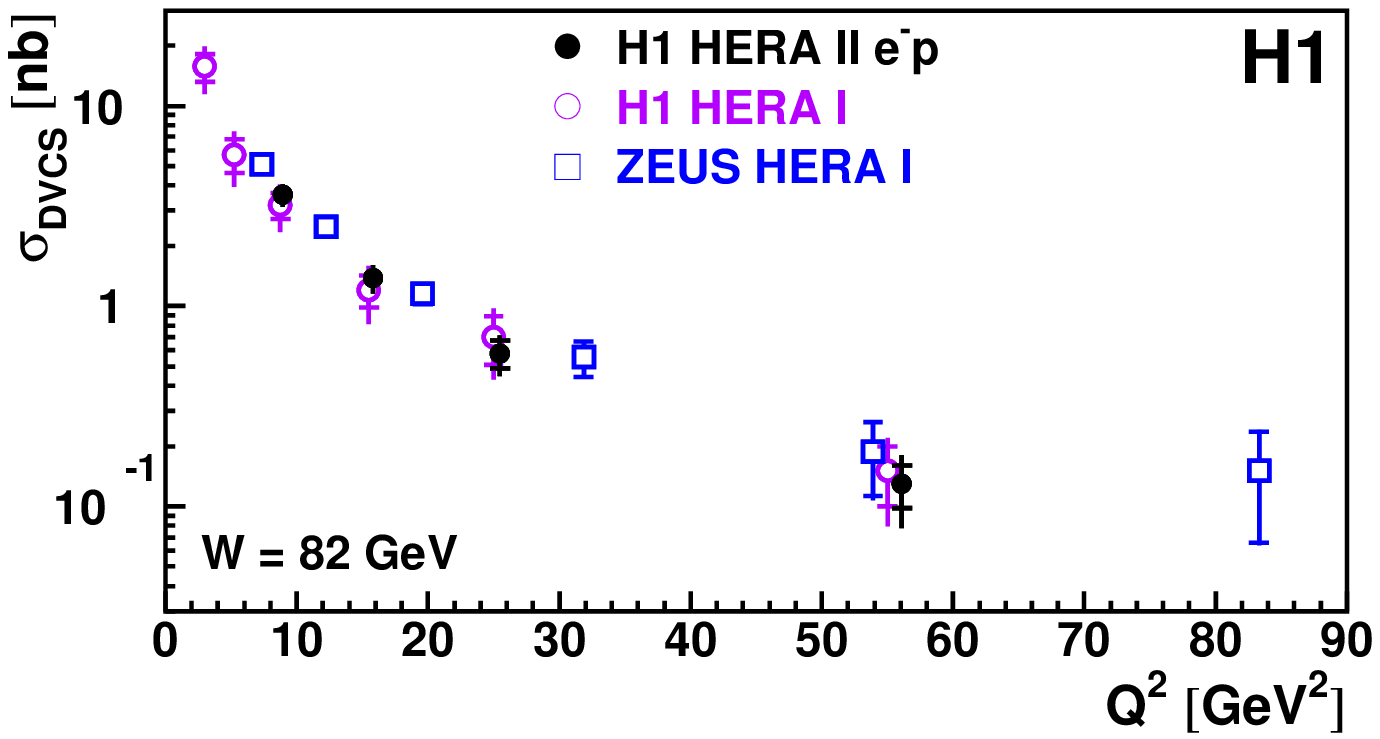,width=7.5cm}
\end{minipage} \hspace*{2mm}
\begin{minipage}[b]{.48\linewidth}
\epsfig{file=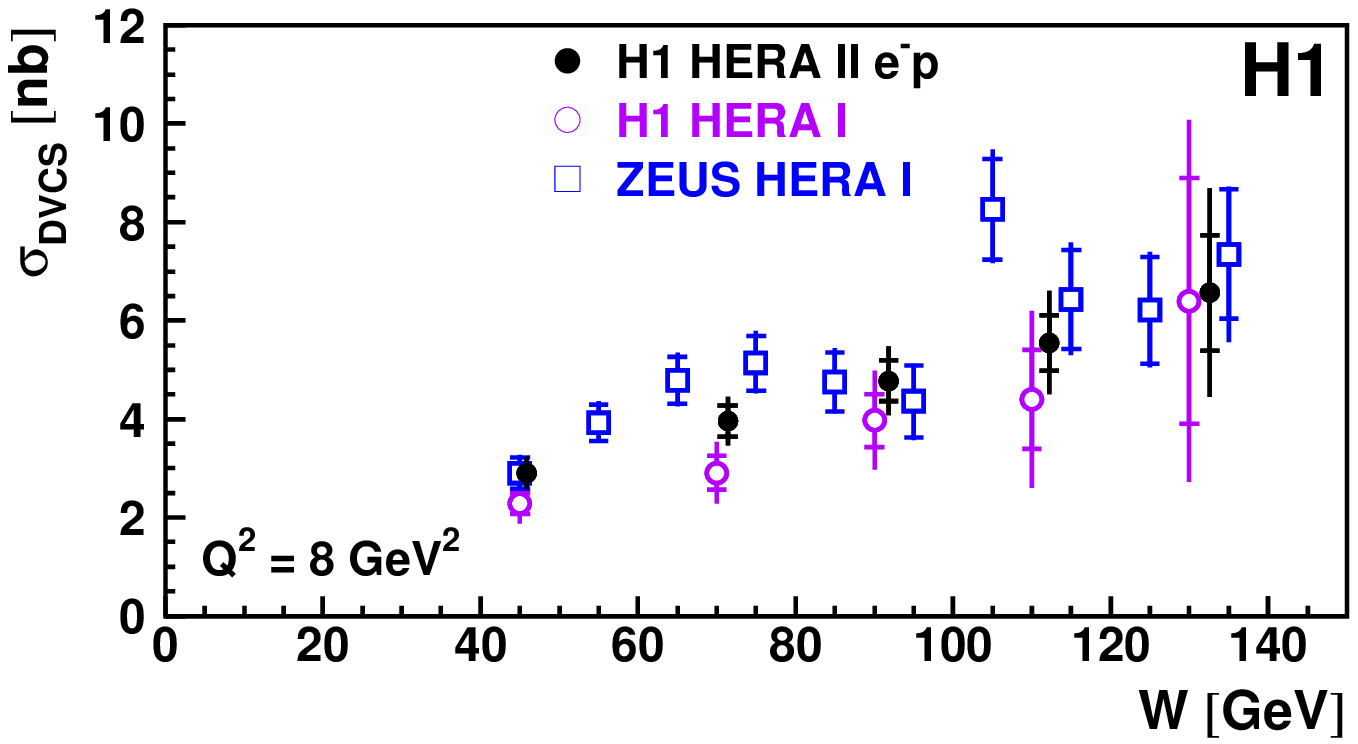,width=7.5cm}
\end{minipage}
\end{center}
\caption{\label{fig:H1+ZeusXsect_vsQ2andW}
Left: $Q^2$-dependence of the differential 
$\gamma^* \, p \rightarrow \gamma \, p$ cross section measured by {\sc H1} 
and {\sc Zeus}. Right: as left panel, but here the $W$-dependence is shown. 
The figures are taken from \cite{Aaron:2007cz}.}
\end{figure}

The  $Q^2$-dependence of the differential 
$\gamma^* \, p \rightarrow \, \gamma \, p$ cross section is shown in the 
left panel of figure~\ref{fig:H1+ZeusXsect_vsQ2andW}, extracted from 
{\sc Hera}-I and {\sc Hera}-II data of {\sc H1}~\cite{Aaron:2007cz} and 
the {\sc Hera}-I data of {\sc Zeus}~\cite{Chekanov:2003ya}, the latter 
based on a somewhat larger event sample. The corresponding $W$ dependence, 
where $W$ is the invariant mass of the system of virtual photon and proton, 
is displayed in the right panel. The virtuality appears high
enough to assign the observed rise with $W$ to the nature of DVCS as 
a hard process, as increasing $W$ implies decreasing $\xbj$, where the 
parton densities in the proton show a rapid rise. 

\begin{figure}[htb]
\begin{center}
\begin{minipage}[b]{.48\linewidth}
\epsfig{file=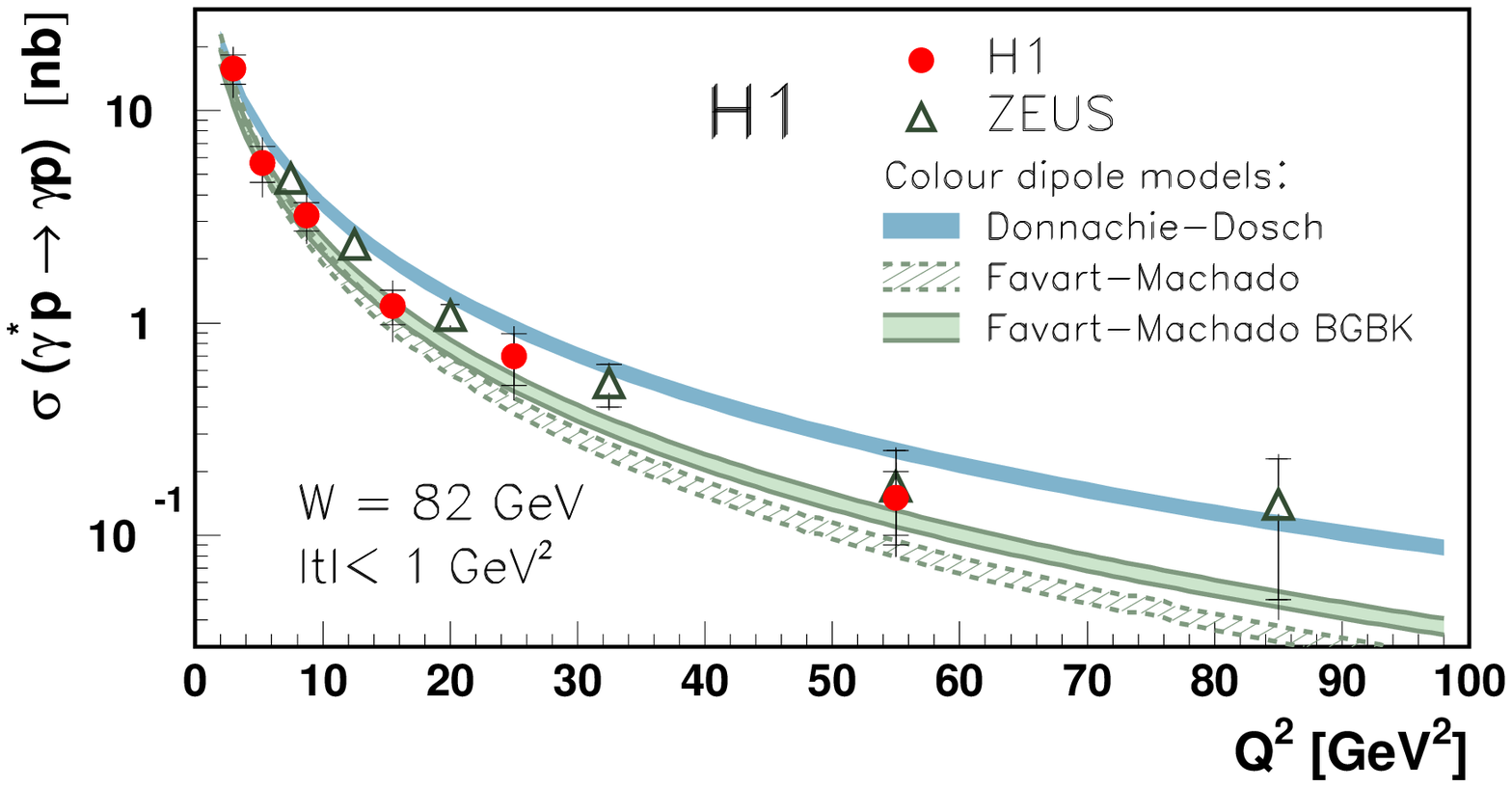,width=7.5cm}
\end{minipage} \hfill
\begin{minipage}[b]{.48\linewidth}
\epsfig{file=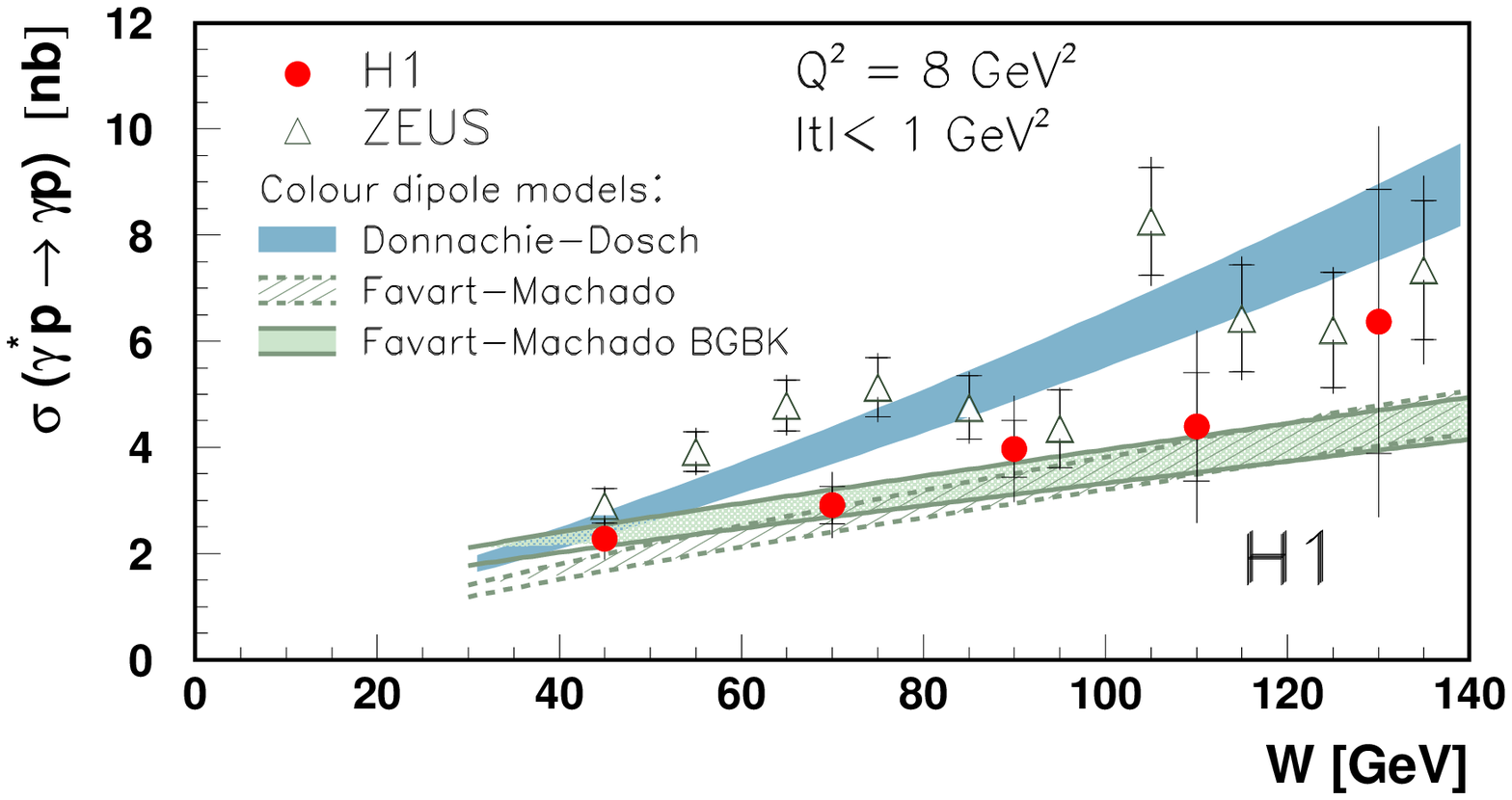,width=7.5cm}
\end{minipage}
\end{center}
\caption{\label{fig:H1+GPDvsQ2andW}
{\sc Hera}-I data shown in the previous figure in comparison to
various Colour Dipole Model calculations (see text). 
The figures are taken from \cite{Aktas:2005ty}.}
\end{figure}
In figure~\ref{fig:H1+GPDvsQ2andW}, various Colour Dipole Model 
calculations~\cite{Favart:2003cu,Favart:2004uv,Donnachie:2000px}
are compared to only the {\sc Hera}-I data of both experiments. 
Presently it is impossible to discrimate between the various 
model versions.

\begin{figure}[htb]
\begin{center}
\epsfig{file=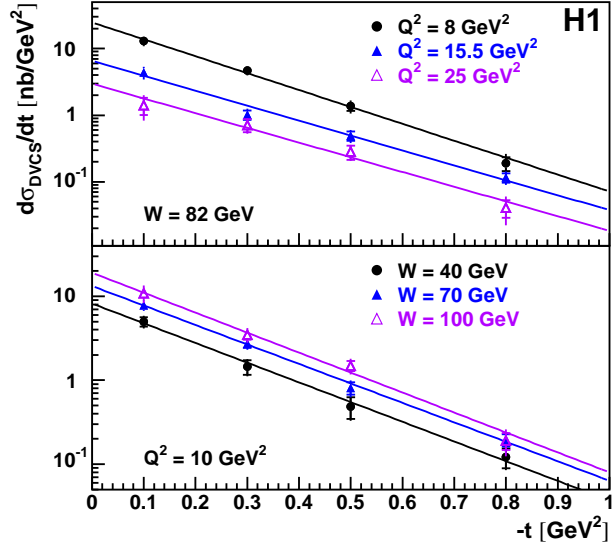,width=8cm}
\end{center}
\caption{\label{fig:H1sigDVCSvs_t}
The $-t$ dependence of the differential 
$\gamma^* \, p \rightarrow \gamma \, p$ cross section, extracted from
{\sc H1} {\sc Hera} II data at 3 values of $\langle Q^2 \rangle$.
The figure is taken from \cite{Aaron:2007cz}.}
\end{figure}
A commonly used ansatz for the $t$ dependence of quark GPDs is an exponential 
decrease $e^{-b \; |t|}$. The slope parameter $b$ was first measured in
DVCS by {\sc H1}~\cite{Aaron:2007cz}, the results being shown in the two 
left panels of figure~\ref{fig:H1sigDVCSvs_t} for $\langle Q^2 \rangle$ = 
8, 15.5 and 25 GeV$^2$ integrated over $W$, and for $\langle W \rangle$ = 
40, 70 and 100 GeV integrated over $Q^2$. Parameterizing a possible $Q^2$ 
dependence as $b(Q^2) = A \; (1-B \cdot {\rm log}(Q^2/2$ GeV$^2))$, the 
fit shows that it is indeed significant with $A = 6.98 \pm 0.54$ Gev$^2$ 
and $B = 0.12 \pm 0.03$. At $Q^2 = 8$ GeV$^2$ the result is 
$b=5.45 \pm 0.19_{stat} \pm 0.34_{syst}$. 
Following \cite{Burkardt:2002hr,Frankfurt:2005mc}, this $t$ slope can be 
converted to an average impact parameter which describes the transverse 
extension of partons. The {\sc H1} result at 
$\langle x \rangle = 1.2 \cdot 10^{-3}$, where sea quarks and gluons 
dominate, is $\sqrt{\langle r^2_T \rangle} = 0.65 \pm 0.02$ fm. This
can be compared to the transverse charge radius of the 
proton~\cite{Thomas:2001kw}:
\be
\sqrt{\langle r^2_T \rangle_{ch}} \equiv 
\sqrt{-4 \; \frac{d}{dQ^2} F_1(Q^2)|_{Q^2=0}} = 0.65 \pm 0.01 \;\; {\rm{fm}.}
\ee

Recently, the convergence of the perturbative treatment of DVCS was 
studied in the low-$\xi$ region typical for collider kinematic conditions, 
in a theoretical study using the predictive power of conformal 
symmetry~\cite{Kumericki:2006xx}. Beyond NLO, further terms in the
perturbation series are found to be small for an input scale 
of a few GeV$^2$. Concerning the scale dependence, larger corrections 
due to evolution may appear only for values of $\xi$ below 
$5 \cdot 10^{-3}$, similar to the situation in DIS~\cite{Vogt:2004mw}.

\begin{figure}[htb]
\begin{center}
\epsfig{file=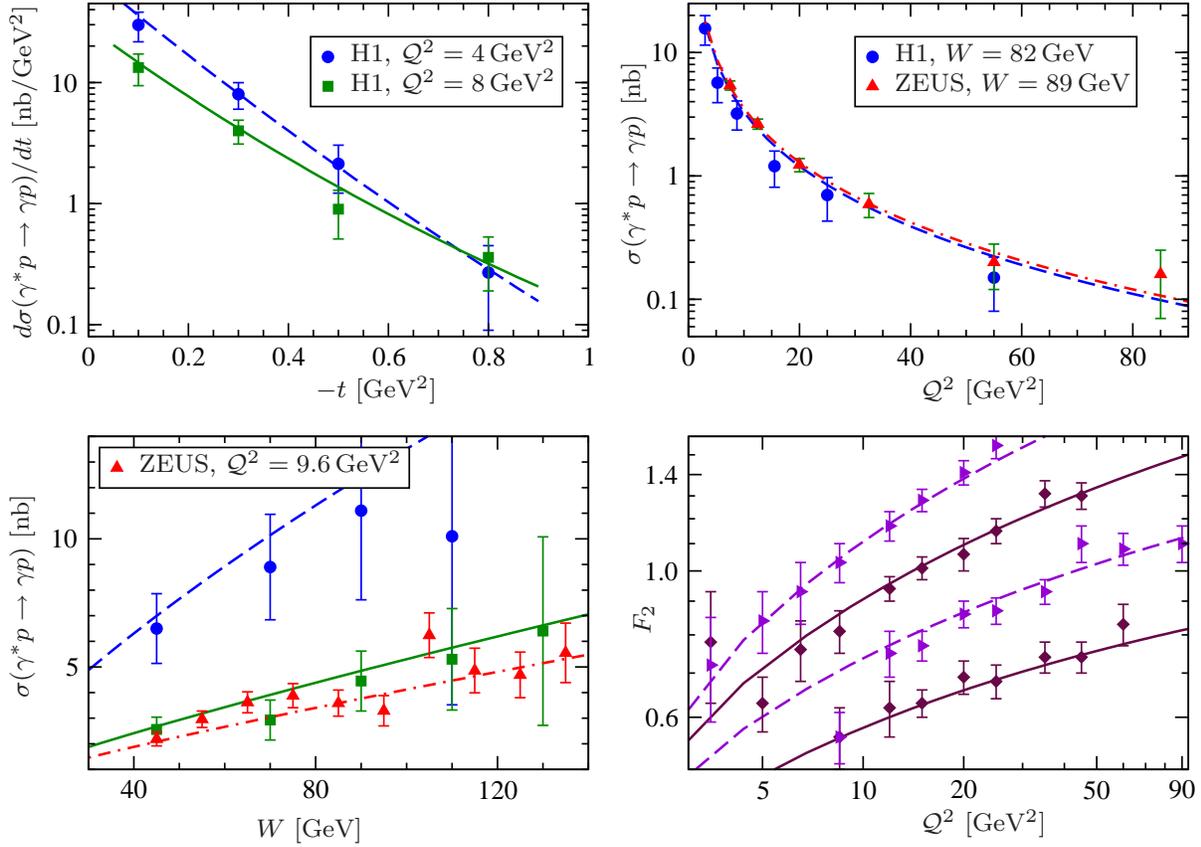,width=16cm}
\end{center}
\caption{\label{fig:Mellin-BarnesFit} Simultaneous fit to the DVCS 
and DIS data in the $\overline{\rm CS}$  scheme at NNLO. Upper left 
panel: DVCS cross section for $Q^2 =4$ GeV$^2$ and $W=71$ GeV 
(circles, dashed line) as well as $Q^2 =8$ GeV$^2$ and $W=82$ GeV
(squares, solid line)~\cite{Aktas:2005ty}. Upper right panel: DVCS
cross section ($-t < -1$ GeV$^2$) versus $Q^2$ for $W=82$ GeV (H1, 
circles, dashed line) and $W=89$ GeV (ZEUS, triangles, dash-dotted 
line)~\cite{Chekanov:2003ya}. Lower left panel: DVCS cross
section versus $W$ for $Q^2 = 4$ GeV$^2$ (H1, circles, dashed line),
$Q^2 = 8$ GeV$^2$ (H1, squares, solid line), and $Q^2 =
9.6$ GeV$^2$ (ZEUS, triangles, dash-dotted line). Lower right panel:
$F_2(\xbj,Q^2)$ versus $Q^2$ for $\xbj=8\cdot 10^{-3}, 3.2\cdot 10^{-3}, 
1.3\cdot 10^{-3}, 5\cdot 10^{-4}$~\cite{Aid:1996au}.
The figure is taken from \cite{Kumericki:2007sa}.}
\end{figure}
Based on this approach, very recently a simultaneous fit was performed 
to the DVCS and DIS collider data, for the first time up to NNLO
accuracy~\cite{Kumericki:2007sa}. The \(\overline{\rm CS}\) scheme was
used which differs from the more commonly used \(\overline{\rm MS}\)
scheme only in the skewness dependence of the conformal moments in the
Mellin-Barnes representation, the latter being used to evaluate the
Compton form factors.
In figure~\ref{fig:Mellin-BarnesFit} the fit results are compared to
the data, showing fair agreement. It is argued in \cite{Kumericki:2007sa} 
that this should be considered satisfactory, as several aspects of the
theoretical approach are not completely settled yet, in particular
regarding the skewness dependence of quarks GPDs.

\subsubsection{DVCS cross sections at low energies}
\label{subsubsec-exp:low-energy-cross-section}

The helicity-dependent DVCS cross sections 
$d^4\sigma^\Erightarrow/dQ^2d\xbj dtd\phi$ 
and $d^4\sigma^\Eleftarrow/dQ^2d\xbj dtd\phi$ were measured at \jlab\
with 5.75 GeV electron beams of positive and negative helicity,
respectively~\cite{Munoz-Camacho:2006hx}. Here the symbol \(\Erightarrow\) 
(\(\Eleftarrow\)) designates positive (negative) beam helicity.
In a three-arm kinematically complete experiment, the scattered 
electron was measured by the high-resolution spectrometer, the produced 
real photon by an electromagnetic calorimeter and the recoil proton by a
scintillator array. Three photon virtualities were selected 
with $\xbj = 0.36$ fixed: $Q^2 =$ 1.5, 1.9, 2.3 GeV$^2$. For each of them, 
four $-t$ values were chosen: 0.17, 0.23, 0.28, 0.33 GeV.

From the two measured single-helicity cross sections, the 
polarized and unpolarized DVCS cross sections were obtained:
\begin{eqnarray}
\frac{d^4\Sigma}{dQ^2d\xbj dtd\phi} & = & \frac{1}{2}
       \left( \frac{d^4\sigma^\Erightarrow}{dQ^2d\xbj dtd\phi} -
       \frac{d^4\sigma^\Eleftarrow}{dQ^2d\xbj dtd\phi} \right) 
       \propto \mathrm{Im\;\; I} \label{eq:SepXsectsDVCS} \\
       \frac{d^4\sigma}{dQ^2d\xbj dtd\phi} & = & \frac{1}{2}
       \left( \frac{d^4\sigma^\Erightarrow}{dQ^2d\xbj dtd\phi} + 
       \frac{d^4\sigma^\Eleftarrow}{dQ^2d\xbj dtd\phi} \right) 
       \propto \mathrm{Re \;\; I,} \label{eq:SepXsectsDVCS2}
\end{eqnarray}
where I is the interference term that consists of linear combinations 
of Compton Form Factors. 
Cross section measurements involve additional experimental systematics, 
as no cancellations of systematic uncertainties occur as in asymmetry 
measurements. Such measurements have the advantage that the cross 
sections in (\ref{eq:SepXsectsDVCS}) and  (\ref{eq:SepXsectsDVCS2})
are directly proportional to Im I and Re I, respectively, while in case 
of asymmetries unwanted Bethe-Heitler terms appear in the denominator.

\begin{figure}[htb]
\begin{center}
\epsfig{file=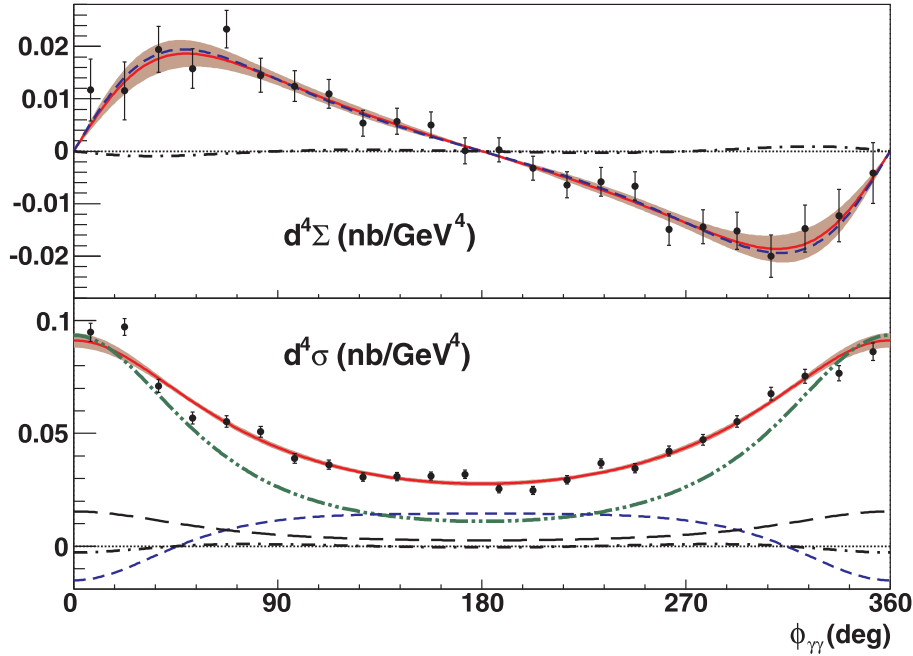,width=12cm} 
\end{center}
\caption{\label{fig:HallA_scalingTest}
\jlab\ data and fit to $d^4\Sigma/dQ^2d\xbj dtd\phi$ and 
$d^4\sigma/dQ^2d\xbj dtd\phi$, shown
as a function of the azimuthal angle $\phi$ at $Q^2=2.3,
-t=-0.28$ at $\langle\xbj\rangle=0.36$. Only statistical
errors are shown.The solid line shows the total fit with the band 
representing one standard deviation. The dot-dot-dashed line is the 
$|$BH$|^2$ squared contribution, twist-2 (twist-3) interaction terms
are shown by a long-dashed (dot-dashed) line. 
The figure is taken from~\cite{Munoz-Camacho:2006hx}.}
\end{figure}
The data for the polarized and unpolarized DVCS cross sections are 
shown in figure~\ref{fig:HallA_scalingTest} together with Fourier
decompositions according to (\ref{eq:ITtoAmpl}).
The polarized cross section exhibits the expected $\sin{\phi}$ 
behaviour. For the unpolarized one, the main contributions
stem from the $|$BH$|^2$ term and the twist-2 part of the 
interference term, while $|$DVCS$|^2$ and twist-3 contributions are 
found to be negligible. 
%
%
\subsubsection{Beam-spin asymmetry}
\label{subsubsec-exp:beam-spin}
It was shown in section~\ref{subsec:GPD_Access} that in hard exclusive 
leptoproduction of a real photon, the
interference of the Bethe-Heitler and Deeply Virtual Compton Scattering
processes allows the extraction of a wealth of information on quark GPDs. 
For the case of an {\em unpolarized} ($U$) proton target,
the {\em beam-spin asymmetry} (BSA) for a longitudinally ($L$) 
{\em polarized beam} is given by:
\be
\label{eq:BSA1}
{\cal{A}}_{LU}(\phi) = \frac
  {d \sigma^{\Erightarrow}(\phi) 
 - d \sigma^{\Eleftarrow}(\phi)}
  {d \sigma^{\Erightarrow}(\phi) 
 + d \sigma^{\Eleftarrow}(\phi)}.
\ee
Here $\Erightarrow$ ($\Eleftarrow$) denotes beam spin parallel (antiparallel)
to the beam direction.

The interference term I can be Fourier expanded in the azimuthal angle $\phi$.
Evaluating the above asymmetries to leading power in $1/Q$ in each 
contribution to this expansion and to leading order in $\alpha_S$, 
only the $\sin \phi$  ($\cos \phi$) term remains in the numerator of the 
beam-spin (beam-charge) asymmetry. To the extent that the leading BH-term 
dominates the denominator, the azimuthal dependence of the beam-spin
(beam-charge) asymmetry is reduced to $\sin \phi$ ($\cos \phi$):
\be
\label{eq:BSA2}
{\cal{A}}_{LU}(\phi) \propto \mathrm{Im} \; \widehat{M}_U \, \sin \phi. 
\ee
The BSA is sensitive to the linear combination $\widehat{M}_U$ of Compton 
Form Factors, which describes an {\em unpolarized} proton 
{\em target}~\cite{Belitsky:2001ns} (see (\ref{eq:Munpol})). Generally,
the quark GPDs $H^f$ are expected to dominate expression (\ref{eq:Munpol}), 
because i) the second term is suppressed by at least a factor of 10, as 
$\xi$ is usually not larger than 0.2 even in fixed-target kinematics,
and the unpolarized
\begin{figure}[htb] 
\begin{center}
\vspace*{-15mm} 
\begin{minipage}[h]{.48\linewidth}
\vspace*{20mm}\epsfig{file=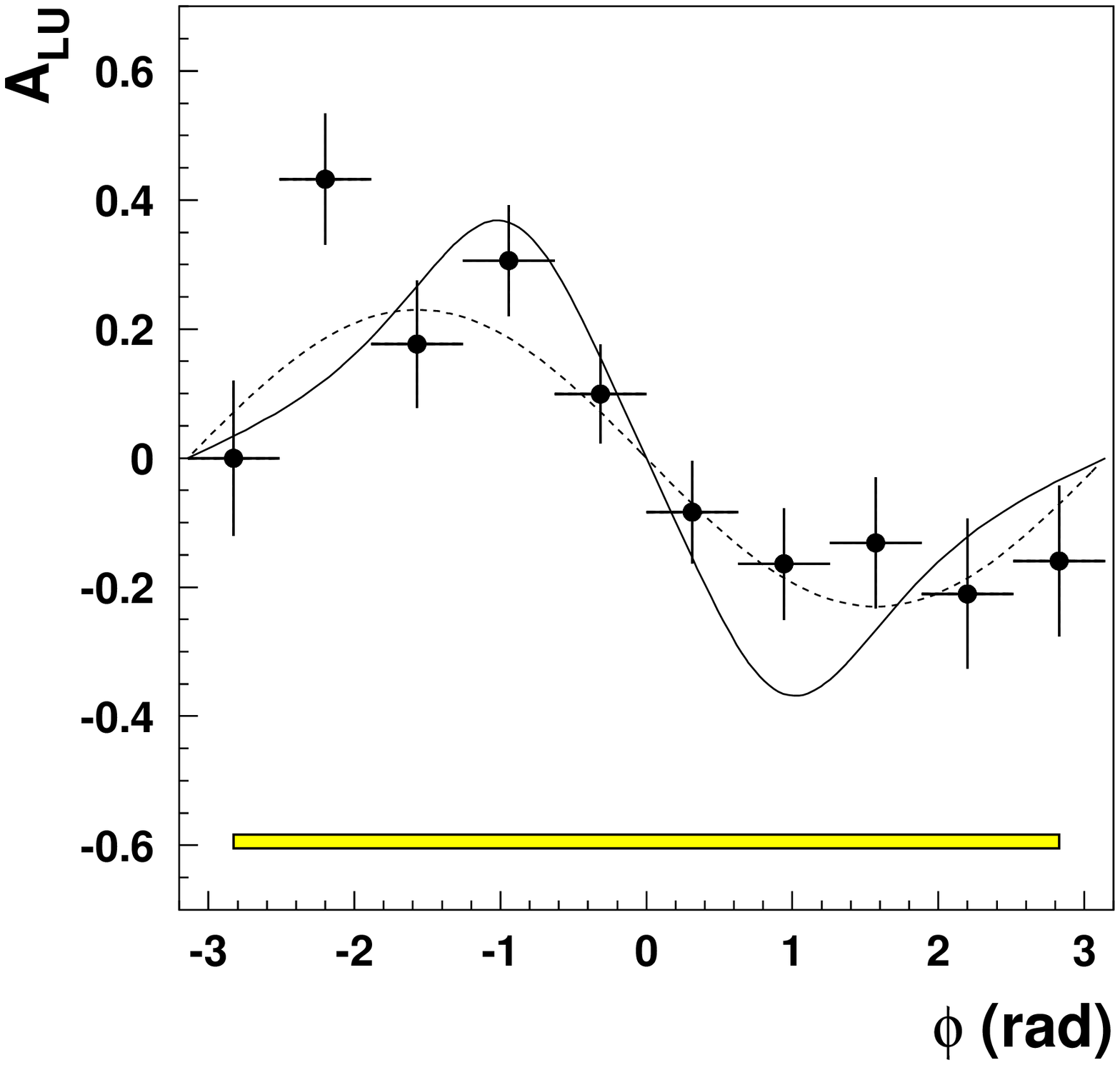,width=7cm}
\end{minipage} \hfill
\begin{minipage}[h]{.48\linewidth}
\epsfig{file=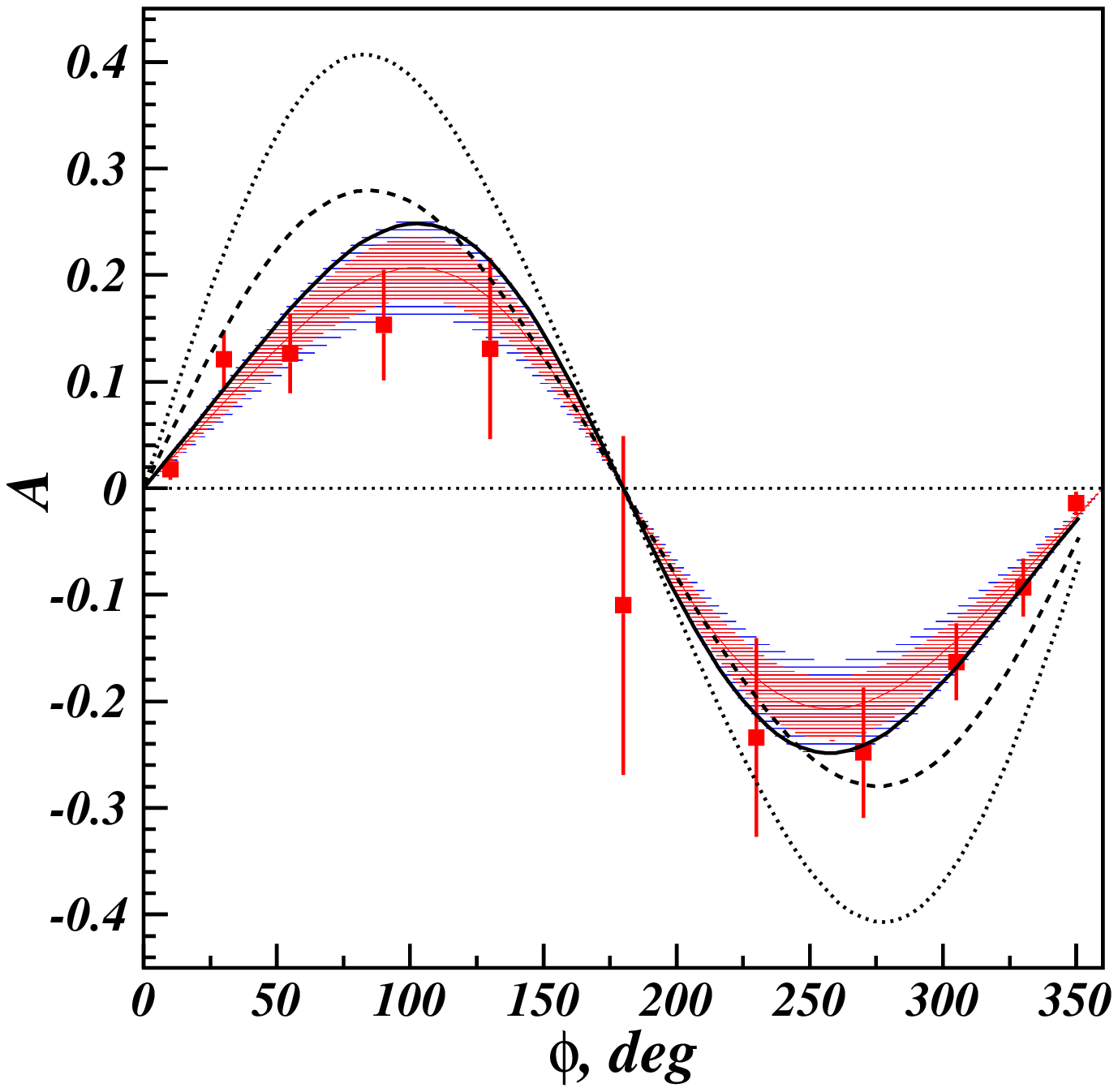,width=7cm}
\end{minipage}
\end{center}
\vspace*{-10mm}
\caption{\label{fig:BSAvsPhi}
Azimuthal dependence of the beam-spin asymmetry.
Left: \hermes\ proton data taken with 27.6~GeV positrons.
Right: \clas\ proton data taken with 4.25 GeV electrons.
The figures are taken from \cite{Airapetian:2001yk,Stepanyan:2001sm}.} 
\end{figure}
contribution $\mathcal{H}$ is expected to dominate the polarized
one $\widetilde{\mathcal{H}}$, in analogy to the forward case;
ii) the third term is $t$-suppressed, by about a factor of 25 for 
typical $t$-values of about 0.15~GeV$^2$.
For scattering on the proton, the quark GPD $H^u$ will yield the major
contribution to $\widehat{M}_U$ because $u$ quarks i) are more abundant 
in the proton and ii) more readily absorb virtual photons because of 
their 4 times larger charge-squared factor.

The first published GPD-related experimental results were 
beam-spin asymmetries measured in DVCS on the proton by the 
fixed-target experiments {\sc Hermes} at {\sc Hera}~\cite{Airapetian:2001yk} 
with a positron beam and by {\sc Clas} at Jefferson 
Laboratory~\cite{Stepanyan:2001sm} with an electron beam (see 
figure~\ref{fig:BSAvsPhi}).
Note that opposite beam charges imply opposite signs of the measured 
beam spin asymmetries (this is not immediately apparent in 
figure~\ref{fig:BSAvsPhi} because of different $\phi$-ranges shown 
for \hermes\ and \clas). In both experiments significant sinusoidal 
modulations have been observed, as expected from 
the theoretical considerations that were briefly explained above.

DVCS on the proton was recently measured at \jlab\ with a 5.77 GeV 
polarised electron beam in a kinematically complete experiment using the 
\begin{figure}[htb]
\begin{center}
\epsfig{file=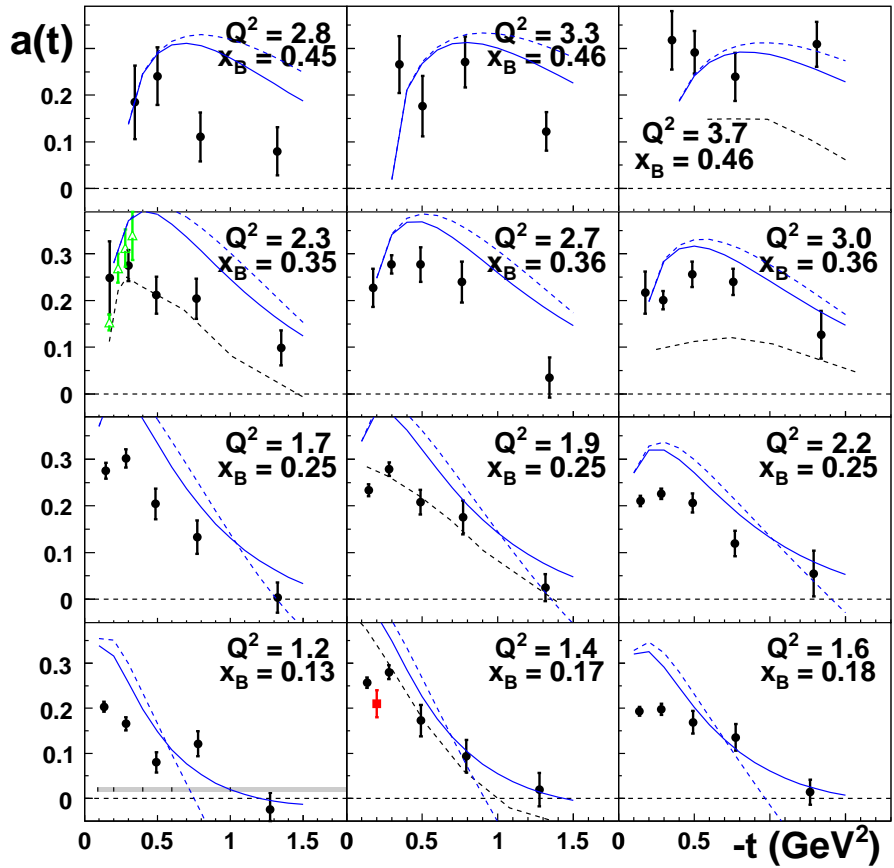,width=12cm} 
\end{center}
\caption{\label{fig:HallA_A_LU_vs_t}
Beam-spin asymmetry measured at \jlab\ as a function of $t$ and shown 
at $\phi = 90^o$ in several
$(\xbj,Q^2)$ bins. Systematic uncertainties and bin limits are shown by the 
(grey) band in the lowest left panel. (Black) circles are from \cite{:2007jq},
the (red) square is earlier data~\cite{Stepanyan:2001sm} and the 
open (green) triangle is the cross section data~\cite{Munoz-Camacho:2006hx}
described in the previous section. The curves are described in the text. The
figure is taken from \cite{:2007jq}.}
\end{figure}
large-acceptance \clas\ spectrometer. Beam-spin asymmetry results were shown
in a simultaneous binning in $x,t$ and $Q^2$ (62 bins in total) (see 
figure~\ref{fig:HallA_A_LU_vs_t}~\cite{:2007jq}). 
The background from undetected
exclusive $\pi^0$'s decaying asymmetrically was estimated using measured
symmetric decays and their relative acceptance function determined by Monte
Carlo techniques. It was subtracted in every kinematic bin, over which it 
varies from 1 to 25\% (5\% on average).
Calculations using the earlier described VGG model including only the 
contribution of the quark GPD $H$, shown as two neighbouring lines,
solid for twist-2 and dashed for twist-3, clearly overshoot most of the data. 
Calculations in a Regge model~\cite{Laget:2007qm}, shown for comparison in a 
few panels as the lower dashed curve, describe the data well in certain bins but not in others.

For a few particular bins, the azimuthal dependence of the same data was 
compared in \cite{Polyakov:2008xm} to the VGG and the dual-parameterization 
models that were described in section~\ref{subsec:GPD_Models}. For both 
models, the same
\begin{figure}[htb]
\begin{center}
\epsfig{file=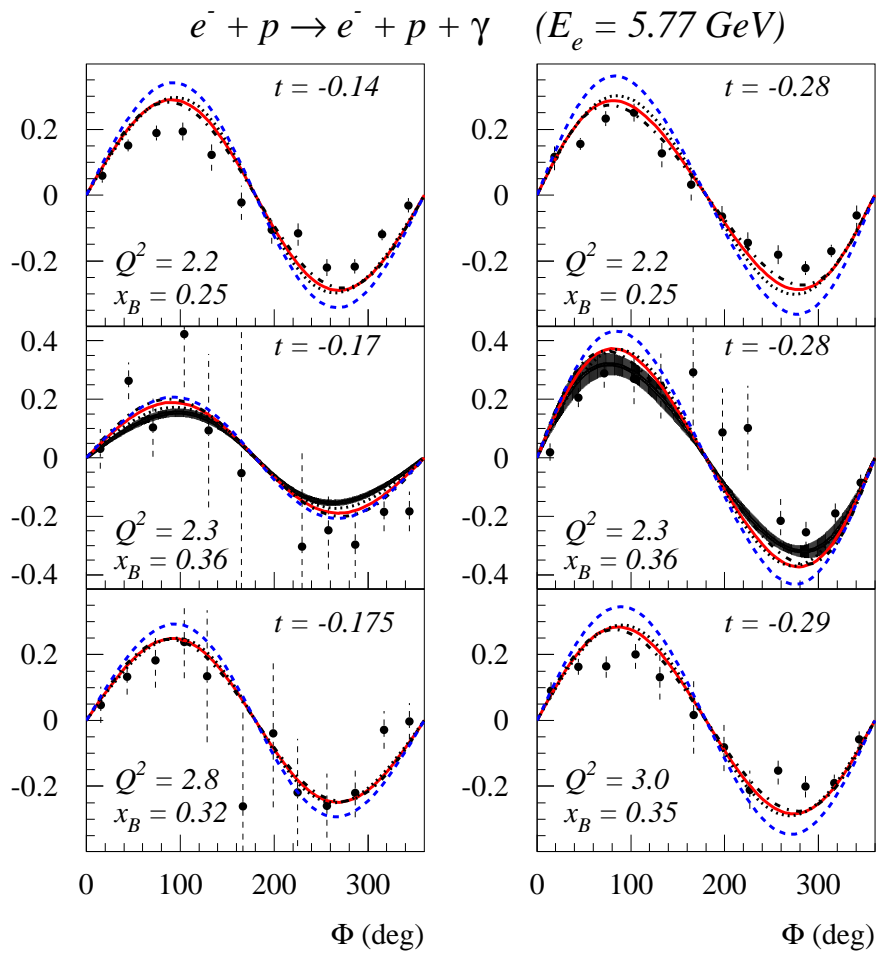,width=12cm} 
\end{center}
\caption{\label{fig:Polyakov+Vand_A_LU_JLab_data}
Azimuthal dependence of the beam-spin asymmetry in DVCS in several kinematic
bins. The black bands in the two middle panels are \jlab\ data from
\cite{Munoz-Camacho:2006hx}, while the data points are \jlab\ data from
\cite{:2007jq}. The curves are explained in the text.
The figure is taken from~\cite{Polyakov:2008xm}.}
\end{figure}
unfactorised ansatz (\ref{eq:unfactorized_Ansatz}) was used. For simplicity, 
only the BSA is shown in figure~\ref{fig:Polyakov+Vand_A_LU_JLab_data}, 
although in \cite{Polyakov:2008xm} also the polarised and unpolarised cross 
sections are discussed. The VGG calculation, for which $b=1$ was used as 
profile parameter, overshoots the data as already discussed in connection with
the previous figure. The dual-parameterization calculation, done with only the
forward function $Q_0$, also overshoots the data, although less severely. 
Adding some estimate for the non-forward function $Q_2$ does not 
improve the situation. Using the values $0,\pm \frac{4}{3}$ for the 
D-term form factor ${\cal{D}}(t)|_{t=0}$ that was mentioned after 
(\ref{eq:singularinverse}) does not yield a simultaneous consistent 
description of the $\phi$ dependence of the unpolarised and polarised cross 
sections and the BSA~\cite{Polyakov:2008xm}. Such a description can
not be achieved by any of the presently existing GPD models.
A similar conclusion is reached in \cite{Guidal:2008jv} after attempting a 
combined fit of the two \jlab\ data sets~\cite{Munoz-Camacho:2006hx,:2007jq}
to variants of the VGG model. It has not been tried to estimate the possible 
influence of higher-order effects for either of the two models.
%
%
%
\subsubsection{Beam-charge asymmetry}
\label{subsubsec-exp:beam-charge}
The measurement of a {\em beam-charge asymmetry} (BCA)
\begin{equation}
\label{eq:BCA1}
{\cal{A}}_{C}(\phi) = \frac
  {d \sigma^+(\phi) - d \sigma^-(\phi)}
  {d \sigma^+(\phi) + d \sigma^-(\phi)},
\end{equation}
where the superscripts $+$ and $-$ denote the lepton beam charge,
requires data for both beam charges. Alternatively, a BCA can also be
derived from the $\cos \phi$ dependence of the cross section for one beam 
charge in an hermetic detector of known efficiency. Using the same 
approximations as in the previous section, the azimuthal dependence of 
the beam-charge asymmetry is reduced to $\cos \phi$:
\be
\label{eq:BCA2}
{\cal{A}}_{C}(\phi)  \propto \mathrm{Re} \; \widehat{M}_U \, \cos \phi.
\end{equation}
The BCA is sensitive to the real part of the {\em same} linear combination 
$\widehat{M}_U$ (\ref{eq:Munpol}), of which the imaginary part determines
the BSA described in the previous section.
\begin{figure}[htb] 
\begin{center} 
\epsfig{file=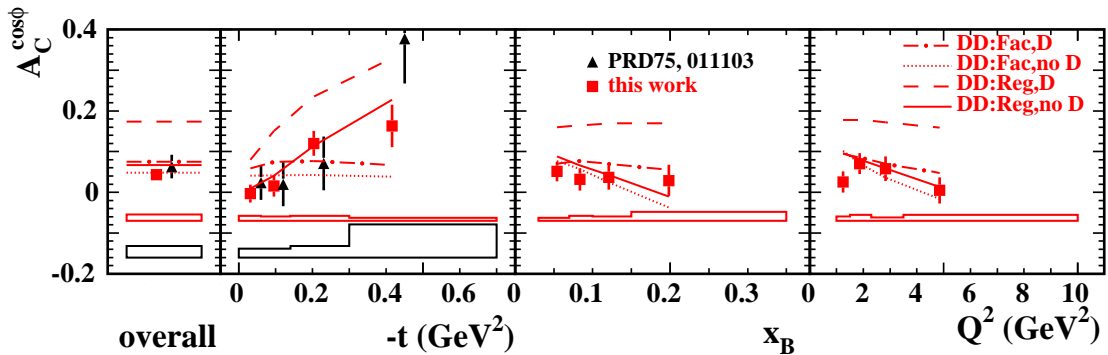,width=15cm}
\end{center}
\caption{\label{fig:BCA-hermes}
Beam-charge asymmetry measured on the proton by {\sc Hermes}:
azimuthal amplitude describing the dependence of the 
interference term on the beam charge ($A_C$). The triangles (shifted 
right for visibility) represent previous results~\cite{Airapetian:2006zr}, 
while most recent data~\cite{:2008jga} are represented by squares. The 
error bars (bands) represent the statistical (systematic) uncertainties. 
The curves labelled 'DD' are calculations of variants of a 
double-distribution GPD model~\cite{Vanderhaeghen:1999xj,Goeke:2001tz} using 
$b_v=\infty$ and $b_s=1$ as profile parameters for valence and sea quarks.
This figure is extracted from figure 5 of \cite{:2008jga}.}
\end{figure}  

{\sc Hera} was the only multi-GeV accelerator providing 
both electron and positron beams. It offered the additional flexibility of 
inverting every few weeks, for the same charge of the beam, the direction 
of its polarization to reduce systematic effects. 
The first measurement of a beam-charge asymmetry was published by
\hermes~\cite{Airapetian:2006zr}, based on the analysis of parts of 
their electron and positron beam data sets. Somewhat later, BCA results were 
published based on the analysis of another much larger \hermes\ data sample, 
taken with transverse target polarisation~\cite{:2008jga}. Here, beam-charge
and transverse-target-spin asymmetries were extracted simultaneously (see 
section \ref{subsubsec-exp:trans-target-spin}). In figure~\ref{fig:BCA-hermes},
the experimental data on the azimuthal asymmetry amplitude $A_C^{\cos \phi}$ 
are compared to calculations using the model of 
\cite{Vanderhaeghen:1999xj,Goeke:2001tz} that is briefly described in section 
\ref{subsec:GPD_Models}. The model results are 
calculated at \hermes\ kinematics, each at the average kinematics of 
each individual bin. The best description within the VGG model is given by 
the version using an unfactorised Regge-inspired $t$ dependence, no D-term 
and profile parameters $b_{val} = \infty$ (no skewing) and $b_{sea} = 1$ 
(some skewing). 
%
%
%
\subsubsection{Longitudinal target-spin asymmetry}
\label{subsubsec-exp:long-target-spin}
Experimental access to quark GPDs other than $H^f$ becomes possible when using 
an {\em unpolarized beam} (U) and a {\em polarized target}. For 
{\em longitudinal} (L) target polarization, the same $\sin{\phi}$ dependence 
is expected as for the BSA, although the relevant twist-2 Compton 
amplitude $\widehat{M}_L$ (\ref{eq:LTSA}) involves a different combination 
of Compton Form Factors than does $\widehat{M}_U$ (\ref{eq:Munpol}).

The longitudinal target-spin asymmetry is defined as:
\begin{equation}
\label{eq:LTSA_asy}
{\cal{A}}_{UL}(\phi) = \frac
  {d \sigma^{\Nleftarrow}(\phi) 
 - d \sigma^{\Nrightarrow}(\phi)}
  {d \sigma^{\Nleftarrow}(\phi) 
 + d \sigma^{\Nrightarrow}(\phi)},
\end{equation}
where $\Nleftarrow$ ($\Nrightarrow$) denotes target spin antiparallel 
(parallel) to the beam direction. This asymmetry is expected to be most 
sensitive to a combination of $H^f$ and $\widetilde{H}^f$, because the 
kinematic suppression of the first 
term in (\ref{eq:LTSA}), relative to the second, may compensate the 
usual dominance of the unpolarized quark GPDs $H^f$ over their polarized 
counterparts $\widetilde{H}^f$. Hence they might be disentangled by 
combining this measurement with BSAs. For not 
too small values of $t$, there exists also some sensitivity to 
$(\xi \widetilde{\mathcal{E}})$, which is written in this way because 
$\widetilde{\mathcal{E}}$ itself is defined to be inversely 
proportional to $\xi$ (see {\em e.g.} \cite{Diehl:2007jb}).

\begin{figure}[htb] 
\begin{center} 
\epsfig{file=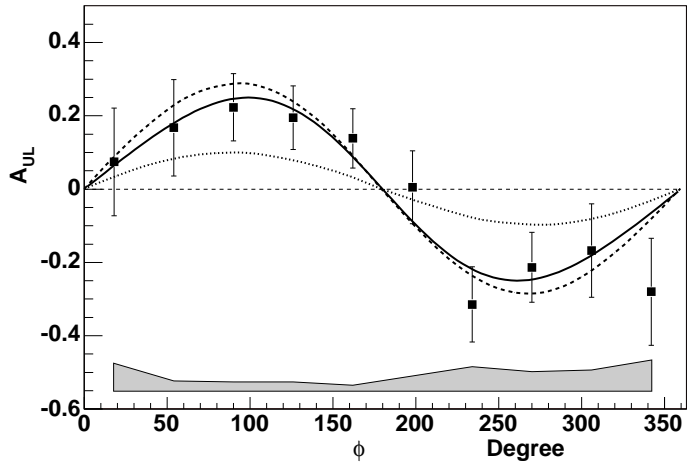,width=10cm}
\end{center}
\caption{\label{fig:A_ULjlab}
Azimuthal dependence of the longitudinal target-spin asymmetry, as 
measured at \jlab~\cite{Chen:2006na} with 5.7 GeV electrons and a
longitudinally polarized NH$_3$ target, shown after subtraction of the
$\pi^0$ background. Error bars represent the statistical uncertainties
and the band at the bottom represents the systematic uncertainties. 
The curves are explained in the text.  
The figure is taken from \cite{Chen:2006na}.} 
\end{figure}
Experimental results on longitudinal target-spin asymmetries were 
obtained at \jlab~\cite{Chen:2006na} and 
\hermes\ \cite{Kopytin:2007zz}. Both experiments observe a $\sin{\phi}$ 
modulation of the yield from the entire acceptance, and scaling of the 
sinusoidal amplitude with $\sqrt{-t}$, in agreement with expectations. 
In figure~\ref{fig:A_ULjlab}, the \jlab\ data~\cite{Chen:2006na} 
are shown. The continous curve represents a fit with the function
$\alpha \sin{\phi} + \beta \sin{2\phi}$ with the parameters
$\alpha = 0.252 \pm 0.042(stat) \pm 0.020(syst)$ and
$\beta = -0.022 \pm 0.045(stat) \pm 0.021(syst)$. The $\sin{2\phi}$ 
term is consistent with zero, indicating that higher-twist contributions
in this kinematic domain are negligible.
The two other curves show calculations using the double-distribution 
model of \cite{Vanderhaeghen:1999xj}, based on MRST02 PDFs, with the
$\xi$ dependence $b_{val}=b_{sea}=1$, including leading-twist 
terms only, and target-mass corrections applied. The nucleon-spin-flip 
quark GPDs $E$ and $\widetilde{E}$ are set to zero. The dashed (dotted) line 
shows the asymmetry when the polarized quark GPD $\widetilde{H}$ is included
(excluded), so that the necessity of $\widetilde{H}$ for the description
of the longitudinally target-spin asymmetry is clearly demonstrated.
%
%
%
\subsubsection{Transverse target-spin asymmetry}
\label{subsubsec-exp:trans-target-spin}
In the case of target polarization {\em transverse} (T) to the direction 
of the virtual photon, the polarization can be resolved into two 
independent components, `normal' to and `sideways' in the production 
plane~\cite{Diehl:2005pc}, defined by the directions of the virtual and
real photons. The DVCS cross section depends only on $\widehat{M}_{T}$, the 
normal component of the twist-2 Compton amplitude (see (\ref{eq:TTSA_N})).
This is the {\em only} amplitude that embodies a combination of CFFs in 
which the contributions
of the quark GPDs $E^f$ are not kinematically suppressed relative to those of 
$H^f$. Hence DVCS measurements on a transversely polarized proton target 
appear to be crucial for the evaluation of $J_u$ and $J_d$ through 
the Ji relation (\ref{eq:Ji_SumRule}). A complication inherent to the
exploitation of this relation lies in the fact that both quark GPDs $H^f$ 
and $E^f$ need to be determined in the limit $t \rightarrow 0$, whereas
at small $t$ the relevant asymmetry is suppressed by a factor of 
$\sqrt{t_0-t}/2m$ when extracting the GPDs $H^f$ and even by a factor of 
$(t_0-t)/4m^2$ when extracting the GPDs $E^f$ (which is apparent from a 
comparison of (\ref{eq:Munpol}) and (\ref{eq:TTSA_N})). The extraction of 
model-dependent constraints on $J_u$ and $J_d$ from experimental data
is discussed in section~\ref{subsec-exp:QuarkTotalAngularMomenta}.

From data taken with both beam charges and transverse target polarization 
$S_\perp$, the beam-charge asymmetry ${\cal{A}}_C(\phi)$ 
(see (\ref{eq:BCA1})) can be extracted simultaneously with the transverse 
target-spin asymmetries arising from the $|\tau_{DVCS}|^2$ and interference 
terms (see (\ref{eq:DVCS-BH-amplitude})), respectively:
\bea
\!\!\!\!\!\!\!\!\!\!\!\!\!\!\!\!\!\!\!\!\!\!\!\!\!\!\!\!\!\!\!\!\!\!\!
{\cal{A}}_{UT}^{DVCS}(\phi,\phi_S) &\equiv&
\frac{1}{S_\perp}\cdot\frac{d\sigma^+(\phi,\phi_S)-d\sigma^+(\phi,\phi_S+\pi)
+d\sigma^-(\phi,\phi_S)-d\sigma^-(\phi,\phi_S+\pi)}{d\sigma^
+(\phi,\phi_S)+d\sigma^+(\phi,\phi_S+\pi)+d\sigma^-(\phi,\phi_S)+d\sigma^
-(\phi,\phi_S+\pi)} \\
\!\!\!\!\!\!\!\!\!\!\!\!\!\!\!\!\!\!\!\!\!\!\!\!\!\!\!\!\!\!\!\!\!\!\!
{\cal{A}}_{UT}^{I}(\phi,\phi_S) &\equiv&\frac{1}{S_\perp}\cdot
\frac{d\sigma^+(\phi,\phi_S)-d\sigma^+(\phi,\phi_S+\pi)
-d\sigma^-(\phi,\phi_S)+d\sigma^-(\phi,\phi_S+\pi)}{d\sigma^+(\phi,\phi_S)+
d\sigma^+(\phi,\phi_S+\pi)+d\sigma^-(\phi,\phi_S)+d\sigma^-(\phi,\phi_S+\pi)}.
\eea
The azimuthal angle $\phi_S$ of the target polarization vector was explained
in section~\ref{sec:SSA}. The subscripts on the $\mathcal{A}$'s represent the 
dependence on beam Charge (C) or Transverse (T) target polarization, with an 
Unpolarized (U) beam, and the superscripts $\pm$ stand for the lepton beam 
charge. All three asymmetries are expanded in terms of various harmonics in
$\phi$ and $\phi_S$ which respective amplitudes are hereafter called 
`(effective) azimuthal (asymmetry) amplitudes'. 

\begin{figure}[htb] 
\begin{center} 
\epsfig{file=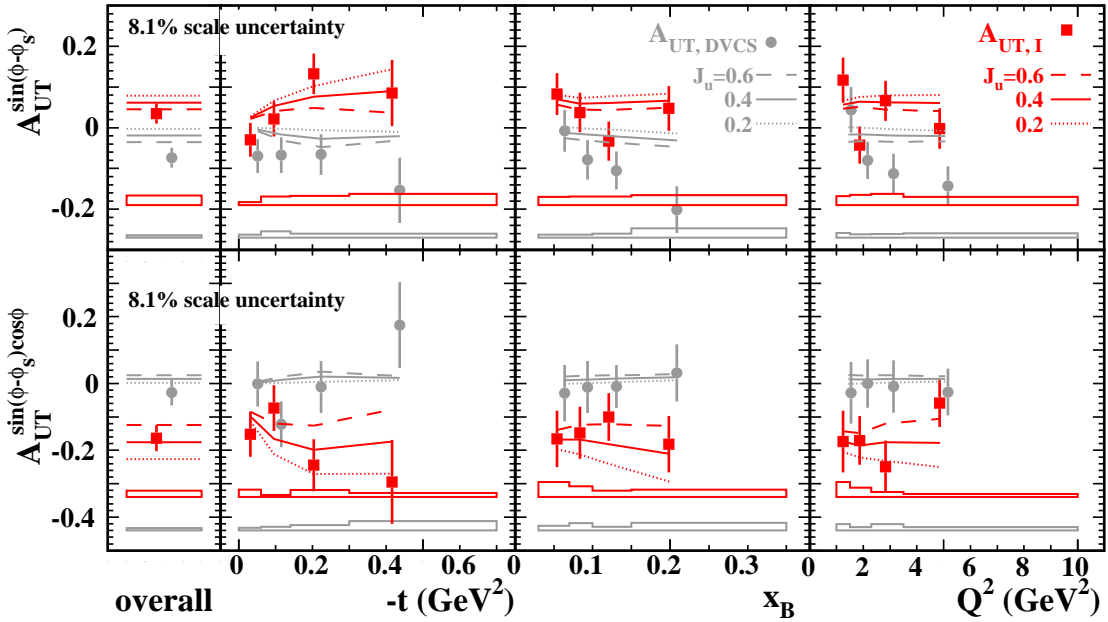,width=15cm}
\end{center}
\caption{\label{fig:AsymsTTSApaper}
Asymmetry amplitudes describing the dependence of the 
squared DVCS amplitude (circles, $A_{UT,DVCS}$) and the interference term 
(squares, $A_{UT,I}$) on the transverse target polarization.
The circles (squares) are shifted right (left) for visibility. 
The error bars represent the statistical uncertainties, while the top 
(bottom) bands denote the systematic uncertainties for $A_{UT,I}$ 
($A_{UT,DVCS}$), excluding an 8.1~\% scale uncertainty from the target 
polarization measurement. The curves in the upper panel are calculations of 
the GPD model variant (Reg, no D) with 
three different values for the $u$-quark total angular momentum $J_u$ 
and fixed $d$-quark total angular momentum $J_d=0$~\cite{Ellinghaus:2005uc}.
This figure is based on figure 6 of \cite{:2008jga}.}
\end{figure}

Only a few azimuthal amplitudes are expected to show substantial sensitivity 
to $J_u$ and $J_d$~\cite{Ellinghaus:2005uc}. At \hermes\, all harmonics
of physical significance were measured with unpolarized beam and a 
transversely polarized proton target~\cite{:2008jga}. 
The kinematic dependences of two asymmetry amplitudes of particular
interest are shown in figure~\ref{fig:AsymsTTSApaper} in comparison with
calculations based on two different types of GPD models. The data is
clearly able to discriminate among these GPD models. The model variants
that agree best with the data, are used in the next section 
to derive constraints from the data on the total angular momentum of 
$u$ quarks. Sensitivity exists, as can be seen from the top panel of the 
figure.


\subsection{Quark total angular momenta}
\label{subsec-exp:QuarkTotalAngularMomenta}

In the theoretical description of the DVCS process, experimental observables 
like cross section differences or asymmetries are related to various 
combinations of the Compton Form Factors
$\mathcal{H},\mathcal{E},\widetilde{\mathcal{H}},\widetilde{\mathcal{E}}$, 
which can be calculated from the respective GPDs 
$H^f,\widetilde{H}^f, E^f,\widetilde{E}^f$ 
(see section~\ref{subsec:GPD_Access}). Parameterizations of the spin-flip 
GPD $E^f$ embody explicitly or implicitly the quark total angular momenta 
$J_u$ and $J_d$~\cite{Goeke:2001tz}.  
\begin{figure}[htb] 
\begin{center} 
\epsfig{file=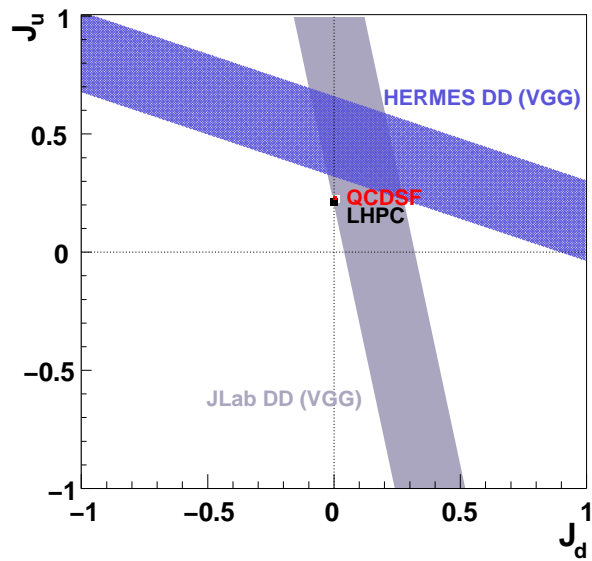,width=8cm}
\end{center}
\caption{\label{fig:Ju_vs_Jd}
Model-dependent constraints on $u$-quark total angular momentum $J_u$ vs 
$d$-quark total angular momentum $J_d$, obtained by comparing DVCS
experimental results and theoretical calculations using the double-distribution
GPD model of \cite{Vanderhaeghen:1999xj,Goeke:2001tz}. 
The constraints based on the \hermes\ data~\cite{:2008jga} for the azimuthal 
amplitudes $A_{\rm UT,\rm I}^{\sin{(\phi-\phi_S)}\cos{\phi}}$ and 
$A_{\rm UT,\rm I}^{\sin(\phi-\phi_S)}$ are labelled \hermes\ DD (VGG). Those
based on the \jlab\ neutron cross section data~\cite{:2007vj} are labelled
\jlab\ DD (VGG). Also shown 
as small (overlapping) rectangles are results from lattice gauge theory
by the QCDSF~\cite{Gockeler:2006ui} and LHPC~\cite{Hagler:2007xi} 
collaborations, as well as a result for only the valence-quark contribution
(DFJK) based on zero-skewness GPDs extracted from nuclear form factor 
data~\cite{Diehl:2004cx,Kroll:2007wn}. The sizes of the small rectangles 
represent the statistical uncertainties of the lattice results and the 
parameter range for which a good DFJK fit to the nucleon form factor 
data was achieved. Theoretical uncertainties are unavailable. The figure 
is taken from \cite{:2008jga}.}
\end{figure}
Hence model-dependent constraints on these two parameters can be derived by 
fitting them to experimental observables~\cite{Ellinghaus:2005uc}. The 
double-distribution model~\cite{Vanderhaeghen:1999xj,Goeke:2001tz} was
thereby fitted to azimuthal amplitudes of 
beam-charge and transverse target-spin asymmetries measured on the proton 
at \hermes\ (see previous section). The resulting constraint~\cite{:2008jga} 
is shown in figure~\ref{fig:Ju_vs_Jd} as sloped band, which in units of 
$\hbar$ can be represented as 
\begin{equation}
\label{eq:constr-DD-H}
J_u+J_d/2.8=0.49\pm0.17(\mathrm{exp_{tot}}).
\end{equation}

Combining DVCS measurements on deuteron and proton targets, the 
beam-spin difference for the neutron was determined at \jlab~\cite{:2007vj}.
Also in this case sensitivity to the GPDs $E^f$ can be found, as for the 
neutron the prefactor $F_1(t)$ in (\ref{eq:Munpol}) is small but $F_2(t)$ 
is not. Similarly as described above, a model-dependent constraint on 
($J_u, J_d$) was derived by comparison to a calculation of the 
double-distribution model. 
The slope of the resulting band differs from that of the band derived 
from proton data (see figure~\ref{fig:Ju_vs_Jd}) due to the different quark 
contents of proton and neutron:
\begin{equation}
\label{eq:constr-DD-JLAB}
J_u+J_d/5.0=0.18\pm0.14(\mathrm{exp_{tot}}).
\end{equation}
These results show that existing data have the potential to provide 
quantitative information about the quark total angular momenta 
$J_u$ and $J_d$. However, as discussed in the previous sections, the
double-distribution model of \cite{Vanderhaeghen:1999xj,Goeke:2001tz}
cannot explain all existing DVCS data.

%
%
%
\section{The spin budget of the nucleon}
\label{sec:SpinBudget}
%
\subsection{Decomposing the Nucleon Spin}
\label{subsec:Decompose}
While the total angular momentum of an isolated system is uniquely
defined, ambiguities arise when decomposing the total angular
momentum of an interacting multi-constituent system into
contributions from various constituents. Moreover, in a gauge 
theory, switching the gauge may result in shuffling angular momentum
between matter and gauge degrees of freedom.
In the context of nucleon structure, this gives rise to subtleties
in defining these quantities that are more fundamental than those
subtleties associated with the choice of factorization scheme,
mentioned in section \ref{subsec:Dq-incl}. 

In section \ref{subsec:GPD-GPDs}, it was discussed how GPDs 
can be used to determine the quark total angular momentum 
contribution $J_f$ to the nucleon spin  
(see (\ref{eq:Ji_SumRule})) in a gauge invariant decomposition of 
the nucleon's spin $\frac{1}{2}$. The 
expectation value of the $\hat{z}$ component of the quark
orbital angular momentum operator can then
be evaluated indirectly as the difference,
\be
{L}_q =\sum_f L_f=\sum_f\left[
J_f-\frac{1}{2}\left(\Delta q_f +\Delta \bar{q}_f\right)\right]
=J_q-\frac{1}{2}\Delta\Sigma
,
\label{eq:Lf}
\ee
which is otherwise presently inaccessible experimentally. 
Similarily one can indirectly evaluate the expectation value of the
operator representing the gluon total
angular momentum as the difference between the nucleon's spin
$\frac{1}{2}$ and $J_q$:
\be
J_\g=\frac{1}{2} - J_q .
\ee
One might consider going even further and use experimental results
for $\Dg$ to determine the gluon orbital angular
momentum as the difference between $J_\g$ and $\Dg$,
providing what appears to be a complete understanding for 
the nucleon spin budget. 
However, the difference between these two operators has no fundamental
connection to orbital angular momentum other than being the 
difference between $J_\g$ and $\Dg$. 

Equation (\ref{eq:Lf}) is a special case of
Ji's decomposition \cite{Ji:1996ek}
\be
{\vec J} = \sum_f \left({\vec S}_f + {\vec L}_f\right) +
{\vec J}_\g \label{eq:JiDeco}
\ee
of the expectation value of the angular momentum operator. Since this 
decomposition of the angular momentum density $M^{\mu \nu \lambda}=
T^{\mu \nu}x^\lambda - T^{\mu \lambda}x^\nu$, where 
$T^{\mu \nu}=T^{\nu \mu}$ is the energy-momentum tensor, in terms of
manifestly gauge invariant local operators (see footnote on page 
\pageref{page:manifestly}) is based on the decomposition of the 
$M^{0ij}$-component ($i,j \in \{x,y,z\}$), it is independent of the 
quantization axis. Here ${\vec S}_f$ is defined through the
expectation value of the axial vector current 
$\psi_f^\dagger{\vec \gamma} \gamma_5 \psi_f$, 
${\vec L}_f$ through the expectation value of 
$\psi_f^\dagger {\vec r}\times\left({\vec p}-g{\vec A}\right)\psi_f$,
and ${\vec J}_\g$ through the expectation value of
${\vec r}\times Tr\left({\vec E}\times{\vec B}\right)$,
where ${\vec E}$ and ${\vec B}$ are the QCD color electric and magnetic 
fields. A further decomposition of ${\vec J}_\g$ into spin 
and orbital  components using manifestly gauge invariant local operators 
has not been found and may be impossible. In general, spin (or 
`intrinsic' angular momentum) is identified with terms in the operator 
that do not depend explicitly on ${\vec r}$, in contradistinction to the 
orbital (or `extrinsic') piece~\cite{Jaffe_PrivComm}.

Being based on manifestly gauge invariant local operators, 
all three terms 
in (\ref{eq:JiDeco}) can be calculated in lattice gauge theory.
As was discussed in previous sections, the expectation values 
${\vec S}_f$ and 
${\vec J}_f$ can be accessed experimentally, and
${\vec J}_\g$ can, at least in principle, be accessed in exclusive
deeply virtual heavy meson production. Beyond leading order in
$\alpha_s$, all three terms in (\ref{eq:JiDeco}) depend on $Q^2$,
but the $Q^2$ dependence cancels in their sum.

The main disadvantage of decomposition (\ref{eq:JiDeco})
is that for no choice of the quantization axis is there a
partonic interpretation as a difference between number densities
for the quark orbital angular momentum ${\vec L}_f$ or the gluon total
angular momentum ${\vec J}_\g$.  Only the quark intrinsic 
angular momentum ${\vec S}_f$ has such an 
interpretation (for the $\hat{z}$ quantization axis).  
This is because the
operator definitions for ${\vec L}_f$ and ${\vec J}_\g$
contain interactions (e.g. ${\vec L}_f$
 contains the gluon field through
the gauge-covariant derivative 
${\vec D}={\vec \partial}-ig{\vec A}$). With quarks and gluons
thereby inextricably intertwined, it is impossible to 
unambigiously identify these terms
with quark orbital angular momentum, and gluon  
angular momentum respectively \cite{Jaffe:2000kr}. 

An alternative `light-cone decomposition'  was proposed by Jaffe and Manohar 
\cite{Jaffe:1989jz} (see also \cite{Harindranath:1998ve,Hagler:1998kg}).
While the Ji-decomposition is based on the $M^{0ij}$ component of the angular 
momentum density, the Jaffe-Manohar decomposition is based on the $M^{+xy}$ 
component and applies only to the $z$ component of $\vec{J}$. It can be 
written as:
\be
\frac{1}{2} = J_z = \frac{1}{2}\Delta \Sigma + \Dg + {\cal L}.  
\label{eq:JaffeDeco}
\ee
The quantity $\Delta\Sigma$ appears also in (\ref{eq:Lf}), the special case of 
(\ref{eq:JiDeco}) in the helicity basis.  The quark-intrinsic piece is 
\emph{the same} in both components: the operator that measures 
this piece in $M^{+12}$ is 
$\epsilon^{+12-}\psi^\dagger\gamma_{-}\gamma_{5}\psi$ and the operator that measures it in $M^{012}$ is 
$\epsilon^{0123}\psi^\dagger\gamma_{3}\gamma_{5}\psi$, i.e. both operators are 
components of the axial vector current and therefore their expectation values 
are proportional to the axial charge, so that
$\Delta\Sigma$ appears in both decompositions.
This is not true for any other term in the decompositions! 
One can understand
this point by analogy with the energy momentum tensor 
$T^{\mu \nu}$~\cite{Jaffe:2000kr}: The 
energy of a single particle state can be expressed in terms of the matrix 
element of $T^{00}$ and decomposed into a quark and gluon piece.  The 
`momentum' in the infinite-momentum frame corresponds to the matrix element of 
$T^{++}$ and can also be decomposed into quark and gluon pieces.  
The quark and gluon pieces of these two objects are unrelated.

In decomposition (\ref{eq:JaffeDeco}), each term has a partonic 
interpretation.
The gluon spin 
contribution $\Dg$ appears explicitly.
It is experimentally accessible (see section \ref{subsec:DeltaG})
and can be defined as the
expectation value of a (nonlocal) manifestly gauge invariant
operator. In light-cone gauge, this operator collapses to a 
local operator (and its expectation value has a partonic 
interpretation).
No direct experimental access to the parton
orbital angular momentum ${\cal L}$ has been identified. Its value 
can be obtained only by subtracting the
quark and gluon spin contributions from the nucleon spin.
Both $\Dg$ and ${\cal L}$ can be defined through matrix 
elements of local operators only in light-cone gauge $A^+=0$.
Explicit definitions for the operators
appearing in both decompositions can be found in 
\cite{Jaffe:2000kr}. 
Since neither one can be represented as the matrix element
of a manifestly gauge invariant local operator, they cannot be 
analytically continued to Euclidean space and are thus inaccessible for
lattice QCD. 

One can decompose ${\cal L}$  further into
contributions from gluons and each quark flavor
\be
{\cal L} = \sum_f{\cal L}_f + {\cal L}_\g,
\label{eq:JaffeSubDeco}
\ee
where for example, the contribution ${\cal L}_f$
from quarks with flavour $f$
can be obtained from the expectation value of
$\psi_{f+}^\dagger \left(xp_y-yp_x\right)\psi_{f+}$ (with
$\psi_{f+}$ being the dynamical or `good' component of the quark field
operator $\psi_f$ in that framework) for a nucleon polarized along the
$\hat{z}$ direction. Each of the terms in (\ref{eq:JaffeSubDeco})
has a partonic interpretation. 
Even a decomposition with respect to the
momentum fraction $x$ is possible, but for none of these terms 
has experimental access been 
identified nor are they accessible in lattice QCD, since their
manifestly gauge invariant forms are nonlocal in Minkowski space
\cite{Bashinsky:1998if}.
Nevertheless, their
expectation values can be easily related to the (light-cone)
wave functions of hadrons, which play an important role in 
phenomenology \cite{Brodsky:2004tq}.

Summarising the above dichotomy, the two decompositions described 
above have only $\Delta \Sigma$ in 
common, which is also manifestly gauge invariant and experimentally
accessible. The experimentally accessible quantities
$J_f$ (and, by subtraction, $L_f$ and $J_\g$) appear in one decomposition, 
while $\Dg$ (and, by subtraction, ${\cal L}$) appear in the other.
Combining terms from these incommensurate decompositions 
would be a sterile exercise.
For example, after determining $J_\g=\frac{1}{2}-\sum_f J_f$ 
by evaluating the right hand side using the Ji relation
for $J_f$, it would be fruitless  to `determine' 
$L_\g$ by subtracting $\Dg$ from $J_\g$
(or equivalently by subtracting $L_q$ from ${\cal L}$). 
In fact, the last step would yield a term that is 
neither directly experimentally accessible nor could it be 
represented by
an operator other than that defined as the difference 
of the two operators. 
{\em Defining} gluon orbital angular momentum
through such a subtraction procedure has no predictive power
(not even in principle). 

Note that an analogous dichotomy (multichotomy) appears when one 
considers the motion of
a charged particle in an electromagnetic field. While the
canonical momentum ${\bm p}\equiv-i{\bm \nabla}$ has a simple 
interpretation in terms of only the charged particle's
wave function, it is not gauge invariant. Evaluation of the gauge 
invariant combination ${\bm p}-e{\bm A}$ requires in addition
knowledge about the vector potential. This combination is directly
accessible experimentally, e.g. on a macroscopic scale by studying 
the trajectory of the particle in a magnetic field.
Similarly, while the canonical
orbital angular momentum ${\bm r}\times{\bm p}\equiv 
-i{\bm r}\times{\bm \nabla}$ has a simple 
interpretation as a property of
the charged particle's wave function alone, 
it is unfortunately not gauge invariant. 
On the other hand, the gauge invariant combination 
${\bm r}\times\left({\bm p}-e{\bm A}\right)$ contains the vector 
potential ${\bm A}$ and its interpretation is therefore more
subtle. Furthermore, only the gauge noninvariant orbital
angular momentum ${\bm r}\times {\bm p}$ satisfies the
familiar angular momentum commutation relations $\left[L_i,L_j\right]
= i \varepsilon^{ijk}L_k$.
In the end, both definitions of the angular momentum
of the charged particle have their merit, and depending on the 
context either may be preferable.

Analogously, in the decomposition of the nucleon spin, 
it also turns out that both
schemes have their respective merits.
In Ji's decomposition both $J_q$ and $J_\g$ are manifestly
gauge invariant and 
calculable in lattice gauge theory (see section
\ref{sec:GPDLattice}). Furthermore, using (\ref{eq:Ji_SumRule})
$J_f$ can be measured, which allows determination of 
$L_f=J_f-\frac{1}{2}\left(\Delta q_f +\Delta \bar{q}_f\right)$. 
However, since it is even more difficult to relate 
$J_\g$ to measurable quantities (\eg using (\ref{eq:Ji_SumRuleGlue}))
than is the case for $J_f$, the 
relation can currently not be `tested',
while it can nevertheless be used to determine 
$J_\g=\frac{1}{2} - J_q$, 
which is interpreted as gluon total angular momentum
in the equal time quantized framework only.

Besides a partonic interpretation of {\em all} terms, the Jaffe-Manohar 
decomposition (\ref{eq:JaffeDeco}) has
the advantage that it contains the experimentally accessible
quantity $\Dg$
(see section \ref{subsec:DeltaG}). Unfortunately, also here
no experimental observable has been found to relate to 
${\cal L}_f$ or ${\cal L}_\g$ even in principle, 
so the light-cone based spin decomposition cannot 
be tested either, but one can still use the sum rule
to determine the net parton orbital angular momentum
${\cal L}$.
Another disadvantage of the light-cone based decomposition 
(\ref{eq:JaffeDeco}) is that
because the operators used to define its terms 
(except $\Delta \Sigma$) 
are not manifestly gauge invariant, 
its terms cannot be calculated in lattice gauge theory.
Nevertheless, although a simple (\ie\ local) operator expression 
for $\Dg$  exists only in 
light-cone gauge, it can be defined through a manifestly
gauge invariant operator. 
In all other gauges, $\Dg$ is described by matrix elements of
a non-local operator rendering its calculation or interpretation
exceedingly difficult.
$\Dg$ has a physical interpretation as the gluon intrinsic angular momentum
in the light-cone framework only. 

The two decompositions of the nucleon spin discussed here are not 
the only possibilities. For
example, another alternative has been proposed in
\cite{Chen:2007ww,Chen:2007wb}. However, it neither has
a partonic interpretation nor does its orbital angular momentum seem
to be 
experimentally accessible and we will not discuss this decomposition 
here.

An analogue may illustrate the dangers in mixing components from
different decompositions: suppose hypothetically that
DIS data on deuterium were available only at $Q^2=100$\,GeV$^2$ and
proton data were available only at $Q^2=10$\,GeV$^2$. The normal 
procedure for extracting neutron data would be to evolve to a common
$Q^2$ value first. However, again hypothetically, suppose there 
existed no evolution equations. One may then ask oneself whether
it would make sense (without first evolving the results from one $Q^2$
to the other value of $Q^2$) to subtract the distribution of$u$ quarks
in the proton at $Q^2=10$\,GeV$^2$ from that in the deuteron 
at $Q^2=100$\,GeV$^2$ and use the results to {\em define} the 
distribution of $u$ quarks in the neutron. 
The main weakness of this analogy is of course that this definition 
is falsifiable 
because experiments may be performed at different values of 
$Q^2$ or QCD evolution may be discovered.
%
%

\subsection{Implications for quark angular momenta}
\label{subsec:Implications}
%
%
\subsubsection{Results using the Ji relation}
\label{subsubsec:JiScheme}
Due to the difficulties in extracting GPDs from DVCS data,
no model-independent experimental constraints on the Ji relation yet
exist. Lattice-based constraints on $J_f$ have been discussed in
section \ref{sec:GPDLattice} and indicate that $L_u+L_d$
is consistent with zero, while $L_u-L_d\approx -0.4\pm 0.05$.
In that subsection it was already emphasized that
this result is counterintuitive, where the intuition derives 
from comparison to constituent quark models and phenomenology
for single-spin asymmetries. Given the complementary and
incommensurate definitions of quark orbital angular momentum,
as discussed in the previous section,
the question arises whether the orbital angular momentum in these
models should be identified with $L_f$ in (\ref{eq:JiDeco})
or with ${\cal L}_f$ in (\ref{eq:JaffeSubDeco}).
If they were to be identified with $L_f$, there would be
additional, and presently unknown, corrections added to account for 
the difference before the orbital angular momentum in the models
could be compared to the lattice QCD results.

\subsubsection{Results in the light-cone decomposition}
\label{subsubsec:JaffeScheme}
The existing information on $ \Dg$ from global fits
of experimental data suggests that its first moment is small 
\cite{DSSV}. 
Neglecting `theoretical' and model uncertainties such as those
associated with the choice of functional forms
used to represent PDFs in the fits while 
interpreting table~\ref{tab:DSSV-1stmoments} leads to 
$\Dg \simeq -0.1 \pm 0.1 (expt.)$ for $\Delta\chi^2=1$.
Using the most precise result 
for $\Delta \Sigma = 0.33 \pm 0.03 (tot.)$ \cite{HERMES_g1pd} the
net parton orbital angular momentum is obtained from 
(\ref{eq:JaffeDeco}) as:
\be
{\cal L}= \sum_f {\cal L}_f+{\cal L}_\g\simeq 0.43\pm 0.1,
\ee
which would be more than twice the intrinsic contribution from quarks. 
It remains open
how gluons and quarks share their orbital contributions.

%
%
%
\section{Conclusions and Outlook}
\label{sec:Summary}

One measure of the vitality of a field is the rapidity with which
its conceptual framework and focus evolve.  By this measure,
the study of the internal spin structure of the proton is indeed
vital.  Its explosive birth in the late 1980's was triggered by
the development of experimental technologies for Deeply Inelastic
Scattering (DIS) of polarized high energy leptons on targets containing
polarized protons.  The pivotal EMC measurement~\cite{EMC:1988,EMC:1989} 
suggested that the helicities 
of the quarks combine to make a net contribution to the helicity of the 
proton that is small compared to naive expectations.  This finding
inspired a flood of theoretical papers and several major new experimental 
efforts at three
laboratories, which confirmed the original finding with much higher
precision, and extended the measurements to the neutron target
to help distinguish the contributions of the various quark flavours. 
While $u$ quarks were found to be strongly polarized in the same
direction as the parent proton, both $d$ quarks and even $s$ 
quarks were found to be negatively polarized.
These measurements of inclusive DIS asymmetries were extended to
semi-inclusive measurements where the type of hadron produced in
the fragmentation of the struck quark tends to `tag' the
flavour of that quark.  The resulting data indicate that
the helicity densities of sea quarks are small in the measured
$x$ range of $0.02<x<0.4$,
but still consistent with fits to inclusive data yielding a negative
first moment for the strange sea, which receives a significant 
negative contribution from $x<0.02$.

Attention soon focused on a possibly substantial role of the helicity
of the gluons, which is experimentally difficult to probe in DIS.
While some of the existing new experiments struggled to extract
constraints on the gluon polarization, yet more new efforts were
launched that were designed specifically to target this problem.
At the time of writing, these are beginning to bear fruit, suggesting 
that the fractional gluon helicity contribution is small, and together
with the quark helicities, do not
exhaust the `spin budget' of the proton.  This leaves the question
of the contribution of parton orbital angular momentum.  In the
light-cone framework for defining this quantity with a partonic interpretation
in the axial gauge, one can estimate it as the difference
${\cal L} = \frac{1}{2} - \frac{1}{2}\Delta\Sigma - \Dg$.
Although $\Delta\Sigma$ is precisely known from measurements of $g_1^d$,
the first {\em moment} $\Dg$ is still so poorly known that
there is still little direct experimental constraint on the contribution of
parton orbital angular momentum to the nucleon spin. 

In the 1990's, dramatic theoretical progress reshaped the field,
identifying new experimental signals for 
nucleon spin structure.  One key issue has been quark transversity,
which differs from helicity because of the relativistic internal
motion of quarks.  It was realized that single-spin asymmetries in semi-inclusive
scattering offered access to this elusive property, through 
a dependence of the fragmentation process on the transverse
polarization of the struck quark---the Collins effect.  While
initial experimental evidence for this phenomenon was ambiguous,
eventually SIDIS measurements with transverse target polarization
in combination with studies of the jet structure of hadrons
from $e^+e^-$ collisions provided a data set with a clear 
interpretation, leading to the first extraction of transversity 
distributions.  As expected, they are found to share the features
of the helicity distributions in that they are positive for $u$ quarks,
and negative and considerably smaller for $d$ quarks.  Their
magnitudes are found be be about half of the Soffer positivity bounds,
which are expressed in terms of the unpolarized and helicity densities.

One of the ambiguities that complicated the interpretation of
early measurements of single-spin asymmetries was a possible
contribution by the Sivers effect, which arises from a `naively T-odd'
correlation between the nucleon spin and the intrinsic transverse momentum
of quarks.  This developed into a fascinating
topic of its own when it was realized that such distributions
raise new questions about key theorems for factorization of
hard and soft dynamics of processes, and the importance of
initial or final-state interactions in the understanding
of parton distributions that depend on intrinsic transverse
momentum.  The above-mentioned SIDIS measurements with transverse
target polarization also provided clear evidence of the Sivers
effect.  Future high-luminosity measurements of the Drell-Yan
process with transversely polarized protons are now anticipated
as a test of the new prediction that the Sivers effect will appear
there with the opposite sign.

The pursuit of parton orbital angular momentum also led to dramatic 
theoretical progress in the late 1990's.  It was realized that
experimental access to this quantity might be provided by hard
exclusive processes that involve hard interactions,
yet leave the target nucleon intact.  Most prominent among these
processes is deeply virtual Compton scattering, the production
of a single energetic photon, because of its fruitful interference
with the well-understood Bethe-Heitler process, providing access
to the phase of the DVCS amplitude.  Such exclusive processes are now
interpreted within the framework of generalized parton distributions
that depend on three kinematic variables and subsume both elastic
form factors as moments and ordinary parton distributions as
special cases.  The excitement centered around the finding that
a second moment of certain GPDs provides the total angular
momentum of partons, including orbital angular momentum.  
Although existing DIS experimental
facilities were not designed to measure exclusive processes,
they managed to make some pioneering measurements, which have
been compared to calculations based on the few available GPD
parameterizations.  While the basic features of the data are described,
none of the existing GPD models are able to simultaneously explain 
all available DVCS data, including cross sections and their
differences or asymmetries with respect to beam charge and helicity, and
target polarization.  Nevertheless, the data for transverse
target polarization were shown to be sensitive to quark total
angular momentum, of which the contribution from quark
intrinsic spin has already been determined as mentioned above.  
Hence the development of more successful parameterizations can be expected 
to thereby result in constraints on quark orbital angular momentum.
Promising efforts already underway based on `conformal moments'
offer a practical means of fitting DVCS data with parameterizations at 
NLO in $\alpha_s$.  Some success with this approach has been achieved 
with DVCS cross sections at collider energies, and work continues to 
extend this to charge and polarization dependences at fixed-target
energies.

A decade ago, progress in the understanding of nonperturbative
QCD was made mainly by comparing experimental data with models that were
constructed to approximately embody some important features of QCD.
Since then, rapid progress has been made in improving the scope and
precision of numerical simulations of QCD on a Euclidean lattice.
Improved algorithms and techniques together with rapidly
increasing computational power have combined to bring this
field to the point that results for a large variety of observables
can now be taken seriously as implications of QCD.  An example
is the spin budget of the proton, where it is computed that the
only substantial contribution by quarks is from the $u$ flavour,
in which the positive contribution from quark intrinsic spin dominates
the negative contribution from orbital angular momentum.  While several
types of important observables will remain inacessible to this
approach for the forseeable future (such as the gluonic
contribution to the spin budget), others that are are calculable
are so difficult to measure experimentally that lattice calculations will 
be the only source of information about them for a considerable time.
Examples here include chiral-odd generalized parton distributions.
Hence calculations on the lattice play a role that is highly 
complementary to both experiment and model building.

The existing experimental facilities for polarized DIS have provided 
a wealth of new data, much of it unanticipated when those experiments
were conceived.  For example, they have provided
high quality data
for inclusive measurements of double-spin asymmetries, determining
the spin structure function $g_1$ of the proton and neutron over
the kinematic range of fixed-target experiments.  However,
other important avenues to nucleon spin structure such as semi-inclusive
DIS and exclusive processes have been exploited with only enough
precision to inspire and guide rapid theoretical progress, but not to
provide detailed information.  Also, the experience
from the HERA $e-p$ collider taught us that the vastly wider kinematic
range available at such a collider can lead to unanticipated insights,
in part because of the much improved access to the region of small $x$.
One clear motivation for planning for a high-luminosity $e-p$ collider
with polarized beams is to obtain data for $g_1$ over a wide range
in hard scale $Q^2$ to allow the precise inference of gluon polarization
from the $Q^2$ evolution of $g_1$, in analogy to the success in
extracting the unpolarized gluon density from HERA data.
The rapidly expanding horizons of the field in the last decade
have provided other compelling motivations.  Only with high luminosity
free of target dilution by unpolarized components and a broad kinematic
range can precise information be produced in the study of transversity,
transverse-momentum-dependent parton distributions and quark
orbital momentum through the constraint of generalized parton
distributions.  The clear need for a new accelerator facility
to make this all possible has inspired specific proposals and design
studies~\cite{EIC}.  Luminosities of order $10^{33}$\,nucleons/(cm$^2$s) 
are considered possible in the energy range $\sqrt{s}=15$--100\,GeV.  
The continuing vitality of this exciting field depends on the determined 
pursuit of this goal.  While the community works toward this new facility, 
the present rate of rapid conceptual progress in the field will depend on 
a continuing flow of new data from \compass~\cite{CompassPlans}, the only 
remaining high-energy lepton DIS facility, as well as from Jefferson 
Laboratory, especially after it achieves its 12\,GeV 
upgrade~\cite{Thomas:1900zz}.

%
%
\section*{Acknowledgements}
The authors would like to warmly thank Markus Diehl, Klaus Rith
and Werner Vogelsang for reading and providing a critique of a draft 
of this paper.  We appreciate their helpful advice and comments, as 
well as those of Bob Jaffe and Christian Weiss. We are thankful to 
Eduard Avetisyan and Michael Engelhardt for technical help, and to 
Ulrike Elschenbroich for Figs.~\ref{fig:SIG3half-SIG1half}
and \ref{fig:Cartoon3}. 
M.B. was supported by the DOE under grant numbers 
DE-FG03-95ER40965 and DE-AC05-06OR23177, (under which Jefferson Science 
Associates, LLC, operates Jefferson Lab).

%
\section*{References}
%

\bibliography{RPPpaper}
\end{document}